\documentclass[manuscript]{acmart}
\usepackage{paralist}
\usepackage{subfiles}
\usepackage{pifont} 
\usepackage{xspace}
\usepackage{xcolor,soul}
\usepackage{tcolorbox}
\usepackage{ifthen}
\usepackage{multibib}

\newcites{PRST}{Primary Studies}

\AtBeginDocument{\providecommand\BibTeX{{\normalfont B\kern-0.5em{\scshape i\kern-0.25em b}\kern-0.8em\TeX}}}

\newcommand\cpaBaseName{combined-program-analysis technique}
\newcommand\cpaTechs{\cpaBaseName s\xspace}
\newcommand\cpaTech{\cpaBaseName\xspace}
\newcommand\CpaTechs{\expandafter\MakeUppercase\cpaBaseName s\xspace}
\newcommand\CpaTech{\expandafter\MakeUppercase\cpaBaseName\xspace}

\usepackage{xcolor}

\setcopyright{acmlicensed}
\copyrightyear{2026}
\acmYear{2026}
\acmDOI{XXXXXXX.XXXXXXX}

\acmJournal{JACM}
\acmVolume{xx}
\acmNumber{x}
\acmArticle{xxx}
\acmMonth{02}

\begin{document}

\title{Combined Program Analysis Techniques: A Systematic Mapping Study}
\titlenote{This work was partially supported by Italian projects PNRR SOP (H73C22000890001) and PRIN 2022  Big Sistah (2022EYX28N).}

\author{Pietro Braione}
\email{pietro.braione@unimib.it}
\orcid{0000-0001-9307-6781}
\affiliation{\institution{University of Milano-Bicocca}
\city{Milano}
\country{Italy}
}

\author{Giovanni Denaro}
\orcid{0000-0002-7566-8051}
\email{giovanni.denaro@unimib.it}
\affiliation{\institution{University of Milano-Bicocca}
\city{Milano}
\country{Italy}
}

\author{Luca Gugliemo}
\email{luca.gugliemo@unimib.it}
\orcid{0000-0002-5580-3623}
\affiliation{\institution{University of Milano-Bicocca}
\city{Milano}
\country{Italy}
}

\author{Elson Kurian}
\email{elson.kurian@unimib.it}
\orcid{0000-0002-6392-0870}
\affiliation{\institution{University of Milano-Bicocca}
\city{Milano}
\country{Italy}
}

\author{Enea Raffaele Ilario Papaleo}
\orcid{0009-0000-9573-9657}
\email{e.papaleo1@campus.unimib.it}
\affiliation{\institution{University of Milano-Bicocca}
\city{Milano}
\country{Italy}
}

\author{Martino Tessaro}
\email{m.tessaro@campus.unimib.it}
\affiliation{\institution{University of Milano-Bicocca}
\city{Milano}
\country{Italy}
}

\renewcommand{\shortauthors}{Braione et al.}

\newcommand\renderingMode[1]{\textcolor{black}{#1}\xspace}
\newcommand\numSynergisticEffects{\renderingMode{19}}
\newcommand\numMappingFunctionStructures{\renderingMode{7}}
\newcommand\numMappingFunctionMechanics{\renderingMode{8}}
\newcommand\numPrimaryStudies{\renderingMode{248}}

\begin{abstract}

\noindent\textbf{Context}. Since the eighties, the combination of program analysis techniques has been increasingly recognized as a promising approach to overcome the limitations of standalone methods. While individual techniques, based on either static or dynamic analysis, address important challenges in software dependability, their integration often yields synergistic effects on  precision, coverage and insights.

\noindent\textbf{Objective}. This paper surveys a significant portion of the modern literature on combining program analysis techniques, consisting of \numPrimaryStudies primary studies, with the aim of cataloging the types of interactions and synergies that were exploited to define \emph{\cpaTechs} so far. 
The goal is to provide a structured understanding of why and how program analysis techniques can be conjoined, and which benefits can arise from their interactions.

\noindent\textbf{Method}. We devise an original taxonomy that classifies \cpaTechs according to their aimed synergistic effects, inter-analysis workflows and interaction schemata (to which we refer to as \emph{mapping functions}). 
We then map the  primary studies to the taxonomy, answering research questions on 
which synergistic effects those studies pursued via the combination of analysis techniques,
which inter-analysis workflows they embodied, and
which types of mapping functions they exploited.

\noindent\textbf{Conclusion}. Our taxonomy and literature mapping reveal the commonalities and the differences, in terms of goals and patterns, in the design of \cpaTechs.
Thereby we provide a  framework of concepts that can foster the ability of researchers and practitioners to reason on existing \cpaTechs, and steer further research on new useful \cpaTechs and analysis frameworks.

\end{abstract}



\maketitle

\newcommand{\synAlarms}{Discriminate true and false alarms}
\newcommand{\synAlarmsDynamic}{Confirm executable alarms}
\newcommand{\synAlarmsStatic}{Identify false alarms}
\newcommand{\synPartitioning}{Exploit state-space partitioning}
\newcommand{\synPartitioningWitness}{Provide partition witnesses}
\newcommand{\synPartitioningCoverage}{Prioritize unexplored partitions}
\newcommand{\synPartitioningDirect}{Direct partner analysis onto partitions}
\newcommand{\synTraversal}{Exploit analysis on problem variants}
\newcommand{\synTraversalSeedInputs}{Convey problem variants by controlling depended artifacts} \newcommand{\synTraversalSeedSink}{Convey problem variants by restricting the analysis scope} \newcommand{\synTraversalTransform}{Convey problem variants by transformation of the target program} \newcommand{\synIntepretability}{Improve interpretability of program semantics}
\newcommand{\synIntepretabilityEntities}{Augment semantics of program entities}
\newcommand{\synIntepretabilityAPI}{Augment semantics of program entities} \newcommand{\synIntepretabilityOracle}{Provide expectations on program states}
\newcommand{\synIntepretabilityArtifacts}{Relate program semantics to software artifacts}
\newcommand{\synRewrite}{Rewrite representations of programs states}
\newcommand{\synRewriteConcrete}{Rewrite with concrete values}
\newcommand{\synRewriteSimulator}{Rewrite with semantically richer data}
\newcommand{\synRefine}{Refine program models}
\newcommand{\synRefineIncorporate}{Incorporate relevant details}
\newcommand{\synRefinePrune}{Prune invalid states}
\newcommand{\synFeature}{Exploit Integrated Feature Spaces}
\newcommand{\synFlow}{Exploit Integrated Program-Flow Data}
\newcommand{\synGUI}{Exploit Integrated GUI Data}
\newcommand{\synReports}{Integrate Analysis Reports}
\newcommand{\workCascade}{Cascade composition}
\newcommand{\workFeedback}{Feedback composition}
\newcommand{\workSidebyside}{Side-by-side composition}
\newcommand{\structPaths}{Program Paths}
\newcommand{\structCFG}{Control-Flow Entities}
\newcommand{\structDF}{Data-Flow Entities}
\newcommand{\structCG}{Call-Graph Entities}
\newcommand{\structModules}{Classes or Modules}
\newcommand{\structGUI}{GUI Entities}
\newcommand{\structProgram}{The Program}
\newcommand{\mechanicAssociation}{Identity}
\newcommand{\mechanicConstraint}{Constraint Solving}
\newcommand{\mechanicInterpolation}{Craig Interpolation}
\newcommand{\mechanicInvariants}{Likely Invariants}
\newcommand{\mechanicMining}{Specification Mining}
\newcommand{\mechanicMerging}{State Merging}
\newcommand{\mechanicSummary}{Function Summaries}
\newcommand{\mechanicML}{Metrics, Data Mining and Machine Learning}
 \section{Introduction}

Program analysis techniques automatically extract information from software artifacts,
to assist in assessing 
the validity or invalidity of dependability properties and dependability characteristics of interest for a software product, or derive data for reasoning on those properties and characteristics~\cite{Pezze:SWTesting:book:2007}. Possible dependability properties and characteristics 
include, for instance, 
the degree of correctness, reliability, maintainability, understandability or security of the software under analysis~\cite{sommerville2011software,DO178C:2011,EN50716_2023}. 
Possible artifacts that can be targets of a program analysis technique 
include specifications, design documents, implementation code, or combinations of those. 

For instance, classical approaches to program analysis encompass techniques for 
model checking, alias analysis and program slicing. 
Model-checking techniques formally demonstrate safety properties of interest by analyzing finite-state representations of the target systems~\cite{Clarke:ModelChecking:1999,Visser:model:asej:2003,jhala2009software}.
Alias analysis techniques extract information about possible aliasing between program variables to support several forms of further reasoning about the semantics of the target programs~\cite{andersen1994program,landi1992safe,diwan1998type}.
Program slicing techniques synthesize portions of the source code that suffice for reasoning on given safety properties~\cite{weiser2009program,Tip:survey:1995}. 
But this is only a limited selection of samples of program analysis techniques. 
Over the years, researchers proposed and explored a plethora of general approaches and specific techniques for program analysis, which get often classified into two main classes, i.e.,  the ones based on \emph{static analysis} of program artifacts and the ones based on \emph{dynamic analysis} of the execution traces monitored at runtime while executing the programs. Other than the program analyses already mentioned above, notable examples of classic static analysis techniques include data flow analysis and symbolic execution~\cite{kildall1973unified,king1976symbolic,clarke1976system}.  Examples of classic dynamic analysis techniques include assertion checking~\cite{rosenblum1995practical} and runtime analysis~\cite{bartocci2018lectures}.
Program analysis techniques empower  software engineering tools, for instance, they have been exploited 
for addressing many problems in test-case generation~\cite{Yang:covtools:AST:2006,McMinn:SBSESurvey:STVR:2004,cadar2013symbolic}.

Starting already in the eighties, several researchers pushed forward the idea that the results of distinct analysis techniques could be combined with each other, aiming to synergistically overcome the limits of using standalone techniques~\cite{osterweil1984integrating,richardson1985partition}.
Since its very inception, this direction of work was  reckoned very promising, as researchers acknowledged the duality of the benefits and limitations of the possible types of analysis means, in particular the duality between  
analysis techniques that pursue their goals by either sampling \emph{reachable} program states with respect to some property, or producing a \emph{program-complete} set of states enjoying some property~\cite{young1989rethinking,ernst:invited:2004}. 
Sampling reachable program states is the typical approach of dynamic analysis and symbolic execution techniques. Runtime analysis and testing identify program failures by displaying some concrete execution sequence from an initial state up to a reachable error state. 
Symbolic execution statically 
analyzes the reachable states along selected program paths, by simulating the execution of a program with respect to symbolic inputs that represent all possible concrete values of the inputs.
Dually, synthesizing a set of abstractly-represented states that enjoy some  properties,  while they subsume all reachable program states (i.e., they are program-complete), is the typical means in which a static analysis can prove program properties. 
If there is no other program state other than the ones already subsumed, then there cannot exist any error state that disproves those properties.  

Since the seminal work of the eighties, exploiting the synergy of results computed with distinct analysis techniques has been gaining increasing momentum. The proposed analysis techniques 
encompassed several ways of composing multiple (two or more) analysis techniques, according to even sophisticated workflows. For instance, the well-known analysis paradigm of counter-example-guided abstraction refinement~\cite{clarke2000counterexample,clarke2003counterexample} combines model checking of abstract program models, with symbolic execution and runtime checking of selected program paths, in a feedback-loop-style workflow. Along the forward direction of the workflow, they aim at increasingly refining the knowledge about the possible error states identified in the abstract models, up to reporting concrete error states successfully confirmed via runtime checking, if any. Along the backward-feedback direction of the workflow, activated when finding that some abstract error states are unreachable, they  improve the precision of the abstract models, up to be eventually  able to demonstrate the absence of error states for correct programs. 

In the light of more than forty years of research on combining program analysis techniques, 
this paper surveys 
a relevant portion of the modern literature on the subject, namely, \numPrimaryStudies primary studies on \emph{\cpaTechs}. Moreover, in the process of conducting the survey, we also devise an original taxonomy that encompasses the synergies and the interactions in those techniques.
The taxonomy captures both  
the classical means of well-acknowledged analysis paradigms, and further ones that emerged from the primary studies considered in our survey.
We then report the mapping between the \cpaTechs from the primary studies
and the entries of the taxonomy. This allows us to gain understanding on which  goals and interaction means have been investigated 
by researchers so far, and reveal the commonalities and the differences, in terms of goals and patterns, in the design of \cpaTechs.

Our taxonomy also aims to instantiate the general statement of
the classic paper of Young et Taylor at ICSE 1989, on devising \cpaTechs~\cite{young1989rethinking}: \emph{it is easiest to exploit the interactions between techniques when the same model schemata is shared between them}.
The taxonomy proposed in this paper specifically indicates 
a set of possible types of those model schemata, which we refer to as  \emph{mapping functions}, by which 
the results of an analysis technique can be mapped into data exploitable in the context of a partner analysis technique. 
We further classify the mapping functions along the dimensions of the \emph{interpretation structure} and the \emph{mechanics} of the mapping that they induce.  
As such two dimensions are orthogonal to each other, each $\langle$\emph{interpretation structure, mechanics}$\rangle$ pair  represents a possible type of mapping function. 

In summary, this paper contributes:
\begin{itemize}
    \item A taxonomy (and thus a catalog) of synergistic effects, workflows and mapping functions by which program analysis techniques can be conjoined with each other to yield \emph{\cpaTechs}. The taxonomy
allows for concise framing of the combined-program-analysis essence of a \cpaTech under consideration.

\item The mapping of \numPrimaryStudies primary studies on \cpaTechs
to the taxonomy, thus answering research questions on 
which are the synergistic effects pursued in \cpaTechs, 
which are the inter-analysis workflows that they exploit, and
which are the mapping functions that they use, along with the representation structures and the mechanics of those mapping functions. 
\end{itemize}

We are aware of a single mapping study published in 2012 that focuses on 51 papers on the combination of static and dynamic quality assurance techniques, encompassing 25 papers specifically focusing on \cpaTechs~\cite{survey:elberzhager:combination_mapping_study}. Other than considering less and less recent papers than our mapping study, their study is limited to classifying high-level goals underlying the considered primary studies, i.e., whether those studies aim at improving effectiveness, efficiency, program coverage or defect coverage, whereas out study addresses the synergies and interactions in the construction of \cpaTechs. Other existing surveys address program analysis techniques in the context of specific application domains, e.g., internet-of-things~\cite{survey:kumar:iot}, distributed software systems~\cite{cai2025survey} or program comprehension~\cite{cornelissen2009systematic}, or do not specifically focus on combined techniques~\cite{wogerer2005survey,gosain2015survey,xu2005brief,kennedy1979survey}. 

The remainder of the paper is organized as follows. Section~\ref{sec:taxonomy} discusses 
a sample of classic \cpaTechs: we aim to both render concrete the concept of \cpaTech, by  introducing paradigmatic examples, and thereby introduce
the main organization of 
our  taxonomy 
with reference to the sample techniques. 
Section~\ref{sec:rqs} presents the research questions that drove our mapping study.
Section~\ref{literature:search} explains how we selected the papers considered for the survey. Section~\ref{sec:survey} develops the taxonomy, and presents the systematic mapping  between the papers in the survey and the entries in the taxonomy. Section~\ref{sec:conclusions} summarizes the results of the mapping study and 
future research directions.

 \section{Combining Program Analysis Techniques}
\label{sec:taxonomy}

Designers of \cpaTechs synergistically integrate 
multiple analysis stages, aiming at ensembles of analysis techniques that overcome the limitations of  relying on those standalone analysis techniques separately. 
The synergy arises because some program analysis techniques can exploit knowledge about the program behavior that they are unable to produce by themselves. 
For instance, as we already commented, reachability analysis techniques
can derive existential information about the program state space, but cannot conclude properties that are universally valid for the target program, 
whereas the scenario is dually inverse in the case of program-complete analysis techniques.
Thus, sharing pieces of information between some reachability analysis stage and some program-complete analysis stage 
may enhance the effectiveness of the exploration at the side of either of those analysis stages, respectively.

A core contribution of this paper is a taxonomy of both the synergistic effects that have been pursued via ensembling program analysis techniques, and the ways in which program analysis techniques have been interconnected to foster their synergies.
In this section, we start by reviewing a sample of \cpaTechs that we use as working examples. We aim at both  exemplifying some paradigmatic techniques, to render concrete the concept of \cpaTech, and illustrating the organization of our taxonomy. 
On purpose,  in this section, we consider examples in which the combination flavors of the techniques are 
well acknowledged in the literature on program analysis. This helps us concentrate on the classification method that we use for framing different types of \cpaTechs. Later, in the subsequent sections of the paper, we extend our mapping study to a larger body of \cpaTechs encompassed in the literature. 

While reviewing a given \cpaTech, we categorize 
\begin{inparaenum}[(i)]
\item 
which \emph{elementary analysis stages}  participate in its realization,
\item 
which \emph{inter-analysis workflow} it exploits to compose the elementary analysis stages into a unified analysis algorithm,  
\item 
which \emph{intermediate results}  get exchanged across those elementary stages
and
\item 
which \emph{synergistic effects} those interactions aim to foster.
\end{inparaenum}
The following example further elaborates on our method, and introduces a first set of entries of our taxonomy.

\begin{example}[Check'n'Crash~\citePRST{csallner:check-n-crash:icse:2005}]\label{sec:example:cnc}
Figure~\ref{fig:checkncrash} graphically illustrates the cascade-style workflow of the technique Check'n'Crash, which aims to automatically detect alarms on executions that make the target program fail due to possible runtime exceptions.
In Check'n'Crash, the inter-analysis workflow can be interpreted  as the coordination between two analysis stages that exploit the combination between static analysis, namely, weakest precondition calculus, as implemented in ESC/Java~\cite{flanagan2002extended}, and random testing, as implemented in JCrasher~\cite{csallner_jcrasher_2004}, respectively. In the static analysis phase (first analysis stage) Check'n'Crash identifies error conditions under which the program can throw runtime exceptions, e.g., due to referencing a null pointer or executing a division by zero. In the random testing phase (second analysis stage), it  dynamically confirms which of those exceptions  actually occur in concrete test cases, and reports only the confirmed alarms.

\begin{figure}[!tb]
\begin{center}
\includegraphics[width=.8\textwidth]{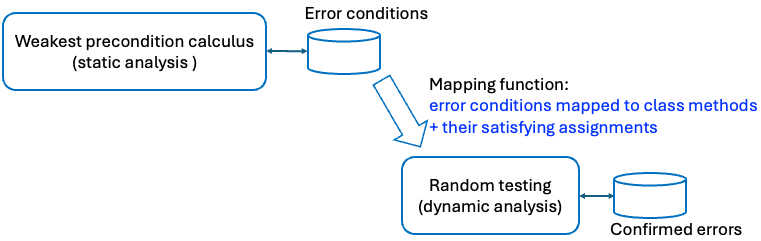}
\end{center}
\parbox{.8\textwidth}{\small 
\textbf{\underline{Identikit}}\\
Name: Check'n'Crash\\
Stage 1: Static analysis, \`a-la weakest precondition calculus\\
Stage 2: Dynamic analysis, \`a-la random testing\\
Workflow: \emph{\workCascade}\\
Interaction Stage 1 $\rightarrow$ Stage 2
\begin{compactitem}[$\circ$]
\item Mapping function: Error conditions associated to class methods (\emph{\structCG}), along with corresponding satisfying assignments computed with \emph{\mechanicConstraint}
\item Synergistic effect:  \emph{\synAlarms}
\end{compactitem}
}
\caption{Combined-analysis workflow of Check'n'Crash}
\label{fig:checkncrash}
\end{figure}

As annotated at the bottom of the figure (\emph{Identikit}), we 
capture this type of workflow under class \emph{cascade composition}: the results of the former stage cascade down to the latter stage that uses those results specifically. 

Figure~\ref{fig:checkncrash} highlights the exchange of intermediate results across the two analysis stages.
The dynamic-analysis stage interprets the statically-detected error conditions by exploiting the knowledge 
that the static-analysis stage has associated each error condition with the class method that might trigger the corresponding error. 

We capture (\emph{Identikit})
this way of sharing the results 
as a \emph{mapping function} with \emph{interpretation structure} that grounds on \emph{call graph entities}.
The call graph entities considered in  Check'n'Crash are class methods.
The interpretation structure of the mapping function allows for the receiver analysis (in this case, the random testing stage of Check'n'Crash) to determine where to plug the results of the companion analysis (in this case, the error conditions computed in the static analysis stage of Check'n'Crash) within its analysis algorithm. 
Knowing the mapping between  the statically-detected error conditions and the corresponding class methods  allows for 
Check'n'Crash
to determine which error conditions shall be paid attention 
while addressing random testing for each specific class method.

Another observation is that, in Check'n'Crash, the random testing phase does not exploit the error conditions directly. Rather it exploits  satisfying assignments, i.e., concrete inputs that satisfy those error conditions, as the overall idea of Check'n'Crash is that the test generation process shall keep those input values fixed, while randomly trying many possible values for the other ones. Check'n'Crash relies on constraint solving technologies~\cite{de2008z3,dutertre2006yices} to compute 
satisfying assignments for the error conditions being considered.

In our taxonomy, we refer such constraint-solving behavior, by which Check'n'Crash maps the results of its static-analysis stage to exploitable data for its dynamic-analysis stage, as  the \emph{mechanics of the mapping function}. 
The mechanics of the mapping function explains how the results fed to the receiver analysis (in this case, the input values fed to the dynamic analysis stage of Check'n'Crash) are mathematically processed (in this case, by means of constraint solving) based on the results computed in the companion analysis (in this case, the static analysis stage of Check'n'Crash).

Finally, for the Check'n'Crash example, we elaborate on the 
the synergistic effect that underlies the idea of combining  two analysis stages as above. In the case of Check'n'Crash, the very goal of combining static and dynamic analysis
is to avoid the possible false alarms that the static analysis of ESC/Java might report if used standalone. Check'n'Crash reports only confirmed alarms, along with supporting evidence in the form of test cases demonstrating the occurrence of those alarms. \emph{\synAlarms} is a common type of synergistic effect that can be pursued by composing program-analysis stages, a goal that Check'n'Crash shares with several other \cpaTechs considered in the mapping study that we discuss in Section~\ref{sec:survey} of this paper.\qed

\end{example}

Figure~\ref{fig:classification} summarizes the classification schema that we exploit for building the taxonomy. 
Our taxonomy renders: 
\begin{inparaenum}[(1)]

\item 
Three main types of inter-analysis workflows, which mimic  classic communication schemes of software-based systems~\cite{lee:embedded:mit:2017}. Beside the \emph{Cascade composition} workflow that we exemplified above with reference to Check'n'Crash, the taxonomy includes \emph{Feedback composition} and \emph{Side-by-side composition}, which we further exemplify in the next examples discussed below in this section; 

\item 
The synergistic effects exploited in \cpaTechs, further classified as \emph{vertical} and \emph{horizontal synergies}.  Figure~\ref{fig:classification}  includes an entry for \emph{Discriminate true and false alarms}, the synergistic effect that we exemplified with Check'n'Crash. This synergistic effect appears under the class of vertical synergies, as the benefits of combining the analysis techniques take the form of exploiting the results of an analysis stage within the algorithm of the partner analysis stage, to improve the capability of the analysis done thereby. On the other hand, horizontal synergies occur as combining the results of the analysis stages out of the boundaries of their respective algorithms. 
The other entries  related to synergistic effects of Figure~\ref{fig:classification} are discussed in the next examples of this section, including a type of horizontal synergy. The list of synergistic-effect entries in the figure is purposely left incomplete, as we will introduce yet further entries, which emerged while surveying the papers considered in our mapping study, and that we will describe while discussing the results of the mapping in Section~\ref{sec:survey}; 

\item 
Our original classification of the types of 
\emph{mapping functions} that could be exploited to share intermediate results across analysis stages, along  
the dimensions of 
\begin{inparaenum}[(3.1)]
\item the \emph{Interpretation structure} and
\item the \emph{Mechanics} 
\end{inparaenum}
on which the mapping functions may ground to accomplish the sharing of intermediate results. We already exemplified the entries \emph{Interpretation structure / Call-graph entities} and \emph{Mechanics / Constraint solving} in the case of  Check'n'Crash, while the other entries included in Figure~\ref{fig:classification} appear in the next examples of this sections, and we will introduce further entries while discussing the results of the mapping in Section~\ref{sec:survey}. 
\end{inparaenum}

\begin{figure}[!tb]
\begin{center}\small
\begin{enumerate}

\item Inter-analysis workflow
\begin{itemize}
\item Cascade composition
\item Feedback composition
\item Side-by-side composition
\end{itemize}

\item Inter-analysis synergistic effects
\begin{itemize}
\item Vertical synergy (improving capabilities of analysis algorithms)  
\begin{itemize}
    \item \synAlarms
    \item \synPartitioning
    \item \synRewrite
    \item \synRefine
    \item\dots
\end{itemize}
\item Horizontal synergy (extending the ranges of  analysis outcomes or working data) 
\begin{itemize}
\item \synFeature \item\dots 
\end{itemize}
\end{itemize}

\item\label{mapfun} Inter-analysis mapping functions
\begin{enumerate}[(\ref{mapfun}.1)]

\item Interpretation structure \label{mapping-schema} 
\begin{itemize}
\item \structPaths
\item \structCG
\item \structProgram 
\item \dots

\end{itemize}

\item Mechanics
\begin{itemize}
\item \mechanicAssociation 
\item \mechanicConstraint
\item \mechanicInterpolation
\item \mechanicML
\item\dots
\end{itemize}

\end{enumerate}

\end{enumerate}

\end{center}
\caption{Dimensions to classify \cpaTechs} 
\label{fig:classification}
\end{figure}

The cascade-composition schema that we exemplified with reference to Check'n'Crash
is generally the main building block for exploiting the synergy of multiple program analysis techniques. 
In fact, cascading two analysis techniques captures the basic idea that an analysis technique $B$ relies on the results from another analysis technique $A$.
The overall analysis algorithm is thus organized in stages, in which the analysis $A$ (former stage) is executed before $B$ (latter stage), and $B$ leverages 
some (intermediate) artifact computed during $A$ as technical means 
to accomplish the further goals.

We remark that, while describing the cascading between analysis techniques, our descriptions in this paper aim to emphasize the main flavours of the combination taking a conceptual standpoint. This means that we dismiss the specific algorithmics that can be part of the practical implementations of the considered program analyses, but is not strictly relevant for describing their combined-analysis flavors. For instance, the representation of  Check'n'Crash in Figure~\ref{fig:checkncrash} is agnostic on whether the second stage is executed 
once for each error condition computed from the former stage, or after completing the analysis of the error conditions, or yet for batches of the error conditions.

The next example discusses a program analysis technique that encompasses a \emph{Feedback composition} workflow, along with
further mapping functions and synergistic effects.
In feedback composition the stages of program analysis iteratively exchange information to enhance each other's effectiveness. Feedback composition involves a bidirectional exchange: Each analysis stage provides insights to subsequent stages and, in turn, receives refined data and information from those stages, steering continuous improvement. 

\begin{example}[Concolic Execution~\cite{sen:cute:esec:2005,godefroid:dart:pldi:2005}]\label{sec:example:concolic}
Figure~\ref{fig:concolic} illustrates the feedback-loop-style
workflow of classic concolic execution that exploits the combination of concrete and symbolic execution. Our taxonomy in Figure~\ref{fig:classification} captures this type of inter-analysis workflow under class \emph{Feedback composition}, as the two analysis stages, dynamic monitoring and symbolic execution, respectively, coordinate between them by feeding results to each other in both directions, respectively. 

\begin{figure}[!tb]
\begin{center}
\includegraphics[width=.6\textwidth]{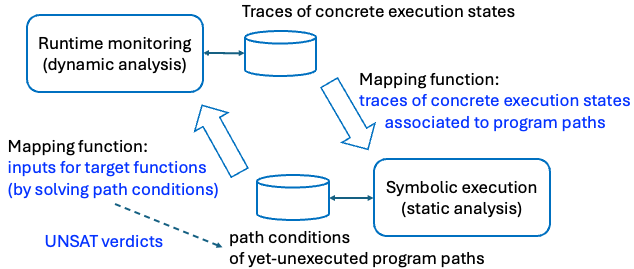}
\end{center}
\parbox{.8\textwidth}{\small 
\textbf{\underline{Identikit}}\\
Name: Concolic execution\\
Stage 1: Dynamic analysis, \`a-la runtime monitoring\\
Stage 2: Static analysis, \`a-la symbolic execution\\
Workflow: \workFeedback\\
Interaction Stage 1 $\rightarrow$ Stage 2
\begin{compactitem}[$\circ$]
\item Mapping function: Execution traces associated (\emph{\mechanicAssociation}) to corresponding \emph{\structPaths}
\item Synergistic effect (i):  \emph{\synPartitioning} to steer progress\\
Synergistic effect (ii): \emph{\synRewrite}  \end{compactitem}
Interaction Stage 2 $\rightarrow$ Stage 1
\begin{compactitem}[$\circ$]
\item Mapping function: Inputs to  execute the target function(s) (\emph{\structCG}), via \emph{\mechanicConstraint}
\item Synergistic effect:  \emph{\synPartitioning} to steer progress
\end{compactitem}
}
\caption{Combined-analysis workflow of concolic execution} 
\label{fig:concolic}

\end{figure}

The aim of concolic execution is to increase the efficiency and the effectiveness of traditional symbolic execution in exploring the path space of a program, usually for the sake of generating test cases that execute the program paths explored thereby. Concolic execution starts with executing some existing (or randomly picked) test case, monitors the execution of the test case at runtime (runtime monitoring stage),
computes the symbolic execution of the program along the program path that corresponds to the execution of the test case (symbolic execution stage), and solves the path conditions of the alternative program paths that originate in path-prefixes of the current program path. Each successful solution allows for instantiating a new test case for a new program path, and to iterate the above process with reference to the new test case. 

For exchanging results from runtime monitoring to symbolic execution, the mapping function refers to the execution paths of the program (Figure~\ref{fig:classification}, \emph{Mapping function / Interpretation structure}: \emph{\structPaths}), associating the concrete execution states visited during the execution of a test case with the symbolic states computed while symbolically executing the program along the program path traversed by the test case. As the symbolic execution stage exploits the association with the concrete states directly, without additional transformations applied to those concrete states, in this case the mapping function works without any special mechanics, transferring the analysis results directly (Figure~\ref{fig:classification}, \emph{Mapping function / Mechanics}: \emph{\mechanicAssociation}).

The synergistic effect of relating the results from runtime monitoring to the symbolically-executed program paths is twofold. 
First, the symbolic execution stage  
can drive its analysis onto the  program paths pinpointed by the concrete execution (Figure~\ref{fig:classification}, \emph{Synergistic effect}: \emph{\synPartitioning}  --in fact,  concolic execution induces a path-coverage-based partitioning of the program state space). Second, it can 
refer to the values available in the concrete states for simplifying the symbolic formulas, e.g., rewriting non-linear constraints to linear counterparts, and granting the satisfiability of the symbolic states that it computes for those program paths (Figure~\ref{fig:classification}, \emph{Synergistic effect}: \emph{\synRewrite}).  

In exchanging results from the symbolic execution stage to the runtime monitoring stage, the mapping function is of the same type as the one that we discussed for the Check'n'Crash technique, as symbolic execution exploits constraint solving against the path conditions of yet-unexecuted program paths, to feed inputs that allow for  further executing the target function(s). The mapping function
refers to the functions of the program (Figure~\ref{fig:classification}, \emph{Mapping function / Interpretation structure}: \emph{\structCG}) and exploits constraint solving (Figure~\ref{fig:classification}, \emph{Mapping function / Mechanics}: \emph{\mechanicConstraint}) to generate inputs for the target function(s). 
The aimed  synergistic effect is to provide test cases for making dynamic analysis progress onto unexecuted program paths (Figure~\ref{fig:classification}, \emph{Synergistic effect}: \emph{\synPartitioning}).\qed

\end{example}

In both of the above examples, the mapping functions with mechanics of type \emph{Constraint solving} illustrated the case of 
using inferential reasoning to
specialize symbolic formulas that represent sets of program inputs, by instantiating concrete  values that satisfy the formulas.
In general, mapping-function mechanics can work by either specializing or generalizing the analysis outcome.
The next example includes a mapping function that fosters synergistic effects by generalizing the outcome of an  analysis stage. 

\begin{example}[Software Model Checking via Counter Example Guided Abstraction Refinement~\citePRST{beyer:blast:isttt:2007,ball:slam:fmcad:2010}]\label{sec:example:cegar} Figure~\ref{fig:cegar} illustrates the workflow of software model checking according to the CEGAR (Counter Example Guided Abstraction Refinement) approach, as defined in the seminal work on the techniques BLAST and SLAM~\citePRST{beyer:blast:isttt:2007,ball:slam:fmcad:2010}. The inter-analysis workflow is  designed in \emph{Feedback composition} style, throughout a stage of model checking of a finite-model representation  of the target program, 
and a stage of symbolic execution. 

In a nutshell, the model checking stage computes a finite model of the program via the technique of predicate abstraction, that is, abstract interpretation of the program with respect to a finite,  representative set of predicates over the program variables, and then unrolls the entire reachability state of the model to identify potential counter-examples for  safety properties of interest. A counter-example indicates a program path that, when executed, may lead the program to violate some safety property. Then, the symbolic execution stage analyzes the program along the counter-example paths, aiming to confirm those counter-examples by generating test cases that show the failures of the program concretely. However, in some cases, symbolic execution may determine that some  of those program paths are infeasible, meaning that the corresponding counter-examples were false alarms, i.e., 
spurious results of the model checking stage.

The identification of false alarms allows for the symbolic execution stage to provide feedback to the model checking stage. The feedback takes the form of additional predicates to be considered in the predicate abstraction process, to refine the precision of the abstract model in order to exclude the spurious counter-examples. The new predicates can be interpolated 
out of the symbolic data that characterize the infeasible program paths. The two stages iterate, aiming at making  the abstraction  precise enough for either reporting a test case that shows a valid counter-example, or demonstrating that there is no counter-example as the program provably enjoys the safety properties of interest.

\begin{figure}[!tb]
\begin{center}
\includegraphics[width=.8\textwidth]{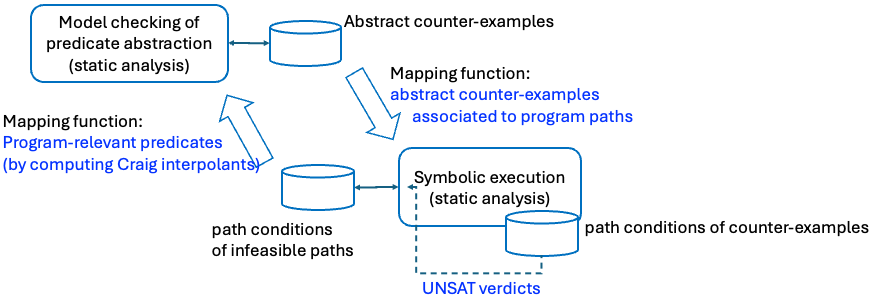}
\end{center}
\parbox{.8\textwidth}{\small 
\textbf{\underline{Identikit}}\\
Name: CEGAR, as embodied in the techniques BLAST and SLAM\\
Stage 1: Static analysis, \`a-la model checking of predicate abstractions of programs\\
Stage 2: Static analysis, \`a-la symbolic execution\\
Workflow: \emph{\workFeedback}\\
Interaction Stage 1 $\rightarrow$ Stage 2
\begin{compactitem}[$\circ$]
\item Mapping function: Abstract counter-examples associated by \emph{\mechanicAssociation} to  \emph{\structPaths} 
\item Synergistic effect:  \emph{\synAlarms}
\end{compactitem}
Interaction Stage 2 $\rightarrow$ Stage 1
\begin{compactitem}[$\circ$]
\item Mapping function: Program-relevant predicates associated to  \emph{\structProgram}, via \emph{\mechanicInterpolation}
\item Synergistic effect:  \emph{\synRefine}  for model checking
\end{compactitem}
}
\caption{Combined-analysis workflow of counter example guided abstract refinement} 
\label{fig:cegar}

\end{figure}

The mapping functions that characterize
the interaction between the two analysis stages can be described as follows. A  mapping function conveys results from the model checking stage to the symbolic execution stage, by
reporting the counter-examples in the form  
of property-violating program paths (Figure~\ref{fig:classification}, \emph{Mapping function / Interpretation structure}: \emph{\structPaths}, \emph{Mapping function / Mechanics}: \emph{\mechanicAssociation}). The symbolic execution stage exploits that information to steer the symbolic analysis exactly along those program paths, aiming to discriminate if the counter-examples are true or false alarms (Figure~\ref{fig:classification}, \emph{Synergistic effect}: \emph{\synAlarms}). 

The symbolic execution stage aims to generate failure-revealing test cases, as the 
output of the overall technique.
It synthesizes program inputs via constraint solving,
in line with the goal of discriminating the true and the false alarms.
A further mapping function  shares feedback from the symbolic execution stage to the model checking stage. 
Specifically, the feedback consists of  predicates that generalize the path conditions of the program states at which the considered counter-examples become unsatisfiable (dashed line labeled as \emph{UNSAT verdicts} in Figure~\ref{fig:cegar}). The mechanics of the mapping relies on the mathematics of Craig-interpolants applied to the unsatisfiability proofs computed via a constraint solver (Figure~\ref{fig:classification}, \emph{Mapping function / Mechanics}: \emph{\mechanicInterpolation}): an interpolant is a predicate that is implied by the path condition of the latest satisfiable state in the program path (that is, the interpolant is more general than the satisfiable formula) while it implies (that is, it is sufficient to state) that the path condition becomes unsatisfiable thereon. 
As we described above, the model checking stage will use the predicates fed from the symbolic execution stage to refine the abstraction that it uses to
finitely model the behavior of the program. In this case, the feedback is associated with the program as a whole (Figure~\ref{fig:classification}, \emph{Mapping function / Interpretation structure}: \emph{\structProgram}).
The aimed synergistic effect is to refine the program abstraction model considered during model checking  (Figure~\ref{fig:classification}, \emph{Synergistic effect}: \emph{\synRefine}).\qed 

\end{example}

The next example discusses a program analysis technique that encompasses a \emph{Side-by-side composition} workflow.
In side-by-side composition the program analysis stages do not influence their  respective algorithms, but rather the data that the analysis stages compute separately are unified into bags of data, which get then exploited as a whole to compute the analysis outcomes thereon.  
These \cpaTechs are part of a class of synergistic effects that we categorize as \emph{horizontal synergies}, as the benefits of combining analysis techniques manifest in expecting increased effectiveness when producing
the analysis outcomes  from intermediate data that range across the different flavors of the results from the elementary stages.

\begin{example}[\textsc{Marvin} malware classifiers~\cite{lindorfer_marvin_2015}]\label{sec:example:marvin}
Figure~\ref{fig:marvin} illustrates the workflow of the malware classification technique \textsc{Marvin}, which combines static and dynamic analysis to collect and exploit different types of static properties and dynamic characteristics of Android apps. 
The inter-analysis workflow of the \textsc{Marvin} classifier is defined in \emph{Side-by-side composition} style (Figure~\ref{fig:classification}, Inter-analysis workflow: Side-by-side composition):  The results computed with either static or dynamic analysis get joined into a feature vector, where each static property and dynamic characteristic
is regarded as a separate feature of the application under analysis; then the entire feature vector is exploited with a supervised classifier to obtain a maliciousness score of the application under analysis. 

The synergistic effect exploited in the \textsc{Marvin} malware classifier derives from  considering a rich feature space (Figure~\ref{fig:classification}, \emph{Synergistic effect}: \emph{\synFeature}), which ranges over the results of multiple analysis techniques, and is thus richer than the considering features computed only with static analysis or only with dynamic analysis. The claim of the \textsc{Marvin} approach is that their classifier can be more effective than a malware classifier that refers only to statically computed features or only to dynamically computed features, respectively. It should be noted that the expected synergy resides in being able to exploit also correlations between static and dynamic characteristics, such that the effectiveness could be arguably higher than the sum of the effectiveness of using the static-analysis and dynamic-analysis feature spaces separately. 

\begin{figure}[!tb]
\begin{center}
\includegraphics[width=.8\textwidth]{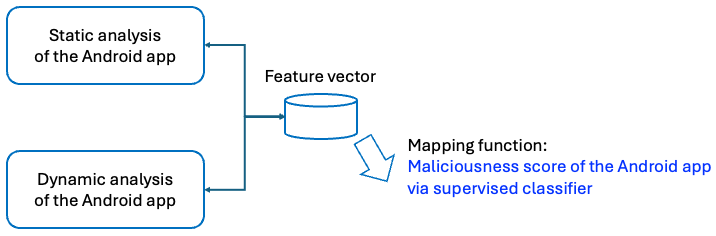}
\end{center}
\parbox{.8\textwidth}{\small 
\textbf{\underline{Identikit}}\\
Name: \textsc{Marvin} malware classifier\\
Stage 1: Static analysis of the Android app under analysis\\
Stage 2: Dynamic analysis of the Android app under analysis\\
Workflow: \emph{\workSidebyside}\\
Interaction Stage 1 $\rightarrow$ Stage 2
\begin{compactitem}[$\circ$]
\item Mapping function: Maliciousness score associated to the Android app under analysis (\emph{\structProgram}), and computed with a supervised classifier (\emph{\mechanicML}) exploited against the feature vector comprised of the results of both static and dynamic analysis 
\item Synergistic effect: \emph{\synFeature} with machine learning 
\end{compactitem}
}
\caption{Combined-analysis workflow of the \textsc{Marvin} malware classifier} 
\label{fig:marvin}
\end{figure}

\textsc{Marvin} exploits the sided static and dynamic features with the mathematics of a supervised classifier. Here, the classification algorithm manipulates the program data, but does not analyze the program or the execution of the program any further. Thus, in our study, we categorize this step of the technique (as well as  similar applications of machine-learning and data-mining algorithms in other techniques)  as a type of mapping-function mechanics, rather than considering it as an additional analysis stage. 
In this case, the mapping function associates the outcome of the classifier, i.e., the maliciousness score, with the application under analysis (Figure~\ref{fig:classification}, \emph{Mapping function / Interpretation structure}: \emph{\structProgram}) according to the mathematics of the supervised classifier\footnote{The mathematics of the classifier includes the need of having trained the supervised classifier against a dataset of reference applications labeled with their known scores.} (Figure~\ref{fig:classification}, \emph{Mapping function / Mechanics}: \emph{\mechanicML}).\qed
\end{example}
\section{Goal and Research Questions}
\label{sec:rqs}

The goal of our mapping study is:

\medskip
\begin{center}
\parbox{.93\textwidth}{
To analyze existing proposals of \cpaTechs, in order to 
describe and explain the types of interactions and synergies that were exploited to define those techniques; and thereby extrapolate a  framework of concepts to foster the ability of researchers and practitioners to reason on existing \cpaTechs, and define new useful \cpaTechs.
}
\end{center}
\medskip

To reach this goal, we analyzed the \cpaTechs proposed in literature, driven by the following research questions:

\begin{enumerate}[\textbf{RQ} 1:]

\item Which synergistic effects do \cpaTechs address? As illustrated in the examples of Section~\ref{sec:taxonomy}, in general, the motivations for combining program analysis techniques link with addressing some type of synergistic effect, to improve on using given program analysis techniques in their standalone embodiments. The possible synergistic effects may concretize in either enhancing the results of given standalone techniques, or addressing analysis goals that  standalone techniques cannot solve on their own.
We will provide an overview of the types of synergistic effects addressed by existing \cpaTechs, including the synergistic effects already introduced in Figure~\ref{fig:classification} and further ones encompassed in the surveyed primary studies. 

\item Which inter-analysis workflows do \cpaTechs exploit? We will characterize the \cpaTechs by considering the type of inter-analysis workflows that comprise them. We classify the inter-analysis workflows as cascade composition, feedback composition or side-by-side composition, as described in Section~\ref{sec:taxonomy}. 

\item Which interpretation structures do the exploited mapping functions rely on? We aim to understand the mapping functions that allow for defining \cpaTechs in terms of the interpretation structures exploited therein.  We will thus provide an overview of the  mapping-function interpretation structures  exploited to define existing \cpaTechs, including the mapping-function interpretation structures already introduced in Figure~\ref{fig:classification} and further ones encompassed in the surveyed primary studies. 

\item Which mechanics  do the exploited mapping functions rely on?
We aim to understand the mapping functions that allow for defining \cpaTechs also in terms of their mechanics.
We will thus provide an overview of the  mapping-function mechanics  exploited to define existing \cpaTechs, including the mapping-function mechanics already introduced in Figure~\ref{fig:classification} and further ones encompassed in the surveyed primary studies. 

\end{enumerate}
 \section{Primary Studies}
\label{literature:search}

In order to answer the research questions introduced in Section~\ref{sec:rqs}, 
we collected a corpus of primary studies on \cpaTechs. 
We followed the guidelines proposed by Kitchenham et al.\ and Petersen et al.~\cite{brereton2007lessons,kitchenham2009systematic,petersen2015guidelines}. The overall process consists of three main activities: \begin{inparaenum}[(i)] \item database search, \item study selection, and \item data extraction. \end{inparaenum}

\subsection{Database Search}
The database search step aims to collect candidate primary studies by searching on scientific databases available online. We relied on the scientific database 
Scopus, 
a comprehensive database of peer-reviewed papers published in well-reputed conference or journal venues on software engineering and testing, 
also suggested in 
guidelines 
on performing
systematic studies in software engineering~\cite{brereton2007lessons,petersen2015guidelines}. 
Scopus offers advanced search features over metadata including title, keywords, publication venue, and subject area.

We defined the search query by iterating the following process: \begin{inparaenum} \item define the candidate keywords and the logical structure of the search query, \item query the Scopus database and \item assess the search results for comprehensiveness.
\end{inparaenum} 

Figure~\ref{fig:searchquery} shows the final search query. 
We aimed to be
comprehensive with respect to research subjects that can be identified with different names, while
reducing the number of irrelevant papers returned because of keywords with ambiguous meaning. As an example, the keywords ``static analysis'' and ``dynamic analysis'' were disjunctively conjoined because we were interested in all the papers describing program analyses that are either static or dynamic. At the same time, they were conjunctively conjoined with the criterion that the publication venue is about  ``software'' to exclude irrelevant papers discussing analyses outside of the software analysis domain, e.g., structural analyses of buildings. To this end, we also paired the choice of keywords of the search string 
with the following
\emph{database search criteria}: 
\begin{inparaenum}[(DSC-1)]
     \item 
 Consider the articles published in scientific journals or in conference proceedings such that the venue  name  contains the term ``software'',
\item 
 Consider the documents that are associated with and only with the Computer Science subject area, 
\item 
 Consider only the documents written in English. 
\end{inparaenum}

\begin{figure}[!tb]
\begin{center}

\begin{tcolorbox}
\scriptsize
\begin{tabular}{p{1.5cm}l}
\S keywords $\triangleright$ & 
( TITLE-ABS-KEY ( "program analysis" ) OR TITLE-ABS-KEY ( "software analysis" ) OR \\
& TITLE-ABS-KEY ( "static analysis" ) OR TITLE-ABS-KEY ( "dynamic analysis" ) OR \\
& TITLE-ABS-KEY ( "software testing" ) OR TITLE-ABS-KEY ( "program testing" ) ) AND \\ 
& ( TITLE-ABS-KEY ( combin* ) OR TITLE-ABS-KEY ( hybrid* ) ) AND\\

\S DSC-1 $\triangleright$ & 
DOCTYPE ( ar OR cp ) AND (SRCTITLE ( *software* ) OR CONFNAME ( *software* ) ) AND\\
\S DSC-2 $\triangleright$
& (LIMIT-TO ( SUBJAREA,"COMP" ) OR\\

& EXCLUDE ( SUBJAREA,"DECI" ) OR   EXCLUDE ( SUBJAREA,"SOCI" ) OR \\
& EXCLUDE ( SUBJAREA,"BUSI" ) OR EXCLUDE ( SUBJAREA,"PHYS" ) OR \\
& EXCLUDE ( SUBJAREA,"ENVI" ) OR EXCLUDE ( SUBJAREA,"ENER" ) OR \\
& EXCLUDE ( SUBJAREA,"MATE" ) OR EXCLUDE ( SUBJAREA,"ARTS" ) OR \\
& EXCLUDE ( SUBJAREA,"MEDI" )) AND\\
\S DSC-3 $\triangleright$ 
& LIMIT-TO ( LANGUAGE,"English" )
\end{tabular}
\end{tcolorbox}
\end{center}
\caption{Keywords and logical structure of the query} 
\label{fig:searchquery}
\end{figure}

We iterated the process while assessing the comprehensiveness of the resulting list of papers. At each iteration, we 
discussed the search results in our research team and checked the inclusion of the relevant primary studies known to us, to the best of our knowledge. 
Eventually we consolidated the  query in the figure. The query was run on the Scopus database on February 20, 2025, and returned a total of 2,776 candidate primary studies.

\subsection{Study Selection}
This step aimed to select the final set of primary studies to consider in the mapping study, 
based on the \emph{study selection criteria} in Figure~\ref{fig:study:selection}. The first criterion (SSC-1) could be partially automated, as we could identify and exclude most short papers based on the metadata available from Scopus. This initial automatic selection  
reduced the candidates to 1,049 papers, which we ordered by similarity of author names, and assigned in equal portions among us, the 6 team members. Each of us finalized the evaluation of the study selection criteria by reviewing the assigned papers, calling for group discussion on need. The ordering by similarity of author names facilitated the evaluation of the criterion (SSC-4) on selecting only one primary study (the most comprehensive paper) per technique, in case of techniques encompassed in multiple papers.
The study selection phase ended up with selecting \numPrimaryStudies primary studies.

\begin{figure}[!tb]
\begin{center}

\begin{tcolorbox}
\scriptsize
\begin{tabular}{p{0.8cm}p{11.2cm}}
SSC-1 $\triangleright$ & 
\textbf{No short papers.} We excluded all studies with length less than 10 pages. \\

SSC-2 $\triangleright$ &
\textbf{Available documents.} We excluded papers that we failed to retrieve online to the best of our means.\\

SSC-3 $\triangleright$ & 
\textbf{Only primary studies.} We excluded the secondary studies, i.e., mapping studies, surveys and literature reviews.\\

SSC-4 $\triangleright$ & 
\textbf{Only a primary study per technique.} 
When two or more papers were about the same technique, we selected only one of those papers as primary study representative of the technique, aiming to  the most comprehensive one. For instance, if there was an "extended version", we preferred the extended version to the original paper.\\

SSC-5 $\triangleright$ & 
\textbf{Only primary studies on \cpaTechs} We excluded the primary studies that, despite of being related to program analysis and testing techniques, did not concern a combination of multiple program analysis techniques.
Even though we had  designed the search query for reflecting this requirement, the initial results included several false positives, such as, studies that did not address actual program analysis techniques, or described analysis techniques based on a single program analysis approach. 
We proceeded as follows: we inspected the workflow of each \cpaTech, identified the workflow steps that corresponded to distinct program-analysis and testing techniques, and selected only the techniques that included multiple steps of this kind. 
We did not count: 
\begin{inparaenum}[(i)]
  \item workflow steps that corresponded to manual activities;
\item workflow steps that did not depend on analyzing or executing the target software. Arguably, these steps do not correspond to static or dynamic program-analysis techniques. An example could be an internal step that computes the centrality degree of the nodes of a call graph. With reference to our examples in Section 2, this type of internal steps correspond to mapping functions with synthesis-style mechanics (e.g., to associate the call-graph nodes with centrality values to be used in a next analysis stage) rather than being themselves  additional program-analysis stages. This consideration applies also for steps that correspond to pure application of  machine learning algorithms or deep neural networks.  
\end{inparaenum}

\end{tabular}
\end{tcolorbox}
\end{center}
\caption{Exclusion criteria} 
\label{fig:study:selection}
\end{figure}

\subsection{Data Extraction}
In this step, we reviewed the selected primary studies in detail, to extract the description of the analysis stages and inter-analysis workflows involved in the \cpaTechs, along with the mapping functions and synergistic effects therein. 
We conducted the work by dividing the primary studies in approximately equal portions among the team members, having the extracted data crosschecked by another team member, and allocating team discussions for the hard cases. We provide our data extraction report as additional material, in Appendix~A of this paper. 

 \section{Results}
\label{sec:survey}

This section presents the data that we extracted from the primary studies, answering the research questions. 

\subsection{Synergistic Effects (RQ~1)}
\label{sec:results:problems}

\begin{figure}[!tb]
\begin{center}
\includegraphics[width=0.9\textwidth,trim=2mm 12mm 2mm 8mm,clip]{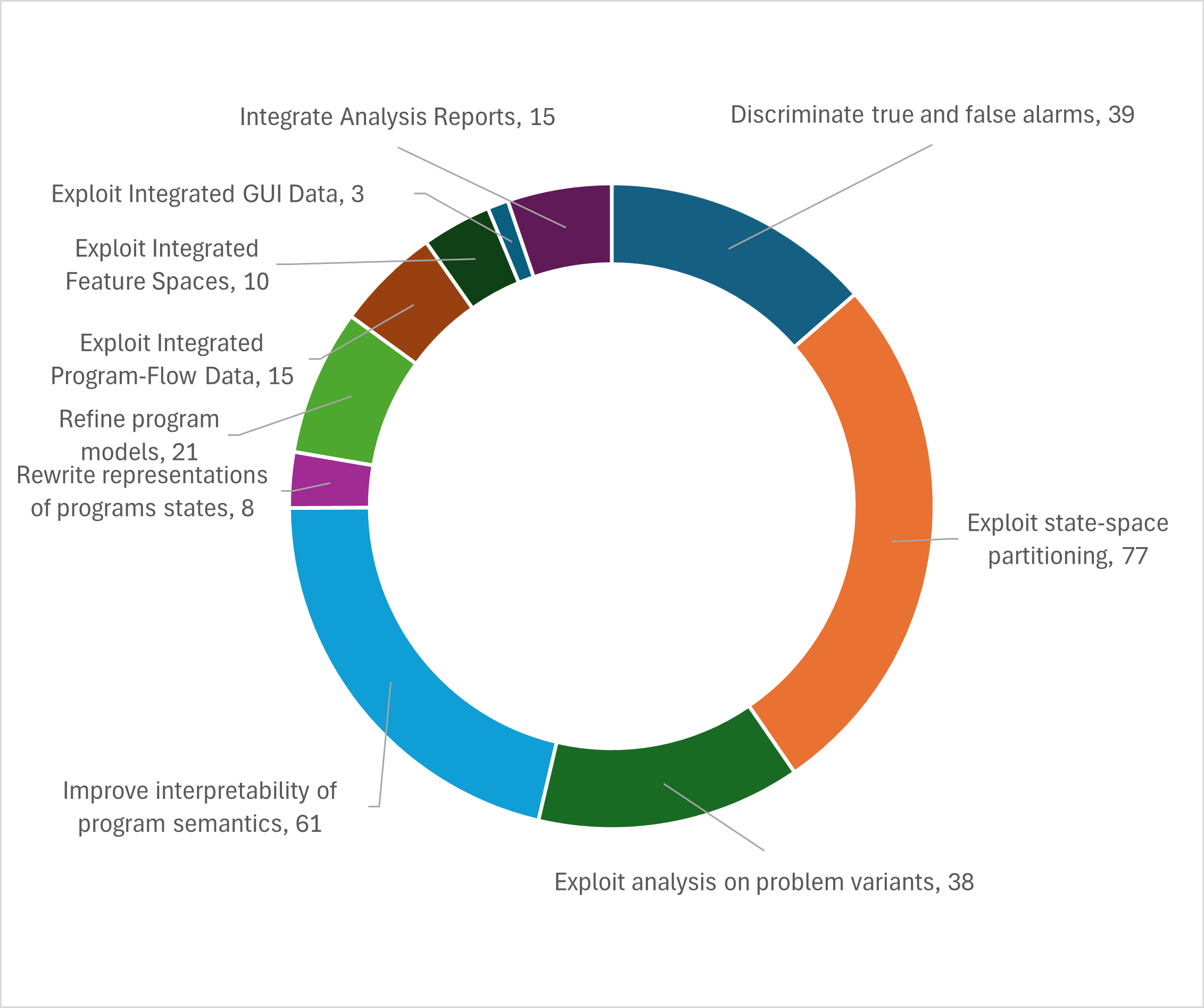}
\end{center}

\caption{Distribution of the synergistic effects across the primary studies} 
\label{fig:results:synergies}
\end{figure}

Figure~\ref{fig:results:synergies}
indicates
the distribution of the 
10 types of synergistic effects that we have classified by surveying the \cpaTechs in the primary studies. 
Below we discuss each type of synergistic effect and the related mapping with the primary studies. As we already did in Figure~\ref{fig:classification}, we  classify the synergistic effects as \emph{vertical} or \emph{horizontal synergies}, if they occur in the context of cascading or side-by-side compositions of analysis techniques, respectively. 
The 10 main types of synergistic effects of Figure~\ref{fig:results:synergies} further classify in a total of
\numSynergisticEffects (sub-)types of synergistic effects, as we describe below.

\subsubsection{Vertical Synergies}

\subsubsection*{\textbf{\synAlarms}}
A common issue of many program analysis techniques is that some reported alarms could be later discovered to be false alarms, at the cost of having wasted precious efforts of software engineers meanwhile.  
A common mitigation is to further analyze the alarms identified with a given program analysis technique in a partner analysis stage that may assist in better discriminating between false and true alarms, respectively. This collaborative way of addressing improved effectiveness characterizes the synergistic effect of various designs of \cpaTechs. In the primary studies, we classified two sub-categories of this type of synergistic effect, according to whether the main objective is to confirm true alarms or remove false alarms, respectively.

$\triangleright$\textbf{\synAlarmsDynamic}.
The dominant construction is the one in which 
the latter analysis stage relies on concrete execution, 
aiming to accept only the true alarms that can be confirmed as they can be concretely observed at runtime~\citePRST{lee_sealant_2017,leesatapornwongsa_flakerepro_2022,liu_tdroid_2018,liu_vd-guard_2023,chen_boosting_2021, Tlili:AMAST:2008, Vorobyov:SEFM:2012,Wang:TSE:2022,Patra:PICSE:2018,burnim_testing_2011,diner_generating_2021,eslamimehr_efficient_2018,fan_efficiently_2018,hu_achyb_2021,huster_using_2017,kim_first_2024,bodden_clara_2012,srivastava_crystallizer_2023,ponta2020detection}.
For instance, this is the case in the technique Check'n'Crash that we discussed in Example~\ref{sec:example:cnc}~\citePRST{csallner:check-n-crash:icse:2005}. Other approaches pursue the same goal by  mapping the alarm reports to program paths, and then analyzing the feasibility those program paths with symbolic execution and constraint solving~\citePRST{ball:slam:fmcad:2010,beyer:blast:isttt:2007,chen_star_2015,Pan:ICSME:2017,garcia_automatic_2017,banerjee_energypatch_2018,busse_combining_2022,chebaro_behind_2014}. 
Often times, the overall \cpaTech will provide test cases witnessing 
the confirmed alarms.
In general, these approaches trade the gains in the precision of the alarm reports for possible losses of recall, because they cannot draw any conclusion for the alarms that they could not confirm, and thus some of the alarms  
discarded due to missing confirmation
could be 
\emph{missed true alarms}.

$\triangleright$\textbf{\synAlarmsStatic}.
In other \cpaTechs, the stages that analyze the alarms  exploit static analysis to filter out the alarms that can be proven to be false alarms~\citePRST{Tripp:ISSTA:2014,lohar_two-phase_2021,gerrard_comprehensive_2017,naik_effective_2009,bodden_clara_2012,goffi_automatic_2016}.
For instance, in this fashion, taint analysis can be exploited to reason on candidate alarms identified as possible source-sink relations between program statements, such as input statements that can be exploited for injection attacks or privacy leakages, in order to discard infeasible source-sink pairs~\citePRST{samhi2022difuzer, Tang:NIVAnalyzer:2017}. 
The already-mentioned techniques that analyze candidate alarms with symbolic execution and constraint solving may
ascertain some false alarms as result of identifying given
sets of infeasible paths~\citePRST{ball:slam:fmcad:2010,beyer:blast:isttt:2007,beyer_explicit-state_2013,su_combining_2015,Wang:TSE:2022}, 
as in the CEGAR algorithm that we discussed in Example~\ref{sec:example:cegar}. 
Some alarms identified by analysis at unit-level can be proven infeasible by a subsequent analysis that discovers that  there are no possible execution contexts that would supposedly trigger those alarms~\citePRST{kim_precise_2018}. 
Bodden et al.\ statically map candidate alarms to feature vectors and leverage supervised machine learning to 
filter out likely false alarms~\citePRST{bodden_finding_2008}. 
Dually to the discussion in the previous paragraph, these approaches
can improve the confidence on the reported alarms by dismissing some false  alarms, though being generally inconclusive about whether the alarms that will be eventually reported are  true or false alarms. 

\subsubsection*{\textbf{\synPartitioning}}
Classically, some program analysis techniques 
may partition the program state space  into distinguishable regions, each enjoying or not enjoying given properties of interest.
In a \cpaTech, 
such type of knowledge, which can be introduced by an analysis stage or emerge incrementally while the overall analysis progresses, can be exploited
as a means to steer the analysis stages, or interrelate their results.
Typically,  
the involved analysis stages shall
suitably adapt their analysis algorithms by mapping
the explored program states onto the corresponding state-space regions.
In the primary studies, we classified three sub-categories of this type of synergistic effect, based on whether the reference state-space partitioning is exploited for interrelating the results across analysis stages, providing guidance within analysis algorithms, or directing the analysis onto specific partitions, respectively.  

$\triangleright$\textbf{\synPartitioningWitness}. Possible synergies may  arise directly from associating the state-space regions with the results of the partner analysis stages, thus  characterizing those state-space regions with the witnesses provided by those analyses. 
For instance,
if a former analysis stage identifies a state-space partition, a partner stage that executes a dynamic analysis technique 
can identify which state-space regions can be associated with concrete observations (or test cases) collected at runtime~\citePRST{liu_general_2021,molina_fuzzing_2022,avancini_comparison_2013,carbin_automatically_2010,gupta_hybrid_1997,ghiduk_reducing_2016,adamsen_analyzing_2016}.\footnote{It can be argued that the synergistic effects on \emph{discriminating true and false alarms} that we discussed above can be also described as discriminating hit and missed property-related states. Indeed marking a set of program states as "possible alarms" is a particular case of assigning a property to those states, and confirming the alarms requires to hit those states in the partner analysis being used. Nonetheless, it is also true that designers of \cpaTechs most often perceive the discrimination of alarms and the coverage of program states as distinct types of design goals, respectively, which is why we decided to assign its own specific semantics to the synergies on discriminating true and false alarms.} 
As another example, MuJava executes an analysis stage that  witnesses weakly killed mutants (arguably representative of corresponding execution-space portions) and a next stage that considers only the weakly killed mutants as candidates for strong mutation analysis~\citePRST{kim_combining_2013}.
At the other side of the spectrum, the properties computed in an analysis stage can be confirmed or disproved by  associating them with corresponding formal proofs statically computed in a partner analysis stage~\citePRST{nguyen_using_2014,balakrishnan_modeling_2012}. The synergy can also derive from coordinating multiple types of witnesses, as it happens in some test generation approaches that combine analysis techniques to provide both, test cases for some test objectives, and infeasibility proofs for other test objectives, thus concomitantly satisfying and refining testing criteria of interest~\citePRST{baluda_bidirectional_2016,gulavani_synergy_2006}.
Ghandehari et al.\  associate combinatorial-testing partitions with test cases to incrementally compute fault-localization indicators~\citePRST{sh_ghandehari_combinatorial_2020}. 

$\triangleright$\textbf{\synPartitioningCoverage}.
The analysis stages of a \cpaTech can draw on the knowledge of a set of state-space regions to 
synergistically prioritize
the exploration of program states that correspond to not-yet-hit regions, thus  steering the progress of the analysis algorithms therein.
For instance, this is the case of many test generation approaches, which combine a former analysis stage, where they identify state-space regions to address as test objectives (e.g., program locations, program paths, data-flow entities,  possible buffer overflows, API calls, concurrency events, GUI states, program mutations, test assertions, and so forth),
with a subsequent test generation stage (e.g., based on fuzzing, search-based algorithms, symbolic execution, constraint solving) where they monitor the incremental coverage of those test objectives during the test generation process~\citePRST{Yu:TSE:2020,wang_could_2019,kechagia_effective_2019,yin_fries_2024,el-serafy_automatic_2015,elyasov_search-based_2018,mahmood_evodroid_2014,mao_sapienz_2016,kahkonen_lightweight_2014,hummer_testing_2013,do_goal-oriented_2015,ValleGomez:IETSW:2022,wang_combodroid_2020,zhou_minerva_2022,zhao_test_2015,nguena_timo_multiple_2019,mirzaei_reducing_2016,su_combining_2015,sabbaghi2019fsct,braione_combining_2017,ahmadi_concolic_2019-1,10.1145/3533767.3534221}. Other techniques coordinate multiple analysis stages to explore incrementally larger sets of target state-space regions, with  later analysis stages focused on extending or refining the set of regions already explored by former analysis stages~\citePRST{LeanBin:Wodiany:ASE:2024,jasper_test_1994,jeng_automatic_1999,noller_badger_2018,noller_hydiff_2020,moukahal_boosting_2021,zhang_discover_2018,baluda_bidirectional_2016,beyer_cooperative_2021,Yi:ASEA:2011,godboley_toward_2021,Xu:ICSME:2020}. 

$\triangleright$\textbf{\synPartitioningDirect}. 
An analysis stage can synergistically direct a partner stage to focus on specific state-space regions. Differently from above, the progress of the partner stage is not pursued by a prioritization policy established in its own analysis algorithm, but rather it is actively directed from the former analysis stage by indicating a specific state-space partition (out of a larger set) to be addressed, or feeding data that make the partner stage analyze specific partitions~\citePRST{yu_descry_2017,eslamimehr_efficient_2018,merriam_measurement_2013,memon_employing_2006,zhang_boundary_2016,jin_f3_2013,dufour_blended_2007,gerrard_conditional_2022,huster_efficient_2015,chen_improving_2018,linares-vasquez_mining_2015,alhanahnah_autompi_2023,hu_semantics-based_2021,song_itree_2014,bertolino_automatic_1994,liu_tdroid_2018,liu_vd-guard_2023}. 
For instance, this is the case in  the feedback loop of concolic execution that we discussed in 
Example~\ref{sec:example:concolic}:
the symbolic execution stage computes test cases that trigger the runtime analysis of specific yet-unexplored program paths, and the program paths observed during runtime analysis direct symbolic execution to analyze those program paths as well~\citePRST{godefroid:dart:pldi:2005,sen:cute:esec:2005}. Many other papers that leverage concolic execution exploit  synergies similar to the ones described in our example~\citePRST{li_human-machine_2023,le_combined_2013,choi_grey-box_2019,10.1145/2491411.2491438,artzi_directed_2010,zhang_heuristic_2017,aquino_worst-case_2018,delahaye_infeasible_2015,dinges_solving_2014,ben_henda_opensaw_2017,cadar_execution_2005,borzacchiello_fuzzing_2021}. 

\subsubsection*{\textbf{\synTraversal}}
Synergistic effects may arise when a \cpaTech
focalizes a partner analysis stage on a (set of) purposely-specialized variant(s) of the considered analysis problem, such that solving the problem variant(s) enables the partner stage to produce the results of interests. 
In the primary studies, we classified three sub-categories of this type of synergistic effect, in which the problem variants to be considered derive from specializing artifacts on which the analysis depends, restricting the scope of the analysis, or manipulating the target program, respectively. 

$\triangleright$\textbf{\synTraversalSeedInputs}.
Some analysis techniques seed their analysis algorithms based on given program artifacts that they take as inputs. Whereas in standalone embodiments of the algorithms, those depended artifacts are set by the users or internally synthesized (e.g., with default or random values), in a \cpaTech they can be controlled by the other analysis stages, aiming at synergistic effects. 
For instance, in the primary studies, a common case occurs with fuzzing techniques, which can be seeded with existing test cases.
Many \cpaTechs integrate  fuzzing-based analysis stages,
by feeding those stages (possibly iteratively) with seed
test cases or inputs computed by other analysis stages or based on data provided by other analysis stages, 
aiming to polarize the execution of the fuzzing stages with respect to the goals of the overall analysis~\citePRST{Vikram:ICSE:2021,mcminn_input_2012,zhang_discover_2018,borzacchiello_fuzzing_2021,zhao_test_2015,aquino_worst-case_2018,hough_revealing_2020,harman_strong_2011,fan_history-driven_2024,Rasthofer:2017:ICST}. Similarly, fault localization can be seeded with passing and failing test cases computed in other analysis stages~\citePRST{gissurarson_propr_2022,Vidziunas:ICSME:2024}.  

$\triangleright$\textbf{\synTraversalSeedSink}.
Restricting the analysis scope of an analysis stage, based on deductions from other analysis stages, can also foster synergistic effects. For instance, the analysis scope can be controlled by injecting constraints over the program inputs~\citePRST{gerrard_comprehensive_2017,holling_profiting_2016}, or focusing the set of program entities (e.g., functions, variables or statements) to be considered~\citePRST{bai_hybrid_2022,dhar_clotho_2015,do_goal-oriented_2015,hui_utilization_2017,huang_discovering_2024,eda_efficient_2019,sun_gptscan_2024,sinha_fault_2009,debroy_combining_2014,Rohatgi:IET:2009,gissurarson_propr_2022,Said:ESE:2020,liu_vd-guard_2023,Vidziunas:ICSME:2024}. 

$\triangleright$\textbf{\synTraversalTransform}.
Another means can be to focalize an analysis stage on a  suitably crafted variant of the program under analysis, which better reflects given goals or properties.
For instance, 
partner analysis stages may pass program variants computed via program slicing algorithms~\citePRST{Vidziunas:ICSME:2024,kwon_static_2012,chebaro_behind_2014,czech_just_2015,chen_improving_2018,liu_tdroid_2018}, by means of fault injection techniques~\citePRST{zhang_adaptive_2023}, or by incorporating the original program with fabricated branches that represent  reachability targets~\citePRST{jahangirova_test_2016,jun-xian_pathwalker_2015}, operators with different numerical precision~\citePRST{tang_software_2017}, runtime mechanisms for fault tolerance~\citePRST{joiner_efficient_2014}, or yet patches generated via program repair techniques~\citePRST{dhar_clotho_2015,banerjee_energypatch_2018,fang_vfix_2024,gissurarson_propr_2022}.  

\subsubsection*{\textbf{\synIntepretability}}

The results of a given analysis stage can convey knowledge on aspects of the program semantics that are not directly available to a partner analysis stage, thus allowing the analysis algorithm executed in the partner stage to interpret the program semantics in the light of those additional pieces of knowledge. 
In the primary studies, we classified three sub-categories of this type of synergistic effect, according to whether the provided knowledge relates with properties of program entities, expectations on the execution states, or relations between the program and other software artifacts.

$\triangleright$\textbf{\synIntepretabilityEntities}.
An analysis stage can characterize the behavior of some
program entities with properties that are not explicitly represented in the code of the program under analysis. It can then feed such additional knowledge to partner analysis stages, allowing for them to draw on those properties.
For instance it is common for \cpaTechs to rely on analysis stages that can compute information on the possible pointer aliases, the possible dynamic types of the program variables, the possible instances of dynamically generated code, possible data races, the precision of floating-point computation, or information (including summaries) on the functions called in the program under analysis~\citePRST{Tiwari:ISSRE:2023, Whelan:CC:2013, pradel2012statically,
Xie:ISSTA:2015,Wei:ISSTA:2013,messaoudi_log-based_2021,Young:TSE:1988,zhang_runtime_2014,zhang_combined_2011,zhang_androidleaker_2017,madsen_practical_2013,lohar_two-phase_2021,logozzo_rata_2010,liu_ipa_2016,liu_efficient_2016,lee_hybridroid_2016,dhar_clotho_2015,zhang_detecting_2012,lameed_staged_2011,kellogg_lightweight_2021,guo_predracer_2024,emmi_rapid_2021,ferrara_tval_2012,
hsu_highly_2023,ivancic_scalable_2015,feng_apposcopy_2014,Yan:SETTA:2021,belevantsev_multilevel_2017,fan_static_2021,tang_separate_1994,santelices_exploiting_2010,schoeberl_worst-case_2010,sochor_grammarforge_2025,souter_tatoo_2001,chen_star_2015,godefroid_proving_2010,backes_regression_2013,asensio_worst-case_2013,cai_hybrid_2018,Rattanasuksun:RRF:2016}.
The provided  
not-directly-available information can also predicate on properties of the application interfaces, such as, properties of the input data structures, concurrency properties, non-functional properties, invocation protocols, information on GUI-level inputs and events~\citePRST{wang_string_2022,mahmood_evodroid_2014,grechanik_preventing_2013,dimovski_quantitative_2022,zhu_dynamic_2021,allwood_high_2011,tanida_automated_2013,bose_columbus_2023}.

$\triangleright$\textbf{\synIntepretabilityOracle}.
Many \cpaTechs feature analysis stages that compute test oracles, assertions, invariants, temporal dependencies or other types of models, to enable partner stages to interpret if the computed program states satisfy given expectations~\citePRST{memon_employing_2006,hahnle_constraint-based_2018,lamela_seijas_model_2018,ma_software_2018,dimovski_computing_2020,armoni_deterministic_2006,schordan_combining_2014}.

$\triangleright$\textbf{\synIntepretabilityArtifacts}.
A partner analysis stage can provide 
knowledge of relevant dependencies between the program under analysis and other software artifacts that comprise the overall software product. For instance, some primary studies exploit static analysis to relate program locations with corresponding  dependencies on configuration files, and then exploit those dependencies to monitor or steer the execution of the program  with respect to the configuration options~\citePRST{zhang2013confdiagnoser,song_itree_2014}. Other \cpaTechs 
exploit the relation between program entities and knowledge from
specification or requirement documents~\citePRST{zhao_framework_2022, brennan_symbolic_2018,bertolino_architectural_2007,boudhiba_model-based_2015}.

\subsubsection*{\textbf{\synRewrite}}

An analysis stage can compute data that allow for a partner analysis to apply useful rewritings at their representation of some program states, 
by exploiting the correspondence between those states and the results of the former analysis. The partner analysis will typically either
replace some entities that belong to the representation of the states with analogous
entities from the corresponding states from the other analysis stages, or extend some properties between mutually corresponding states.
In the primary studies, we classified two sub-categories of this type of synergistic effect, according to whether the rewriting mechanisms rely on either concrete or semantically richer counterparts.

$\triangleright$\textbf{\synRewriteConcrete}. 
An analysis stage can exploit the correspondence with its internal representations of the program states and  concrete data computed by a partner analysis, in order to partially concretize intermediate results therein.  
The general goal can be to strengthen somehow the efficiency of the analysis, though possibly sacrificing either the recall or the precision (or both) of the results. For instance, this type of synergistic effects occur in many primary studies that ground on possible variants of concolic execution~\citePRST{godefroid:dart:pldi:2005,sen:cute:esec:2005,dinges_solving_2014,grech_shooting_2018,schordan_combining_2014}. As we also discussed in   Example~\ref{sec:example:concolic}, concolic execution takes advantage of the concrete values observed during runtime monitoring to  both ``linearize'' the symbolic formulas computed along the corresponding program paths, and assume the satisfiability of the symbolic states that correspond (by construction) to concretely executed program states.
Other techniques improve efficiency  of static analysis by concretely executing given function calls~\citePRST{Park:JMESECSFSE:2021}, or by partially rewriting abstract strings with corresponding concrete string values observed at runtime~\citePRST{Tripp:ISSTA:2014}.

$\triangleright$\textbf{\synRewriteSimulator}.
We found a single primary study in which the rewriting aims 
at providing semantically-richer representations~\citePRST{li_view-based_2008}. After computing the model of an interactive GUI-based application, they rely on the results of a partner analysis stage that extracted the source code of given GUI entities, and replace some abstract GUI entities in their model with the corresponding source code, making the model \emph{actionable} for running simulations. 

\subsubsection*{\textbf{\synRefine}}

The results of an analysis stage may allow for refining the precision of a program model computed in the analysis algorithm of a partner analysis stage, thus allowing the partner stage to improve the precision of its results accordingly. In the primary studies, we classified two sub-categories of this type of synergistic effect, according to whether the refinement of the program model at hand occurs by either incorporating new details or pruning some invalid states, respectively, in the light of the new information available.

$\triangleright$\textbf{\synRefineIncorporate}.
On one hand, the precision of a program model can be refined by incorporating new details based on the data provided by partner analysis stages.
Usually this is obtained by re-executing the analysis that computed the initial model based on the new data avaliable~\citePRST{liu_general_2021,chen_synthesising_2016,dallmeier_generating_2010, chen_have_2009,matias_determining_2020,beyer_explicit-state_2013,camilli_model-based_2020,chen_automatically_2023,howar_hybrid_2013,cousot_refining_1999, srivastava_crystallizer_2023,sun_tlv_2015,merriam_measurement_2013,cai_hybrid_2018,Sadeghi:JMFSE:2017,bruning_complete_2023}. For instance, this is the case in the 
CEGAR technique that we discussed in Example~\ref{sec:example:cegar}, in which the detection and characterization of some false alarms 
allowed for re-executing the model checking stage against a more precise program model.

$\triangleright$\textbf{\synRefinePrune}.
On the other hand, the precision of a program model can be also refined
by pruning away some portions of the current model, which were initially included for conservativeness, but could then be identified as irrelevant based on the information conveyed from another analysis stage~\citePRST{li_calculating_2013,liu_promal_2022, azzopardi_technique_2020,bertolino_automatic_1994,kama_change_2015,howar_hybrid_2013,sun_tlv_2015,cousot_refining_1999}.

\subsubsection{Horizontal Synergies}

\subsubsection*{\textbf{\synFlow}}
Some primary studies compose (side-by-side) analysis stages that compute program-flow data, aiming to
synergistically exploit program-flow data that can mutually complement each other. This can support subsequent analysis stages to better reason on the possible program flows~\citePRST{Tarvo:ASE:2018,Ye:2016:SANER, Rimsa:SPE:2020, Obbink:SANER:2018, Perez:TSE:2019,chen_have_2009,zhang_hybrid_2024,zhang_combined_2011,zhang_hybrid_2018,jones_addressing_2016,lamela_seijas_model_2018,he_python_2023,homaei_athena_2019,sun_combort_2016,Sadeghi:JMFSE:2017}.

\subsubsection*{\textbf{\synFeature}}

Some primary studies exploit synergistic effects by forming feature vectors out of program features (software metrics or classifications) computed by different 
analysis stages, and then handling the feature vectors with either mathematical models or machine learning algorithms to derive program properties~\citePRST{li_fault_2022,krisper_metric_2017,ulrich_experience_2016, golagha_can_2020,greca_comparing_2022,khoshgoftaar_investigating_1995,freitas_scout_2016,yang_multi-objective_2024,alshoaibi_price_2019}. 
For instance, this is the case of the Marvin approach that we discussed in Example~\ref{sec:example:marvin}~\citePRST{lindorfer_marvin_2015}.
This allows for both integrating  the capabilities of distinct types of information and exploiting a richer set of possible correlations.

\subsubsection*{\textbf{\synGUI}}
\CpaTechs that address GUI properties of programs can integrate GUI data and other program data from multiple analysis stages, in order to synergistically support improved reasoning on the GUIs and the GUI properties~\citePRST{linares-vasquez_mining_2015,mao_sapienz_2016,fazzini_enhancing_2023}. 

\subsubsection*{\textbf{\synReports}}
Some primary studies 
directly integrate the results from distinct analysis techniques that work on the same type of problem (e.g., the identification of alarms, smells, issues, regression test cases, and so forth), in order to synergistically improve the delivered analysis reports with respect to precision, coverage, traceability, or supporting evidence~\citePRST{Vasquez:IST:2019, Xu:SANER:2024,Yu:TACAS:2009,zhang_comparing_2022,gergely_differences_2019,ling_essential_2024, neelofar_improving_2017,marin_pinpointing_2008,jannesari_generating_2014,poshyvanyk2007feature,shan_self-hiding_2018,sohn_cement_2022,gerrard_conditional_2022,botella_complementary_2019,saputri_software_2020}.

\subsection{Inter-analysis Workflow (RQ~2)}\label{sec:results:workflows}

\begin{figure}[!tb]
\begin{center}
\includegraphics[width=0.5\textwidth]{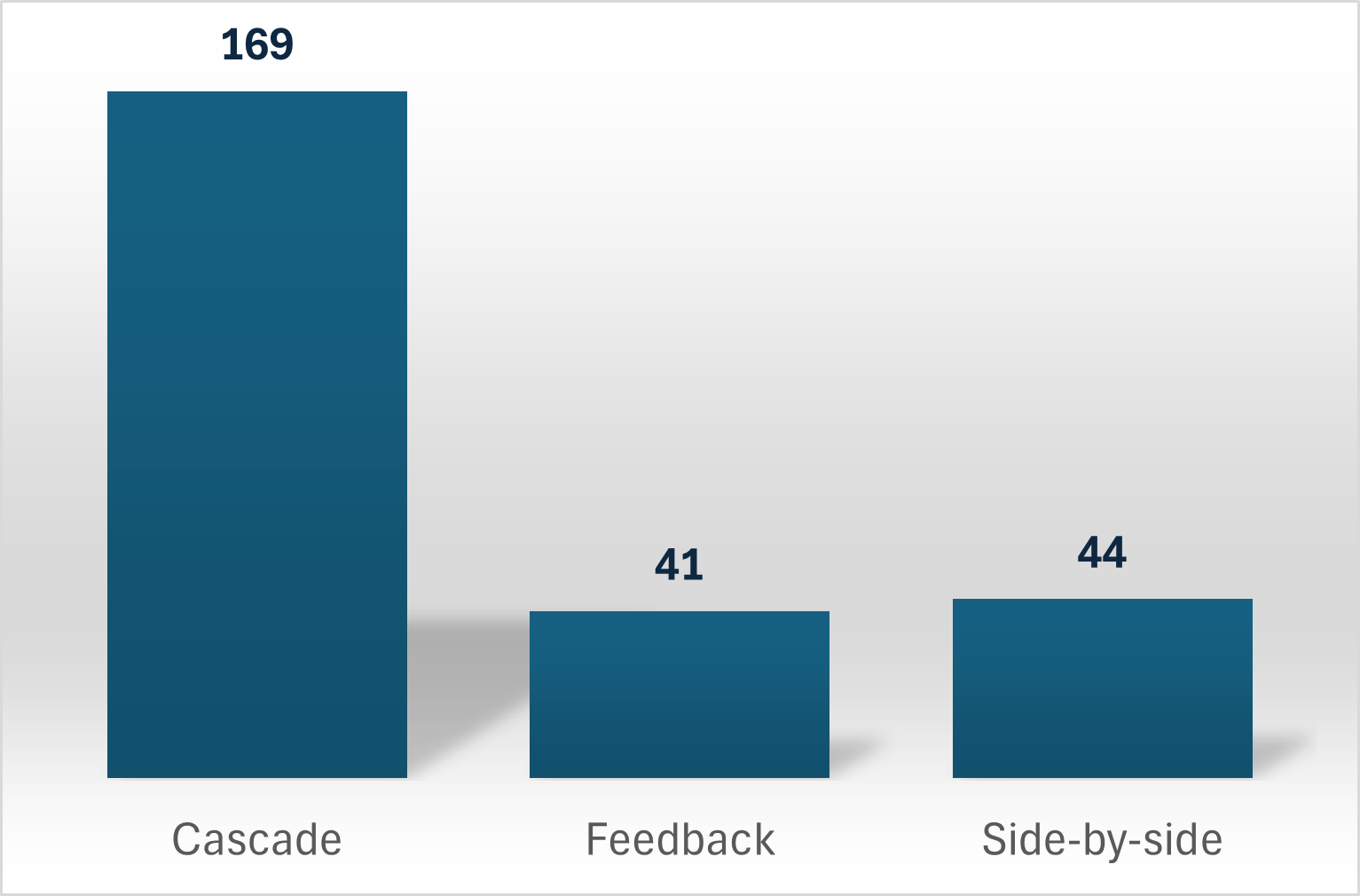}
\end{center}
\caption{Distribution of inter-analysis workflows across the primary studies} 
\label{fig:results:workflows}
\end{figure}

Figure~\ref{fig:results:workflows} illustrates the distribution of the types of inter-analysis workflow of the \cpaTechs in the primary studies. The specific mapping is:

 \textbf{\workCascade}
  \citePRST{pradel2012statically, souter_tatoo_2001,sochor_grammarforge_2025,sinha_fault_2009,lee_sealant_2017,leesatapornwongsa_flakerepro_2022,li_calculating_2013,li_view-based_2008,li_human-machine_2023,linares-vasquez_mining_2015,liu_tdroid_2018,mahmood_evodroid_2014,chen_synthesising_2016,chen_star_2015,chen_boosting_2021,chen_improving_2018,liu_general_2021,czech_just_2015,liu_promal_2022,liu_vd-guard_2023,bose_columbus_2023,debroy_combining_2014, Tiwari:ISSRE:2023, Tlili:AMAST:2008, Tripp:ISSTA:2014, ValleGomez:IETSW:2022, Vidziunas:ICSME:2024, Vikram:ICSE:2021, Vorobyov:SEFM:2012, Wang:TSE:2022, Wei:ISSTA:2013, Whelan:CC:2013, LeanBin:Wodiany:ASE:2024, Xie:ISSTA:2015, Xu:ICSME:2020, Yan:SETTA:2021, 10.1145/3533767.3534221, Yi:ASEA:2011, Rohatgi:IET:2009, Sadeghi:JMFSE:2017, Said:ESE:2020, Pan:ICSME:2017, Patra:PICSE:2018, samhi2022difuzer, Tang:NIVAnalyzer:2017,zhang2013confdiagnoser,bruning_complete_2023,chebaro_behind_2014,matias_determining_2020,mcminn_input_2012,messaoudi_log-based_2021,Young:TSE:1988,Yu:TSE:2020,yu_descry_2017,zhang_runtime_2014,zhang_combined_2011,zhang_androidleaker_2017,zhou_minerva_2022,nguyen_using_2014,nguena_timo_multiple_2019,molina_fuzzing_2022,mirzaei_reducing_2016,memon_employing_2006,mao_sapienz_2016,lohar_two-phase_2021,logozzo_rata_2010,liu_ipa_2016,liu_efficient_2016,lee_hybridroid_2016,ma_software_2018,zhu_dynamic_2021,zhao_framework_2022,alhanahnah_autompi_2023,allwood_high_2011,armoni_deterministic_2006,azzopardi_technique_2020,backes_regression_2013,bai_hybrid_2022,balakrishnan_modeling_2012,bertolino_architectural_2007,bodden_clara_2012,bodden_finding_2008,boudhiba_model-based_2015,ponta2020detection,braione_combining_2017,brennan_symbolic_2018,burnim_testing_2011,busse_combining_2022,cadar_execution_2005,cai_hybrid_2018,camilli_model-based_2020,carbin_automatically_2010,asensio_worst-case_2013,sh_ghandehari_combinatorial_2020,
zhang_boundary_2016,belevantsev_multilevel_2017,garcia_automatic_2017,
wang_combodroid_2020,wang_could_2019,zhang_detecting_2012,zhang_adaptive_2023,
hough_revealing_2020,hsu_highly_2023,hu_achyb_2021,jeng_automatic_1999,jin_f3_2013,joiner_efficient_2014,jun-xian_pathwalker_2015,grech_shooting_2018,grechanik_preventing_2013,gupta_hybrid_1997,kechagia_effective_2019,kellogg_lightweight_2021,kim_first_2024,kim_precise_2018,guo_predracer_2024,kim_combining_2013,jasper_test_1994,chen_automatically_2023,wang_string_2022,yin_fries_2024,harman_strong_2011,holling_profiting_2016,dimovski_quantitative_2022,dimovski_computing_2020,diner_generating_2021,do_goal-oriented_2015,dufour_blended_2007,el-serafy_automatic_2015,elyasov_search-based_2018,emmi_rapid_2021,fan_static_2021,fan_efficiently_2018,fan_history-driven_2024,fang_vfix_2024,feng_apposcopy_2014,ferrara_tval_2012,gerrard_conditional_2022,eda_efficient_2019,hu_semantics-based_2021,sun_gptscan_2024,huang_discovering_2024,hui_utilization_2017,hummer_testing_2013,huster_efficient_2015,huster_using_2017,ivancic_scalable_2015,cousot_refining_1999, Rasthofer:2017:ICST, Rattanasuksun:RRF:2016,santelices_exploiting_2010,schoeberl_worst-case_2010,tang_separate_1994,schordan_combining_2014,srivastava_crystallizer_2023,tang_software_2017,su_combining_2015,tanida_automated_2013,csallner:check-n-crash:icse:2005,ghiduk_reducing_2016,godefroid_proving_2010,goffi_automatic_2016,artzi_directed_2010,kama_change_2015,kwon_static_2012,lameed_staged_2011,adamsen_analyzing_2016,ahmadi_concolic_2019-1,naik_effective_2009}

\textbf{\workFeedback} \citePRST{merriam_measurement_2013,le_combined_2013,choi_grey-box_2019,dallmeier_generating_2010, Park:JMESECSFSE:2021,10.1145/2491411.2491438,noller_badger_2018,noller_hydiff_2020,moukahal_boosting_2021,zhang_discover_2018,madsen_practical_2013,dhar_clotho_2015,avancini_comparison_2013,baluda_bidirectional_2016,zhang_heuristic_2017,zhao_test_2015,aquino_worst-case_2018,gulavani_synergy_2006,hahnle_constraint-based_2018,kahkonen_lightweight_2014,delahaye_infeasible_2015,dinges_solving_2014,eslamimehr_efficient_2018,gerrard_comprehensive_2017,howar_hybrid_2013,cousot_refining_1999, sabbaghi2019fsct,ben_henda_opensaw_2017,beyer_explicit-state_2013,sun_tlv_2015,beyer_cooperative_2021,song_itree_2014,beyer:blast:isttt:2007,ball:slam:fmcad:2010,godefroid:dart:pldi:2005,sen:cute:esec:2005,gissurarson_propr_2022,banerjee_energypatch_2018,bertolino_automatic_1994,borzacchiello_fuzzing_2021,godboley_toward_2021}

\textbf{\workSidebyside} \citePRST{linares-vasquez_mining_2015,lindorfer_marvin_2015,sun_combort_2016,ling_essential_2024,chen_have_2009, Tarvo:ASE:2018, Vasquez:IST:2019, Xu:SANER:2024, Ye:2016:SANER, Rimsa:SPE:2020, Obbink:SANER:2018, Perez:TSE:2019, li_fault_2022, neelofar_improving_2017,Yu:TACAS:2009,zhang_hybrid_2024,zhang_combined_2011,marin_pinpointing_2008,mao_sapienz_2016,zhang_hybrid_2018,jahangirova_test_2016,jannesari_generating_2014,krisper_metric_2017,golagha_can_2020,greca_comparing_2022,jones_addressing_2016,khoshgoftaar_investigating_1995,lamela_seijas_model_2018,yang_multi-objective_2024,he_python_2023,fazzini_enhancing_2023,freitas_scout_2016,gergely_differences_2019,homaei_athena_2019,shan_self-hiding_2018,poshyvanyk2007feature,alshoaibi_price_2019,ulrich_experience_2016,zhang_comparing_2022,sohn_cement_2022,gerrard_conditional_2022,botella_complementary_2019,Sadeghi:JMFSE:2017,saputri_software_2020}

\subsection{Interpretation Structures of Mapping Functions (RQ~3)}\label{sec:results:interpretations}

\begin{figure}[!tb]
\begin{center}
\includegraphics[width=.8\textwidth]{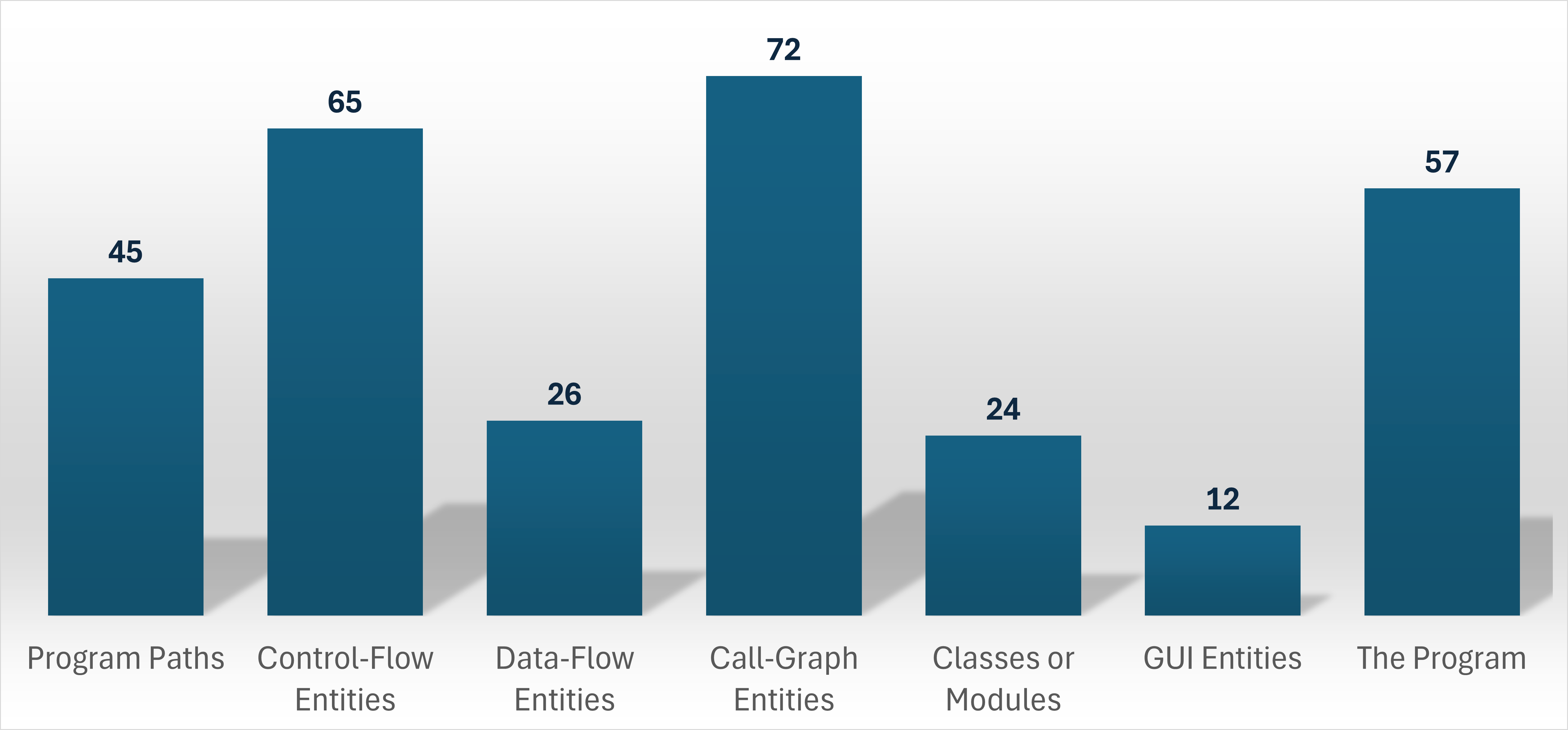}
\end{center}
\caption{Distribution of mapping-function interpretation structures across the primary studies} 
\label{fig:results:interpretation-structures}
\end{figure}

Figure~\ref{fig:results:interpretation-structures} indicates the distribution across the primary studies of the \numMappingFunctionStructures types of mapping-function interpretation structures that we have classified during the survey, and that we describe in detail below.
As introduced in Section~\ref{sec:taxonomy}, the interpretation structure of a mapping function explains how the results of an analysis stage are referred to program entities, such that the partner analysis stages may suitably interpret those results when considering corresponding program entities through their analysis algorithms.

\subsubsection*{\textbf{\structPaths}}
Many program analysis techniques produce results that correspond to possible paths of the program control-flow graph (the program paths), or the program states traversed thereby. Straightforwardly, in \cpaTechs, analysis algorithms that reason on program paths can interpret the results that other analysis stages associated with the corresponding program paths. This type of mapping is very common in \cpaTechs that ground on dynamic analysis, symbolic execution and flow-sensitive static analysis in different ways (as, for instance, we discussed in Example~\ref{sec:example:concolic} and Example~\ref{sec:example:cegar}) due to the program-path-oriented nature of those techniques~\citePRST{
le_combined_2013,
lee_sealant_2017,
choi_grey-box_2019,
Tlili:AMAST:2008,
Wang:TSE:2022,
Wei:ISSTA:2013,
Xie:ISSTA:2015,
Yi:ASEA:2011,
Said:ESE:2020,
sabbaghi2019fsct,
10.1145/2491411.2491438,
Young:TSE:1988, yu_descry_2017,zhang_discover_2018,braione_combining_2017,cadar_execution_2005,asensio_worst-case_2013,
zhang_heuristic_2017,
zhang_boundary_2016,
zhao_test_2015,aquino_worst-case_2018,
hsu_highly_2023,hu_achyb_2021,kim_precise_2018,kahkonen_lightweight_2014,delahaye_infeasible_2015,harman_strong_2011,holling_profiting_2016,emmi_rapid_2021,
huster_efficient_2015,ben_henda_opensaw_2017,srivastava_crystallizer_2023,beyer_explicit-state_2013,dinges_solving_2014,eslamimehr_efficient_2018,gerrard_comprehensive_2017,li_calculating_2013,banerjee_energypatch_2018,fan_efficiently_2018,beyer:blast:isttt:2007,ball:slam:fmcad:2010,godefroid:dart:pldi:2005,sen:cute:esec:2005,busse_combining_2022,bertolino_automatic_1994}. 

\subsubsection*{\textbf{\structCFG}}
The results of some program analysis techniques can be 
transferred to other analysis stages as properties or facts that occur at instructions, decisions, loops or other instruction blocks of the program control flow graph (therefore, with grosser granularity than expressing precise mappings between results and program paths). 
With this type of mapping functions, the results of an analysis can be  exploited in partner analysis stages that inspect the same control-flow entities
in their analysis algorithms~\citePRST{ponta2020detection,sabbaghi2019fsct,
leesatapornwongsa_flakerepro_2022,chen_have_2009,debroy_combining_2014,Vorobyov:SEFM:2012,Whelan:CC:2013,LeanBin:Wodiany:ASE:2024,Xie:ISSTA:2015,Xu:ICSME:2020,Xu:SANER:2024,Yan:SETTA:2021,Rimsa:SPE:2020,samhi2022difuzer,Tang:NIVAnalyzer:2017,zhang2013confdiagnoser,adamsen_analyzing_2016,chebaro_behind_2014,li_fault_2022,neelofar_improving_2017,nguyen_using_2014,merriam_measurement_2013,marin_pinpointing_2008,lohar_two-phase_2021,logozzo_rata_2010,liu_ipa_2016,ma_software_2018,dhar_clotho_2015,zhao_framework_2022,backes_regression_2013,baluda_bidirectional_2016,bodden_clara_2012,bodden_finding_2008,brennan_symbolic_2018,Ye:2016:SANER,zhang_detecting_2012,zhang_adaptive_2023,jahangirova_test_2016,jeng_automatic_1999,joiner_efficient_2014,lameed_staged_2011,gupta_hybrid_1997,kellogg_lightweight_2021,gulavani_synergy_2006,fan_efficiently_2018,jasper_test_1994,do_goal-oriented_2015,el-serafy_automatic_2015,elyasov_search-based_2018,emmi_rapid_2021,fan_static_2021,fang_vfix_2024,ferrara_tval_2012,garcia_automatic_2017,belevantsev_multilevel_2017,srivastava_crystallizer_2023,mcminn_input_2012,Park:JMESECSFSE:2021,sohn_cement_2022,gissurarson_propr_2022,godboley_toward_2021,godefroid_proving_2010,balakrishnan_modeling_2012,aquino_worst-case_2018,Vidziunas:ICSME:2024}.

\subsubsection*{\textbf{\structDF}} 
Data-flow dependencies relate pairs of program locations that may operate on the same (groups of) memory locations. The possible data-flow dependencies  can thus play a similar mapping-function role as we explained for the above case of control-flow entities, letting an analysis stage associate results as facts that relate with given data-flow relations, and transfer those results to partner analysis stages  that consider corresponding data-flow relations within their respective analysis algorithm~\citePRST{chen_boosting_2021,Xie:ISSTA:2015, sinha_fault_2009,Yu:TACAS:2009,Yu:TSE:2020,zhang_androidleaker_2017,Ye:2016:SANER,hough_revealing_2020,he_python_2023,do_goal-oriented_2015,homaei_athena_2019,eda_efficient_2019,huang_discovering_2024,hui_utilization_2017,hummer_testing_2013,huster_using_2017, santelices_exploiting_2010,bai_hybrid_2022,su_combining_2015,cai_hybrid_2018,sun_gptscan_2024,ghiduk_reducing_2016,burnim_testing_2011,carbin_automatically_2010,schordan_combining_2014,Said:ESE:2020}. 

\subsubsection*{\textbf{\structCG}}
A program analysis technique can transfer its results to partner analysis stages in the form of properties or facts that it attaches to the entities of the program call graph, i.e., the program functions that belong to the call graph (the nodes of the call graph) or the call relations between those functions (the edges of the call graph).
The partner analysis stages can suitably interpret those results while visiting the call graph within their analysis algorithms.
In the most common scenarios, some analysis stages produce results that map to program functions or call sites (including the case of inputs for executing some functions),
allowing for a partner analysis to consider those data  as well~\citePRST{chen_synthesising_2016,chen_star_2015, Rattanasuksun:RRF:2016, poshyvanyk2007feature,Tiwari:ISSRE:2023,le_combined_2013,liu_tdroid_2018,liu_general_2021,liu_vd-guard_2023,chen_improving_2018,choi_grey-box_2019, alhanahnah_autompi_2023,Tripp:ISSTA:2014,Vikram:ICSE:2021,Obbink:SANER:2018,Tang:NIVAnalyzer:2017,10.1145/2491411.2491438,noller_badger_2018,noller_hydiff_2020,moukahal_boosting_2021,zhang_discover_2018,molina_fuzzing_2022,zhu_dynamic_2021,zhou_minerva_2022,zhao_test_2015,Tarvo:ASE:2018,wang_could_2019,10.1145/3533767.3534221,Ye:2016:SANER,jannesari_generating_2014,jun-xian_pathwalker_2015,krisper_metric_2017,kwon_static_2012,gupta_hybrid_1997,kechagia_effective_2019,wang_string_2022,yin_fries_2024,delahaye_infeasible_2015,dinges_solving_2014,dufour_blended_2007,ivancic_scalable_2015,saputri_software_2020,srivastava_crystallizer_2023,ulrich_experience_2016,zhang_runtime_2014,madsen_practical_2013,beyer_explicit-state_2013,Sadeghi:JMFSE:2017,sun_combort_2016,sun_gptscan_2024,Park:JMESECSFSE:2021,Rasthofer:2017:ICST,schoeberl_worst-case_2010,shan_self-hiding_2018,hu_semantics-based_2021,tang_separate_1994,beyer:blast:isttt:2007,ball:slam:fmcad:2010,godefroid:dart:pldi:2005,sen:cute:esec:2005,csallner:check-n-crash:icse:2005,goffi_automatic_2016,belevantsev_multilevel_2017,cai_hybrid_2018,ben_henda_opensaw_2017,borzacchiello_fuzzing_2021,Vidziunas:ICSME:2024}. For instance, this is the case in the techniques of
Example~\ref{sec:example:cnc}, Example~\ref{sec:example:concolic} and Example~\ref{sec:example:cegar}.
In other cases, some analysis stages may contribute to the overall analysis by decorating the call graph with further call relations, which
may occur between some functions,
but might 
have been missed during the construction of the ``initial'' call graph,\footnote{ 
At the simplest extreme of the spectrum,
the very activity of computing the call graph can be itself framed as a (preliminary) static analysis step that delivers its results to other analysis stages, exploiting a mapping function grounded on call graph entities. In our mapping study, we purposely avoid into this extremely fine-grained
anatomy of the analysis techniques, 
at least as long as those call-graph-computation steps occur with straightforward traversals or the program code. 
} e.g., 
call relations related to event-oriented message dispatching~\citePRST{Perez:TSE:2019,mahmood_evodroid_2014} or inter-component  flows~\citePRST{feng_apposcopy_2014,lee_hybridroid_2016,bose_columbus_2023}.

\subsubsection*{\textbf{\structModules}}
Other types of mapping functions may abstract from the intra-procedural or inter-procedural execution flows, which characterize the mappings above, grounding on the modular structuring of the code. This includes associating the results of an analysis stage with 
the classes of an object-oriented program, or any other modular decomposition (e.g., package structures, components) of the program, or modules that comprise crosscutting concerns (e.g., bug-fixing patches or test cases), aiming to exploit those results in a partner analysis stage that considers those classes or modules for further analysis~\citePRST{pradel2012statically, dallmeier_generating_2010,liu_general_2021,Vasquez:IST:2019, Rohatgi:IET:2009, Pan:ICSME:2017,Patra:PICSE:2018,zhang_hybrid_2024,zhang_combined_2011,molina_fuzzing_2022,zhang_hybrid_2018,golagha_can_2020,jones_addressing_2016,kama_change_2015,kim_first_2024,guo_predracer_2024, alshoaibi_price_2019,sun_tlv_2015,greca_comparing_2022,sh_ghandehari_combinatorial_2020,freitas_scout_2016,gergely_differences_2019,matias_determining_2020,naik_effective_2009}.

\subsubsection*{\textbf{\structGUI}}
Programs that interact with users via graphical-user-interfaces are often analyzed by hypothesizing  functional decompositions grounded on the structure of the GUIs, for instance with reference to the possible GUI screens and widgets, as the GUI entities can be often identified directly in the code. It is then straightforward for analysis algorithms that  specifically consider GUI entities to integrate results that other analysis stages logically associated with corresponding GUI entities~\citePRST{mahmood_evodroid_2014,li_view-based_2008,li_human-machine_2023,linares-vasquez_mining_2015,liu_promal_2022,mirzaei_reducing_2016,memon_employing_2006,mao_sapienz_2016,chen_automatically_2023,fazzini_enhancing_2023,tanida_automated_2013,Sadeghi:JMFSE:2017}. 

\subsubsection*{\textbf{\structProgram}}
The mapping functions with grossest granularity  track 
properties and facts that hold for the
the  program under analysis as a whole, 
transferring those results to other analysis stages as general, program-level data. 
Program models, e.g., state machines, Petri nets or program abstractions (as the ones of Example~\ref{sec:example:cegar}), can be computed in an analysis stage and then transferred to help other analysis stages~\citePRST{beyer:blast:isttt:2007,ball:slam:fmcad:2010,cousot_refining_1999,nguena_timo_multiple_2019,hahnle_constraint-based_2018,lamela_seijas_model_2018,wang_combodroid_2020,diner_generating_2021,howar_hybrid_2013,bertolino_architectural_2007,souter_tatoo_2001,grechanik_preventing_2013,beyer_cooperative_2021,sochor_grammarforge_2025,song_itree_2014,gissurarson_propr_2022,camilli_model-based_2020,ahmadi_concolic_2019-1,azzopardi_technique_2020,boudhiba_model-based_2015,bruning_complete_2023,schordan_combining_2014}. Similarly, program variants, such as slices, mutants and patched versions, can be transferred by some analysis stages as alter ego versions of the program under analysis and subsequently analyzed in place of the original program~\citePRST{czech_just_2015,liu_tdroid_2018,Vidziunas:ICSME:2024,eda_efficient_2019,chebaro_behind_2014,ValleGomez:IETSW:2022,armoni_deterministic_2006,kim_combining_2013,dhar_clotho_2015,fang_vfix_2024,chen_improving_2018,tang_software_2017,dimovski_quantitative_2022,banerjee_energypatch_2018}. Flow-insensitive analyses may generalize locally observed data, e.g., pointer aliases observed at any instruction or string literals used in the code, 
to the whole program, allowing for partner analysis stages to  rely on that knowledge~\citePRST{liu_efficient_2016,mao_sapienz_2016,grech_shooting_2018,ling_essential_2024}. Other program-level mappings occur with program-level invariants or constraints~\citePRST{dimovski_computing_2020,gerrard_conditional_2022,gerrard_comprehensive_2017,beyer_explicit-state_2013}, test data and test suites~\citePRST{fan_history-driven_2024,zhang_comparing_2022, liu_vd-guard_2023,jin_f3_2013,linares-vasquez_mining_2015,allwood_high_2011,avancini_comparison_2013,artzi_directed_2010,botella_complementary_2019}, logging data~\citePRST{messaoudi_log-based_2021}, program-level measurements and predictions (as in Example~\ref{sec:example:marvin})~\citePRST{khoshgoftaar_investigating_1995,lindorfer_marvin_2015,yang_multi-objective_2024}.

\subsection{Mechanics of Mapping Functions (RQ~4)}\label{sec:results:mechanics}

\begin{figure}[!tb]
\begin{center}
\includegraphics[width=.85\textwidth]{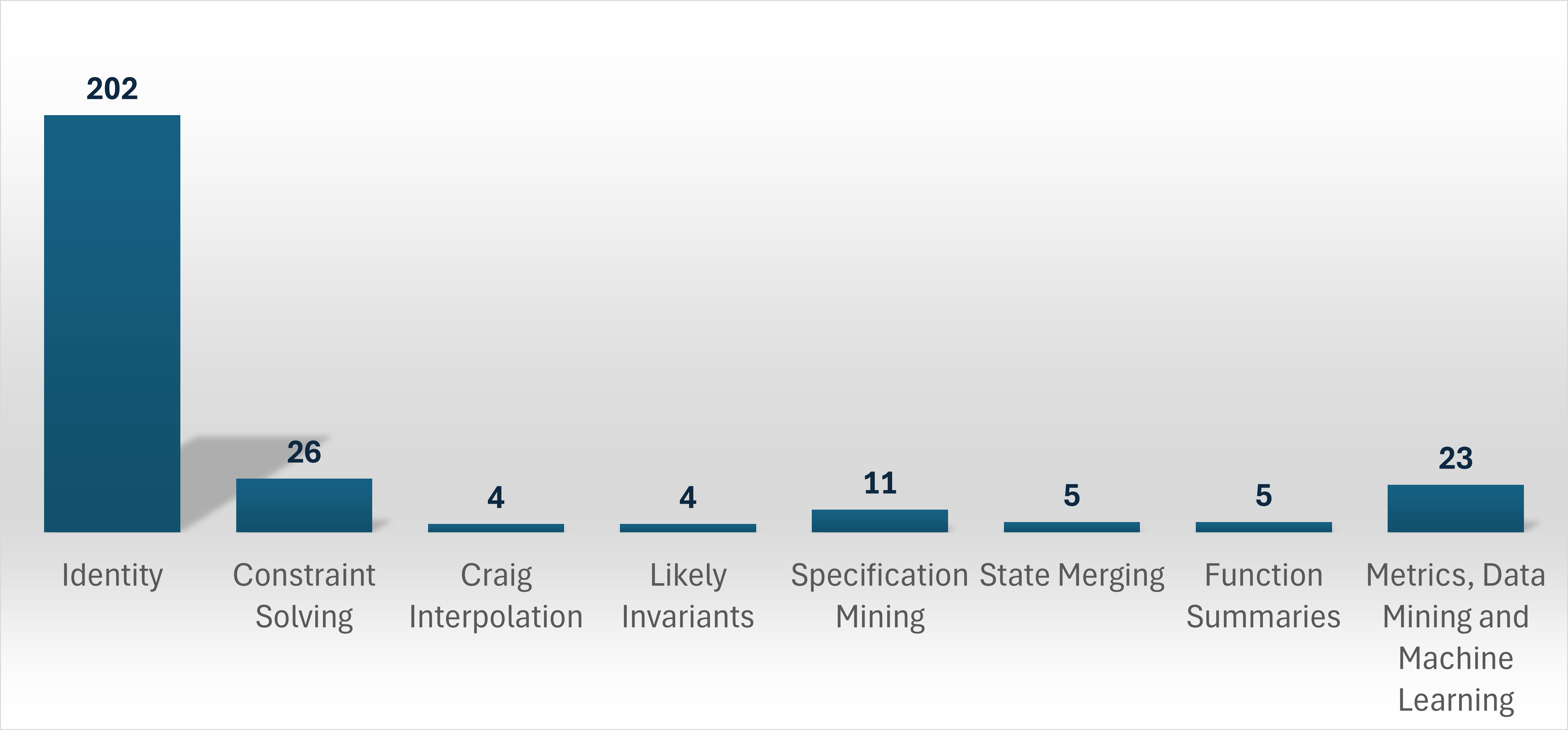}
\end{center}
\caption{Distribution of mapping-function mechanics across the primary studies} 
\label{fig:results:mechanics}
\end{figure}

Figure~\ref{fig:results:mechanics} indicates the distribution across the primary studies of the \numMappingFunctionMechanics types of mapping-function mechanics that we have classified during the survey, and that we describe in detail below.
As introduced in Section~\ref{sec:taxonomy}, the mechanics of a mapping function explains how the results of an analysis stage get further processed, typically through program-agnostic, mathematical means, to turn them in the final format in which they are transferred to the partner analysis stages. 

\subsubsection*{\textbf{\mechanicAssociation}}
The identity mechanics captures the baseline case in which the results of an analysis are transferred to other analysis stages without 
additional processing,
plainly associating those results to corresponding program entities as  
we discussed in the previous section. We discussed some mapping functions with identity mechanics in Example~\ref{sec:example:concolic} and example~\ref{sec:example:cegar}. 
Unsurprisingly, the identity mechanics occurs in many mapping functions encompassed in the primary studies~\citePRST{
naik_effective_2009,
le_combined_2013,
lee_sealant_2017,
sinha_fault_2009,
tang_separate_1994,
leesatapornwongsa_flakerepro_2022,
sohn_cement_2022,
li_calculating_2013,li_view-based_2008,
sochor_grammarforge_2025,
li_human-machine_2023,
linares-vasquez_mining_2015,
song_itree_2014,
ling_essential_2024,liu_tdroid_2018,
mahmood_evodroid_2014,
chen_synthesising_2016,chen_star_2015,
chen_have_2009,chen_improving_2018,
choi_grey-box_2019,liu_general_2021,
dallmeier_generating_2010,
liu_promal_2022,
liu_vd-guard_2023,
chen_boosting_2021,
sun_gptscan_2024,
sun_combort_2016,
czech_just_2015,alhanahnah_autompi_2023,
Tlili:AMAST:2008,
Tripp:ISSTA:2014,
Vasquez:IST:2019,
ValleGomez:IETSW:2022,
Vidziunas:ICSME:2024,
Vikram:ICSE:2021,
souter_tatoo_2001,
Vorobyov:SEFM:2012,
Wang:TSE:2022,
Wei:ISSTA:2013,
Whelan:CC:2013,
LeanBin:Wodiany:ASE:2024,
Xu:SANER:2024,
10.1145/3533767.3534221,
Ye:2016:SANER,
Yi:ASEA:2011,
Rohatgi:IET:2009,
Sadeghi:JMFSE:2017,
Patra:PICSE:2018,
samhi2022difuzer,
Tang:NIVAnalyzer:2017,
Xie:ISSTA:2015,
Xu:ICSME:2020,
zhang2013confdiagnoser,
10.1145/2491411.2491438,
matias_determining_2020,
mcminn_input_2012,
messaoudi_log-based_2021,
neelofar_improving_2017,
noller_badger_2018,
noller_hydiff_2020,
moukahal_boosting_2021,
Young:TSE:1988,
Yu:TACAS:2009,
Yu:TSE:2020,
yu_descry_2017,
fazzini_enhancing_2023,
zhang_discover_2018,
zhang_hybrid_2024,
zhang_runtime_2014,
zhang_combined_2011,
zhang_androidleaker_2017,
zhou_minerva_2022,
nguena_timo_multiple_2019,
molina_fuzzing_2022,
mirzaei_reducing_2016,
merriam_measurement_2013,
memon_employing_2006,
mao_sapienz_2016,
madsen_practical_2013,
lohar_two-phase_2021,
logozzo_rata_2010,
liu_ipa_2016,
liu_efficient_2016,
dhar_clotho_2015,
zhu_dynamic_2021,allwood_high_2011,avancini_comparison_2013,asensio_worst-case_2013,
zhang_heuristic_2017,
zhang_boundary_2016,
zhao_test_2015,
wang_could_2019,
zhang_comparing_2022,
zhang_hybrid_2018,
hough_revealing_2020,
hsu_highly_2023,
hu_achyb_2021,
jahangirova_test_2016,
jannesari_generating_2014,
jeng_automatic_1999,
jin_f3_2013,
kwon_static_2012,
lameed_staged_2011,
grechanik_preventing_2013,
gupta_hybrid_1997,
jones_addressing_2016,
kama_change_2015,
kechagia_effective_2019,
kellogg_lightweight_2021,
tanida_automated_2013,
kim_first_2024,
kim_precise_2018,
guo_predracer_2024,
elyasov_search-based_2018,
kahkonen_lightweight_2014,
kim_combining_2013,
jasper_test_1994,
chen_automatically_2023,
delahaye_infeasible_2015,
el-serafy_automatic_2015,
harman_strong_2011,
holling_profiting_2016,
dimovski_quantitative_2022,
diner_generating_2021,
dinges_solving_2014,
do_goal-oriented_2015,
dufour_blended_2007,
eslamimehr_efficient_2018,
emmi_rapid_2021,
fan_static_2021,
fan_efficiently_2018,
Rattanasuksun:RRF:2016,
fang_vfix_2024,
feng_apposcopy_2014,
ferrara_tval_2012,
gerrard_conditional_2022,
gerrard_comprehensive_2017,
eda_efficient_2019,
hu_semantics-based_2021,
hummer_testing_2013,
garcia_automatic_2017,
huster_efficient_2015,
huster_using_2017,
schoeberl_worst-case_2010,
santelices_exploiting_2010,
cousot_refining_1999,
Pan:ICSME:2017,
ponta2020detection,
sabbaghi2019fsct,
chebaro_behind_2014,
saputri_software_2020,
sh_ghandehari_combinatorial_2020,
schordan_combining_2014,
aquino_worst-case_2018,
balakrishnan_modeling_2012,Said:ESE:2020,
srivastava_crystallizer_2023,
su_combining_2015,bai_hybrid_2022,armoni_deterministic_2006,bodden_clara_2012,bodden_finding_2008,sun_tlv_2015,cai_hybrid_2018,banerjee_energypatch_2018,joiner_efficient_2014,jun-xian_pathwalker_2015,yin_fries_2024,huang_discovering_2024,hui_utilization_2017,backes_regression_2013,beyer_cooperative_2021,
shan_self-hiding_2018,tang_software_2017,
beyer:blast:isttt:2007,
ball:slam:fmcad:2010,godefroid:dart:pldi:2005,sen:cute:esec:2005,ghiduk_reducing_2016,gissurarson_propr_2022,godboley_toward_2021,godefroid_proving_2010,braione_combining_2017,belevantsev_multilevel_2017,burnim_testing_2011,busse_combining_2022,camilli_model-based_2020,debroy_combining_2014,fan_history-driven_2024,adamsen_analyzing_2016,azzopardi_technique_2020,ben_henda_opensaw_2017,bertolino_automatic_1994,borzacchiello_fuzzing_2021,bose_columbus_2023,boudhiba_model-based_2015,brennan_symbolic_2018,carbin_automatically_2010,ahmadi_concolic_2019-1,
Tiwari:ISSRE:2023,
lee_hybridroid_2016}.

\subsubsection*{\textbf{\mechanicConstraint}}
Constraint solving is the process of automatically computing satisfying assignments for a set of constraints predicated on 
variables of given types~\cite{de2008z3,dutertre2006yices}. For example, the variables might represent integer numbers, a constraint may predicate logical facts over expressions formed with arithmetic and comparison operators with respect to those variables, and the satisfying assignments (also called solutions) will then be suitable integer values of the variables that make the constraint satisfied. In some cases, the constraint solving process may also be able to conclude that given constraints are unsatisfiable, meaning that there exists no solution. 
In \cpaTechs, constraint solving is classically paired with
symbolic analysis stages (as, for instance, in Example~\ref{sec:example:cnc}, Example~\ref{sec:example:concolic} and Example~\ref{sec:example:cegar}), in order 
to feed partner analysis stages with solutions and satisfiability outcomes of the constraints computed with symbolic analysis~\citePRST{
le_combined_2013,
choi_grey-box_2019,
Wang:TSE:2022,
10.1145/2491411.2491438,artzi_directed_2010,avancini_comparison_2013,baluda_bidirectional_2016,botella_complementary_2019,cadar_execution_2005,
zhang_heuristic_2017,delahaye_infeasible_2015,eslamimehr_efficient_2018,Rasthofer:2017:ICST,aquino_worst-case_2018,ben_henda_opensaw_2017,beyer_explicit-state_2013,dinges_solving_2014,beyer:blast:isttt:2007,ball:slam:fmcad:2010,godefroid:dart:pldi:2005,sen:cute:esec:2005,csallner:check-n-crash:icse:2005,banerjee_energypatch_2018,chebaro_behind_2014,gulavani_synergy_2006,lohar_two-phase_2021}. 

\subsubsection*{\textbf{\mechanicInterpolation}}
A Craig interpolant is a logical formula that captures the reason why two formulas are mutually inconsistent~\cite{craig1957three}. Formally, it means that given two mutually inconsistent formulas $F_1$ and $F_2$ (i.e., $F_1 \wedge F_2$ is unsatisfiable), then an interpolant $I$ is a formula on the variables common to  $F_1$ and $F_2$, with the property that $F_1 \implies I$, meaning that the interpolant is more general than $F_1$, while it maintains that $I \wedge F_2$ is  unsatisfiable.
In \cpaTechs, Craig interpolants have been exploited for refining program abstractions by characterizing infeasible regions of the program execution space~\citePRST{beyer_explicit-state_2013,godboley_toward_2021,beyer:blast:isttt:2007,ball:slam:fmcad:2010}. We discussed the use of interpolants in Example~\ref{sec:example:cegar}. 

\subsubsection*{\textbf{\mechanicInvariants}}
Likely invariants are the results of an inference process over a set of observations of the behavior of a program under analysis~\cite{ernst2001dynamically}.
In a nutshell, the overall process boils down to considering a set of possible symbolically-represented properties over program variables, and then determine the subset of those properties for which it is possible to reject the null-hypothesis that they are contradicted by the available observations. As the observations indicate that those properties may generally hold, the inference process concludes that they likely can be invariants for the program under analysis. In \cpaTechs, the computation of likely invariants is a possible way of post-processing data collected with runtime monitoring, aiming to support symbolic reasoning thereafter~\citePRST{nguyen_using_2014,molina_fuzzing_2022}, or  
steering the exploration of \emph{diverse} state-space regions that contradict the invariants synthesized up to a given point of the analysis~\citePRST{ma_software_2018,dimovski_computing_2020}. 

\subsubsection*{\textbf{\mechanicMining}}
Specification mining is an inference process to automatically generalize a set of observations for modeling the execution protocols of given program entities, most often in the form of finite state machines~\cite{ammons2002mining}. Specification mining can support various tasks in program understanding, testing, debugging and program analysis. 
\CpaTechs may include analysis stages that synthesize the analysis results 
based on specification mining algorithms, to feed partner analysis stages with the resulting execution protocols. In the primary studies, the partner analysis stages aimed to either  mutually complement  mined specifications~\citePRST{dallmeier_generating_2010,Tarvo:ASE:2018,wang_combodroid_2020,Perez:TSE:2019,wang_string_2022,howar_hybrid_2013,bertolino_architectural_2007,bruning_complete_2023} or identify alarms by searching for violations of those specifications~\citePRST{hahnle_constraint-based_2018,lamela_seijas_model_2018,pradel2012statically}. 

\subsubsection*{\textbf{\mechanicMerging}}
State merging consists in joining multiple program states computed through an analysis into a single state representation that generalizes those states, typically aiming to increase the degree of conciseness of the results~\cite{kuznetsov2012efficient}.
In the primary studies, \cpaTechs exploited state merging to  
generalize dynamically observed or statically observed states~\citePRST{grech_shooting_2018,homaei_athena_2019,Obbink:SANER:2018,Rimsa:SPE:2020,Park:JMESECSFSE:2021}.

\subsubsection*{\textbf{\mechanicSummary}}
In program analysis, function summaries indicate intensional models of the essential behavior of some program functions, e.g., in terms of pre-conditions and  post-conditions, which an analyzer can use as helpers for 
relieving the analysis process from the burden of explicitly analyzing some depended functions~\cite{godefroid2007compositional}.
In the primary studies, some analysis stages synthesized their results as functions summaries and fed partner analysis stages with those summaries, typically aiming to increasing efficiency and scalability~\citePRST{belevantsev_multilevel_2017,ivancic_scalable_2015,Tiwari:ISSRE:2023,chen_star_2015,goffi_automatic_2016}.

\subsubsection*{\textbf{\mechanicML}}
Software metrics, possibly further processed with data-mining or machine-learning algorithms,  can be means for numerically summarizing execution data and static characteristics of programs~\cite{chidamber1994metrics,fenton2014software,witten2002data}. In the primary studies, the results of some analysis stages can be delivered as software metrics~\citePRST{marin_pinpointing_2008,zhao_test_2015,krisper_metric_2017,greca_comparing_2022,freitas_scout_2016, poshyvanyk2007feature,ulrich_experience_2016,Yan:SETTA:2021,song_itree_2014,gerrard_conditional_2022}, or indicators derived with data mining and machine-learning (as, for instance, in Example~\ref{sec:example:marvin})~\citePRST{zhao_framework_2022,zhang_detecting_2012,zhang_adaptive_2023,yang_multi-objective_2024,he_python_2023,golagha_can_2020,alshoaibi_price_2019,gergely_differences_2019,lindorfer_marvin_2015,li_fault_2022,khoshgoftaar_investigating_1995,gissurarson_propr_2022,saputri_software_2020}.

 \section{Discussion and Conclusions}
\label{sec:conclusions}

This systematic mapping study examined a large body of literature that spans four decades of research on combined program analysis techniques, motivated by the wide recognition that integrating static and dynamic analyses can overcome the limitations inherent to the standalone approaches.
From an initial pool of 2,776 papers retrieved from the Scopus scientific database, we systematically selected and surveyed \numPrimaryStudies primary studies, thereby documenting the synergistic effects, the workflows and the inter-analysis mapping functions, which underlie the combined-program-analysis techniques reported in those studies. We organized the concepts that emerged during the survey into a comprehensive taxonomy of synergistic effects, inter-analysis workflows, mapping-function interpretation structures and mapping-function mechanics, thus devising a structured lens through which we can describe common motivations and interactions patterns across the techniques, capturing commonalities and differences with respect to the synergies they exploit, and how they exchange and transform the analysis results throughout their workflows.

Our taxonomy and the mapping study contribute novel dimensions to systematize the landscape of combined-program-analysis techniques.
In particular, other than just describing the inter-analysis workflows, we originally classified the techniques in the primary studies along three complementary dimensions: synergistic effects, mapping-function interpretation structures and mapping-function mechanics. 
Out of the surveyed literature, we documented 10 distinct families of synergistic effects (which classify 19 types of synergistic effects in total), capturing both vertical interplays (where an analysis provides results that enhance the capabilities of another analysis) and horizontal interplays (which extend the range of outcomes by combining the outcomes of individual analyses) in the surveyed techniques. We further documented 7 types of mapping-function interpretation structures, i.e., program structures that enabled downstream analysis stages to contextualize their analysis algorithms with respect to the intermediate results produced by upstream analysis stages.  
Finally, we documented 7 types of mapping-function  mechanics (different from the identity-style mechanics), each describing a program-agnostic, mathematically-grounded mechanism for post-processing analysis results, to make those results actionable for  partner techniques or users.

By classifying a large and sparse body of \numPrimaryStudies primary studies according to this taxonomy, our mapping study exposes both the recurring patterns that characterize established combined-analysis paradigms and the rich diversity of interaction schemes that have emerged over time. This structured classification makes it possible to appreciate common design principles across otherwise heterogeneous approaches, while also highlighting the breadth of innovative combinations explored in the literature.

Other than providing a structured picture over the past work on combined program analysis, we believe that this study has the potential to beneficially influence future research on the subject. In particular,  by providing a unified taxonomy and a systematic mapping of existing work, this study can lay the foundation for both designing new combined-program-analysis techniques and steering future research toward more powerful, flexible and scalable analysis frameworks.  On one hand, the identification of the possible synergistic effects and mapping functions can be exploited, possibly also further refined, by researches in both industry and academia to better establish the relation between their problems at hand and the approaches explored in the past, favoring the reuse of known designs while eliciting new inventions. On the other hand, our conceptualization of the inter-analysis interactions in terms of mapping functions can enable the construction of analysis frameworks for automating the composition between different analysis components, where the analysis framework 
standardizes and manages the interpretation structures used by the analysis stages to exchange data with each other (in the same spirit in which SMT-LIB initiative standardized the input and output formats of constraint solvers), and provides APIs and off-the-shelf components to plug mapping-function mechanics and analysis algorithms into \cpaTechs.

\section{Acknowledgments}
This work has been partially supported by the  PNRR project SOP, H73C22000890001 (part of SERICS PE00000014), and by the PRIN 2022 project Big Sistah, 2022EYX28N. We thank Mojtaba Daryabari for his collaboration with the team in the initial phases of the mapping study.

\bibliographystylePRST{abbrv}
\bibliographyPRST{combined_analyses_survey_unimib,papersSLR}
\bibliographystyle{abbrv}
\renewcommand\refname{Further References}
\bibliography{references}

\begin{thebibliography}{100}

\bibitem{adamsen_analyzing_2016}
C.~Adamsen, G.~Mezzetti, and A.~Møller.
\newblock Analyzing test completeness for dynamic languages.
\newblock pages 142--153. Association for Computing Machinery, Inc, 2016.

\bibitem{ahmadi_concolic_2019-1}
R.~Ahmadi and J.~Dingel.
\newblock Concolic testing for models of state-based systems.
\newblock In {\em Proceedings of the 2019 27th {ACM} {Joint} {Meeting} on
  {European} {Software} {Engineering} {Conference} and {Symposium} on the
  {Foundations} of {Software} {Engineering}}, pages 4--15, Tallinn Estonia,
  Aug. 2019. ACM.

\bibitem{alhanahnah_autompi_2023}
M.~Alhanahnah, S.~Ma, A.~Gehani, G.~Ciocarlie, V.~Yegneswaran, S.~Jha, and
  X.~Zhang.
\newblock {autoMPI}: {Automated} {Multiple} {Perspective} {Attack}
  {Investigation} {With} {Semantics} {Aware} {Execution} {Partitioning}.
\newblock {\em IEEE Transactions on Software Engineering}, 49(4):2761--2775,
  2023.
\newblock Publisher: Institute of Electrical and Electronics Engineers Inc.

\bibitem{allwood_high_2011}
T.~Allwood, C.~Cadar, and S.~Eisenbach.
\newblock High coverage testing of {Haskell} programs.
\newblock pages 375--385, 2011.

\bibitem{alshoaibi_price_2019}
D.~Alshoaibi, K.~Hannigan, H.~Gupta, and M.~Mkaouer.
\newblock {PRICE}: {Detection} of {Performance} {Regression} {Introducing}
  {Code} {Changes} {Using} {Static} and {Dynamic} {Metrics}.
\newblock volume 11664 LNCS, pages 75--88. Springer Verlag, 2019.

\bibitem{aquino_worst-case_2018}
A.~Aquino, G.~Denaro, and P.~Salza.
\newblock Worst-{Case} {Execution} {Time} {Testing} via {Evolutionary}
  {Symbolic} {Execution}.
\newblock volume 2018-October, pages 76--87. IEEE Computer Society, 2018.

\bibitem{armoni_deterministic_2006}
R.~Armoni, D.~Korchemny, A.~Tiemeyer, M.~Vardi, and Y.~Zbar.
\newblock Deterministic dynamic monitors for linear-time assertions.
\newblock volume 4262 LNCS, pages 163--177. Springer Verlag, 2006.

\bibitem{artzi_directed_2010}
S.~Artzi, J.~Dolby, F.~Tip, and M.~Pistoia.
\newblock Directed test generation for effective fault localization.
\newblock pages 49--59, 2010.

\bibitem{asensio_worst-case_2013}
E.~Asensio, I.~Lafoz, A.~Coombes, and J.~Navas.
\newblock Worst-case execution time analysis approach for safety-critical
  airborne software.
\newblock volume 7896 LNCS, pages 161--176, 2013.

\bibitem{avancini_comparison_2013}
A.~Avancini and M.~Ceccato.
\newblock Comparison and integration of genetic algorithms and dynamic symbolic
  execution for security testing of cross-site scripting vulnerabilities.
\newblock {\em Information and Software Technology}, 55(12):2209--2222, 2013.
\newblock Publisher: Elsevier B.V.

\bibitem{azzopardi_technique_2020}
S.~Azzopardi, C.~Colombo, and G.~Pace.
\newblock A {Technique} for {Automata}-based {Verification} with {Residual}
  {Reasoning}.
\newblock pages 237--248. Science and Technology Publications, Lda, 2020.

\bibitem{backes_regression_2013}
J.~Backes, S.~Person, N.~Rungta, and O.~Tkachuk.
\newblock Regression verification using impact summaries.
\newblock volume 7976 LNCS, pages 99--116. Springer Verlag, 2013.

\bibitem{bai_hybrid_2022}
J.-J. Bai, Q.-L. Chen, Z.-M. Jiang, J.~Lawall, and S.-M. Hu.
\newblock Hybrid {Static}-{Dynamic} {Analysis} of {Data} {Races} {Caused} by
  {Inconsistent} {Locking} {Discipline} in {Device} {Drivers}.
\newblock {\em IEEE Transactions on Software Engineering}, 48(12):5120--5135,
  2022.
\newblock Publisher: Institute of Electrical and Electronics Engineers Inc.

\bibitem{balakrishnan_modeling_2012}
G.~Balakrishnan, N.~Maeda, S.~Sankaranarayanan, F.~Ivančić, A.~Gupta, and
  R.~Pothengil.
\newblock Modeling and analyzing the interaction of {C} and {C}++ strings.
\newblock volume 7421 LNCS, pages 67--85, 2012.

\bibitem{ball:slam:fmcad:2010}
T.~Ball, E.~Bounimova, R.~Kumar, and V.~Levin.
\newblock Slam2: Static driver verification with under 4\% false alarms.
\newblock In {\em Formal Methods in Computer Aided Design}, pages 35--42. IEEE,
  2010.

\bibitem{baluda_bidirectional_2016}
M.~Baluda, G.~Denaro, and M.~Pezzè.
\newblock Bidirectional {Symbolic} {Analysis} for {Effective} {Branch}
  {Testing}.
\newblock {\em IEEE Transactions on Software Engineering}, 42(5):403--426,
  2016.
\newblock Publisher: Institute of Electrical and Electronics Engineers Inc.

\bibitem{banerjee_energypatch_2018}
A.~Banerjee, L.~Chong, C.~Ballabriga, and A.~Roychoudhury.
\newblock {EnergyPatch}: {Repairing} {Resource} {Leaks} to {Improve}
  {Energy}-{Efficiency} of {Android} {Apps}.
\newblock {\em IEEE Transactions on Software Engineering}, 44(5):470--490,
  2018.
\newblock Publisher: Institute of Electrical and Electronics Engineers Inc.

\bibitem{belevantsev_multilevel_2017}
A.~Belevantsev.
\newblock Multilevel static analysis for improving program quality.
\newblock {\em Programming and Computer Software}, 43(6):321--336, 2017.
\newblock Publisher: Maik Nauka Publishing / Springer SBM.

\bibitem{ben_henda_opensaw_2017}
N.~Ben~Henda, B.~Johansson, P.~Lantz, K.~Norrman, P.~Saarinen, and
  O.~Segersvärd.
\newblock {OpenSAW}: {Open} security analysis workbench.
\newblock volume 10202 LNCS, pages 321--337. Springer Verlag, 2017.

\bibitem{bertolino_automatic_1994}
A.~Bertolino and M.~Marré.
\newblock Automatic {Generation} of {Path} {Covers} {Based} on the {Control}
  {Flow} {Analysis} of {Computer} {Programs}.
\newblock {\em IEEE Transactions on Software Engineering}, 20(12):885--899,
  1994.

\bibitem{bertolino_architectural_2007}
A.~Bertolino, H.~Muccini, and A.~Polini.
\newblock Architectural verification of black-box component-based systems.
\newblock volume 4401 LNCS, pages 98--113. Springer Verlag, 2007.

\bibitem{beyer:blast:isttt:2007}
D.~Beyer, T.~A. Henzinger, R.~Jhala, and R.~Majumdar.
\newblock The software model checker {B}last: Applications to software
  engineering.
\newblock {\em International Journal on Software Tools for Technology
  Transfer}, 9:505--525, 2007.

\bibitem{beyer_cooperative_2021}
D.~Beyer and M.-C. Jakobs.
\newblock Cooperative verifier-based testing with {CoVeriTest}.
\newblock {\em International Journal on Software Tools for Technology
  Transfer}, 23(3):313--333, 2021.
\newblock Publisher: Springer Science and Business Media Deutschland GmbH.

\bibitem{beyer_explicit-state_2013}
D.~Beyer and S.~Löwe.
\newblock Explicit-state software model checking based on {CEGAR} and
  interpolation.
\newblock volume 7793 LNCS, pages 146--162, 2013.

\bibitem{bodden_clara_2012}
E.~Bodden and L.~Hendren.
\newblock The {Clara} framework for hybrid typestate analysis.
\newblock {\em International Journal on Software Tools for Technology
  Transfer}, 14(3):307--326, 2012.

\bibitem{bodden_finding_2008}
E.~Bodden, P.~Lam, and L.~Hendren.
\newblock Finding programming errors earlier by evaluating runtime monitors
  ahead-of-time.
\newblock pages 36--47, 2008.

\bibitem{borzacchiello_fuzzing_2021}
L.~Borzacchiello, E.~Coppa, and C.~Demetrescu.
\newblock Fuzzing symbolic expressions.
\newblock pages 711--722. IEEE Computer Society, 2021.

\bibitem{bose_columbus_2023}
P.~Bose, D.~Das, S.~Vasan, S.~Mariani, I.~Grishchenko, A.~Continella,
  A.~Bianchi, C.~Kruegel, and G.~Vigna.
\newblock Columbus: {Android} {App} {Testing} {Through} {Systematic} {Callback}
  {Exploration}.
\newblock pages 1381--1392. IEEE Computer Society, 2023.

\bibitem{botella_complementary_2019}
J.~Botella, J.-F. Capuron, F.~Dadeau, E.~Fourneret, B.~Legeard, and F.~Schadle.
\newblock Complementary test selection criteria for model-based testing of
  security components.
\newblock {\em International Journal on Software Tools for Technology
  Transfer}, 21(4):425--448, 2019.
\newblock Publisher: Springer Verlag.

\bibitem{boudhiba_model-based_2015}
I.~Boudhiba, C.~Gaston, P.~Gall, and V.~Prevosto.
\newblock Model-based testing from input output symbolic transition systems
  enriched by program calls and contracts.
\newblock volume 9447, pages 35--51. Springer Verlag, 2015.

\bibitem{braione_combining_2017}
P.~Braione, G.~Denaro, A.~Mattavelli, and M.~Pezze.
\newblock Combining symbolic execution and search-based testing for programs
  with complex heap inputs.
\newblock pages 90--101. Association for Computing Machinery, Inc, 2017.

\bibitem{brennan_symbolic_2018}
T.~Brennan, S.~Saha, T.~Bultan, and C.~Pǎsǎreanu.
\newblock Symbolic path cost analysis for side-channel detection.
\newblock pages 27--37. Association for Computing Machinery, Inc, 2018.

\bibitem{bruning_complete_2023}
F.~Brüning, M.~Gleirscher, W.-L. Huang, N.~Krafczyk, J.~Peleska, and
  R.~Sachtleben.
\newblock Complete {Property}-{Oriented} {Module} {Testing}.
\newblock volume 14131 LNCS, pages 183--201. Springer Science and Business
  Media Deutschland GmbH, 2023.

\bibitem{burnim_testing_2011}
J.~Burnim, K.~Sen, and C.~Stergiou.
\newblock Testing concurrent programs on relaxed memory models.
\newblock pages 122--132, 2011.

\bibitem{busse_combining_2022}
F.~Busse, P.~Gharat, C.~Cadar, and A.~Donaldson.
\newblock Combining static analysis error traces with dynamic symbolic
  execution (experience paper).
\newblock pages 568--579. Association for Computing Machinery, Inc, 2022.

\bibitem{cadar_execution_2005}
C.~Cadar and D.~Engler.
\newblock Execution generated test cases: {How} to make systems code crash
  itself.
\newblock volume 3639, pages 2--23. Springer Verlag, 2005.

\bibitem{cai_hybrid_2018}
H.~Cai.
\newblock Hybrid {Program} {Dependence} {Approximation} for {Effective}
  {Dynamic} {Impact} {Prediction}.
\newblock {\em IEEE Transactions on Software Engineering}, 44(4):334--364,
  2018.
\newblock Publisher: Institute of Electrical and Electronics Engineers Inc.

\bibitem{camilli_model-based_2020}
M.~Camilli and B.~Russo.
\newblock Model-{Based} {Testing} {Under} {Parametric} {Variability} of
  {Uncertain} {Beliefs}.
\newblock volume 12310 LNCS, pages 175--192. Springer Science and Business
  Media Deutschland GmbH, 2020.

\bibitem{carbin_automatically_2010}
M.~Carbin and M.~Rinard.
\newblock Automatically identifying critical input regions and code in
  applications.
\newblock pages 37--47, 2010.

\bibitem{chebaro_behind_2014}
O.~Chebaro, P.~Cuoq, N.~Kosmatov, B.~Marre, A.~Pacalet, N.~Williams, and
  B.~Yakobowski.
\newblock Behind the scenes in {SANTE}: {A} combination of static and dynamic
  analyses.
\newblock {\em Automated Software Engineering}, 21(1):107--143, 2014.
\newblock Publisher: Kluwer Academic Publishers.

\bibitem{chen_synthesising_2016}
H.-Y. Chen, C.~David, D.~Kroening, P.~Schrammel, and B.~Wachter.
\newblock Synthesising interprocedural bit-precise termination proofs.
\newblock pages 53--64. Institute of Electrical and Electronics Engineers Inc.,
  2016.

\bibitem{chen_star_2015}
N.~Chen and S.~Kim.
\newblock {STAR}: {Stack} trace based automatic crash reproduction via symbolic
  execution.
\newblock {\em IEEE Transactions on Software Engineering}, 41(2):198--220,
  2015.
\newblock Publisher: Institute of Electrical and Electronics Engineers Inc.

\bibitem{chen_have_2009}
Q.~Chen, L.~Wang, Z.~Yang, and S.~Stoller.
\newblock Have: {Detecting} atomicity violations via integrated dynamic and
  static analysis.
\newblock volume 5503, pages 425--439, 2009.

\bibitem{chen_automatically_2023}
S.~Chen, L.~Fan, C.~Chen, and Y.~Liu.
\newblock Automatically {Distilling} {Storyboard} {With} {Rich} {Features} for
  {Android} {Apps}.
\newblock {\em IEEE Transactions on Software Engineering}, 49(2):667--683,
  2023.
\newblock Publisher: Institute of Electrical and Electronics Engineers Inc.

\bibitem{chen_boosting_2021}
T.~Chen, K.~Heo, and M.~Raghothaman.
\newblock Boosting static analysis accuracy with instrumented test executions.
\newblock pages 1154--1165. Association for Computing Machinery, Inc, 2021.

\bibitem{chen_improving_2018}
Z.~Chen, H.-F. Guo, and M.~Song.
\newblock Improving regression test efficiency with an awareness of refactoring
  changes.
\newblock {\em Information and Software Technology}, 103:174--187, 2018.
\newblock Publisher: Elsevier B.V.

\bibitem{choi_grey-box_2019}
J.~Choi, J.~Jang, C.~Han, and S.~Cha.
\newblock Grey-{Box} {Concolic} {Testing} on {Binary} {Code}.
\newblock volume 2019-May, pages 736--747. IEEE Computer Society, 2019.

\bibitem{cousot_refining_1999}
P.~Cousot and R.~Cousot.
\newblock Refining {Model} {Checking} by {Abstract} {Interpretation}.
\newblock {\em Automated Software Engineering}, 6(1):69--95, 1999.
\newblock Publisher: Springer Netherlands.

\bibitem{csallner:check-n-crash:icse:2005}
C.~Csallner and Y.~Smaragdakis.
\newblock Check'n'crash: combining static checking and testing.
\newblock In {\em Proceedings of the 27th International Conference on Software
  Engineering (ICSE 2005)}, pages 422--431, 2005.

\bibitem{czech_just_2015}
M.~Czech, M.-C. Jakobs, and H.~Wehrheim.
\newblock Just test what you cannot verify!
\newblock volume 9033, pages 100--114. Springer Verlag, 2015.

\bibitem{dallmeier_generating_2010}
V.~Dallmeier, N.~Knopp, C.~Mallon, S.~Hack, and A.~Zeller.
\newblock Generating test cases for specification mining.
\newblock In {\em Proceedings of the 19th International Symposium on Software
  Testing and Analysis}, pages 85--95, 2010.

\bibitem{debroy_combining_2014}
V.~Debroy and W.~Wong.
\newblock Combining mutation and fault localization for automated program
  debugging.
\newblock {\em Journal of Systems and Software}, 90(1):45--60, 2014.

\bibitem{delahaye_infeasible_2015}
M.~Delahaye, B.~Botella, and A.~Gotlieb.
\newblock Infeasible path generalization in dynamic symbolic execution.
\newblock {\em Information and Software Technology}, 58:403--418, 2015.
\newblock Publisher: Elsevier B.V.

\bibitem{dhar_clotho_2015}
A.~Dhar, R.~Purandare, M.~Dhawan, and S.~Rangaswamy.
\newblock {CLOTHO}: {Saving} programs from malformed strings and incorrect
  string-handling.
\newblock pages 555--566. Association for Computing Machinery, Inc, 2015.

\bibitem{dimovski_quantitative_2022}
A.~Dimovski.
\newblock Quantitative {Program} {Sketching} using {Lifted} {Static}
  {Analysis}.
\newblock volume 13241 LNCS, pages 102--122. Springer Science and Business
  Media Deutschland GmbH, 2022.

\bibitem{dimovski_computing_2020}
A.~Dimovski and A.~Legay.
\newblock Computing program reliability using forward-backward precondition
  analysis and model counting.
\newblock volume 12076 LNCS, pages 182--202. Springer, 2020.

\bibitem{diner_generating_2021}
D.~Diner, G.~Fraser, S.~Schweikl, and A.~Stahlbauer.
\newblock Generating {Timed} {UI} {Tests} from {Counterexamples}.
\newblock volume 12740 LNCS, pages 53--71. Springer Science and Business Media
  Deutschland GmbH, 2021.

\bibitem{dinges_solving_2014}
P.~Dinges and G.~Agha.
\newblock Solving complex path conditions through heuristic search on induced
  polytopes.
\newblock volume 16-21-November-2014, pages 425--436. Association for Computing
  Machinery, 2014.

\bibitem{do_goal-oriented_2015}
T.~Do, S.-C. Khoo, A.~Fong, R.~Pears, and T.~Quan.
\newblock Goal-oriented dynamic test generation.
\newblock {\em Information and Software Technology}, 66:40--57, 2015.
\newblock Publisher: Elsevier B.V.

\bibitem{dufour_blended_2007}
B.~Dufour, B.~Ryder, and G.~Sevitsky.
\newblock Blended analysis for performance understanding of framework-based
  applications.
\newblock {\em 2007 ACM International Symposium on Software Testing and
  Analysis, ISSTA'07}, pages 118--128, 2007.

\bibitem{eda_efficient_2019}
R.~Eda and H.~Do.
\newblock An efficient regression testing approach for {PHP} {Web} applications
  using test selection and reusable constraints.
\newblock {\em Software Quality Journal}, 27(4):1383--1417, 2019.
\newblock Publisher: Springer.

\bibitem{el-serafy_automatic_2015}
A.~El-Serafy, C.~Salama, and A.~Wahba.
\newblock Automatic test data generation targeting hybrid coverage criteria.
\newblock volume 532, pages 149--160. Springer Verlag, 2015.

\bibitem{elyasov_search-based_2018}
A.~Elyasov, I.~Prasetya, and J.~Hage.
\newblock Search-{Based} {Test} {Data} {Generation} for {JavaScript}
  {Functions} that {Interact} with the {DOM}.
\newblock volume 2018-October, pages 88--99. IEEE Computer Society, 2018.

\bibitem{emmi_rapid_2021}
M.~Emmi, L.~Hadarean, R.~Jhala, L.~Pike, N.~Rosner, M.~Schäf, A.~Sengupta, and
  W.~Visser.
\newblock {RAPID}: {Checking} {API} usage for the cloud in the cloud.
\newblock pages 1416--1426. Association for Computing Machinery, Inc, 2021.

\bibitem{eslamimehr_efficient_2018}
M.~Eslamimehr, M.~Lesani, and G.~Edwards.
\newblock Efficient detection and validation of atomicity violations in
  concurrent programs.
\newblock {\em Journal of Systems and Software}, 137:618--635, 2018.
\newblock Publisher: Elsevier Inc.

\bibitem{fan_static_2021}
G.~Fan, T.~Chen, B.~Yin, L.~Chen, T.~Wang, and J.~Wang.
\newblock Static {Bound} {Analysis} of {Dynamically} {Allocated} {Resources}
  for {C} {Programs}.
\newblock volume 2021-October, pages 390--400. IEEE Computer Society, 2021.

\bibitem{fan_efficiently_2018}
L.~Fan, G.~Meng, T.~Su, Y.~Liu, G.~Pu, S.~Chen, and L.~Xu.
\newblock Efficiently manifesting asynchronous programming errors in android
  apps.
\newblock pages 486--497. Association for Computing Machinery, Inc, 2018.

\bibitem{fan_history-driven_2024}
Z.~Fan, G.~Ye, T.~Hu, and Z.~Tang.
\newblock History-driven {Compiler} {Fuzzing} via {Assembling} and {Scheduling}
  {Bug}-{Triggering} {Code} {Segments}.
\newblock pages 331--342. IEEE Computer Society, 2024.

\bibitem{fang_vfix_2024}
P.~Fang, P.~Gao, Y.~Peng, Q.~Zhang, T.~Xie, D.~Song, P.~Mittal, S.~Kulkarni,
  Z.~Liu, and X.~Xiao.
\newblock {VFIX}: {Facilitating} {Software} {Maintenance} of {Smart}
  {Contracts} via {Automatically} {Fixing} {Vulnerabilities}.
\newblock pages 13--24. Institute of Electrical and Electronics Engineers Inc.,
  2024.

\bibitem{fazzini_enhancing_2023}
M.~Fazzini, K.~Moran, C.~Bernal-Cardenas, T.~Wendland, A.~Orso, and
  D.~Poshyvanyk.
\newblock Enhancing {Mobile} {App} {Bug} {Reporting} via {Real}-{Time}
  {Understanding} of {Reproduction} {Steps}.
\newblock {\em IEEE Transactions on Software Engineering}, 49(3):1246--1272,
  2023.
\newblock Publisher: Institute of Electrical and Electronics Engineers Inc.

\bibitem{feng_apposcopy_2014}
Y.~Feng, S.~Anand, I.~Dillig, and A.~Aiken.
\newblock Apposcopy: {Semantics}-based detection of android malware through
  static analysis.
\newblock volume 16-21-November-2014, pages 576--587. Association for Computing
  Machinery, 2014.

\bibitem{ferrara_tval_2012}
P.~Ferrara, R.~Fuchs, and U.~Juhasz.
\newblock {TVAL}+ : {TTTVLA} and value analyses together.
\newblock volume 7504 LNCS, pages 63--77, 2012.

\bibitem{freitas_scout_2016}
E.~Freitas, C.~Camilo-Junior, and A.~Vincenzi.
\newblock {SCOUT}: {A} {Multi}-objective {Method} to {Select} {Components} in
  {Designing} {Unit} {Testing}.
\newblock pages 36--46. IEEE Computer Society, 2016.

\bibitem{garcia_automatic_2017}
J.~Garcia, M.~Hammad, N.~Ghorbani, and S.~Malek.
\newblock Automatic generation of inter-component communication exploits for
  android applications.
\newblock volume Part F130154, pages 661--671. Association for Computing
  Machinery, 2017.

\bibitem{gergely_differences_2019}
T.~Gergely, G.~Balogh, F.~Horvath, B.~Vancsics, A.~Beszedes, and T.~Gyimothy.
\newblock Differences between a static and a dynamic test-to-code traceability
  recovery method.
\newblock {\em Software Quality Journal}, 27(2):797--822, 2019.
\newblock Publisher: Springer New York LLC.

\bibitem{gerrard_conditional_2022}
M.~Gerrard, M.~Borges, M.~Dwyer, and A.~Filieri.
\newblock Conditional {Quantitative} {Program} {Analysis}.
\newblock {\em IEEE Transactions on Software Engineering}, 48(4):1212--1227,
  2022.
\newblock Publisher: Institute of Electrical and Electronics Engineers Inc.

\bibitem{gerrard_comprehensive_2017}
M.~Gerrard and M.~Dwyer.
\newblock Comprehensive failure characterization.
\newblock pages 365--376. Institute of Electrical and Electronics Engineers
  Inc., 2017.

\bibitem{ghiduk_reducing_2016}
A.~Ghiduk.
\newblock Reducing the number of higher-order mutants with the aid of data
  flow.
\newblock {\em E-Informatica Software Engineering Journal}, 10(1):31--49, 2016.
\newblock Publisher: Wroclaw University of Science and Technology.

\bibitem{gissurarson_propr_2022}
M.~Gissurarson, L.~Applis, A.~Panichella, A.~Van~Deursen, and D.~Sands.
\newblock {PROPR}: {Property}-{Based} {Automatic} {Program} {Repair}.
\newblock volume 2022-May, pages 1768--1780. IEEE Computer Society, 2022.

\bibitem{godboley_toward_2021}
S.~Godboley, J.~Jaffar, R.~Maghareh, and A.~Dutta.
\newblock Toward optimal mc/dc test case generation.
\newblock pages 505--516. Association for Computing Machinery, Inc, 2021.

\bibitem{godefroid_proving_2010}
P.~Godefroid and J.~Kinder.
\newblock Proving memory safety of floating-point computations by combining
  static and dynamic program analysis.
\newblock pages 1--11, 2010.

\bibitem{godefroid:dart:pldi:2005}
P.~Godefroid, N.~Klarlund, and K.~Sen.
\newblock {DART}: directed automated random testing.
\newblock In {\em Proceedings of the ACM SIGPLAN 2005 Conference on Programming
  Language Design and Implementation (PLDI 2005)}, pages 213--223, 2005.

\bibitem{goffi_automatic_2016}
A.~Goffi, A.~Gorla, M.~Ernst, and M.~Pezzè.
\newblock Automatic generation of oracles for exceptional behaviors.
\newblock pages 213--224. Association for Computing Machinery, Inc, 2016.

\bibitem{golagha_can_2020}
M.~Golagha, A.~Pretschner, and L.~Briand.
\newblock Can {We} {Predict} the {Quality} of {Spectrum}-based {Fault}
  {Localization}?
\newblock pages 4--15. Institute of Electrical and Electronics Engineers Inc.,
  2020.

\bibitem{greca_comparing_2022}
R.~Greca, B.~Miranda, M.~Gligoric, and A.~Bertolino.
\newblock Comparing and {Combining} {File}-based {Selection} and
  {Similarity}-based {Prioritization} towards {Regression} {Test}
  {Orchestration}.
\newblock pages 115--125. Institute of Electrical and Electronics Engineers
  Inc., 2022.

\bibitem{grech_shooting_2018}
N.~Grech, G.~Fourtounis, A.~Francalanza, and Y.~Smaragdakis.
\newblock Shooting from the heap: {Ultra}-scalable static analysis with heap
  snapshots.
\newblock pages 198--208. Association for Computing Machinery, Inc, 2018.

\bibitem{grechanik_preventing_2013}
M.~Grechanik, B.~Mainul~Hossain, U.~Buy, and H.~Wang.
\newblock Preventing database deadlocks in applications.
\newblock pages 356--366. Association for Computing Machinery, 2013.

\bibitem{gulavani_synergy_2006}
B.~Gulavani, T.~Henzinger, Y.~Kannan, A.~Nori, and S.~Rajamani.
\newblock {SYNERGY}: {A} new algorithm for property checking.
\newblock pages 117--127, 2006.

\bibitem{guo_predracer_2024}
X.~Guo, X.~Qi, Y.~Li, and C.~Wu.
\newblock {PredRacer}: {Predictively} {Detecting} {Data} {Races} in {Android}
  {Applications}.
\newblock pages 239--249. Institute of Electrical and Electronics Engineers
  Inc., 2024.

\bibitem{gupta_hybrid_1997}
R.~Gupta, M.~Soffa, and J.~Howard.
\newblock Hybrid {Slicing}: {Integrating} {Dynamic} {Information} with {Static}
  {Analysis}.
\newblock {\em ACM Transactions on Software Engineering and Methodology},
  6(4):370--397, 1997.
\newblock Publisher: Association for Computing Machinery (ACM).

\bibitem{harman_strong_2011}
M.~Harman, Y.~Jia, and W.~Langdon.
\newblock Strong higher order mutation-based test data generation.
\newblock pages 212--222, 2011.

\bibitem{he_python_2023}
X.~He, X.~Liu, and L.~Xu.
\newblock Python {API} {Misuse} {Mining} and {Classification} {Based} on
  {Hybrid} {Analysis} and {Attention} {Mechanism}.
\newblock {\em International Journal of Software Engineering and Knowledge
  Engineering}, 33(10):1567--1597, 2023.
\newblock Publisher: World Scientific.

\bibitem{holling_profiting_2016}
D.~Holling, A.~Hofbauer, A.~Pretschner, and M.~Gemmar.
\newblock Profiting from {Unit} {Tests} for {Integration} {Testing}.
\newblock pages 353--363. Institute of Electrical and Electronics Engineers
  Inc., 2016.

\bibitem{homaei_athena_2019}
H.~Homaei and H.~Shahriari.
\newblock Athena: {A} framework to automatically generate security test oracle
  via extracting policies from source code and intended software behaviour.
\newblock {\em Information and Software Technology}, 107:112--124, 2019.
\newblock Publisher: Elsevier B.V.

\bibitem{hough_revealing_2020}
K.~Hough, G.~Welearegai, C.~Hammer, and J.~Bell.
\newblock Revealing injection vulnerabilities by leveraging existing tests.
\newblock pages 284--296. IEEE Computer Society, 2020.

\bibitem{howar_hybrid_2013}
F.~Howar, D.~Giannakopoulou, and Z.~Rakamarić.
\newblock Hybrid learning: {Interface} generation through static, dynamic, and
  symbolic analysis.
\newblock pages 268--279, 2013.

\bibitem{hsu_highly_2023}
M.-Y. Hsu, F.~Hetzelt, D.~Gens, M.~Maitland, and M.~Franz.
\newblock A {Highly} {Scalable}, {Hybrid}, {Cross}-{Platform} {Timing}
  {Analysis} {Framework} {Providing} {Accurate} {Differential} {Throughput}
  {Estimation} via {Instruction}-{Level} {Tracing}.
\newblock pages 821--831. Association for Computing Machinery, Inc, 2023.

\bibitem{hu_semantics-based_2021}
Y.~Hu, H.~Wang, Y.~Zhang, B.~Li, and D.~Gu.
\newblock A {Semantics}-{Based} {Hybrid} {Approach} on {Binary} {Code}
  {Similarity} {Comparison}.
\newblock {\em IEEE Transactions on Software Engineering}, 47(6):1241--1258,
  2021.
\newblock Publisher: Institute of Electrical and Electronics Engineers Inc.

\bibitem{hu_achyb_2021}
Y.~Hu, W.~Wang, C.~Hunger, R.~Wood, S.~Khurshid, and M.~Tiwari.
\newblock {ACHyb}: {A} hybrid analysis approach to detect kernel access control
  vulnerabilities.
\newblock pages 316--327. Association for Computing Machinery, Inc, 2021.

\bibitem{huang_discovering_2024}
Z.~Huang, S.~Ravi, and C.~Wang.
\newblock Discovering {Likely} {Program} {Invariants} for {Persistent}
  {Memory}.
\newblock pages 1795--1807. Association for Computing Machinery, Inc, 2024.

\bibitem{hui_utilization_2017}
Z.~Hui.
\newblock Utilization of {Dependence} and {Weight} to {Improve} {Fault}
  {Localization} {Method} of {Regression} {Test} {Cases}.
\newblock {\em International Journal of Software Engineering and Knowledge
  Engineering}, 27(3):423--447, 2017.
\newblock Publisher: World Scientific Publishing Co. Pte Ltd.

\bibitem{hummer_testing_2013}
W.~Hummer, O.~Raz, O.~Shehory, P.~Leitner, and S.~Dustdar.
\newblock Testing of data-centric and event-based dynamic service compositions.
\newblock volume~23, pages 465--497, 2013.
\newblock Issue: 6.

\bibitem{huster_efficient_2015}
S.~Huster, S.~Burg, H.~Eichelberger, J.~Laufenberg, J.~Ruf, T.~Kropf, and
  W.~Rosenstiel.
\newblock Efficient testing of different loop paths.
\newblock volume 9276, pages 117--131. Springer Verlag, 2015.

\bibitem{huster_using_2017}
S.~Huster, J.~Ströbele, J.~Ruf, T.~Kropf, and W.~Rosenstiel.
\newblock Using robustness testing to handle incomplete verification results
  when combining verification and testing techniques.
\newblock volume 10533 LNCS, pages 54--70. Springer Verlag, 2017.

\bibitem{hahnle_constraint-based_2018}
R.~Hähnle and B.~Steffen.
\newblock Constraint-based behavioral consistency of evolving software systems.
\newblock volume 11026 LNCS, pages 205--218. Springer Verlag, 2018.

\bibitem{ivancic_scalable_2015}
F.~Ivančić, G.~Balakrishnan, A.~Gupta, S.~Sankaranarayanan, N.~Maeda,
  T.~Imoto, R.~Pothengil, and M.~Hussain.
\newblock Scalable and scope-bounded software verification in {Varvel}.
\newblock {\em Automated Software Engineering}, 22(4):517--559, 2015.
\newblock Publisher: Kluwer Academic Publishers.

\bibitem{jahangirova_test_2016}
G.~Jahangirova, D.~Clark, M.~Harman, and P.~Tonella.
\newblock Test oracle assessment and improvement.
\newblock pages 247--258. Association for Computing Machinery, Inc, 2016.

\bibitem{jannesari_generating_2014}
A.~Jannesari, N.~Koprowski, J.~Schimmel, and F.~Wolf.
\newblock Generating classified parallel unit tests.
\newblock volume 8570 LNCS, pages 117--133. Springer Verlag, 2014.

\bibitem{jasper_test_1994}
R.~Jasper, M.~Brennan, K.~Williamson, B.~Currier, and D.~Zimmerman.
\newblock Test data generation and feasible path analysis.
\newblock pages 95--107. Association for Computing Machinery, Inc, 1994.

\bibitem{jeng_automatic_1999}
B.~Jeng and I.~Forgács.
\newblock An automatic approach of domain test data generation.
\newblock {\em Journal of Systems and Software}, 49(1):97--112, 1999.
\newblock Publisher: Elsevier Inc.

\bibitem{jin_f3_2013}
W.~Jin and A.~Orso.
\newblock F3: {Fault} localization for field failures.
\newblock pages 213--223, 2013.

\bibitem{joiner_efficient_2014}
R.~Joiner, T.~Reps, S.~Jha, M.~Dhawan, and V.~Ganapathy.
\newblock Efficient runtime-enforcement techniques for {Policy} weaving.
\newblock volume 16-21-November-2014, pages 224--234. Association for Computing
  Machinery, 2014.

\bibitem{jones_addressing_2016}
A.~Jones.
\newblock Addressing the regression test problem with change impact analysis
  for {Ada}.
\newblock volume 9695, pages 61--77. Springer Verlag, 2016.

\bibitem{jun-xian_pathwalker_2015}
Z.~Jun-Xian, L.~Zhou-Jun, and Z.~Xian-Chen.
\newblock {PathWalker}: {A} dynamic symbolic execution tool based on {LLVM}
  byte code instrumentation.
\newblock volume 9409, pages 227--242. Springer Verlag, 2015.

\bibitem{kama_change_2015}
N.~Kama, S.~Ismail, K.~Kamardin, N.~Zainuddin, A.~Azmi, and W.~Zainuddin.
\newblock A change impact analysis tool: {Integration} between static and
  dynamic analysis techniques.
\newblock volume 532, pages 413--424. Springer Verlag, 2015.

\bibitem{kechagia_effective_2019}
M.~Kechagia, X.~Devroey, A.~Panichella, G.~Gousios, and A.~Van~Deursen.
\newblock Effective and efficient {API} misuse detection via exception
  propagation and search-based testing.
\newblock pages 192--203. Association for Computing Machinery, Inc, 2019.

\bibitem{kellogg_lightweight_2021}
M.~Kellogg, N.~Shadab, M.~Sridharan, and M.~Ernst.
\newblock Lightweight and modular resource leak verification.
\newblock pages 181--192. Association for Computing Machinery, Inc, 2021.

\bibitem{khoshgoftaar_investigating_1995}
T.~Khoshgoftaar and R.~Szabo.
\newblock Investigating {ARIMA} models of software system quality.
\newblock {\em Software Quality Journal}, 4(1):33--48, 1995.
\newblock Publisher: Kluwer Academic Publishers.

\bibitem{kim_first_2024}
D.~Kim, T.-H. Chen, and J.~Yang.
\newblock A {First} {Look} at the {Inheritance}-{Induced} {Redundant} {Test}
  {Execution}.
\newblock pages 1397--1408. IEEE Computer Society, 2024.

\bibitem{kim_combining_2013}
S.-W. Kim, Y.-S. Ma, and Y.-R. Kwon.
\newblock Combining weak and strong mutation for a noninterpretive {Java}
  mutation system.
\newblock {\em Software Testing Verification and Reliability}, 23(8):647--668,
  2013.

\bibitem{kim_precise_2018}
Y.~Kim, Y.~Choi, and M.~Kim.
\newblock Precise {Concolic} {Unit} {Testing} of {C} {Programs} using
  {Extended} {Units} and {Symbolic} {Alarm} {Filtering}.
\newblock volume 2018-January, pages 315--326. IEEE Computer Society, 2018.

\bibitem{krisper_metric_2017}
M.~Krisper, J.~Iber, C.~Kreiner, and M.~Quaritsch.
\newblock A metric for evaluating residual complexity in software.
\newblock volume 748, pages 138--149. Springer Verlag, 2017.

\bibitem{kwon_static_2012}
T.~Kwon and Z.~Su.
\newblock Static detection of unsafe component loadings.
\newblock volume 7210 LNCS, pages 122--143, 2012.

\bibitem{kahkonen_lightweight_2014}
K.~Kähkönen and K.~Heljanko.
\newblock Lightweight state capturing for automated testing of multithreaded
  programs.
\newblock volume 8570 LNCS, pages 187--203. Springer Verlag, 2014.

\bibitem{lameed_staged_2011}
N.~Lameed and L.~Hendren.
\newblock Staged static techniques to efficiently implement array copy
  semantics in a {MATLAB} {JIT} compiler.
\newblock volume 6601 LNCS, pages 22--41, 2011.

\bibitem{lamela_seijas_model_2018}
P.~Lamela~Seijas, S.~Thompson, and M.~Francisco.
\newblock Model extraction and test generation from {JUnit} test suites.
\newblock {\em Software Quality Journal}, 26(4):1519--1552, 2018.
\newblock Publisher: Springer New York LLC.

\bibitem{le_combined_2013}
A.~Le, T.~Quan, N.~Huynh, P.~Nguyen, and N.-V. Le.
\newblock Combined {Constraint}-{Based} {Analysis} for {Efficient} {Software}
  {Regression} {Detection} in {Evolving} {Programs}.
\newblock volume 303, pages 108--120. Springer Verlag, 2013.

\bibitem{lee_hybridroid_2016}
S.~Lee, J.~Dolby, and S.~Ryu.
\newblock {HybriDroid}: {Static} analysis framework for android hybrid
  applications.
\newblock pages 250--261. Association for Computing Machinery, Inc, 2016.

\bibitem{lee_sealant_2017}
Y.~Lee, J.~Bang, G.~Safi, A.~Shahbazian, Y.~Zhao, and N.~Medvidovic.
\newblock A sealant for inter-app security holes in android.
\newblock pages 312--323. Institute of Electrical and Electronics Engineers
  Inc., 2017.

\bibitem{leesatapornwongsa_flakerepro_2022}
T.~Leesatapornwongsa, X.~Ren, and S.~Nath.
\newblock {FlakeRepro}: automated and efficient reproduction of
  concurrency-related flaky tests.
\newblock pages 1509--1520. Association for Computing Machinery, Inc, 2022.

\bibitem{li_calculating_2013}
D.~Li, S.~Hao, W.~Halfond, and R.~Govindan.
\newblock Calculating source line level energy information for {Android}
  applications.
\newblock pages 78--89, 2013.

\bibitem{li_view-based_2008}
P.~Li and E.~Wohlstadter.
\newblock View-based maintenance of graphical user interfaces.
\newblock pages 156--167, 2008.

\bibitem{li_human-machine_2023}
Y.~Li, Y.~Feng, R.~Hao, and Z.~Chen.
\newblock Human-{Machine} {Collaborative} {Testing} for {Android}
  {Applications}.
\newblock pages 440--451. Institute of Electrical and Electronics Engineers
  Inc., 2023.

\bibitem{li_fault_2022}
Y.~Li, S.~Wang, and T.~Nguyen.
\newblock Fault localization to detect co-change fixing locations.
\newblock pages 659--671. Association for Computing Machinery, Inc, 2022.

\bibitem{linares-vasquez_mining_2015}
M.~Linares-Vásquez, M.~White, C.~Bernal-Cárdenas, K.~Moran, and
  D.~Poshyvanyk.
\newblock Mining android app usages for generating actionable {GUI}-based
  execution scenarios.
\newblock volume 2015-August, pages 111--122. IEEE Computer Society, 2015.

\bibitem{lindorfer_marvin_2015}
M.~Lindorfer, M.~Neugschwandtner, and C.~Platzer.
\newblock {MARVIN}: {Efficient} and {Comprehensive} {Mobile} {App}
  {Classification} through {Static} and {Dynamic} {Analysis}.
\newblock volume~2, pages 422--433. IEEE Computer Society, 2015.

\bibitem{ling_essential_2024}
Y.~Ling, Y.~Hao, Y.~Wang, K.~Wang, G.~Bai, and J.~Dong.
\newblock Essential or {Excessive}? {MINDAEXT}: {Measuring} {Data}
  {Minimization} {Practices} among {Browser} {Extensions}.
\newblock pages 964--975. Institute of Electrical and Electronics Engineers
  Inc., 2024.

\bibitem{liu_general_2021}
C.~Liu.
\newblock A {General} {Framework} to {Detect} {Design} {Patterns} by
  {Combining} {Static} and {Dynamic} {Analysis} {Techniques}.
\newblock {\em International Journal of Software Engineering and Knowledge
  Engineering}, 31(1):21--54, 2021.
\newblock Publisher: World Scientific.

\bibitem{liu_promal_2022}
C.~Liu, H.~Wang, T.~Liu, D.~Gu, Y.~Ma, H.~Wang, and X.~Xiao.
\newblock {PROMAL}: {Precise} {Window} {Transition} {Graphs} for {Android} via
  {Synergy} of {Program} {Analysis} and {Machine} {Learning}.
\newblock volume 2022-May, pages 1755--1767. IEEE Computer Society, 2022.

\bibitem{liu_efficient_2016}
F.~Liu, B.~Li, and R.~Nasre.
\newblock Efficient online cycle detection technique combining with
  {Steensgaard} points-to information.
\newblock {\em Software - Practice and Experience}, 46(5):601--623, 2016.
\newblock Publisher: John Wiley and Sons Ltd.

\bibitem{liu_tdroid_2018}
J.~Liu, D.~Wu, and J.~Xue.
\newblock {TDroid}: {Exposing} app switching attacks in android with control
  flow specialization.
\newblock pages 236--247. Association for Computing Machinery, Inc, 2018.

\bibitem{liu_ipa_2016}
P.~Liu, O.~Tripp, and X.~Zhang.
\newblock {IPA}: {Improving} predictive analysis with pointer analysis.
\newblock pages 59--69. Association for Computing Machinery, Inc, 2016.

\bibitem{liu_vd-guard_2023}
Y.~Liu, S.~Chen, Y.~Xie, Y.~Wang, L.~Chen, B.~Wang, Y.~Zeng, Z.~Xue, and P.~Su.
\newblock {VD}-{Guard}: {DMA} {Guided} {Fuzzing} for {Hypervisor} {Virtual}
  {Device}.
\newblock pages 1676--1687. Institute of Electrical and Electronics Engineers
  Inc., 2023.

\bibitem{logozzo_rata_2010}
F.~Logozzo and H.~Venter.
\newblock {RATA}: {Rapid} atomic type analysis by abstract interpretation -
  {Application} to {JavaScript} optimization.
\newblock volume 6011 LNCS, pages 66--83, 2010.

\bibitem{lohar_two-phase_2021}
D.~Lohar, C.~Jeangoudoux, J.~Sobel, E.~Darulova, and M.~Christakis.
\newblock A {Two}-{Phase} {Approach} for {Conditional} {Floating}-{Point}
  {Verification}.
\newblock volume 12652 LNCS, pages 43--63. Springer Science and Business Media
  Deutschland GmbH, 2021.

\bibitem{ma_software_2018}
L.~Ma and Z.~Ding.
\newblock Software {Bug} {Localization} {Based} on {Key} {Range} {Invariants}.
\newblock volume 11293 LNCS, pages 20--32. Springer Verlag, 2018.

\bibitem{madsen_practical_2013}
M.~Madsen, B.~Livshits, and M.~Fanning.
\newblock Practical static analysis of {JavaScript} applications in the
  presence of frameworks and libraries.
\newblock pages 499--509, 2013.

\bibitem{mahmood_evodroid_2014}
R.~Mahmood, N.~Mirzaei, and S.~Malek.
\newblock {EvoDroid}: {Segmented} evolutionary testing of {Android} apps.
\newblock volume 16-21-November-2014, pages 599--609. Association for Computing
  Machinery, 2014.

\bibitem{mao_sapienz_2016}
K.~Mao, M.~Harman, and Y.~Jia.
\newblock Sapienz: {Multi}-objective automated testing for android
  applications.
\newblock pages 94--105. Association for Computing Machinery, Inc, 2016.

\bibitem{marin_pinpointing_2008}
G.~Marin and J.~Mellor-Crummey.
\newblock Pinpointing and exploiting opportunities for enhancing data reuse.
\newblock pages 115--126, 2008.

\bibitem{10.1145/2491411.2491438}
P.~D. Marinescu and C.~Cadar.
\newblock Katch: high-coverage testing of software patches.
\newblock In {\em Proceedings of the 2013 9th Joint Meeting on Foundations of
  Software Engineering}, ESEC/FSE 2013, page 235–245, New York, NY, USA,
  2013. Association for Computing Machinery.

\bibitem{matias_determining_2020}
T.~Matias, F.~Correia, J.~Fritzsch, J.~Bogner, H.~Ferreira, and A.~Restivo.
\newblock Determining microservice boundaries: {A} case study using static and
  dynamic software analysis.
\newblock volume 12292 LNCS, pages 315--332. Springer Science and Business
  Media Deutschland GmbH, 2020.

\bibitem{mcminn_input_2012}
P.~McMinn, M.~Harman, K.~Lakhotia, Y.~Hassoun, and J.~Wegener.
\newblock Input domain reduction through irrelevant variable removal and its
  effect on local, global, and hybrid search-based structural test data
  generation.
\newblock {\em IEEE Transactions on Software Engineering}, 38(2):453--477,
  2012.

\bibitem{memon_employing_2006}
A.~Memon.
\newblock Employing user profiles to test a new version of a {GUI} component in
  its context of use.
\newblock {\em Software Quality Journal}, 14(4):359--377, 2006.
\newblock Publisher: Kluwer Academic Publishers.

\bibitem{merriam_measurement_2013}
N.~Merriam, P.~Gliwa, and I.~Broster.
\newblock Measurement and tracing methods for timing analysis: {Independently}
  and in combination with modelling methods.
\newblock {\em International Journal on Software Tools for Technology
  Transfer}, 15(1):9--28, 2013.
\newblock Publisher: Springer Verlag.

\bibitem{messaoudi_log-based_2021}
S.~Messaoudi, D.~Shin, A.~Panichella, D.~Bianculli, and L.~Briand.
\newblock Log-based slicing for system-level test cases.
\newblock pages 517--528. Association for Computing Machinery, Inc, 2021.

\bibitem{mirzaei_reducing_2016}
N.~Mirzaei, J.~Garcia, H.~Bagheri, A.~Sadeghi, and S.~Malek.
\newblock Reducing combinatorics in {GUI} testing of android applications.
\newblock volume 14-22-May-2016, pages 559--570. IEEE Computer Society, 2016.

\bibitem{molina_fuzzing_2022}
F.~Molina, M.~D'Amorim, and N.~Aguirre.
\newblock Fuzzing {Class} {Specifications}.
\newblock volume 2022-May, pages 1008--1020. IEEE Computer Society, 2022.

\bibitem{moukahal_boosting_2021}
L.~Moukahal, M.~Zulkernine, and M.~Soukup.
\newblock Boosting {Grey}-box {Fuzzing} for {Connected} {Autonomous} {Vehicle}
  {Systems}.
\newblock pages 516--527. Institute of Electrical and Electronics Engineers
  Inc., 2021.

\bibitem{naik_effective_2009}
M.~Naik, C.-S. Park, K.~Sen, and D.~Gay.
\newblock Effective static deadlock detection.
\newblock pages 386--396, 2009.

\bibitem{neelofar_improving_2017}
N.~Neelofar, L.~Naish, J.~Lee, and K.~Ramamohanarao.
\newblock Improving spectral-based fault localization using static analysis.
\newblock {\em Software - Practice and Experience}, 47(11):1633--1655, 2017.
\newblock Publisher: John Wiley and Sons Ltd.

\bibitem{nguena_timo_multiple_2019}
O.~Nguena~Timo, D.~Prestat, and A.~Rollet.
\newblock Multiple {Mutation} {Testing} for {Timed} {Finite} {State} {Machine}
  with {Timed} {Guards} and {Timeouts}.
\newblock volume 11812 LNCS, pages 104--120. Springer, 2019.

\bibitem{nguyen_using_2014}
T.~Nguyen, D.~Kapur, W.~Weimer, and S.~Forrest.
\newblock Using dynamic analysis to generate disjunctive invariants.
\newblock pages 608--619. IEEE Computer Society, 2014.
\newblock Issue: 1.

\bibitem{noller_badger_2018}
Y.~Noller, R.~Kersten, and C.~Pǎsǎreanu.
\newblock Badger: {Complexity} analysis with fuzzing and symbolic execution.
\newblock pages 322--332. Association for Computing Machinery, Inc, 2018.

\bibitem{noller_hydiff_2020}
Y.~Noller, C.~Pasareanu, M.~Bohme, Y.~Sun, H.~Nguyen, and H.~Nguyen.
\newblock Hydiff: {Hybrid} differential software analysis.
\newblock pages 1273--1285. IEEE Computer Society, 2020.

\bibitem{Obbink:SANER:2018}
N.~G. Obbink, I.~Malavolta, G.~L. Scoccia, and P.~Lago.
\newblock An extensible approach for taming the challenges of javascript dead
  code elimination.
\newblock In {\em 2018 IEEE 25th International Conference on Software Analysis,
  Evolution and Reengineering (SANER)}, pages 291--401, 2018.

\bibitem{Pan:ICSME:2017}
J.~Pan and X.~Mao.
\newblock Detecting dom-sourced cross-site scripting in browser extensions.
\newblock In {\em 2017 IEEE International Conference on Software Maintenance
  and Evolution (ICSME)}, pages 24--34. IEEE, 2017.

\bibitem{Park:JMESECSFSE:2021}
J.~Park, J.~Park, D.~Youn, and S.~Ryu.
\newblock Accelerating javascript static analysis via dynamic shortcuts.
\newblock In {\em Proceedings of the 29th ACM Joint Meeting on European
  Software Engineering Conference and Symposium on the Foundations of Software
  Engineering}, pages 1129--1140, 2021.

\bibitem{Patra:PICSE:2018}
J.~Patra, P.~N. Dixit, and M.~Pradel.
\newblock Conflictjs: finding and understanding conflicts between javascript
  libraries.
\newblock In {\em Proceedings of the 40th international conference on software
  engineering}, pages 741--751, 2018.

\bibitem{Perez:TSE:2019}
D.~D. Perez and W.~Le.
\newblock Specifying callback control flow of mobile apps using finite
  automata.
\newblock {\em IEEE Transactions on Software Engineering}, 47(2):379--392,
  2019.

\bibitem{ponta2020detection}
S.~E. Ponta, H.~Plate, and A.~Sabetta.
\newblock Detection, assessment and mitigation of vulnerabilities in open
  source dependencies.
\newblock {\em Empirical Software Engineering}, 25(5):3175--3215, 2020.

\bibitem{poshyvanyk2007feature}
D.~Poshyvanyk, Y.-G. Gu{\'e}h{\'e}neuc, A.~Marcus, G.~Antoniol, and V.~Rajlich.
\newblock Feature location using probabilistic ranking of methods based on
  execution scenarios and information retrieval.
\newblock {\em IEEE Transactions on Software Engineering}, 33(6):420--432,
  2007.

\bibitem{pradel2012statically}
M.~Pradel, C.~Jaspan, J.~Aldrich, and T.~R. Gross.
\newblock Statically checking api protocol conformance with mined multi-object
  specifications.
\newblock In {\em 2012 34th International Conference on Software Engineering
  (ICSE)}, pages 925--935. IEEE, 2012.

\bibitem{Rasthofer:2017:ICST}
S.~Rasthofer, S.~Arzt, S.~Triller, and M.~Pradel.
\newblock Making malory behave maliciously: Targeted fuzzing of android
  execution environments.
\newblock In {\em 2017 IEEE/ACM 39th International Conference on Software
  Engineering (ICSE)}, pages 300--311, 2017.

\bibitem{Rattanasuksun:RRF:2016}
S.~Rattanasuksun, T.~Yu, W.~Srisa-An, and G.~Rothermel.
\newblock Rrf: A race reproduction framework for use in debugging process-level
  races.
\newblock In {\em 2016 IEEE 27th International Symposium on Software
  Reliability Engineering (ISSRE)}, pages 162--172, 2016.

\bibitem{Rimsa:SPE:2020}
A.~Rimsa, J.~Nelson~Amaral, and F.~M.~Q. Pereira.
\newblock Practical dynamic reconstruction of control flow graphs.
\newblock {\em Software: Practice and Experience}, 51(2):353--384, 2021.

\bibitem{Rohatgi:IET:2009}
A.~Rohatgi, A.~Hamou-Lhadj, and J.~Rilling.
\newblock Approach for solving the feature location problem by measuring the
  component modification impact.
\newblock {\em IET software}, 3(4):292--311, 2009.

\bibitem{sabbaghi2019fsct}
A.~Sabbaghi, H.~R. Kanan, and M.~R. Keyvanpour.
\newblock Fsct: A new fuzzy search strategy in concolic testing.
\newblock {\em Information and Software Technology}, 107:137--158, 2019.

\bibitem{Sadeghi:JMFSE:2017}
A.~Sadeghi, R.~Jabbarvand, and S.~Malek.
\newblock Patdroid: permission-aware gui testing of android.
\newblock In {\em Proceedings of the 2017 11th Joint Meeting on Foundations of
  Software Engineering}, pages 220--232, 2017.

\bibitem{Said:ESE:2020}
W.~Said, J.~Quante, and R.~Koschke.
\newblock Mining understandable state machine models from embedded code.
\newblock {\em Empirical Software Engineering}, 25(6):4759--4804, 2020.

\bibitem{samhi2022difuzer}
J.~Samhi, L.~Li, T.~F. Bissyand{\'e}, and J.~Klein.
\newblock Difuzer: Uncovering suspicious hidden sensitive operations in android
  apps.
\newblock In {\em Proceedings of the 44th International Conference on Software
  Engineering}, pages 723--735, 2022.

\bibitem{santelices_exploiting_2010}
R.~Santelices and M.~Harrold.
\newblock Exploiting program dependencies for scalable multiple-path symbolic
  execution.
\newblock pages 195--205, 2010.

\bibitem{saputri_software_2020}
T.~Saputri and S.-W. Lee.
\newblock Software {Analysis} {Method} for {Assessing} {Software}
  {Sustainability}.
\newblock {\em International Journal of Software Engineering and Knowledge
  Engineering}, 30(1):67--95, 2020.
\newblock Publisher: World Scientific Publishing Co. Pte Ltd.

\bibitem{schoeberl_worst-case_2010}
M.~Schoeberl, W.~Puffitsch, P.~Ulslev, and B.~Huber.
\newblock Worst-case execution time analysis for a {Java} processor.
\newblock {\em Software - Practice and Experience}, 40(6):507--542, 2010.

\bibitem{schordan_combining_2014}
M.~Schordan and A.~Prantl.
\newblock Combining static analysis and state transition graphs for
  verification of event-condition-action systems in the {RERS} 2012 and 2013
  challenges.
\newblock {\em International Journal on Software Tools for Technology
  Transfer}, 16(5):493--505, 2014.
\newblock Publisher: Springer Verlag.

\bibitem{sen:cute:esec:2005}
K.~Sen, D.~Marinov, and G.~Agha.
\newblock {CUTE}: a concolic unit testing engine for {C}.
\newblock In {\em Proceedings of the 10th European software engineering
  conference held jointly with 13th ACM SIGSOFT international symposium on
  Foundations of software engineering (ESEC/FSE-13)}, pages 263--272, 2005.

\bibitem{sh_ghandehari_combinatorial_2020}
L.~Sh.~Ghandehari, Y.~Lei, R.~Kacker, R.~Kuhn, T.~Xie, and D.~Kung.
\newblock A {Combinatorial} {Testing}-{Based} {Approach} to {Fault}
  {Localization}.
\newblock {\em IEEE Transactions on Software Engineering}, 46(6):616--645,
  2020.
\newblock Publisher: Institute of Electrical and Electronics Engineers Inc.

\bibitem{shan_self-hiding_2018}
Z.~Shan, I.~Neamtiu, and R.~Samuel.
\newblock Self-hiding behavior in {Android} apps: {Detection} and
  characterization.
\newblock pages 728--739. IEEE Computer Society, 2018.

\bibitem{sinha_fault_2009}
S.~Sinha, H.~Shah, C.~Görg, S.~Jiang, M.~Kim, and M.~Harrold.
\newblock Fault localization and repair for {Java} runtime exceptions.
\newblock pages 153--163. Association for Computing Machinery, Inc, 2009.

\bibitem{sochor_grammarforge_2025}
H.~Sochor, F.~Ferrarotti, and R.~Wille.
\newblock {GrammarForge}: {Learning} {Program} {Input} {Grammars} for {Fuzz}
  {Testing}.
\newblock volume 15280 LNCS, pages 272--289. Springer Science and Business
  Media Deutschland GmbH, 2025.

\bibitem{sohn_cement_2022}
J.~Sohn and M.~Papadakis.
\newblock {CEMENT}: {On} the {Use} of {Evolutionary} {Coupling} {Between}
  {Tests} and {Code} {Units}. {A} {Case} {Study} on {Fault} {Localization}.
\newblock volume 2022-October, pages 133--144. IEEE Computer Society, 2022.

\bibitem{song_itree_2014}
C.~Song, A.~Porter, and J.~Foster.
\newblock {ITree}: {Efficiently} discovering high-coverage configurations using
  interaction trees.
\newblock {\em IEEE Transactions on Software Engineering}, 40(3):251--265,
  2014.
\newblock Publisher: Institute of Electrical and Electronics Engineers Inc.

\bibitem{souter_tatoo_2001}
A.~Souter, T.~Wong, S.~Shindo, and L.~Pollock.
\newblock {TATOO}: {Testing} and analysis tool for object-oriented software.
\newblock volume 2031 LNCS, pages 389--403. Springer Verlag, 2001.

\bibitem{srivastava_crystallizer_2023}
P.~Srivastava, F.~Toffalini, K.~Vorobyov, F.~Gauthier, A.~Bianchi, and
  M.~Payer.
\newblock Crystallizer: {A} {Hybrid} {Path} {Analysis} {Framework} to {Aid} in
  {Uncovering} {Deserialization} {Vulnerabilities}.
\newblock pages 1586--1597. Association for Computing Machinery, Inc, 2023.

\bibitem{su_combining_2015}
T.~Su, Z.~Fu, G.~Pu, J.~He, and Z.~Su.
\newblock Combining symbolic execution and model checking for data flow
  testing.
\newblock volume~1, pages 654--665. IEEE Computer Society, 2015.

\bibitem{sun_tlv_2015}
J.~Sun, H.~Xiao, Y.~Liu, S.-W. Lin, and S.~Qin.
\newblock {TLV}: {Abstraction} through testing, learning, and validation.
\newblock pages 698--709. Association for Computing Machinery, Inc, 2015.

\bibitem{sun_combort_2016}
X.~Sun, X.~Peng, H.~Leung, and B.~Li.
\newblock {ComboRT}: {A} {New} {Approach} for {Generating} {Regression} {Test}
  {Cases} for {Evolving} {Programs}.
\newblock {\em International Journal of Software Engineering and Knowledge
  Engineering}, 26(6):1001--1026, 2016.
\newblock Publisher: World Scientific Publishing Co. Pte Ltd.

\bibitem{sun_gptscan_2024}
Y.~Sun, D.~Wu, Y.~Xue, H.~Liu, H.~Wang, Z.~Xu, X.~Xie, and Y.~Liu.
\newblock {GPTScan}: {Detecting} {Logic} {Vulnerabilities} in {Smart}
  {Contracts} by {Combining} {GPT} with {Program} {Analysis}.
\newblock pages 2048--2060. IEEE Computer Society, 2024.

\bibitem{tang_software_2017}
E.~Tang, X.~Zhang, N.~Muller, Z.~Chen, and X.~Li.
\newblock Software {Numerical} {Instability} {Detection} and {Diagnosis} by
  {Combining} {Stochastic} and {Infinite}-{Precision} {Testing}.
\newblock {\em IEEE Transactions on Software Engineering}, 43(10):975--994,
  2017.
\newblock Publisher: Institute of Electrical and Electronics Engineers Inc.

\bibitem{Tang:NIVAnalyzer:2017}
J.~Tang, X.~Cui, Z.~Zhao, S.~Guo, X.~Xu, C.~Hu, T.~Ban, and B.~Mao.
\newblock Nivanalyzer: A tool for automatically detecting and verifying
  next-intent vulnerabilities in android apps.
\newblock In {\em 2017 IEEE International Conference on Software Testing,
  Verification and Validation (ICST)}, pages 492--499, 2017.

\bibitem{tang_separate_1994}
Y.~Tang and P.~Jouvelot.
\newblock Separate abstract interpretation for control-flow analysis.
\newblock volume 789 LNCS, pages 224--243. Springer Verlag, 1994.

\bibitem{tanida_automated_2013}
H.~Tanida, M.~Prasad, S.~Rajan, and M.~Fujita.
\newblock Automated {System} {Testing} of {Dynamic} {Web} {Applications}.
\newblock volume 303, pages 181--196. Springer Verlag, 2013.

\bibitem{Tarvo:ASE:2018}
A.~Tarvo and S.~P. Reiss.
\newblock Automatic performance prediction of multithreaded programs: a
  simulation approach.
\newblock {\em Automated Software Engineering}, 25:101--155, 2018.

\bibitem{Tiwari:ISSRE:2023}
A.~Tiwari, J.~Prakash, and C.~Hammer.
\newblock Demand-driven information flow analysis of {WebView} in {Android}
  hybrid apps.
\newblock In {\em 2023 IEEE 34th International Symposium on Software
  Reliability Engineering (ISSRE)}, pages 415--426, 2023.

\bibitem{Tlili:AMAST:2008}
S.~Tlili, Z.~Yang, H.~Z. Ling, and M.~Debbabi.
\newblock A hybrid approach for safe memory management in {C}.
\newblock In J.~Meseguer and G.~Ro{\c{s}}u, editors, {\em Algebraic Methodology
  and Software Technology}, pages 377--391, Berlin, Heidelberg, 2008. Springer
  Berlin Heidelberg.

\bibitem{Tripp:ISSTA:2014}
O.~Tripp, P.~Ferrara, and M.~Pistoia.
\newblock Hybrid security analysis of web javascript code via dynamic partial
  evaluation.
\newblock In {\em Proceedings of the 2014 International Symposium on Software
  Testing and Analysis}, ISSTA 2014, pages 49--59, New York, NY, USA, 2014.
  Association for Computing Machinery.

\bibitem{ulrich_experience_2016}
A.~Ulrich and A.~Votintseva.
\newblock Experience report: {Formal} verification and testing in the
  development of embedded software.
\newblock pages 293--302. Institute of Electrical and Electronics Engineers
  Inc., 2016.

\bibitem{ValleGomez:IETSW:2022}
K.~J. Valle\space{}Gómez, A.~García\space{}Domínguez,
  P.~Delgado\space{}Pérez, and I.~Medina\space{}Bulo.
\newblock Mutation-inspired symbolic execution for software testing.
\newblock {\em IET Software}, 16(5):478--492, 2022.

\bibitem{Vidziunas:ICSME:2024}
L.~Vidziunas, D.~Binkley, and L.~Moonen.
\newblock The impact of program reduction on automated program repair.
\newblock In {\em 2024 IEEE International Conference on Software Maintenance
  and Evolution (ICSME)}, pages 337--349, 2024.

\bibitem{Vikram:ICSE:2021}
V.~Vikram, R.~Padhye, and K.~Sen.
\newblock Growing a test corpus with bonsai fuzzing.
\newblock In {\em 2021 IEEE/ACM 43rd International Conference on Software
  Engineering (ICSE)}, pages 723--735, 2021.

\bibitem{Vorobyov:SEFM:2012}
K.~Vorobyov, P.~Krishnan, and P.~Stocks.
\newblock A low-overhead, value-tracking approach to information flow security.
\newblock In G.~Eleftherakis, M.~Hinchey, and M.~Holcombe, editors, {\em
  Software Engineering and Formal Methods}, pages 367--381, Berlin, Heidelberg,
  2012. Springer Berlin Heidelberg.

\bibitem{Vasquez:IST:2019}
H.~Vázquez, A.~Bergel, S.~Vidal, J.~{Díaz Pace}, and C.~Marcos.
\newblock Slimming javascript applications: An approach for removing unused
  functions from javascript libraries.
\newblock {\em Information and Software Technology}, 107:18--29, 2019.

\bibitem{wang_combodroid_2020}
J.~Wang, Y.~Jiang, C.~Xu, C.~Cao, X.~Ma, and J.~Lu.
\newblock Combodroid: {Generating} high-quality test inputs for android apps
  via use case combinations.
\newblock pages 469--480. IEEE Computer Society, 2020.

\bibitem{wang_string_2022}
M.~Wang, B.~Cui, J.~Yan, J.~Yan, and J.~Zhang.
\newblock String {Test} {Data} {Generation} for {Java} {Programs}.
\newblock volume 2022-October, pages 251--262. IEEE Computer Society, 2022.

\bibitem{Wang:TSE:2022}
Y.~Wang, F.~Gao, L.~Wang, T.~Yu, J.~Zhao, and X.~Li.
\newblock Automatic detection, validation, and repair of race conditions in
  interrupt-driven embedded software.
\newblock {\em IEEE Transactions on Software Engineering}, 48(1):346--363,
  2022.

\bibitem{wang_could_2019}
Y.~Wang, M.~Wen, R.~Wu, Z.~Liu, S.~Tan, Z.~Zhu, H.~Yu, and S.-C. Cheung.
\newblock Could i {Have} a {Stack} {Trace} to {Examine} the {Dependency}
  {Conflict} {Issue}?
\newblock volume 2019-May, pages 572--583. IEEE Computer Society, 2019.

\bibitem{Wei:ISSTA:2013}
S.~Wei and B.~G. Ryder.
\newblock Practical blended taint analysis for javascript.
\newblock In {\em Proceedings of the 2013 International Symposium on Software
  Testing and Analysis}, ISSTA 2013, pages 336--346, New York, NY, USA, 2013.
  Association for Computing Machinery.

\bibitem{Whelan:CC:2013}
R.~Whelan, T.~Leek, and D.~Kaeli.
\newblock Architecture-independent dynamic information flow tracking.
\newblock In R.~Jhala and K.~De~Bosschere, editors, {\em Compiler
  Construction}, pages 144--163, Berlin, Heidelberg, 2013. Springer Berlin
  Heidelberg.

\bibitem{LeanBin:Wodiany:ASE:2024}
I.~Wodiany, A.~Pop, and M.~Luj\'{a}n.
\newblock Leanbin: Harnessing lifting and recompilation to debloat binaries.
\newblock In {\em Proceedings of the 39th IEEE/ACM International Conference on
  Automated Software Engineering}, ASE '24, page 1434–1446, New York, NY,
  USA, 2024. Association for Computing Machinery.

\bibitem{Xie:ISSTA:2015}
X.~Xie, Y.~Liu, W.~Le, X.~Li, and H.~Chen.
\newblock S-looper: automatic summarization for multipath string loops.
\newblock In {\em Proceedings of the 2015 International Symposium on Software
  Testing and Analysis}, ISSTA 2015, page 188–198, New York, NY, USA, 2015.
  Association for Computing Machinery.

\bibitem{Xu:ICSME:2020}
X.~Xu, C.~Zou, and J.~Xue.
\newblock Every mutation should be rewarded: Boosting fault localization with
  mutated predicates.
\newblock In {\em 2020 IEEE International Conference on Software Maintenance
  and Evolution (ICSME)}, pages 196--207, 2020.

\bibitem{Xu:SANER:2024}
Y.~Xu, M.~Zhou, Q.~Gao, S.~Zhang, and Z.~Wu.
\newblock Swat4j: Generating system call allowlist for java container attack
  surface reduction.
\newblock In {\em 2024 IEEE International Conference on Software Analysis,
  Evolution and Reengineering (SANER)}, pages 929--939, 2024.

\bibitem{Yan:SETTA:2021}
Y.~Yan, S.~Jiang, S.~Zhang, and Y.~Huang.
\newblock Csfl: Fault localization on real software bugs based on the
  combination of context and spectrum.
\newblock In S.~Qin, J.~Woodcock, and W.~Zhang, editors, {\em Dependable
  Software Engineering. Theories, Tools, and Applications}, pages 219--238.
  Springer International Publishing, 2021.

\bibitem{yang_multi-objective_2024}
M.~Yang, S.~Yang, and W.~Wong.
\newblock Multi-{Objective} {Software} {Defect} {Prediction} via
  {Multi}-{Source} {Uncertain} {Information} {Fusion} and {Multi}-{Task}
  {Multi}-{View} {Learning}.
\newblock {\em IEEE Transactions on Software Engineering}, 50(8):2054--2076,
  2024.
\newblock Publisher: Institute of Electrical and Electronics Engineers Inc.

\bibitem{10.1145/3533767.3534221}
S.~Yang, Z.~Zeng, and W.~Song.
\newblock {PermDroid}: automatically testing permission-related behaviour of
  android applications.
\newblock In {\em Proceedings of the 31st ACM SIGSOFT International Symposium
  on Software Testing and Analysis}, ISSTA 2022, page 593–604, New York, NY,
  USA, 2022. Association for Computing Machinery.

\bibitem{Ye:2016:SANER}
J.~Ye, C.~Zhang, L.~Ma, H.~Yu, and J.~Zhao.
\newblock Efficient and precise dynamic slicing for client-side javascript
  programs.
\newblock In {\em 2016 IEEE 23rd International Conference on Software Analysis,
  Evolution, and Reengineering (SANER)}, volume~1, pages 449--459, 2016.

\bibitem{Yi:ASEA:2011}
Q.~Yi, J.~Liu, and W.~Shen.
\newblock Efficient loop-extended model checking of data structure methods.
\newblock In T.-h. Kim, H.~Adeli, H.-k. Kim, H.-j. Kang, K.~J. Kim, A.~Kiumi,
  and B.-H. Kang, editors, {\em Software Engineering, Business Continuity, and
  Education}, pages 237--249, Berlin, Heidelberg, 2011. Springer Berlin
  Heidelberg.

\bibitem{yin_fries_2024}
X.~Yin, Y.~Feng, Q.~Shi, Z.~Liu, H.~Liu, and B.~Xu.
\newblock {FRIES}: {Fuzzing} {Rust} {Library} {Interactions} via {Efficient}
  {Ecosystem}-{Guided} {Target} {Generation}.
\newblock pages 1137--1148. Association for Computing Machinery, Inc, 2024.

\bibitem{Young:TSE:1988}
M.~Young and R.~N. Taylor.
\newblock Combining static concurrency analysis with symbolic execution.
\newblock {\em IEEE Transactions on Software Engineering}, 14(10):1499--1511,
  1988.

\bibitem{Yu:TACAS:2009}
F.~Yu, T.~Bultan, and O.~H. Ibarra.
\newblock Symbolic string verification: Combining string analysis and size
  analysis.
\newblock In S.~Kowalewski and A.~Philippou, editors, {\em Tools and Algorithms
  for the Construction and Analysis of Systems}, pages 322--336, Berlin,
  Heidelberg, 2009. Springer Berlin Heidelberg.

\bibitem{Yu:TSE:2020}
T.~Yu, Z.~Huang, and C.~Wang.
\newblock Contesa: Directed test suite augmentation for concurrent software.
\newblock {\em IEEE Transactions on Software Engineering}, 46(4):405--419,
  2020.

\bibitem{yu_descry_2017}
T.~Yu, T.~Zaman, and C.~Wang.
\newblock {DESCRY}: {Reproducing} system-level concurrency failures.
\newblock volume Part F130154, pages 694--704. Association for Computing
  Machinery, 2017.

\bibitem{zhang_discover_2018}
B.~Zhang, C.~Feng, A.~Herrera, V.~Chipounov, G.~Candea, and C.~Tang.
\newblock Discover deeper bugs with dynamic symbolic execution and
  coverage-based fuzz testing.
\newblock {\em IET Software}, 12(6):507--519, 2018.
\newblock Publisher: Institution of Engineering and Technology.

\bibitem{zhang_detecting_2012}
C.~Zhang and Y.~Chen.
\newblock Detecting infeasible paths via mining branch correlations.
\newblock {\em Journal of Software Engineering}, 6(4):65--78, 2012.

\bibitem{zhang_hybrid_2024}
G.~Zhang, L.~Liu, Z.~Chen, and J.~Wang.
\newblock Hybrid {Regression} {Test} {Selection} by {Integrating} {File} and
  {Method} {Dependences}.
\newblock pages 1557--1569. Association for Computing Machinery, Inc, 2024.

\bibitem{zhang_comparing_2022}
J.~Zhang, Y.~Liu, M.~Gligoric, O.~Legunsen, and A.~Shi.
\newblock Comparing and {Combining} {Analysis}-{Based} and {Learning}-{Based}
  {Regression} {Test} {Selection}.
\newblock pages 17--28. Institute of Electrical and Electronics Engineers Inc.,
  2022.

\bibitem{zhang_hybrid_2018}
L.~Zhang.
\newblock Hybrid {Regression} {Test} {Selection}.
\newblock volume 2018-January, pages 199--209. IEEE Computer Society, 2018.

\bibitem{zhang_runtime_2014}
L.~Zhang and C.~Wang.
\newblock Runtime prevention of concurrency related type-state violations in
  multithreaded applications.
\newblock pages 1--12. Association for Computing Machinery, Inc, 2014.

\bibitem{zhang2013confdiagnoser}
S.~Zhang and M.~D. Ernst.
\newblock Automated diagnosis of software configuration errors.
\newblock In {\em 2013 35th International Conference on Software Engineering
  (ICSE)}, pages 312--321, 2013.

\bibitem{zhang_combined_2011}
S.~Zhang, D.~Saff, Y.~Bu, and M.~Ernst.
\newblock Combined static and dynamic automated test generation.
\newblock pages 353--363, 2011.

\bibitem{zhang_adaptive_2023}
W.~Zhang, Y.~Hu, B.~Tan, X.~Shi, and J.~Jiang.
\newblock Adaptive {Tracing} and {Fault} {Injection} based {Fault} {Diagnosis}
  for {Open} {Source} {Server} {Software}.
\newblock pages 729--740. Institute of Electrical and Electronics Engineers
  Inc., 2023.

\bibitem{zhang_heuristic_2017}
X.-Z. Zhang, Y.-Z. Gong, and Y.-W. Wang.
\newblock Heuristic guided selective path exploration for loop structure in
  coverage testing.
\newblock {\em International Journal of Open Source Software and Processes},
  8(2):59--75, 2017.
\newblock Publisher: IGI Global.

\bibitem{zhang_androidleaker_2017}
Z.~Zhang and X.~Feng.
\newblock {AndroidLeaker}: {A} hybrid checker for collusive leak in android
  applications.
\newblock volume 10606 LNCS, pages 164--180. Springer Verlag, 2017.

\bibitem{zhang_boundary_2016}
Z.~Zhang, T.~Wu, and J.~Zhang.
\newblock Boundary value analysis in automatic white-box test generation.
\newblock pages 239--249. Institute of Electrical and Electronics Engineers
  Inc., 2016.

\bibitem{zhao_test_2015}
R.~Zhao, Z.~Li, and Q.~Wang.
\newblock Test {Generation} for {Programs} with {Binary} {Tree} {Structure} as
  {Input}.
\newblock {\em International Journal of Software Engineering and Knowledge
  Engineering}, 25(7):1129--1151, 2015.
\newblock Publisher: World Scientific Publishing Co. Pte Ltd.

\bibitem{zhao_framework_2022}
Y.~Zhao, G.~Yi, F.~Liu, Z.~Hui, and J.~Zhao.
\newblock A {Framework} for {Scanning} {Privacy} {Information} based on
  {Static} {Analysis}.
\newblock volume 2022-December, pages 1135--1145. Institute of Electrical and
  Electronics Engineers Inc., 2022.

\bibitem{zhou_minerva_2022}
C.~Zhou, Q.~Zhang, M.~Wang, L.~Guo, J.~Liang, Z.~Liu, M.~Payer, and Y.~Jiang.
\newblock Minerva: browser {API} fuzzing with dynamic mod-ref analysis.
\newblock pages 1135--1147. Association for Computing Machinery, Inc, 2022.

\bibitem{zhu_dynamic_2021}
S.~Zhu, N.~Alawar, M.~Erez, and M.~Gligoric.
\newblock Dynamic {Generation} of {Python} {Bindings} for {HPC} {Kernels}.
\newblock pages 92--103. Institute of Electrical and Electronics Engineers
  Inc., 2021.

\end{thebibliography}

\appendix
\section{Primary Studies: Data Extraction}

\subsection{Analyzing test completeness for dynamic languages \cite{adamsen_analyzing_2016}}

\subsubsection*{Summary}
This paper presents a hybrid of lightweight static analysis and dynamic execution of test suites to determine when test suites have sufficient coverage to guarantee type-related correctness properties, which is particularly challenging for program code with overloading and value-dependent types. 
The static analysis has two parts: a dependence analysis and a type analysis (technically, a points-to analysis). It is context- and path-insensitive, and thereby scales to large programs, and it is relatively easy to implement; notably, it requires simpler modeling of native functions than what would be required by a fully static analysis approach.
The analysis combined abstract interpretation of value and type dependencies, with runtime monitoring of a test suite, to establish the type completeness property of the test suite, evaluated as coverage of the abstract states produced by the type analysis from Section 5.

\subsubsection*{Synergistic Effects}
\synPartitioning~/~\synPartitioningWitness
Because, dynamic execution is used to confirm 
coverage of the abstract states produced by the type analysis.
\subsubsection*{Inter-Analysis Workflow}
\workCascade: the analysis starts with abstract interpretation of value and type dependencies, then it monitors a test suite at runtime to establish the type completeness property of the test suite, evaluated as coverage of the abstract states produced by the type analysis.
\subsubsection*{Mapping-Function Interpretation Structure}
\structCFG: They abstract interpretation layer compute the relevant set of abstract states (type-analysis results) per program point.

\subsubsection*{Mapping-Function Mechanics}
\mechanicAssociation: 

\subsection{Concolic testing for models of state-based systems \cite{ahmadi_concolic_2019-1}}

\subsubsection*{Summary}
This paper presents a novel approach and tool (mCUTE: Model-level Concolic Unit Testing Engine) to support automatic unit testing of models of real-time embedded systems by conducting concolic testing, a hybrid testing technique based on concrete and symbolic execution. In the first phase, the model is isolated from its environment, is transformed to a testable model, and is integrated with a test harness. In the second phase, the harness tests the model concolically and reports the test execution results.
So the proposed technique transforms UML-RT models by instrumenting action code to collect path constraints and synchronizing with a test harness that dynamically generates inputs. This enables side-by-side executions of concrete and symbolic paths, ensuring systematic branch coverage. 
\subsubsection*{Synergistic Effects}
\begin{itemize}
\item \synPartitioning~/~\synPartitioningCoverage:
    The overall approach aims to generate test cases (in the style of concolic testing) for all execution oaths of a UML-RT model.
\end{itemize}
\subsubsection*{Inter-Analysis Workflow}
\workCascade: First they make model testable (by integrating it with the test harness and instrumentation) and then they use concolic testing.  
\subsubsection*{Mapping-Function Interpretation Structure}
\begin{itemize}
\item \structProgram: the test harness is associated with the program model to be tested.
\end{itemize}
\subsubsection*{Mapping-Function Mechanics}
\mechanicAssociation

\subsection{autoMPI: Automated Multiple Perspective Attack Investigation With Semantics Aware Execution Partitioning \cite{alhanahnah_autompi_2023}}

\subsubsection*{Summary}
This paper introduces autoMPI, a hybrid analysis approach to performing a differential analysis based on crafted inputs, and the static analysis is conducted to identify the annotation sites within the application code afterward automatically.

1) Dynamic trace analysis in which manually crafted test cases are used to derive a set of program traces. Differential analysis is performed to identify which program regions (e.g. callback functions) contain indicator and channel variables that are not common between the different traces.
2) Static liveness analysis receives the previously derived program regions and analyzes the variables therein contained to check which ones are killed in the program region. These are the relevant channel and indicator variables that will be annotated in the following steps of the technique (not relevant since they do not perform program analyses).
\subsubsection*{Synergistic Effects}
\synPartitioning~/~\synPartitioningDirect:
Combined analysis aims to determine the relevant program regions by dynamic trace analysis and the selected regions used for static liveness analysis.
\subsubsection*{Inter-Analysis Workflow}
\workCascade: first identifies candidate regions dynamically, then refines them statically.
\subsubsection*{Mapping-Function Interpretation Structure}
\structCG: tool relies on function calls and execution traces of the functions.
\subsubsection*{Mapping-Function Mechanics}
\mechanicAssociation

\subsection{High coverage testing of Haskell programs \cite{allwood_high_2011}}

\subsubsection*{Summary}
Static type information to retrieve constructors and functions, to be used as supporting data for generating test data.
The supporting data are used to construct program inputs.
while dynamically executing expressions to detect errors and measure coverage of Haskell programs. 
\subsubsection*{Synergistic Effects}
\synIntepretability~/~\synIntepretabilityEntities
Static type information on constants, constructors, and functions allows for interpreting how to build the proper inputs to instantiate the expressions during the evaluation of the Haskell program under test.  

\subsubsection*{Inter-Analysis Workflow}
\workCascade

\subsubsection*{Mapping-Function Interpretation Structure}
\structProgram: The inferred static type information on constants, constructors, and functions is associated with the program under analysis, to be used in the next phase.

\subsubsection*{Mapping-Function Mechanics}
\mechanicAssociation: see above.

\subsection{PRICE: Detection of Performance Regression Introducing Code Changes Using Static and Dynamic Metrics \cite{alshoaibi_price_2019}}

\subsubsection*{Summary}
Optimization based approach for detecting performance regression, 
by combining metrics computed with static and dynamic analysis. Static metrics, such as the number of deleted or added functions, and dynamic metrics like runtime overhead of frequently called functions are collected from git commits. These metrics are used to train a multi-object evolutionary algorithm that generated detections rules maximizing hit and dismiss rate. 
By integrating static code structure metrics with dynamic runtime profiling metrics, the approach prioritizes performance testing for commits that are most likely to cause regressions, reducing testing overhead.
\subsubsection*{Synergistic Effects}
\synFeature: the static and dynamic metrics are used as features to train and exploit a multi-object evolutionary algorithm that generated detection rules.
Static metrics capture structural changes, while dynamic metrics validate runtime impact, enabling accurate detection of performance regressions. This dual analysis optimizes hit and dismiss rates, reducing false positives and negatives.

\subsubsection*{Inter-Analysis Workflow}
\workSidebyside: static and dynamic metrics are separately collected and then combined.

\subsubsection*{Mapping-Function Interpretation Structure}
\structModules: the metrics are associated with the commit patches under analysis.

\subsubsection*{Mapping-Function Mechanics}
\mechanicML: the metrics are processed with a multi-object evolutionary algorithm that generated detections rules.

\subsection{Worst-Case Execution Time Testing via Evolutionary Symbolic Execution \cite{aquino_worst-case_2018}}

\subsubsection*{Summary}
This work presents a
hybrid approach combining symbolic execution and evolutionary algorithms to generate worst-case execution time test cases. Symbolic execution indicates and identifies feasible paths (and their path conditions) that can be further analyzed to identify
worst-case executions, while evolutionary algorithms guide the search by providing pre-conditions to be used in the next iteration of symbolic execution.

\subsubsection*{Synergistic Effects}
\begin{itemize}
    \item \synPartitioning~/~\synPartitioningDirect: Symbolic execution indicates feasible paths where to run  worst-case cost analysis.
    \item \synTraversal~/~\synTraversalSeedInputs: The evolutionary algorithm (via crossover and mutation) seeds relevant pre-conditions.
\end{itemize}

\subsubsection*{Inter-Analysis Workflow}
\workFeedback: the symbolic execution stage informs the evolutionary search stage of feasible paths where to run  worst-case cost analysis, and then evolved conditions are fed back to symbolic execution to identify new candidate program paths iteratively.

\subsubsection*{Mapping-Function Interpretation Structure}
\begin{itemize}
    \item \structPaths: the symbolic execution stage identifies sets of feasible paths. \item \structCFG: the evolutionary search stage identifies preconditions for further analysis of the target function. 
\end{itemize}

\subsubsection*{Mapping-Function Mechanics}
\begin{itemize}
    \item \mechanicConstraint: for symbolic execution to provide test cases for worst-case execution analysis.
    \item \mechanicAssociation: the evolutionary search stage associates the target function with preconditions that allow to focalize the next steps of the analysis.
\end{itemize}

\subsection{Deterministic dynamic monitors for linear-time assertions \cite{armoni_deterministic_2006}}

\subsubsection*{Summary}
The paper introduce a framework that integrates formal specs into a program or executable model, in order to enable runtime verification of the specified properties.
This approach aims to achieving semantic consistency by compiling assertions into deterministic monitors for runtime verification. 

\subsubsection*{Synergistic Effects}
\synIntepretability~/~\synIntepretabilityOracle: Once integrated with the formal specs (assertions) the program is able to interpret the validity of the properties at runtime.

\subsubsection*{Inter-Analysis Workflow}
\workCascade

\subsubsection*{Mapping-Function Interpretation Structure}
\structProgram: The monitors are integrated by instrumentation in the program.

\subsubsection*{Mapping-Function Mechanics}
\mechanicAssociation: The monitors are integrated by instrumentation in the program.

\subsection{Directed test generation for effective fault localization \cite{artzi_directed_2010}}

\subsubsection*{Summary}

They introduce an approach for generating test suites to increase fault localization effectiveness when there are no prior test suites of few failing test cases. This leverages concolic execution.
The proposed directed test generation uses path constraint similarity to create tests resembling failing executions. Then spectrum-based fault localization is applied.

\subsubsection*{Synergistic Effects}
\begin{itemize}
    \item \synPartitioning~/~\synPartitioningDirect
\end{itemize}

The tool Apollo uses concrete values from concrete execution to direct the analysis of the SBFL layer.
\subsubsection*{Inter-Analysis Workflow}
\workCascade: concolic execution (driven by similarity with failing execution) then SBFL 
\subsubsection*{Mapping-Function Interpretation Structure}
\structProgram: test cases.

\subsubsection*{Mapping-Function Mechanics}
\mechanicConstraint: constraint solving to produce results of concolic execution.

\subsection{Worst-case execution time analysis approach for safety-critical airborne software \cite{asensio_worst-case_2013}}

\subsubsection*{Summary}
A hybrid WCET analysis approach for safety-critical airborne software by integrating measurement based timing analysis with static structural modeling. The RapiTime tool is used to collect execution time measurements from instrumented source code running on target hardware is combined with a static code model to predict worst case execution paths.

In the dynamic analysis phase, they instrument the code and then execute it to find the execution paths and time.
In the static analysis phase, the collected execution paths and time combine their data across all paths, to find the worst case execution time by recombining executed paths.
\subsubsection*{Synergistic Effects}
\synIntepretability~/~\synIntepretabilityEntities:
Dynamically executed the instrumented code provide runtime-collected semantics of the timing data associated to the executed program paths.
This informs the static analysis that works by combining the information from ther execution paths, to predict worst case execution paths.

\subsubsection*{Inter-Analysis Workflow}
\workCascade
\subsubsection*{Mapping-Function Interpretation Structure}
\structPaths
\subsubsection*{Mapping-Function Mechanics}
\mechanicAssociation

\subsection{Comparison and integration of genetic algorithms and dynamic symbolic execution for security testing of cross-site scripting vulnerabilities \cite{avancini_comparison_2013}}

\subsubsection*{Summary}
This paper introduces an approach to automatically generate security test cases by combining genetic algorithms and concrete symbolic execution. Which means, explore input space heuristically for some iterations, then use concolic execution based on the current test cases form the genetic algorithms, then include the new test cases in the population and iterate.
\subsubsection*{Synergistic Effects}
\synPartitioning~/~\synPartitioningWitness:
    they used the target branches that are derived from taint analysis as adequacy criteria to partition the state space for the test generation.

\subsubsection*{Inter-Analysis Workflow}
\workFeedback:
alternate between search based testing and concolic execution.
\subsubsection*{Mapping-Function Interpretation Structure}
\structProgram: the test suites computed by  search based testing and concolic execution, respectively.
\subsubsection*{Mapping-Function Mechanics}
\begin{itemize}
    \item \mechanicConstraint: used by CSE
    \item \mechanicAssociation: used by GA
\end{itemize}

\subsection{A Technique for Automata-based Verification with Residual Reasoning \cite{azzopardi_technique_2020}}

\subsubsection*{Summary}
Aiming to reduce the resources required by verification, this paper introduces how analysis techniques at
decreasing levels of abstraction  can be combined in a complementary manner through residual analysis, where any useful partial information discovered at a high-level is used to reduce the verification problem, leaving an easier residual problem for lower-level analyses. 
When full verifications are infeasible, they use residual analysis to silence control-flow-automata events, remove dynamic-event-automata transitions, and weaken dynamic-event-automata guards. 
\subsubsection*{Synergistic Effects}
\synRefine~/~\synRefinePrune:
residual analysis prunes the program models, by
identifying parts of the property respected by the program and parts of the program that cannot violate the property.
\subsubsection*{Inter-Analysis Workflow}
\workCascade:  
static analysis first, then runtime verification on the residual problem.
\subsubsection*{Mapping-Function Interpretation Structure}
\structProgram:
Residual analysis transfer the pruned program models to the next stage.
\subsubsection*{Mapping-Function Mechanics}
\mechanicAssociation

\subsection{Regression verification using impact summaries \cite{backes_regression_2013}}

\subsubsection*{Summary}
The paper introduces regression verification techniques by combining static analysis and symbolic execution. It combines the static change impact analysis with symbolic execution to improve scalability when checking the equivalence between program versions or summaries.

\subsubsection*{Synergistic Effects}
\synIntepretability~/~\synIntepretabilityEntities:
    The overall synergy consists in enabling symbolic execution to interpret impacted and 
    unimpacted constraints, to prune the unimpacted ones from path conditions to provide symbolic summaries representative of impact analysis.

\subsubsection*{Inter-Analysis Workflow}
\workCascade: static analysis runs first to compute the impacted set, then symbolic executions use this info to generate summaries
\subsubsection*{Mapping-Function Interpretation Structure}
\structCFG: The former analysis stage computes impacted statements
\subsubsection*{Mapping-Function Mechanics}
\mechanicAssociation

\subsection{Hybrid Static-Dynamic Analysis of Data Races Caused by Inconsistent Locking Discipline in Device Drivers \cite{bai_hybrid_2022}}

\subsubsection*{Summary}
The paper introduce SDILP, a hybrid static-dynamic analysis for detect data races in Linux drivers caused by inconsistent locking  discipline. SDILP integrates static taint analysis (to identify possibly-shared variable accesses, to reduce the number of accesses to be monitored at runtime) and dynamic lockset analysis (to detect races by monitoring locksets and concurrency contexts). The it relies on a further phase of static lockset analysis, to analyze the variables accesses that have the same contexts as the found data races, namely they access the same variables, occur in the same function and hold the same locks, but these accesses are not covered by dynamic analysis. 

\subsubsection*{Synergistic Effects}
\synTraversal~/~\synTraversalSeedSink:
The SDILP applies dynamic lockset analysis only to the variables that static taint analysis identified as candidates for possibly-shared variable accesses, and applies static lockset analysis only to the variables contexts that belong to the outcomes of the dynamic analysis. 

\subsubsection*{Inter-Analysis Workflow}
\workCascade: static taint analysis $\rightarrow$ dynamic lockset  analysis $\rightarrow$ static lockset  analysis 

\subsubsection*{Mapping-Function Interpretation Structure}
\structDF

\subsubsection*{Mapping-Function Mechanics}
\mechanicAssociation

\subsection{Modeling and analyzing the interaction of C and C++ strings \cite{balakrishnan_modeling_2012}}

\subsubsection*{Summary}
The paper introduces heap-aware memory models for C and C++ programs to emphasizing the interaction between both C and C++ null-terminated buffers as strings. Here it combines the abstract interpretation (for scalable property proofs) and SAT-based bounded model checking (for precise bug detection). Via abstract interpretation, the model is analyzed in an attempt to prove that the assertions are never violated, and the safe assertions are removed from the model. Then the remaining model and properties is considered for model checking. 

\subsubsection*{Synergistic Effects}
\synPartitioning~/~\synPartitioningWitness: from the program model construction, using abstract interpretation, the model is analyzed in an attempt to prove that the assertions are never violated, and the safe assertions are removed from the model. Then the remaining sliced model is considered for model checking. 

\subsubsection*{Inter-Analysis Workflow}
\workCascade: abstract interpretation, then Model checking

\subsubsection*{Mapping-Function Interpretation Structure}
\structCFG

\subsubsection*{Mapping-Function Mechanics}
\mechanicAssociation

\subsection{Bidirectional Symbolic Analysis for Effective Branch Testing \cite{baluda_bidirectional_2016}}

\subsubsection*{Summary}
This work combines symbolic execution (forward analysis) and symbolic reachability analysis (backward analysis), aiming to improve branch coverage. Symbolic execution explores feasible paths forward, while symbolic reachability analysis computes weakest preconditions backward to identify infeasible branches and rare execution conditions.

\subsubsection*{Synergistic Effects}
\synPartitioning~/~\synPartitioningWitness:
By combining witness test cases obtained with forward and unreachability proofs obtained with backward analysis, the overall technique aims to confirm feasibility or infeasibility of the program branches.\\
\synPartitioning~/~\synPartitioningCoverage: The backward and forward analysis stages coordinate with each other to extend and refine each other's exploration by mapping their results to the program branches.

\subsubsection*{Inter-Analysis Workflow}
\workFeedback: The technique alternates forward and backward analysis iteratively. 

\subsubsection*{Mapping-Function Interpretation Structure}

\structCFG: The reachability conditions and test cases computed during the analysis are mapped to the program branches. 

\subsubsection*{Mapping-Function Mechanics}

\mechanicConstraint: drives the mapping of reachability conditions and test cases to the branches of the program.

\subsection{EnergyPatch: Repairing Resource Leaks to Improve Energy-Efficiency of Android Apps \cite{banerjee_energypatch_2018}}

\subsubsection*{Summary}
The EnergyPatch tool by Banerjee et al. detects resource leaks in Android applications to improve their energy efficiency. 
To this end, it exploits a first stage of dynamic analysis to build a model of the application based on its event-based structure (called an event-flow graph, EFG), and then it applies on it a static analysis based on the abstract interpretation framework to identify pairs of program locations where a resource is acquired and released, respectively. Then they analyze the program with symbolic execution to detect possible paths in which a resource can be acquired and never released (which means that energy is wasted by keeping the resource active). 
The exploration is guided by the constructed event-flow graph and each of the transitions in the event-flow graph are made symbolic so that in case of event branches all of them are explored.

\subsubsection*{Synergistic Effects}
\synAlarms~/~\synAlarmsDynamic: The approach exploit symbolic execution and runtime analysis onto the program paths that abstract interpretation identified as the ones with potential energy bugs.

\synTraversal~/~\synTraversalTransform: The approach fixes energy bugs incrementally .

\subsubsection*{Inter-Analysis Workflow}
\workFeedback: the approach iterates after fixing the energy bugs incrementally.

\subsubsection*{Mapping-Function Interpretation Structure}
\structPaths: energy bugs and test cases associated to program paths;\\
\structProgram: energy patched programs associated to the original program

\subsubsection*{Mapping-Function Mechanics}
\mechanicAssociation,
\mechanicConstraint

\subsection{Multilevel static analysis for improving program quality \cite{belevantsev_multilevel_2017}}

\subsubsection*{Summary}
This paper proposes a multilevel static analysis system. 
L1 checks the program under analysis by using abstract syntax tree AST walks and intraprocedural dataflow. Other than detecting defects at intraprocedural locations, L1 builds a memory model that tracks pointer operations and integer values using abstract domains and point-to sets.
L2 is interprocedural summary-based analysis, which detects further (and more critical) defects provides  summaries of the functions based on the intraprocedural results of L1;
L3 is a path sensitive analysis, by constructing and solving reachability formulas for program points.

\subsubsection*{Synergistic Effects}

\synIntepretability~/~\synIntepretabilityEntities: L1 collect intraprocedural information and L2 builds function summaries, improving interpretability of program semantics at the next level;

\subsubsection*{Inter-Analysis Workflow}
\workCascade: First and second levels build memory model and function summaries, respectively. Then they use the information to do path sensitive analysis.

\subsubsection*{Mapping-Function Interpretation Structure}

\structCFG: L1 provide intraprocedural information with value analysis and points-to analysis

\structCG: L2 computes function summaries

\subsubsection*{Mapping-Function Mechanics}
\mechanicAssociation: L1 to L2,\\
\mechanicSummary: L2 to L3

\subsection{OpenSAW: Open security analysis workbench \cite{ben_henda_opensaw_2017}}

\subsubsection*{Summary}
The paper introduces OpenSAW, an open, flexible, and scalable framework for dynamic test generation. OpenSAW combines runtime analysis, to compute branch priorities, with dynamic symbolic execution to generate new test cases (which will then be passed to runtime analysis).
\subsubsection*{Synergistic Effects}
\begin{itemize}
\item \synPartitioning~/~\synPartitioningDirect:
This is a generalized analysis that combines dynamic and static analysis. 
Stage 1: Dynamic analysis directs stage 2 by path selection strategies based on branch-priority heuristics,
Stage 2: Static analysis directs stage 1 by indicating new inputs for runtime analysis

\end{itemize}
\subsubsection*{Inter-Analysis Workflow}
\workFeedback 
\subsubsection*{Mapping-Function Interpretation Structure}
\begin{itemize}
    \item \structCG: inputs associated with program functions,
    \item \structPaths: branch priorities associated with program paths. 
\end{itemize}

\subsubsection*{Mapping-Function Mechanics}
\mechanicAssociation,\\
\mechanicConstraint:
new inputs are generated by brach conditions and solving path constraints derived from symbolic execution.

\subsection{Automatic Generation of Path Covers Based on the Control Flow Analysis of Computer Programs \cite{bertolino_automatic_1994}}

\subsubsection*{Summary}
The paper introduces a recursive algorithm called FTPS for automatic generation of path covers for branch testing base on the ddgraph (a reduced flowgraph). The algorithm can be combined with test generation by directing the generation of test cases for the selected path covers. As a feedback, it maintains a table which stores desired or unsuitable combinations of arcs and can be updated interactively as a deeper knowledge
of program functioning is acquired.

\subsubsection*{Synergistic Effects}
\synPartitioning~/~\synPartitioningDirect: FTPS directs the generation of test cases for the selected path covers.

\synRefine~/~\synRefinePrune:
FTPS addresses the infeasible paths by refining the model during recursive path construction. \subsubsection*{Inter-Analysis Workflow}
\workFeedback
\subsubsection*{Mapping-Function Interpretation Structure}
\structPaths
\subsubsection*{Mapping-Function Mechanics}
\mechanicAssociation

\subsection{Architectural verification of black-box component-based systems \cite{bertolino_architectural_2007}}

\subsubsection*{Summary}
This paper introduce a technique that integrates runtime monitoring and model checking, for verifying component-based system at the architectural level. Monitoring collects execution traces from black-box components via middleware instrumentation, addressing observability challenges without access the source code. Those traces allow to represent the behavioral semantics of the program and are then validated using  SPIN for compliance with architectural requirements.  

\subsubsection*{Synergistic Effects}
\synIntepretability~/~\synIntepretabilityArtifacts:
Runtime monitoring allows for relating the behavioral semantics with the architectural requirements, to enable the model checking step.

\subsubsection*{Inter-Analysis Workflow}
\workCascade

\subsubsection*{Mapping-Function Interpretation Structure}
\structProgram

\subsubsection*{Mapping-Function Mechanics}
\mechanicMining

\subsection{Cooperative verifier-based testing with CoVeriTest \cite{beyer_cooperative_2021}}

\subsubsection*{Summary}
The paper proposes a hybrid technique for test-suite generation using cooperative combinations of verification approaches. CoVeriTest iteratively applies different conditional model checkers and allows users to adjust the level of cooperation and to configure individual time limits for each conditional model checker. It interleaves different reachability analyses and exchanges various types of analysis information between analyses. In contrast to existing  approaches, CoVeriTest allows analysists to configure the analyses that will be combined and the level of cooperation, i.e., which information is exchanged. CoVeriTest iteratively executes a configurable sequence of reachability analyses. In each iteration, the analyses are run sequentially, and each analysis in the sequence is limited to its individual but configurable time limit. Moreover, one can configure CoVeriTest to exchange different types of information gained during a reachability analysis, e.g., which paths are infeasible or have already been explored or which abstraction level to use. \subsubsection*{Synergistic Effects}
\synPartitioning~/~\synPartitioningCoverage:
CoVeriTest's entire workflow is built around satisfying the analysis goals incrementally and steering the analysis towards unexplored program paths.

\subsubsection*{Inter-Analysis Workflow}
\workFeedback cycles through a sequence of analysis runs. Each analysis gets information on the Abstract Reachability Graphs (ARGs) built at previous cycles.
\subsubsection*{Mapping-Function Interpretation Structure}
\structProgram: model a program by Abstract Reachability Graphs and conditions (intended as automata that accept the verified program paths)

\subsubsection*{Mapping-Function Mechanics}
\mechanicAssociation

\subsection{Explicit-state software model checking based on CEGAR and interpolation \cite{beyer_explicit-state_2013}}

\subsubsection*{Summary}
The paper proposes an approach that integrates abstraction and interpolation-based refinement into an explicit-value analysis for software model checking (CEGAR).
\subsubsection*{Synergistic Effects}

\synAlarms~/~\synAlarmsStatic:
Refine the program model used for model checking-Incorporate relevant details

\synRefine~/~\synRefineIncorporate: To refine the model
\subsubsection*{Inter-Analysis Workflow}
\workFeedback

\subsubsection*{Mapping-Function Interpretation Structure}
\structPaths,
\structCG,
\structProgram
\subsubsection*{Mapping-Function Mechanics}
\mechanicInterpolation,
\mechanicConstraint

\subsection{The Clara framework for hybrid typestate analysis \cite{bodden_clara_2012}}

\subsubsection*{Summary}
Clara, a framework for hybrid typestate analysis that combines static and dynamic analysis to verify typestate properties in java programs. The typestate properties define permissible operations on objects based on their internal states. Traditional static analyses provide states 
Clara brides the gap by integrating AspectJ-based runtime monitors with static typestate analyses through annotated dependency state machine. These capture the monitored property as finite state machine, to reason the instrumentation points. 
Clara uses three static analyses --Quick Check, Orphan-shadows, and Nop-shadows.

\subsubsection*{Synergistic Effects}
\synAlarms~/~\synAlarmsStatic:
This approach leverages static typestate analysis to eliminate irrelevant instrumentation points, up to possibly confirming property compliance statically, or at least reducing runtime overhead. 

\synAlarms~/~\synAlarmsDynamic:
By runtime checks, Clara can report confirmed alarms on typestate properties.

\subsubsection*{Inter-Analysis Workflow}
\workCascade: executes analysis sequentially.

\subsubsection*{Mapping-Function Interpretation Structure}
\structCFG: use intra-procedural control flow graphs.

\subsubsection*{Mapping-Function Mechanics}
\mechanicAssociation: analysis computes equivalence relations between states based on continuation semantics.

\subsection{Finding programming errors earlier by evaluating runtime monitors ahead-of-time \cite{bodden_finding_2008}}

\subsubsection*{Summary}
This paper introduces a hybrid approach to detect violations in programs. Using Tracematches, it first specifies runtime properties and use flow-insensitive whole program analysis to find possible failure points in the program (to be instrumented for runtime checking). Then it applies static flow-sensitive analysis and decision trees 
to prune a first set of false positives, and statically maps the remaining alarms to feature vectors to apply decision trees trained on a label dataset of false positives, to filter out the remaining false positives.  

\subsubsection*{Synergistic Effects}
\synAlarms~/~\synAlarmsStatic: Flow-sensitive static analysis and decision trees (ML) are used to filter out false positives. 

\subsubsection*{Inter-Analysis Workflow}
\workCascade: flow-insensitive analysis, followed by flow-sensitive analysis augmented with machine learning 

\subsubsection*{Mapping-Function Interpretation Structure}
\structCFG: analysis results mapped to failure points in the program.

\subsubsection*{Mapping-Function Mechanics}
\mechanicAssociation

\subsection{Fuzzing symbolic expressions \cite{borzacchiello_fuzzing_2021}}

\subsubsection*{Summary}

The paper combines symbolic execution and coverage-guided fuzzing to improve scalability and coverage. It introduces FUZZY-SAT, an approximate constraint solver that leverages fuzzing-mutations to solve symbolic constraints that are generated during concolic execution. 

\subsubsection*{Synergistic Effects}
\synTraversal~/~\synTraversalSeedInputs:
FUZZY-SAT uses symbolic expressions and the input related to the path prefix that led concolic execution to generate each symbolic expression, to guide fuzzing mutations for each branch condition. 
\synPartitioning~/~\synPartitioningDirect: the results from FUZZY-SAT steer concolic execution to progress

\subsubsection*{Inter-Analysis Workflow}
\workFeedback: concolic analysis feeds input-condition pairs for FUZZY-SAT, while the inputs generated with FUZZY-SAT steer concolic execution to progress.

based reasoning
\subsubsection*{Mapping-Function Interpretation Structure}
\structCG: FUZZY-SAT associates new inputs to program functions. Concolic execution associates input-condition pairs to program function.

\subsubsection*{Mapping-Function Mechanics}
\mechanicAssociation

\subsection{Columbus: Android App Testing Through Systematic Callback Exploration \cite{bose_columbus_2023}}

\subsubsection*{Summary}
The paper proposes a callback-driven testing technique for Android apps that automatically identifies callbacks and generates valid inputs. The tool COLUMBUS, a callback-driven testing technique that employs two strategies to eliminate the need for human involvement: (i) it automatically identifies callbacks by simultaneously analyzing both the Android framework and the app under test; (ii) it uses a combination of under-constrained symbolic execution (primitive arguments) and type-guided dynamic heap introspection (object arguments) to generate valid and effective inputs, by symbolizing
the primitive arguments of a callback, and performing an under-constrained symbolic execution to generate the possible values of those arguments.

\subsubsection*{Synergistic Effects}
\synIntepretability~/~\synIntepretabilityAPI:
COLUMBUS augments the semantics of APIs by statically identifying which program functions correspond  to callback from the Android framework.
\subsubsection*{Inter-Analysis Workflow}
\workCascade: statically analysis for identifying callbacks and dependencies, and then generate arguments via under-constrained symbolic execution.
\subsubsection*{Mapping-Function Interpretation Structure}
\structCG: the initial, static analysis stage identifies callbacks.

\subsubsection*{Mapping-Function Mechanics}
\mechanicAssociation

\subsection{Complementary test selection criteria for model-based testing of security components \cite{botella_complementary_2019}}

\subsubsection*{Summary}
This paper introduces model-based testing (MBT) methodology for validating security components by combining static structural coverage with dynamic criteria. The static criterion ensures behavioral coverage of UML-OCL models, while dynamic criteria focus on TOCL (temporal OCL) and Test Purposes.

\subsubsection*{Synergistic Effects}

\synReports: the test cases from the two criteria complement each other to provide a more thorough test suite.

\subsubsection*{Inter-Analysis Workflow}
\workSidebyside: the two criteria are used to generate complementary test cases.

\subsubsection*{Mapping-Function Interpretation Structure}
\structProgram: Test cases for the program under test

\subsubsection*{Mapping-Function Mechanics}
\mechanicConstraint

\subsection{Model-based testing from input output symbolic transition systems enriched by program calls and contracts \cite{boudhiba_model-based_2015}}

\subsubsection*{Summary}

This paper proposes extending the IOSTS (Input Output Symbolic Transition System) framework by integrating the calls for the actual program under test in the IOSTS. Then they can generate test cases with symbolic execution and compute new contracts for the target program. 

\subsubsection*{Synergistic Effects}
\synIntepretability~/~\synIntepretabilityArtifacts: they relate the target program with the IOSTS model, aiming to improve test generation and computation of contracts.

\subsubsection*{Inter-Analysis Workflow}
\workCascade

\subsubsection*{Mapping-Function Interpretation Structure}
\structProgram: the extend IOTS model is associated with the program under analysis

\subsubsection*{Mapping-Function Mechanics}
\mechanicAssociation

\subsection{Complete Property-Oriented Module Testing \cite{bruning_complete_2023}}

\subsubsection*{Summary}
The paper presents a novel approach to complete property-oriented white box module testing using fuzzing, model learning, and model checking to verify LTL properties.
The approach starts with fuzzing, to learn a set of equivalence classes and test them, aiming to construct a first version of the model (with an adaptation of the L* algorithm). Then it iterates 
through model learning and model checking until either 1) revealing a failure wrt the target LTL property, or 2) proving the conformance between the implementation and the model learned so far (which supports the conformance with the target LTL property). 

\subsubsection*{Synergistic Effects}
\synRefineIncorporate: The results from fuzzing allow for learning refined  first version of the model (with an adaptation of the L* algorithm) used in the model learning phase.

\subsubsection*{Inter-Analysis Workflow}
\workCascade: fuzzing allows for identifying equivalence classes, and build a model to get the model learning phase started. 
\subsubsection*{Mapping-Function Interpretation Structure}
\structProgram: the learned model is passed to the model learning analysis stage.

\subsubsection*{Mapping-Function Mechanics}
\mechanicMining: model learning via adaptation of the L* algorithm 

\subsection{Combining symbolic execution and search-based testing for programs with complex heap inputs \cite{braione_combining_2017}}

\subsubsection*{Summary}
This paper introduces SUSHI, a tool to automatically generate test cases for programs with complex data structures as inputs by combining symbolic execution and search-based testing. Symbolic execution is used to compute path conditions that characterize the dependencies between the program paths and the input structures and convert the path conditions to optimization problems that they solve with search-based techniques to produce sequences of method calls that instantiate those inputs.
\subsubsection*{Synergistic Effects}
\synPartitioning~/~\synPartitioningCoverage:
Use symbolic execution to model a sufficient set of program paths to be tested for accomplishing branch coverage.

\subsubsection*{Inter-Analysis Workflow}
\workCascade
\subsubsection*{Mapping-Function Interpretation Structure}
\structPaths:  path conditions associated to program paths
\subsubsection*{Mapping-Function Mechanics}
\mechanicAssociation: path conditions associated to program paths

\subsection{Symbolic path cost analysis for side-channel detection \cite{brennan_symbolic_2018}}

\subsubsection*{Summary}
The paper proposed a tool called CoCo-Channel a technique for detecting side-channel vulnerabilities by combining static analysis and symbolic execution.
Static taint analysis allows for annotating control-flow structures that depend (or not depend) on input secrets and have given costs (according a cost model). Then, symbolic execution exploit this information to formulate queries (to be solved with a constraint solver) on the presence of side channels.
\subsubsection*{Synergistic Effects}
\synIntepretability~/~\synIntepretabilityArtifacts:
The first stage of static taint analysis makes the control-flow structures interpretable with respect to dependencies on external information, that is, input secrets and cost models.

\subsubsection*{Inter-Analysis Workflow}
\workCascade: static taint analysis for annotating control-flow structures is followed by symbolic execution to formulate queries (to be solved with a constraint solver) on the presence of side channels.

\subsubsection*{Mapping-Function Interpretation Structure}
\structCFG:  They map symbolic cost expressions and secret dependencies to control-flow-graph components.

\subsubsection*{Mapping-Function Mechanics}
\mechanicAssociation

\subsection{Testing concurrent programs on relaxed memory models \cite{burnim_testing_2011}}

\subsubsection*{Summary}
The tool called RELAXER integrates predictive dynamic analysis with active testing to uncover concurrency bugs under relaxed memory models. First, it analyses a sequentially consistent execution to predict potential happens-before cycles using dynamic race detections. These cycles indicate possible violations of sequential consistency.
Then, the tool actively controls thread scheduling and simulates operational semantics of memory models to confirm prediction violations.
\subsubsection*{Synergistic Effects}
\synAlarms~/~\synAlarmsDynamic: the tool predicts potential violations and confirms them through controlled execution and reporting real bugs.
First, predictive dynamic analysis identifies potential happens-before cycles, while active testing confirms them under controlled executions. 
\subsubsection*{Inter-Analysis Workflow}
\workCascade: \subsubsection*{Mapping-Function Interpretation Structure}
\structDF: Potential data races.
\subsubsection*{Mapping-Function Mechanics}
\mechanicAssociation.

\subsection{Combining static analysis error traces with dynamic symbolic execution (experience paper) \cite{busse_combining_2022}}

\subsubsection*{Summary}
This paper explores a technique that integrates static analysis error traces with dynamic symbolic executions to confirm potential bugs. By static analysis, generate partial traces annotated with conditions that lead to potential bugs and then use KLEE for symbolic execution by using the partial paths and the associated conditions.

\subsubsection*{Synergistic Effects}
\synAlarms~/~\synAlarmsDynamic
\subsubsection*{Inter-Analysis Workflow}
\workCascade.
\subsubsection*{Mapping-Function Interpretation Structure}
\structPaths: potential bugs are associated with (partial) paths with annotated conditions
\subsubsection*{Mapping-Function Mechanics}
\mechanicAssociation

\subsection{Execution generated test cases: How to make systems code crash itself \cite{cadar_execution_2005}}

\subsubsection*{Summary}
This paper introduces Execution Generated Testing (EGT), a technique that does symbolic execution, while generating test cases to detect defects on the executed paths.
\subsubsection*{Synergistic Effects}
\synPartitioning~/~\synPartitioningDirect: symbolic execution generates input-space partitions and direct test execution on those partitions.

\subsubsection*{Inter-Analysis Workflow}
\workCascade.
\subsubsection*{Mapping-Function Interpretation Structure}
\structPaths.
\subsubsection*{Mapping-Function Mechanics}
\mechanicConstraint.

\subsection{Hybrid Program Dependence Approximation for Effective Dynamic Impact Prediction \cite{cai_hybrid_2018}}

\subsubsection*{Summary}
This paper introduces Diver, a hybrid dynamic impact analysis technique that combines static program dependence graphs and dynamic method-execution events to identify runtime method-level dependencies. 
Dynamic analysis of available test cases detects exceptions that are not handled in methods. 
Static analysis builds a dependency program-level dependency graph by computing data dependencies and control dependencies, considering the unhandled exceptions observed in the previous phase. 
Then, at runtime, the program-level dependency graph allows to better report the dynamic dependencies between the executed methods.
\subsubsection*{Synergistic Effects}
\synRefine~\~\synRefineIncorporate: refines the program-level dependency graph based on incorporating exceptions flow observed at runtime.

\synIntepretability~\~\synIntepretabilityEntities: the incorporate program-level dependency graph allows for interpreting the dependencies observed at runtime.

\subsubsection*{Inter-Analysis Workflow}
\workCascade: 

\subsubsection*{Mapping-Function Interpretation Structure}
\structCG: unhandled exceptions associated to methods.
    
    \structDF: the program-level dependency graph represents data-flow dependency relations
    
\subsubsection*{Mapping-Function Mechanics}
\begin{itemize}
    \item \mechanicAssociation.
\end{itemize}

\subsection{Model-Based Testing Under Parametric Variability of Uncertain Beliefs \cite{camilli_model-based_2020}}

\subsubsection*{Summary}
This paper introduces a methodology that integrates parametric model checking and online model-based testing algorithms to handle uncertainty in software systems. At design-time (static), parametric model checking analysis uses a Markov Decision Processes (MDP) model to pre-compute verification conditions for PCTL requirement satisfaction. At run-time (dynamic), the MBT algorithm explores the SUT and applies Bayesian inference to refine parameter estimates.
\subsubsection*{Synergistic Effects}
\synRefine~/~\synRefineIncorporate: the design time, parametric model checking computes hyper-rectangles satisfying PCTL requirements, while dynamic Bayesian inference refines these models with runtime evidence.
\subsubsection*{Inter-Analysis Workflow}
\workCascade.
\subsubsection*{Mapping-Function Interpretation Structure}
\structProgram: the inferred model is associated to the program.
\subsubsection*{Mapping-Function Mechanics}
\mechanicAssociation.

\subsection{Automatically identifying critical input regions and code in applications \cite{carbin_automatically_2010}}

\subsubsection*{Summary}
The paper proposes an approach that automatically groups subsets of input bytes into fields and classifies the fields (and corresponding regions of code) as critical or forgiving (non-critical).
It starts by monitoring a set of representative executions to learn a baseline for normal behavior and dependencies between input fields and code regions.
Then is exploits fuzzing based on the learned fields and analyzed the execution traces, aiming to establish which traces differ from normal behavior and thus pinpoint critical fields and regions.
\subsubsection*{Synergistic Effects}
\synPartitioning~/~\synPartitioningWitness:
Snap allows for the fuzzing stage to associate execution results with the input fields and related code dependencies identified by the initial runtime analysis stage, and exploit such knowledge to distinguish critical and non-critical brhaviors.

\subsubsection*{Inter-Analysis Workflow}
\workCascade

\subsubsection*{Mapping-Function Interpretation Structure}
\structDF: relations between input fields (subsets of input bytes) and regions in the code.

\subsubsection*{Mapping-Function Mechanics}
\mechanicAssociation

\subsection{Behind the scenes in SANTE: A combination of static and dynamic analyses \cite{chebaro_behind_2014}}

\subsubsection*{Summary}
The paper presents a tool SANTE (Static ANalysis and TEsting) that combines static and dynamic analyses for verification of C programs. 
They showed how several tools based on heterogeneous techniques such as abstract interpretation, dependency analysis, program slicing, constraint solving and test generation can be combined within one tool. The technique works as follows:
1. It asks the VALUE plugin to perform the value analysis on the analyzed program. The results of this analysis are threatening statements in the programs. 
2. It partitions the alarms into subsets, each to be used as slicing criterion. Dependency analysis is called if an advanced slicing option (taking into account alarm dependencies) was provided. Dependency analysis uses directly the results of value analysis without calling it again. The SLICING plugin is then asked to simplify the program with respect to each subset of alarms. 
3. Each sliced program is then analyzed with PATHCRAWLER. The diagnostic is constructed.

\subsubsection*{Synergistic Effects}
\synAlarms~/~\synAlarmsDynamic:
The core synergistic effect of SANTE line in its ability to confirm static-analysis alarms through dynamic test generation, thereby reducing false positives and improving diagnostic precision. Test generation allows SANTE to confirm alarms that correspond to feasible paths, and discard false alarms if all-path test generation terminates without activating the given alarm.

\synTraversal~/~\synTraversalTransform: program slicing reduces the program to focalize the concolic traversals on the statically identified alarms. 

\subsubsection*{Inter-Analysis Workflow}
\workCascade: perform static value analysis first, then slicing and finally dynamic test generation in a sequential pipeline.

\subsubsection*{Mapping-Function Interpretation Structure}
\structCFG: alarms are associated with program instructions.

\structProgram: sliced programs are associated with the program under analysis.

\subsubsection*{Mapping-Function Mechanics}
\mechanicAssociation, \mechanicConstraint

 \subsection{Synthesising interprocedural bit-precise termination proofs \cite{chen_synthesising_2016}}
\subsubsection*{Summary}
This paper proposes a new technique to compute sufficient preconditions, which can then be used to identify pre-states that guarantee procedure termination. To do so, they propose an algorithm based on a cascading combination of two types of analysis, namely a forward and a backward analysis. The proposed algorithm can then be used to analyze any non-recursive program. The first step is an over-approximating forward analysis focused on determining the over-approximating call context and loop invariants for every procedure that is called from the program's entry point and its callees, analyzing them recursively. After that, an under-approximating backward analysis is started, again from the program's entry point. During this analysis, again for every function that is invoked by the entry point or one of its callees, considering them recursively, an under-approximating call context is determined. Both the under-approximating and over-approximating call contexts, along with other information, are then used to determine the sufficient preconditions that guarantee program termination. Finally, the result comes from picking the preconditions associated with the program's entry point. The study then verifies effectiveness of the technique by using a prototype developed to test C programs, noting that the results have been deemed satisfactory with programs with up to 5k LoC.

\subsubsection*{Synergistic Effects}
\synRefine~/~\synRefineIncorporate: The Synergistic Effect of the combination is that it allows analysis of programs with a much higher LoC than existing tools, along with being able to identify more cases of terminating programs. The obtained definitions for the formal mathematical formulas that help with identifying termination conditions leverage, in fact, both over-approximating and under-approximating information, which likely leads to them being more refined.

\subsubsection*{Inter-Analysis Workflow}
\workCascade: The results of the forward analysis in the form of summaries and invariants are used in the backward analysis phase.

\subsubsection*{Mapping-Function Interpretation Structure}
\structCG: The forward analysis stage augments the information of the various functions that are part of the call graph by adding two additional properties: summaries and invariants.

\subsubsection*{Mapping-Function Mechanics}
\mechanicAssociation: Every function is essentially associated directly with itself.

\subsection{STAR: Stack trace based automatic crash reproduction via symbolic execution \cite{chen_star_2015}}
\subsubsection*{Summary}
The paper presents STAR, a framework that can automatically reproduce crashes based on stack traces and generate test cases to exercise them. STAR initially processes the bug report to extract the stack trace and other information about the crash; then, the stack trace is processed in order to infer information about the conditions that lead to a crash; at this point, a backward symbolic execution algorithm is used to determine how to trigger the crash conditions at the method entry points, essentially determining a set of preconditions that will trigger the target crash; finally, the computed preconditions are used as inputs for a test case generation framework, focused on generating test inputs that will trigger the crash when given to the method under examination. In particular, this last step is made up of more internal steps: initially forward symbolic execution is used to compute intra-procedural method summaries (i.e., the method semantics represented in propositional logic), then they are combined in inter-procedural method summaries, and finally a triggering method sequence is deduced by confronting the collected summaries with the preconditions computed previously.

\subsubsection*{Synergistic Effects}
\begin{itemize}
\item \synAlarms~/~\synAlarmsDynamic: The first step computes the weakest preconditions for a given stack trace, which are then handed over to the second step in order to generate test cases that verify the preconditions and can replicate the detected crash.
\item \synIntepretability~/~\synIntepretabilityEntities: The second stage's inner components compute function summaries to aid the interpretation of the behavior.
\end{itemize}

\subsubsection*{Inter-Analysis Workflow}
\workCascade: The results of the backward symbolic execution are used to determine the preconditions that are then required by the test engine to generate the target test cases. The forward symbolic execution, on the other hand, determines method summaries which the test engine will then use to generate the target test cases.

\subsubsection*{Mapping-Function Interpretation Structure}
\structCG: The backward symbolic execution communicates with the test engine by providing preconditions associated to the target functions that need to be verified in order for the crash to be triggered.

\subsubsection*{Mapping-Function Mechanics}
\begin{itemize}
\item \mechanicAssociation: There is no further mapping being done between the first two stages, as each operate on essentially the same \structCG,
\item \mechanicSummary: The second stage computes method summaries which are then used to interpret the semantics of the methods during test generation.
\end{itemize}

\subsection{Have: Detecting atomicity violations via integrated dynamic and static analysis \cite{chen_have_2009}}

\subsubsection*{Summary}
HAVE is a hybrid approach that combines static and dynamic analysis in order to detect otherwise hard to identify atomicity violations in software. In particular, HAVE is made up of five components: a static analyzer, a dynamic monitor, an instrumentation tool, a speculator, and a detector. The source code is first analyzed statically through the use of a static analyzer to build an intra-procedural static summary tree. The source code is then instrumented to intercept certain specific events (such as synchronization monitors), which are then communicated to the dynamic monitor during the execution phase of the program. The dynamic monitor then uses these events to generate dynamic trees. Finally both static trees and dynamic trees are merged into a single type of tree, dubbed the hybrid tree. It is to be noted that all of these trees are essentially control-flow graphs that have been extended with additional information (such as synchronization lock information). The hybrid trees are then used by the speculator, which performs a speculative execution (i.e. a theoretical execution with concrete values) of the branches of the hybrid tree that have no associated dynamic events, recording the concrete events that result from this. Finally the detector runs an algorithm devoted to identifying \textit{conflict-edge}s, meaning edges connecting nodes that might conflict with each other. These edges are then used to identify if any atomic violations might occur.

\subsubsection*{Synergistic Effects}
\begin{enumerate}
\item \synFlow: SA + DM: The two analysis phases are merged to obtain a hybrid tree.
\item \synRefine~/~\synRefineIncorporate: SA + DM $\rightarrow$ S: The speculator aims to incorporate new details in the target model, in the form of concrete events that can be associated to the various elements of the hybrid tree, even if no associated dynamic event has been observed.
\end{enumerate}

\subsubsection*{Inter-Analysis Workflow}
\begin{enumerate}
\item \workSidebyside: SA + DM: The two analyses are executed independently of one another and the results are then combined by the detector into a hybrid tree.
\item \workCascade: SA + DM $\rightarrow$ S: The output of two analyses that are merged are then provided to the speculative execution.
\end{enumerate}

\subsubsection*{Mapping-Function Interpretation Structure}
\begin{enumerate}
\item \structCFG: SA + DM: The two analyses communicate through a tree structure representing in both cases the summary tree of the executed method, one being the static one, and the other being the one with dynamic events.
\item \structCFG: SA + DM $\rightarrow$ S: The tree structure derived from the two techniques is provided to the speculative execution. 
\end{enumerate}

\subsubsection*{Mapping-Function Mechanics}
\begin{enumerate}
\item \mechanicAssociation: SA + DM: The events are mapped directly between the two trees in order to build the hybrid tree.
\item \mechanicAssociation: SA + DM $\rightarrow$ S: The speculative step is enriching the hybrid tree obtained from the previous two techniques.
\end{enumerate}

\subsection{Automatically Distilling Storyboard With Rich Features for Android Apps \cite{chen_automatically_2023}}

\subsubsection*{Summary}
The paper proposes a technique, called StoryDistiller, with the goal of automatically generating Android app storyboards from a combination of static and dynamic analysis. In particular, an initial static analysis step is performed to extract an Activity transition graph (ATG) from the CFG. Following this, every activity in the graph is launched and its UI analyzed to augment the ATG obtained from the previous step. Finally the resulting ATG is then used for feature extraction without further program analysis techniques.

\subsubsection*{Synergistic Effects}
\synRefine~/~\synRefineIncorporate: The combination of the two techniques aims to incorporate details related to which components can trigger which activity into the model initially obtained by the static analysis phase.

\subsubsection*{Inter-Analysis Workflow}
\workCascade: The two analysis steps are executed sequentially without feedback loops.

\subsubsection*{Mapping-Function Interpretation Structure}
\structGUI: The communication between the two techniques occurs at the level of the activities that make up the GUI of the targeted application.

\subsubsection*{Mapping-Function Mechanics}
\mechanicAssociation: The GUI entities are shared with identity-style mechanics.

\subsection{Boosting static analysis accuracy with instrumented test executions \cite{chen_boosting_2021}}
\subsubsection*{Summary}
The paper proposes Dynaboost as a technique that uses dynamic information gained from test case executions to prioritize bug reports obtained by a static analyzer, aiming to reduce the case of false positives and limitations of two already existing tools: Sparrow and Bingo. Initially, the code is statically analyzed, gathering identified alarms, in the form of dataflow behaviours, and intermediate conclusions, along with modeling a derivation graph connecting the various alarms, and a Bayesian network is built based on these inferences. Then the program is instrumented based on the static analysis output through a tool called SDTransfer, which is then executed on a set of test cases. The results of this execution are examined by another tool, called DSTransfer, which determines an initial feedback representing the dataflow facts that can be observed. Finally the various alarms are reported to the user in an interactive fashion, allowing each of them to be marked as correct or invalid. The initial and final step are said to be common in other works, making the paper's relevant contribution the introduction of SDTransfer and DSTransfer.

\subsubsection*{Synergistic Effects}
\synAlarms~/~\synAlarmsDynamic: The Synergistic Effect is strengthening the output of the two analysis processes, by leveraging their complementary characteristics. In fact, the static analysis is useful to provide an upper bound through thorough exploration of the graph, whereas the dynamic portion focuses more on actual program behavior that can be verified through executions.

\subsubsection*{Inter-Analysis Workflow}
\workCascade: The information from the static analysis is passed to the dynamic analysis but not vice versa.

\subsubsection*{Mapping-Function Interpretation Structure}
\structDF: The communication between the static and dynamic analysis is made up of a graph of alarms represented by the paths needed to trigger them, which is then used to instrument the code so that the dynamic component is able to empirically verify them.

\subsubsection*{Mapping-Function Mechanics}
\mechanicAssociation: The static analysis is used to instrument the original source code, and the instrumented program is then used to execute a given test suite and extract dynamic information from it.

\subsection{Improving regression test efficiency with an awareness of refactoring changes \cite{chen_improving_2018}}

\subsubsection*{Summary}
The paper proposes RIT with the goal of making validation changes more efficient. In particular, the technique can be divided into three steps, and requires two versions of the program to be examined: the original version and the one that underwent refactoring.

In the first step, the available tests are executed against the original version of the program to dynamically extract the call graph referencing which methods are invoked by the tests. These call graphs are then correlated with the refactoring changes to identify a subset of tests that are deemed to be potentially affected by the refactoring process.

In the second step, data-flow analysis is executed on each of the selected tests to isolate the statements that can impact each of the assertions identified in the tests, using this result to build so-called test slices (i.e. the equivalent of program slices, except applied to the test cases themselves). The resulting output is thus a set of minimized test cases focused specifically on verifying assertions that were potentially impacted by the refactoring effort.

Finally, each test slice is executed in the last step of the technique, once again to dynamically obtain the resulting call graph. This second call graph is compared with the one obtained in the first step, identifying the atomic changes that lead to differences in behavior between the two suites, which are reported to the developer for analysis.

\subsubsection*{Synergistic Effects}
\begin{enumerate}
\item \synPartitioning~/~\synPartitioningDirect: Dynamic $\rightarrow$ Data-Flow Analysis:
The call graph obtained by the first dynamic analysis stage is used to determine which tests of the given test suite must be subjected to data-flow analysis.
\item \synTraversal~/~\synTraversalTransform: Dynamic + Data-Flow Analysis $\rightarrow$ Dynamic: The test slice obtained from the partner analysis stage is provided for execution to the third stage, while the initial call graph is provided for comparison.
\end{enumerate}

\subsubsection*{Inter-Analysis Workflow}
\begin{enumerate}
\item \workCascade: Dynamic $\rightarrow$ Data-Flow Analysis: The call graph is given to the partner analysis stage without means of feedback.
\item \workCascade: Dynamic + Data-Flow Analysis $\rightarrow$ Dynamic: The sliced test and the original call graph are provided to the second dynamic analysis stage in cascading fashion.
\end{enumerate}

\subsubsection*{Mapping-Function Interpretation Structure}
\begin{enumerate}
\item \structCG: Dynamic $\rightarrow$ Data-Flow Analysis: The test methods themselves and their statements are used to communicate between the two techniques.
\item \structCG + \structProgram: Dynamic + Data-Flow Analysis $\rightarrow$ Dynamic: The sliced test and the original call graph are used in their entirety to communicate between the partner analysis stages.
\end{enumerate}

\subsubsection*{Mapping-Function Mechanics}
\begin{enumerate}
\item \mechanicAssociation: Dynamic $\rightarrow$ Data-Flow Analysis: The call graph is provided to the partner analysis stage with identity-style mechanics.
\item \mechanicAssociation: Dynamic + Data-Flow Analysis $\rightarrow$ Dynamic: The full program and call graph are provided as-is to the partner analysis stage.
\end{enumerate}

\subsection{Grey-Box Concolic Testing on Binary Code \cite{choi_grey-box_2019}}
\subsubsection*{Summary}
This paper proposes a variant of concolic testing dubbed "grey-box concolic testing", with the goal of combining the best elements of both white-box testing, of which concolic testing is an example, and grey-box testing. In particular, they state the behavior of the proposed technique is similar to the usual concolic testing behavior, except for the fact that they do not need to rely on SMT solving. In particular, they note that the difference resides on the fact that they maintain a subset of approximate path constraints for every byte present in the input, which allows them to compare distinct input values even if they take the same execution path. It is worth noting that the paper does not really mention the techniques used explicitly, likely because the difference is essentially in the way the second interaction is done, rather than on the technique itself.

\subsubsection*{Synergistic Effects}
\synPartitioning~/~\synPartitioningDirect: The aim of concolic execution is to increase the efficiency and the effectiveness of traditional symbolic execution in exploring the path space of a program, usually for the sake of generating test cases that execute the program paths explored thereby.

\subsubsection*{Inter-Analysis Workflow}
\workFeedback: Each time a new test case is generated it can be used to derive new constraints that are used to potentially generate new combinations of path conditions, that will once again lead to the creation of new test cases. This is the classic concolic execution behavior.

\subsubsection*{Mapping-Function Interpretation Structure}
\begin{enumerate}
\item \structPaths: D $\rightarrow$ S: Execution traces associated to corresponding program paths.
\item \structCG: D $\leftarrow$ S: The custom algorithm is used to determine inputs that must be fed to the target function.
\end{enumerate}

\subsubsection*{Mapping-Function Mechanics}
\begin{itemize}
\item \mechanicAssociation: D $\rightarrow$ S: Execution traces associated to corresponding program paths.
\item \mechanicConstraint: D $\leftarrow$ S: An approximate-path-solving algorithm is used to identify valid input ranges that can then be used to trigger not-yet-executed program paths.
\end{itemize}

\subsection{Refining Model Checking by Abstract Interpretation \cite{cousot_refining_1999}}

\subsubsection*{Summary}
This work combines forward abstract interpretation, backward abstract interpretation and model checking. Forward abstract interpretation and backward abstract interpretation interact to compute a suitable finite abstraction of the target program, to enable model checking to prove properties conclusively. The model in incrementally built and new states are built step by step. When abstraction is unsound the combined abstract interpretation techniques are used in order to reduce; on-the-fly; the concrete state space to be searched by model-checking. 

\subsubsection*{Synergistic Effects}
\synRefine~/~\synRefineIncorporate~/~\synRefinePrune: Forward abstract interpretation and backward abstract interpretation collaborate to compute a suitable finite abstraction of the target program, enabling effective model checking based on the computed abstraction. 

\subsubsection*{Inter-Analysis Workflow}
\begin{enumerate}
    \item \workFeedback: between forward abstract interpretation and backward abstract interpretation.
    \item \workCascade: as model checking uses the abstraction computed with abstract interpretation.
\end{enumerate}

\subsubsection*{Mapping-Function Interpretation Structure}
\structProgram: an abstract model of the program is propagated through the analysis stages.

\subsubsection*{Mapping-Function Mechanics}
\mechanicAssociation: the computed abstract model represents the (suitably abstracted) behavior of the program. 

\subsection{Just test what you cannot verify! \cite{czech_just_2015}}

\subsubsection*{Summary}
This paper proposes a technique that aims to combine the verification and testing activities in an effective way, by ensuring that testing is executed solely on the parts of the programs that cannot be verified formally. In particular, they call this the \emph{residual program}. Initially they perform conditional model checking, where the program is run through a verifier to obtain both a partial verification and a condition, meaning information on what the verifier checked against. This condition is then used to obtain the residual program and this can be done in one of two different ways, which the paper describes: (1) use the condition directly to eliminate path conditions that only lead to executions that have been proven safe, essentially obtaining a new control flow automaton that can execute only unsafe paths; (2) use program slicing with unproven assertions as the slicing criteria in order to obtain slices that can exercise those assertions. After that, the residual program is tested with any additional testing technique. 
The relevant program analysis for our survey refers to the  employment of program slicing in order to obtain a new program that can exercise only the unverified assertions. The obtained program slices are then used to derive formal assertions that can exercise them.

\subsubsection*{Synergistic Effects}
\synPartitioning~/~\synTraversalTransform: 
They use slicing to obtain a second piece of software sliced with respect to the subset of assertions that could not be proven verified through formal verification, and feed the sliced program to the stage of the technique that can derive the missing formal assertions.

\subsubsection*{Inter-Analysis Workflow}
\workCascade.

\subsubsection*{Mapping-Function Interpretation Structure}
\structProgram: 
The first analysis stage propagates a program slice that can exercise the unverified assertions.

\subsubsection*{Mapping-Function Mechanics}
\mechanicAssociation.

\subsection{Generating test cases for specification mining \cite{dallmeier_generating_2010}}
\subsubsection*{Summary}
This paper proposes a technique that aims to generate test cases based on what they call \emph{specification mining}, meaning extracting high-level software specifications based on the behavior of the software under test. In particular, the technique is based around a two step approach. First off, they leverage a previously self-developed tool to mine type-state specifications, which are represented as FSA indicating the possible object states along with their transitions. The details of this first technique are described in other papers. Then, this initial FSA containing only \emph{observed} transitions gets augmented by generating test cases aiming to build transitions from every state to every other state and then executing them, gathering the resulting transition and states into an augmented FSA. In particular, every valid state is represented along with its transitions, whereas any transition that causes an exception to be raised is identified as an exceptional transition and a special \emph{exceptional} state is used to mark it. 

\subsubsection*{Synergistic Effects}
\synRefine~/~\synRefineIncorporate: Transitions to additional states are added along with exceptional states information.

\subsubsection*{Inter-Analysis Workflow}
\workFeedback: The technique can be summarized by a dynamic phase executed in a feedback loop with a static analysis technique (model-based testing).

\subsubsection*{Mapping-Function Interpretation Structure}
\structModules: The states the program assumes along their transitions are tracked.

\subsubsection*{Mapping-Function Mechanics}
\begin{enumerate}
\item \mechanicMining: Dynamic $\rightarrow$ Static: Specifications are mined based on the results of the dynamic analysis phase.
\item \mechanicAssociation: Dynamic $\leftarrow$ Static.
\end{enumerate}

\subsection{Combining mutation and fault localization for automated program debugging \cite{debroy_combining_2014}}

\subsubsection*{Summary}
This paper proposes a new technique aimed at fixing program faults through the use of mutations. In particular, they identify the technique as three steps, of only which two are covered in the paper. The first step is fault localization, in which they reuse previously made work in the form of Tarantula (as they describe in the technique) or Ochiai (which they state they use in the experiments along the first) to identify the statements that might cause faults and rank them accordingly. Following that, they generate mutants aiming to fix the identified faults through mutation of the target statements. In particular, they identify the correct operators, then pick the highest ranking statement, generate various mutants, and then verify their effectiveness by simply running the target test suite and verify if the presence of the mutant causes tests to pass rather than fail. In case of a failure the mutant is discarded, otherwise it is kept as being a possible fix. Finally, the last step (briefly mentioned) is prompting the programmer to know if the mutant is actually a correct fix or not. The techniques that are interacting are Tarantula/Ochiai and mutants generation. 

\subsubsection*{Synergistic Effects}
\synTraversal~/~\synTraversalSeedSink:
The first step fault localization technique augments the statements that might cause faults by ranking them in order of relevance with respect to the faults. The mutation generation step tries to fix the faults by mutating the statements starting from the highest ranked ones.

\subsubsection*{Inter-Analysis Workflow}
\workCascade: The results of the first stage are passed directly to the second stage.

\subsubsection*{Mapping-Function Interpretation Structure}
\structCFG: The communication occurs by augmenting the data at the level of statements, which are part of the control-flow graph.

\subsubsection*{Mapping-Function Mechanics}
\mechanicAssociation: The data of the statements is given as is to the mutants generation stage.

\subsection{Infeasible path generalization in dynamic symbolic execution \cite{delahaye_infeasible_2015}}

\subsubsection*{Summary}
This paper proposes a technique aimed at augmenting concolic execution, here referred to as Dynamic Symbolic Execution, through the generalization of path infeasibility conditions. In particular, if a path is deemed as infeasible, then the conditions that determine infeasibility are extracted and generalized to a whole category of paths, which are then excluded from the possible ones the concolic execution can execute. This step does not rely on analysis techniques, meaning that the work reduces to basic concolic execution.

\subsubsection*{Synergistic Effects}
\synPartitioning~/~\synPartitioningDirect: The objective of concolic execution is to increase the efficiency and the effectiveness of traditional symbolic execution in exploring the path space of a program, usually for the sake of generating test cases that execute the program paths explored thereby.

\subsubsection*{Inter-Analysis Workflow}
\workFeedback: Each time a new test case is generated it can be used to derive new constraints that are used to potentially generate new combinations of path conditions, that will once again lead to the creation of new test cases. This is the classic concolic execution behavior.

\subsubsection*{Mapping-Function Interpretation Structure}
\begin{enumerate}
\item \structPaths: D $\rightarrow$ S: Execution traces associated to corresponding program paths.
\item \structCG: D $\leftarrow$ S: Symbolic execution feeds inputs to
further execute the target function.
\end{enumerate}

\subsubsection*{Mapping-Function Mechanics}
\begin{enumerate}
\item \mechanicAssociation: D $\rightarrow$ S.
\item \mechanicConstraint: D $\leftarrow$ S: Constraint solving is used to identify input values that satisfy the path conditions that symbolic execution identified as executability conditions of yet-unvisited program paths.
\end{enumerate}

\subsection{CLOTHO: Saving programs from malformed strings and incorrect string-handling \cite{dhar_clotho_2015}}

\subsubsection*{Summary}
This paper introduces CLOTHO, a technique that can identify program locations that are vulnerable to faults related to string handling and automatically perform repairs on them. To do so, the tool operates in various distinct steps according to a hybrid approach that employs both static and dynamic techniques.

In the first step, CLOTHO tries to identify the points in the program that may trigger faults. This is done via static analysis through the use of taint analysis, identifying "sensitive sources" and "sensitive sinks" of strings. From here, taint analysis analyzes the Java bytecode of the program and builds a CFG which is then used to identify the paths connecting sensitive sources to sensitive sinks, marking all string objects that lie on these paths as not requiring repairs. Furthermore, CLOTHO obtains a call graph and uses it to determine if any exceptions are caught by the method or one of its callers, excluding additional points if these conditions are met. Finally, reaching definitions are then used to identify additional areas that should not be patched due to existing earlier in the chain. In the end, the results are points in the program that should be repaired.

In the second step, CLOTHO uses the information gathered previously to compute a series of static constraints, i.e. constraints on the values that the string can assume to avoid a program crash based on statements that exist in the code (such as if statements that check the length of the string and enter an error path if it matches or differs). This is done using a custom algorithm that operates on a set of conditional statements. This information is used to update a constraint store, whose information is then used to generate a series of patches with the goal of "Object repairing" and "Parameter twerking" (essentially adding try-catches around code blocks that may throw exceptions, and then wrapping dangerous parameters in code that ensures they are always bounded properly).

In the third and last step, CLOTHO instruments the program with additional collection statements and then runs the program under test, so that additional information on constraints that cannot be computed statically (such as method calls on strings that are user-dependent) can be added to the constraint store. Patch generation is then repeated. This operation is then repeated until some  termination criteria is reached.

\subsubsection*{Synergistic Effects}
\begin{enumerate}
\item \synTraversal~/~\synTraversalSeedSink: the taint analysis (and subsequent pruning steps) identify the points where unsafe operations are executed on strings, so that they can be the subject of further analysis;
\item \synIntepretability~/~\synIntepretabilityEntities: the static/dynamic analysis steps are used to inspect the code and identify the constraints that strings must follow;
\item \synTraversal~/~\synTraversalTransform: the patching process is used to obtain a new program with the fix applied to it.
\end{enumerate}

\subsubsection*{Inter-Analysis Workflow}
\workFeedback: The patches are used to both fix the software and to instrument it so that the dynamic analysis stage can gather more information; once that happens, the information is then fed back to the patching algorithm to update patches and possibly trigger further dynamic execution steps.

\subsubsection*{Mapping-Function Interpretation Structure}
\begin{enumerate}
\item \structCFG: From taint analysis to the rest: \textemdash The nodes of the control-flow graph are used to indicate which elements should be inspected by the static/dynamic analysis technique as it follows from the output of the taint analysis
\item \structProgram: The feedback loop: \textemdash The entire program along with its patches are used to communicate discoveries during the feedback loop construct (so static/dynamic to patching and then back).
\end{enumerate}

\subsubsection*{Mapping-Function Mechanics}
\mechanicAssociation: The first two stages communicate directly the nodes that should be inspected, while the second and third loop by sending the entire program back and forth, essentially.

\subsection{Quantitative Program Sketching using Lifted Static Analysis \cite{dimovski_quantitative_2022}}
\subsubsection*{Summary}
The paper presents an approach to realize programs from program sketches, i.e. programs that are missing some numerical parameters which must be configured, which respect certain constraints and execution times. The approach proposed by the work treats these programs as part of a program family in which a member is called a variant.

Initially, a forward numerical lifted analysis is used to determine which program variants are "correct", meaning the programs satisfy the constraints given as input to the technique.

Following this, a backward termination lifted analysis is performed on the variants that have been previously identified, to obtain an upper-bound on the amount of steps that are required for the program to terminate.

The results are then ranked based on the computed metric, and then the ones with minimal values are reported as the result of the technique. From this, the necessary inputs to complete the program sketch can be extracted and used in the program sketch itself.

\subsubsection*{Synergistic Effects}
\synIntepretability~/~\synIntepretabilityAPI: 
The synergy is on augmenting the initial program sketches with candidate values, such that the resulting candidate program can be evaluated (in the second step) to determine the execution step upper-bounds. 

\subsubsection*{Inter-Analysis Workflow}
\workCascade: The output of the first lifted analysis is given as an input to the second without any form of feedback.

\subsubsection*{Mapping-Function Interpretation Structure}
\structProgram: The communication between the two analysis steps happens through the definition of the candidate programs (with values filling the "holes" of the initial program sketch), to be analyzed in the second step.

\subsubsection*{Mapping-Function Mechanics}
\mechanicAssociation: The candidate programs are associated to the original program sketch, and are analyzed in the second step.

\subsection{Computing program reliability using forward-backward precondition analysis and model counting \cite{dimovski_computing_2020}}

\subsubsection*{Summary}
This paper proposes a technique to identify the necessary preconditions on inputs that can determine whether a certain assertion is satisfied or, conversely, violated. The technique relies on abstract interpretation.

Starting from the entire program source, the first step employs a forward analysis to automatically infer an over-approximation of the program's invariants.

Following this, a backward analysis step is executed, integrating the results of the first analysis, in order to determine an over-approximation for the necessary preconditions. In particular, the backward analysis is made up of two components that can be executed in parallel, one determining the necessary preconditions to satisfy invariants, and the other the necessary preconditions to violate invariants; these results are then integrated and treated as the output of the backward analysis step.

\subsubsection*{Synergistic Effects}
\synIntepretability~/~\synIntepretabilityOracle: In particular, the technique aims to identify which inputs lead to invariants being satisfied or violated, with those inputs acting as potential oracles.

\subsubsection*{Inter-Analysis Workflow}
\workCascade: The results of the forward analysis are used by the backward analysis step to refine its results.

\subsubsection*{Mapping-Function Interpretation Structure}
\structProgram: The communication between the two analysis steps occurs via the definition of the invariants with respect to the program in its entirety.

\subsubsection*{Mapping-Function Mechanics}
\mechanicInvariants:  The forward analysis stage determines the invariants which are then given as an input to the backward analysis stage.

\subsection{Generating Timed UI Tests from Counterexamples \cite{diner_generating_2021}}

\subsubsection*{Summary}
This paper proposes a custom testing technique focused on verifying the behavior of SCRATCH programs through error witnesses, which are sequences of user mouse or keyboard inputs that lead the program into an error state, while also attempting to communicate these faults to the user visually by means of counterexamples. To do so, the paper employs a combination of static and dynamic analysis.

First, the program and its specifications are given as an input to Bastet, which is a tool that performs static analysis to build an abstract reachability graph, which determines which program states are reachable by which other states. Bastet, in particular, performs an over-approximation of the state space. Following that, an operator determines the properties that a state violates, which represents the error witness.

Following that, the results are confirmed through the use of a dynamic analysis tool, in this case Whisker, which runs the Scratch program according to the inputs indicated by the error witness to properly confirm the fault that has been identified. This results in only witnesses that have been confirmed.
Finally, the resulting witnesses are visualized to the user.

\subsubsection*{Synergistic Effects}
\synAlarms~/~\synAlarmsDynamic: The static analysis and the following operator identify sequences of inputs (called witnesses) that lead the program towards a fault, represented by a state in which some properties are violated; these witnesses are then provided to the dynamic analysis which can verify their correctness.

\subsubsection*{Inter-Analysis Workflow}
\workCascade: The results of one analysis are given to the next without any feedback being present.

\subsubsection*{Mapping-Function Interpretation Structure}
\structProgram: The static analysis communicates the inputs to lead the program states that are to be entered for the error to be identified by the dynamic analysis.

\subsubsection*{Mapping-Function Mechanics}
\mechanicAssociation: The results of the static analysis are given as is to the dynamic analysis, together with the sequence of inputs that need to be provided to reach the desired state.

\subsection{Solving complex path conditions through heuristic search on induced polytopes \cite{dinges_solving_2014}}

\subsubsection*{Summary}
This paper introduces the Concolic Walk algorithm, with the goal of trying to solve the inherent limitation of regular solvers, i.e. their inability to handle general non-linear integer constraints. For this reason, this paper proposes a variation on concolic execution which aids itself of two types of solvers depending on the type of constraint. In particular, the technique splits constraints into linear constraints and non-linear constraints. Linear constraints are solved as per regular concolic execution with the help of an SMT solver, whereas non-linear constraints use a meta-heuristic, in particular tabu-search adapted into an adaptive search algorithm. Essentially, in case of non-linear constraints, they carry out search-based testing.

\subsubsection*{Synergistic Effects}
\begin{enumerate}
\item \synRewrite~/~\synRewriteConcrete: Dynamic $\rightarrow$ Static: The execution results from the dynamic component of concolic executions are used to properly handle opaque library methods without requiring further simplification;
\item \synPartitioning~/~\synPartitioningDirect: Static $\rightarrow$ Dynamic: The results of the SMT solving and the meta-heuristic approach are used to steer the execution onto specific program paths that have to yet be explored.
\end{enumerate}

\subsubsection*{Inter-Analysis Workflow}
\workFeedback: This is the classic concolic execution approach, with the two analysis stages communicating results.

\subsubsection*{Mapping-Function Interpretation Structure}
\begin{enumerate}
\item \structPaths: Dynamic $\rightarrow$ Static: The paths executed during dynamic execution are communicated to the partner analysis stage.
\item \structCG: Static $\rightarrow$ Dynamic: The results map to inputs that can then be used to execute the target function.
\end{enumerate}

\subsubsection*{Mapping-Function Mechanics}
\begin{enumerate}
\item \mechanicAssociation: Dynamic $\rightarrow$ Static: The path taken by the execution is given as-is to the partner analysis stage.
\item \mechanicConstraint: Static $\rightarrow$ Dynamic: The SMT solver/tabu search combination computes valid inputs for the target function via the resolution of constraints.
\end{enumerate}

\subsection{Goal-oriented dynamic test generation \cite{do_goal-oriented_2015}}

\subsubsection*{Summary}
This paper proposes an improvement on dynamic symbolic execution (aka concolic testing) with the goal of improving its exploration process. In particular, they have devised a dynamic symbolic execution-based buffer overflow testing framework called Sebo, which has three phases.

In its first phase, Sebo uses Deputy to statically analyze the code and identify potential buffer memory violations, which are then instrumented by Deputy first and then reprocessed by Sebo itself to inject error conditions, which are then used by test targets in the second phase.

In the second phase, a series of event sequences are determined through the "extended chaining mechanism", a technique that leverages data dependence analysis to collect both direct and indirect data dependencies to build a sequence of events, which identify the statements that must be executed to reach a target test goal

In the final phase, Sebo performs concolic execution via another tool named Crest. Each sequence of events identified in the previous step is explored through the use of concolic execution, which adjusts the given test sequences to ensure the test goals are reached. Thusly, the concolic execution is then tasked to generate test cases to verify buffer overflows.

\subsubsection*{Synergistic Effects}
\begin{enumerate}
\item \synTraversal~/~\synTraversalSeedSink: Deputy $\rightarrow$ Chaining: This step identifies the places where unsafe buffer accesses might occur.
\item \synPartitioning~/~\synPartitioningCoverage: Chaining $\rightarrow$ Concolic: The objective of the analysis is to steer the concolic analysis towards the test goals, which are reached by using the event sequences as coverage criteria.
\end{enumerate}

\subsubsection*{Inter-Analysis Workflow}
\begin{enumerate}
\item \workCascade: Deputy $\rightarrow$ Chaining: The output of Deputy is used, albeit with some modifications, as an input for the chaining approach.
\item \workCascade: Chaining $\rightarrow$ Concolic: The concolic feedback loop is fully self-contained and does not communicate back to the previous stage to create new event sequences. \end{enumerate}

\subsubsection*{Mapping-Function Interpretation Structure}
\begin{enumerate}
\item \structCFG: Deputy $\rightarrow$ Chaining: The communication occurs at the level of program instructions that might trigger out-of-bounds buffer accesses.
\item \structDF: Chaining $\rightarrow$ Concolic: The event sequences represent relationships between variable usages that must be satisfied to reach the target test goals.
\end{enumerate}

\subsubsection*{Mapping-Function Mechanics}
\begin{enumerate}
\item \mechanicAssociation: Deputy $\rightarrow$ Chaining: Deputy, along with the modifications carried out by Sebo, passes information that can be 
injected into the original program to identify buffer overflows, which represent the test goals for the rest of the technique.
\item \mechanicAssociation: Chaining $\rightarrow$ Concolic: The event sequences are given as-is to the concolic execution engine.
\end{enumerate}

\subsection{Blended analysis for performance understanding of framework-based applications \cite{dufour_blended_2007}}

\subsubsection*{Summary}
The paper proposes a technique called blended program analysis with the goal of analyzing the performance of a particular execution. In particular, the paper focuses on an instantiation of this analysis called \emph{blended escape analysis}.

The first step is to execute the target program and then analyze its execution trace to build the execution call graph limited only to the calls observed in that specific run of the program.  The call graph is subsequently edited to remove elements that are irrelevant for static analysis (such as calls to \texttt{\textless clinit\textgreater}).

Finally, the modified call graph is operated on by the desired static analysis. In the case of the example of blended escape analysis, the paper presents an analysis variation that keeps track for every object of its escape state.

\subsubsection*{Synergistic Effects}
\synPartitioning~/~\synPartitioningDirect: The dynamic analysis creates a partition by identifying the relevant function calls performed by the program, which then steer the static analysis towards analyzing only that path.

\subsubsection*{Inter-Analysis Workflow}
\workCascade: The output of the dynamic phase are given directly to the static phase.

\subsubsection*{Mapping-Function Interpretation Structure}
\structCG: The communication happens at the call graph level, computed by the dynamic analysis.

\subsubsection*{Mapping-Function Mechanics}
\mechanicAssociation: The call graph, albeit with a few elements removed, is provided directly to the partner stage.

\subsection{An efficient regression testing approach for PHP Web applications using test selection and reusable constraints \cite{eda_efficient_2019}}

\subsubsection*{Summary}
This paper focuses on a technique that allows to reduce the cost of regression testing by identifying which subset of test cases that exercise the modified code paths may be executed with the same input data and which ones cannot. To do so, the process is split into steps.

The first step of the technique derives a slice of the program (a set of affected paths) which is obtained from the diff between the original and modified source code at the level of the control flow.

The second step consists in deriving the data-flow dependency graphs of the original and modified source code and identify which are the input variables that are affected by the changes in the code. For the variables that are not affected, the technique can re-use the same test input values.

The third step takes the program slice, the list of affected variables, and previously gathered code coverage statistics to determine which are the test cases that can be re-used.

\subsubsection*{Synergistic Effects}
\synTraversal~/~\synTraversalSeedSink: The slice of the program and the unaffected variables, which represents a restricted scope of the program, are provided to the third stage analysis that extracts the test cases that can be reused.

\subsubsection*{Inter-Analysis Workflow}
\workCascade: The outputs of the slicing and data-flow stage are provided to the third analysis stage.

\subsubsection*{Mapping-Function Interpretation Structure}
\begin{enumerate}
\item \structProgram: Slice $\rightarrow$ Test Selection: The entire sliced program is used for the communication stage.
\item \structDF: Definition-Use Analysis $\rightarrow$ Test Selection: The data flow graph is used to provide information regarding the reusable variables to the partner analysis stage.
\end{enumerate}

\subsubsection*{Mapping-Function Mechanics}
\begin{enumerate}
\item \mechanicAssociation: Slice $\rightarrow$ Test Selection: The slice is provided to the partner stage with identity-style mechanics.
\item \mechanicAssociation: Definition-Use Analysis $\rightarrow$ Test Selection: The unaffected input variables are provided as-is to the partner analysis stage.
\end{enumerate}

\subsection{Automatic test data generation targeting hybrid coverage criteria \cite{el-serafy_automatic_2015}}

\subsubsection*{Summary}
This paper proposes an approach based around attempting to increase MC/DC and BVA coverage criteria by means of a Genetic Algorithm.

In particular, they first parse the the code to build its Abstract Syntax Tree (AST for short), then they apply instrumentation to the code to be able to track its execution path during execution. Following this, they statically analyze the AST to identify the branches that must be covered for the MC/DC test coverage criterion, which they then employ to obtain the desired test cases.

Once that is done, both the instrumented code and the MC/DC test cases are fed as an input to the genetic algorithm to perform search-based testing with the goal of reaching the desired MC/DC coverage target.

\subsubsection*{Synergistic Effects}
\synPartitioning~/~\synPartitioningCoverage: The static analysis stage is responsible for identifying the branches that must be covered along with determining input criteria that must be followed to satisfy MC/DC, with the dynamic analysis step needing to actually generate the relevant test cases.

\subsubsection*{Inter-Analysis Workflow}
\workCascade: The branch detection and following MC/DC test cases are given as an input to the search-based testing phase.

\subsubsection*{Mapping-Function Interpretation Structure}
\structCFG: The branches that must be covered are provided to the dynamic analysis stage.

\subsubsection*{Mapping-Function Mechanics}
\mechanicAssociation: The branch entities are provided as-is to the target analysis.

\subsection{Search-Based Test Data Generation for JavaScript Functions that Interact with the DOM \cite{elyasov_search-based_2018}}

\subsubsection*{Summary}
The paper presents JEDI, which is a testing framework focused on testing JavaScript code that interacts with the DOM. In particular, the provided analysis is fully centered around the use of a genetic algorithm to perform search-based testing starting from the CFG extracted from the original code.

\subsubsection*{Synergistic Effects}
\synPartitioning~/~\synPartitioningCoverage: The static analysis stage extracts the CFG which is then provided to the search-based testing phase in order to address the target of branch coverage.

\subsubsection*{Inter-Analysis Workflow}
\workCascade: The output of the static analysis is given to the dynamic one without feedback.

\subsubsection*{Mapping-Function Interpretation Structure}
\structCFG: The communication occurs at the level of the branches of the control flow graph.

\subsubsection*{Mapping-Function Mechanics}
\mechanicAssociation: The branches are shared directly between the two partner analysis stages.

\subsection{RAPID: Checking API usage for the cloud in the cloud \cite{emmi_rapid_2021}}

\subsubsection*{Summary}
The paper presents RAPID, which is an analysis technique meant to ensure correct usage of cloud APIs. The technique is divided into three phases: the first one creates code partitions to allow for parallelization of the analysis, the second one is the analysis proper, and the third one is simply reporting.

Initially, RAPID statically creates a call graph through a Class Hierarchy Analysis (CHA), which is then used to derive an inter-procedural control flow graph, which is also augmented by exceptional edges (i.e. edges that are followed in case an exception is thrown either explicitly or implicitly).

Following this, a local typestate analysis (LTA) is performed to determine the states the object can assume as the target method is executed. At the same time, the analysis also determines that the states the various objects assume are consistent with what a set of properties given as an input to the technique dictate, and that no object can escape the scope of the method it is initialized in via e.g. field assignment.

Finally, a global backward reaching definition analysis (GRDA) is executed on the same data structures, to verify further properties that depend on the values that the arguments of the target method assume; as before any violation is then flagged to be reported.

\subsubsection*{Synergistic Effects}
\synIntepretability~/~\synIntepretabilityEntities: 
CHA $\rightarrow$ LTA $\rightarrow$ GRDA:
Implicit behavior of API methods is derived by the combination of Class Hierarchy Analysis and local typestate analysis. This information on the the implicit behavior is then used by the global backward reaching definition analysis to verify additional properties.

\subsubsection*{Inter-Analysis Workflow}
\begin{enumerate}
\item \workCascade: CHA $\rightarrow$ LTA.
\item \workCascade: LTA $\rightarrow$ GRDA.
\end{enumerate}

\subsubsection*{Mapping-Function Interpretation Structure}
\begin{enumerate}
\item \structCFG: CHA $\rightarrow$ LTA: The interprocedural control-flow graph with exceptional edges is used to communicate between the two partner analysis steps.
\item \structPaths: LTA $\rightarrow$ GRDA: The communication occurs at the level of the states that the objects can assume during code execution, which can be assumed to imply the relevant program paths.
\end{enumerate}

\subsubsection*{Mapping-Function Mechanics}
\begin{enumerate}
\item \mechanicAssociation: CHA $\rightarrow$ LTA: The CFG is used as-is for communication.
\item \mechanicAssociation: LTA $\rightarrow$ GRDA: The program paths are used to communicate between the two stages without additional computation operations.
\end{enumerate}

\subsection{Efficient detection and validation of atomicity violations in concurrent programs \cite{eslamimehr_efficient_2018}}

\subsubsection*{Summary}
The paper proposes a technique named AtomChase, which has as its objective the ability to direct dynamic analysis towards situations in which a three-access pattern (TAP hereinafter) arises. In particular, they describe an approach that is built up of two nested feedback loop cycles preceded by a preparation step, which are described in the text that follows.

Initially, the technique identifies TAP candidates in which atomicity guarantees may be violated. For every candidate that is found, a concolic analysis is performed. This is made up of two components: a concolic execution engine using an SMT solver for the static analysis component and a "Plan Synthesizer".

The concolic execution engine attempts to execute the TAP according to a plan indicating a desired order (for the first execution, no order is provided, so any order is assumed to be fine).

Once the TAP is executed, the relevant execution trace is communicated to the plan synthesizer, which attempts to modify the given execution trace into a similar one which can lead the concolic engine towards the TAP candidate. In particular, this is done by constructing constraints based on the execution trace and then attempting a resolution via the use of a SAT solver.

\subsubsection*{Synergistic Effects}
\begin{enumerate}
\item \synPartitioning~/~\synPartitioningDirect: Concolic $\rightarrow$ PS: The concolic execution provides the execution traces to the plan synthetizer so that it can verify and determine the order in which the events should be executed to reach the TAP.
\item \synAlarms~/~\synAlarmsDynamic: Concolic $\leftarrow$ PS: The execution traces are given to the concolic engine to attempt to verify the alarm in the form of the TAP candidate is confirmed.
\item \synAlarms~/~\synAlarmsDynamic: The global objective of the technique combination is to determine if a given TAP candidate can lead to atomicity violations.
\end{enumerate}

\subsubsection*{Inter-Analysis Workflow}
\workFeedback: The concolic engine and the Path Synthetizer communicate in a feedback structure to assess the paths to hit the TAP candidates.

\subsubsection*{Mapping-Function Interpretation Structure}
\begin{enumerate}
\item \structPaths: Concolic $\rightarrow$ PS: The execution sequence that has reached the target ordering objective is communicated to the Plan Synthesizer.
\item \structPaths: Concolic $\leftarrow$ PS: The Plan Synthetizer hands over an execution sequence that might lead to a desired sequence of events which must be verified and potentially edited by the concolic engine to verify the target TAP candidate's behavior.
\end{enumerate}

\subsubsection*{Mapping-Function Mechanics}
\begin{enumerate}
\item \mechanicConstraint: Concolic $\rightarrow$ PS: The given program paths are used to obtain constraints which can then be solved by an SMT solver to obtain an updated path.
\item \mechanicAssociation: Concolic $\leftarrow$ PS: The desired execution trace as obtained by the Plan Synthesizer is given to the partner analysis stage as is.
\end{enumerate}

\subsection{Static Bound Analysis of Dynamically Allocated Resources for C Programs \cite{fan_static_2021}}

\subsubsection*{Summary}
This paper presents an analysis technique to determine the amount of resources a C program requires statically, both tracking its peak usage and current usage as the program executes.
In particular, the technique consists of two static techniques that are applied one after the other: pointer analysis and numerical value analysis. Both these techniques rely on an abstract interpretation framework, by defining two abstract domains, $\mathcal{B}^\#$ for pointer analysis and $\mathcal{N}^\#$ for numerical value analysis.

As a first step, the program is instrumented by adding both new integer variables and statements using these variables to track resource usage, i.e. the places in the program where resources are allocated and released. Following this, pointer analysis is executed with the goal of determining what every pointer operation does.

Finally, numerical value analysis is executed. In this last step, the results of the pointer analysis are used whenever pointer arithmetic is found, to properly compute the amount of memory that is either allocated or released during the course of the program. At the end of the analysis step, the peak resource usage is reported, along with the resource usage at every point in the program.

\subsubsection*{Synergistic Effects}
\synIntepretability~/~\synIntepretabilityEntities: The pointer analysis stage provides further information with respect to which pointers are being operated upon at a given point in the program, which can then be used by the numerical value analysis stage to properly track the peak resource usage even in the presence of multiple pointer aliases.

\subsubsection*{Inter-Analysis Workflow}
\workCascade: The output of pointer analysis is used by the numerical value analysis.

\subsubsection*{Mapping-Function Interpretation Structure}
\structCFG: The analysis communicate via the locations of the control flow graph representing the behavior of the method.

\subsubsection*{Mapping-Function Mechanics}
\mechanicAssociation: The properties of the pointer analysis are communicated as-is to the partner stage.

\subsection{Efficiently manifesting asynchronous programming errors in android apps \cite{fan_efficiently_2018}}

\subsubsection*{Summary}
The paper proposes APEChecker, a technique that automatically identifies async programming errors (APE) through the use of both static and dynamic UI analysis. The technique consists of three stages.

The first stage represents a static analysis stage, where the call graph of the application is statically extracted from the APK and the locations of possible APEs is determined, in the form of a list of candidate methods and statements within each method.

Each of these pairs is then fed as an input to the second analysis stage, which uses a modified backwards symbolic execution to determine the program traces that can lead to the discovery of the APE.

Finally, these pieces of information are then used in the final step to generate a proper sequence of events and an environment (e.g. certain permissions being granted or revoked) that can lead to the discovery of the APE, to then execute the APK and recreate the situation that can lead to the APE. In particular, instrumentation is injected into the application, and then the test is executed; if the crash is triggered the crash report is dumped to allow for bug fixing at a later point in time.

\subsubsection*{Synergistic Effects}
\synAlarms~/~\synAlarmsDynamic: The general synergy of the three techniques is to discriminate actual alarms out of an initial set of potential alarms (the APE candidates).

\subsubsection*{Inter-Analysis Workflow}
\workCascade: The three analysis stages are executed sequentially.

\subsubsection*{Mapping-Function Interpretation Structure}
\begin{enumerate}
\item \structCFG: The code locations of the possible APEs are passed to the backward symbolic execution stage,
\item \structPaths: The program traces indicated by the backward symbolic execution are used to generate event sequences to execute the APKs to verify the presence or absence of the APE.
\end{enumerate}

\subsubsection*{Mapping-Function Mechanics}
\mechanicAssociation: The control flow graph and paths are given to the subsequent analysis stages with identity-style mechanics.

\subsection{History-driven Compiler Fuzzing via Assembling and Scheduling Bug-Triggering Code Segments \cite{fan_history-driven_2024}}

\subsubsection*{Summary}
This paper proposes a technique called ASMFuzz with the goal of identifying compiler bugs via a history-driven approach, meaning through a process of fuzzing that uses as a starting test cases that identified previously fixed bugs.

In particular, the technique proposed by this paper initially scrapes bug repositories for historical bugs and patches in the form of code that has been used to identify the bug in test cases. The patches are then analyzed using a static analysis technique, with the goal of extracting changes to variables and code blocks. These changes, called segments, are then stored in a database, together with a set of seed programs generated by COMFUZZ. The seeds and segments are then combined into a single code by injecting the latter into the former.
The code obtained is then executed on different compilers (fuzzying) to achieve differential testing, with any errors identified leading to a bug report.

\subsubsection*{Synergistic Effects}
\synTraversal~/~\synTraversalSeedInputs: The segments identified by the static analysis allow for seeding the fuzzing stage 

\subsubsection*{Inter-Analysis Workflow}
\workCascade: The output of the static analysis to select the "segments" is fed to the dynamic fuzzing stage. 

\subsubsection*{Mapping-Function Interpretation Structure}
\structProgram: The communication occurs at the program level, as the "segments" relate to effective tests for program under test.

\subsubsection*{Mapping-Function Mechanics}
\mechanicAssociation: The "segments" are associated to program under test.

\subsection{VFIX: Facilitating Software Maintenance of Smart Contracts via Automatically Fixing Vulnerabilities \cite{fang_vfix_2024}}

\subsubsection*{Summary}
This paper proposes VFIX, a technique to validate and automatically fix vulnerabilities in smart contracts by generating patches. In particular, the technique is divided into three phases.

In the first phase, the smart contract is parsed to obtain its AST and at the same time a static verification pass is executed on the contract to verify the presence of vulnerabilities that can be automatically fixed (in particular, this is done three times with three different tools, which then vote). The static verification can employ different types of static analysis (e.g., intra-procedural data flow analysis, pointer analysis, method summaries for inter-procedural analysis).

The second phase is focused on patch generation, in which the output of the various analysis steps are used to transform the contract's AST to fix the vulnerabilities that have been found and output the new source code containing the patched fixes.

Finally, the last phase involves the verification of the smart contract, which is once again done statically with the same techniques adopted in the first phase. The contract is considered fixed if this additional step does not result in any alarms being triggered from the verification step.

\subsubsection*{Synergistic Effects}
\begin{enumerate}
\item \synTraversal~/~\synTraversalTransform: Phase 1 $\rightarrow$ Phase 2: The various analysis steps are used to identify the points in which unsafe operations that lead to vulnerabilities in the smart contract are located.
\item \synTraversal~/~\synTraversalTransform: Phase 2 $\rightarrow$ Phase 3: The patching process executed by Phase 2 is used to obtain a new program that can be further verified not to carry the observed vulnerabilities.
\end{enumerate}

\subsubsection*{Inter-Analysis Workflow}
\workCascade: The three analysis steps are executed in succession, without looping.

\subsubsection*{Mapping-Function Interpretation Structure}
\begin{enumerate}
\item \structCFG: Phase 1 $\rightarrow$ Phase 2: The locations on the AST where the smart contract exhibits issues that need to be patched are used as the communication method between the two analysis steps.
\item \structProgram: Phase 2 $\rightarrow$ Phase 3: The entire patched program is used to communicate between the two phases.
\end{enumerate}

\subsubsection*{Mapping-Function Mechanics}
\begin{enumerate}
\item \mechanicAssociation: Phase 1 $\rightarrow$ Phase 2: The first two stages do not perform any operations on the target nodes.
\item \mechanicAssociation: Phase 2 $\rightarrow$ Phase 3: The patch is provided as-is to the next stage.
\end{enumerate}

\subsection{Enhancing Mobile App Bug Reporting via Real-Time Understanding of Reproduction Steps \cite{fazzini_enhancing_2023}}

\subsubsection*{Summary}
This paper proposes EBug, a tool aimed at helping users submit bug reports by providing autocompletion for future reproduction steps. In particular, the tool collects this information from an analysis of the application behavior. The tool leverages three different analysis steps executed in parallel to obtain a predictive model, which can then be used by the bug reporting stage to provide a real-time bug report analysis. In particular, EBug analyzes the program through a static GUI analysis, a dynamic GUI analysis, and a program trace analysis: they will be covered in the text that follows.

The static GUI analysis analyzes the application's code to determine the elements that characterize the GUIs along with the interactions that can be executed on them, obtaining a GUI model, represented as a directed graph with screens and elements as nodes, and both containing and transition edges.

The dynamic GUI analysis has the same goal, except the application is executed and a depth-first traversal is executed by clicking on every input and interacting with every text field.

The program trace analysis, on the other hand, takes as an input user interaction traces obtained through real-world usage by real application users (e.g. beta testers) and the traces so obtained are split into action and element traces, building an action prediction and an element prediction model respectively.

Finally, the first and second analysis are then combined and then their result is further combined with the third analysis in the second component, which therefore acts as the collector and merger of the models obtained by the various analysis stages.

\subsubsection*{Synergistic Effects}
\synGUI: The various analysis steps all combine towards building a more comprehensive model of the GUI that can then be used to predict user interactions.

\subsubsection*{Inter-Analysis Workflow}
\workSidebyside: The three analysis steps are executed in parallel and integrated only in the final stage by the bug reporting component.

\subsubsection*{Mapping-Function Interpretation Structure}
\structGUI: The various analysis steps communicate by sharing the GUI model and interactions that can be executed on the various elements of the GUI.

\subsubsection*{Mapping-Function Mechanics}
\mechanicAssociation: The models are all used directly by the bug reporting component and combined in it, without the use of another program analysis or transformation step in-between.

\subsection{Apposcopy: Semantics-based detection of android malware through static analysis \cite{feng_apposcopy_2014}}

\subsubsection*{Summary}
The paper presents an approach dubbed Apposcopy, which aims to provide a way to identify malicious Android applications through the use of a series of analysis techniques. The approach first requires a description of the characteristics of malware applications, which must be described in a custom language. Following this, a series of analysis steps are executed on the target applications.

Initially, a static analysis step consisting of pointer analysis is executed to determine which heap objects may be pointed to by which variable; this step is also paired with the construction of the call graph. 

Following this, these pieces of information are used to build the inter-component call graph (i.e. a graph where the nodes are application components and the edges are annotated with actions and data types of \texttt{Intent} objects), through the use of a forward data flow analysis.

Finally, taint analysis is executed statically on the ICCG, to determine whether any of the data flow queries expressed by the custom language mentioned before apply to the application. This determines whether the application is categorized as malware or not, along with its type.

\subsubsection*{Synergistic Effects}
\synIntepretability~/~\synIntepretabilityEntities: The overall synergy is to ground on improved semantics of the inter-component dependencies, in order to enable taint analysis to effectively identify malware in Android applications.

\subsubsection*{Inter-Analysis Workflow}
\workCascade: The three analysis steps form a chain without any feedback loops.

\subsubsection*{Mapping-Function Interpretation Structure}
\structCG: The various analysis stages communicate with the usage of the call graph along with the Intent data types required to move between the various components of an application.

\subsubsection*{Mapping-Function Mechanics}
\mechanicAssociation: The communication occurs with identity-style mechanics.

\subsection{TVAL+ : TTTVLA and value analyses together \cite{ferrara_tval_2012}}

\subsubsection*{Summary}
The paper proposes a new technique named TVAL+, which is an extension of a TVLA-based heap analysis, meaning a heap analysis based around three-valued logic. In particular, the paper proposes an extension to the already existing generic analyzer SAMPLE.

The described approach sees SAMPLE compute an abstract state for every program point, with a heap analysis stage powered by TVLA. To address the limitation of node naming of the original technique, namely that they are arbitrary, the various states recognized by TVLA are augmented with name predicates to keep the names consistent: this augmentation leads to TVAL+. Finally, the output of this technique is provided to a value analysis stage.

\subsubsection*{Synergistic Effects}
\synIntepretability~/~\synIntepretabilityEntities: The technique provides additional properties to the value analysis namely in the form of heap accesses and variable names for easier associations.

\subsubsection*{Inter-Analysis Workflow}
\workCascade: The output of TVLA+ is provided to the value analysis stage.

\subsubsection*{Mapping-Function Interpretation Structure}
\structCFG: The two analysis stages communicate with the use of the control flow graph.

\subsubsection*{Mapping-Function Mechanics}
\mechanicAssociation: The two techniques work based on information around heap accesses.

\subsection{SCOUT: A Multi-objective Method to Select Components in Designing Unit Testing \cite{freitas_scout_2016}}

\subsubsection*{Summary}
The paper proposes SCOUT, which is a technique that uses a combination of static and dynamic analysis to select the components to be covered by a unit testing suite. SCOUT works in two primary steps: metric extraction and then multi-objective optimization.

Metric extraction is executed by combining into the same database data obtained through static analysis, dynamic analysis, and additional sources (such as business value) according to real-world Android user interactions. In particular, the static analysis part leverages JHawk to compute various software metrics (in particular Halstead effort and Halstead bugs) aiming to identify the cost of future maintenance.

On the other hand, the dynamic analysis component gathers the call frequency through profiling and fault risk through spectrum-based fault localization techniques.

\subsubsection*{Synergistic Effects}
\synFeature: The various analysis steps all compute numerical information relative to the various component applications, which are then integrated through a database for a multi-objective optimization technique aimed at indicating the components that should be subjected to testing. 

\subsubsection*{Inter-Analysis Workflow}
\workSidebyside: The various metrics are computed separately and essentially in parallel with one another and combined only at the end.

\subsubsection*{Mapping-Function Interpretation Structure}
\structModules: The components selected represent code units, which in Android are represented as classes.

\subsubsection*{Mapping-Function Mechanics}
\mechanicML: The various analysis gather software metrics of various kinds that can then be combined for selection purposes.

\subsection{Automatic generation of inter-component communication exploits for android applications \cite{garcia_automatic_2017}}

\subsubsection*{Summary}
The paper proposes an approach called LetterBomb, with the goal of automatically generating exploits for Android applications through the use of a combination of analysis techniques.

The technique takes the application as an input and  analyzes it to identify vulnerabilities by means of static backward data-flow analysis and backward symbolic execution. 
Ultimately, it generates malicious intents which can exploit the vulnerabilities of the application under test. 

\subsubsection*{Synergistic Effects}
\synTraversal~/~\synAlarmsDynamic: The first static analysis step aims to find vulnerable statements, 
which can then be further analyzed in the second analysis step with symbolic execution to
identify how to craft malicious intents for the target application.

\subsubsection*{Inter-Analysis Workflow}
\workCascade: The output of the first analysis is provided directly to the second.

\subsubsection*{Mapping-Function Interpretation Structure}
\structCFG: The communication occurs at the level of the single vulnerable statement.

\subsubsection*{Mapping-Function Mechanics}
\mechanicAssociation: The statement is provided as-is to the partner analysis stage.

\subsection{Differences between a static and a dynamic test-to-code traceability recovery method \cite{gergely_differences_2019}}

\subsubsection*{Summary}
This paper proposes a semi-automatic method to recover test-to-code traceability links, based on a combination of static and dynamic analysis combined together and a manual operator to resolve conflicts occurring from the combination of the analysis results.

The static analysis step is used to record the structure of the Java code itself, so its organization into classes and packages.

At the same time, the test suite is executed dynamically to obtain coverage data.

The output of these two analysis steps are used as inputs for the traceability analysis, with the static data being used for package-based clustering and the dynamic data for coverage-based clustering. The clustering results are then merged into a cluster similarity graph with any discrepancy being signaled, so that an operator can manually fix the traceability issues.

\subsubsection*{Synergistic Effects}
\synReports: The two analysis stages are integrated into a report to show issues related to traceability, which must then be fixed by an operator.

\subsubsection*{Inter-Analysis Workflow}
\workSidebyside: The two analysis are executed in parallel, and the results are then combined at the end in the cluster similarity graph after traceability analysis.

\subsubsection*{Mapping-Function Interpretation Structure}
\structModules: The information between the two partner analysis stages are integrated at the level of the class during the clustering phase to determine proper traceability.

\subsubsection*{Mapping-Function Mechanics}
\mechanicML: The algorithm used is clustering.

\subsection{Conditional Quantitative Program Analysis \cite{gerrard_conditional_2022}}

\subsubsection*{Summary}
This paper proposes a program analysis framework called CQA (Conditional Quantitative Analysis), with the goal of combining evidence from various analysis stages to determine the probability of failures. The technique described is based around three steps, dubbed \texttt{generate\_intervals}, \texttt{estimate}, and \texttt{quantify\_in\_bounds}; note that the general description only offers what these steps are supposed to do but does not describe any way for them to accomplish it, rather the paper then presents a possible instantiation using already existing static analysis techniques: for this survey, we will focus on this instantiation.

The authors propose the usage of alternating conditional analysis
for \texttt{generate\_intervals}, which is a type of analysis that takes the program as an input and then alternates between over-approximating and under-approximating steps to identify the areas for which a program \textit{may} or \textit{must} satisfy a given property.

Following this, \texttt{estimate} does not rely on any program analysis technique, limiting itself to estimating the probability of satisfying a property that \textit{must} be satisfied that was identified by the previous technique.

Finally, \texttt{quantity\_in\_bounds} gets executed for every property that \textit{may} be satisfied and is instantiated as symbolic execution (in either its probabilistic or statistical fashion).

The results of both these last two steps are mathematically combined to provide a lower and upper bound on the quantity of inputs that can lead a program to reach a state holding the desired property.

\subsubsection*{Synergistic Effects}
\begin{enumerate}
\item \synPartitioning~/~\synPartitioningDirect: Interval generation is used to detect the partitions on which the in-bound quantification function must operate.
\item \synReports: the output of the estimate step, which is derived from the analysis done in the \texttt{generate\_intervals} step, is mathematically combined with the output of the \texttt{quantity\_in\_bounds}.
\end{enumerate}

\subsubsection*{Inter-Analysis Workflow}
\begin{enumerate}
\item \workCascade: The output of interval generation is used as inputs for the in-bound quantification function, in particular the ones representing may properties.
\item \workSidebyside: The estimation step, which is itself derived from the output of the interval generation step, is executed independently of the quantification function, with the results being combined at the end through mathematical means.
\end{enumerate}

\subsubsection*{Mapping-Function Interpretation Structure}
\structProgram: The two functions communicate with each other merely by providing the program and input bounds that can then be leveraged for symbolic execution.

\subsubsection*{Mapping-Function Mechanics}
\begin{enumerate}
    \item \mechanicAssociation: Communication of the intervals between \texttt{generate\_intervals} and \texttt{quantity\_in\_bounds} occurs via identity-style mechanics.
    \item \mechanicML: The output of the \texttt{generate\_intervals} step is processed in the \texttt{estimate} step and transformed in a metric (probability) that can be combined with the output of the \texttt{quantity\_in\_bounds} step.
\end{enumerate}

\subsection{Comprehensive failure characterization \cite{gerrard_comprehensive_2017}}

\subsubsection*{Summary}
This paper  focuses on a technique aimed at identifying bounds on the input that can lead the program to a failure state, through an iterative framework in which over-approximating and under-approximating program analysis steps are executed.

Initially, CPAchecker is used to obtain an over-approximation (OA) and thus detect a possible program failure based on the upper-bound of the failure interval that has been computed so far (or a default value if this is the first iteration); the result is either a proof that there are no further failures in the program, in which case the analysis terminates, or some new possible failure.

In the latter case, CIVL is used to compute an under-approximation (UA) aimed at determining whether the fault that has been identified is valid or spurious, thus representing definite proof. 
The conditions identified wth CIVL are fed back, to either remove spurious warnings or further refine the failure space,
and the loop begins anew.

\subsubsection*{Synergistic Effects}
\begin{enumerate}
\item \synAlarms~/~\synAlarmsStatic: OA $\rightarrow$ UA: The under-approximating stage aims at verifying whether the possible fault identified through the over-approximating analysis represents a real failure or not.
\item \synTraversal~/~\synTraversalSeedSink: UA $\rightarrow$ OA: The over-approximating analysis operates on a constrained scope, according to the conditions that confirmed spurious traces or failures.
\end{enumerate}

\subsubsection*{Inter-Analysis Workflow}
\workFeedback: The over-approximating and under-approximating techniques communicate with each other the identified failures in a loop.

\subsubsection*{Mapping-Function Interpretation Structure}
\begin{enumerate}
\item \structPaths: OA $\rightarrow$ UA: reported failure paths.
\item \structProgram: UA $\rightarrow$ OA: program-level constraints over the program inputs.  
\end{enumerate}

\subsubsection*{Mapping-Function Mechanics}
\mechanicAssociation.

\subsection{Reducing the number of higher-order mutants with the aid of data flow \cite{ghiduk_reducing_2016}}

\subsubsection*{Summary}
The proposed technique combines data-flow analysis and mutation analysis to generate higher-order mutants. It utilizes a data-flow analysis to decrease the number of mutation points through the program under test and consequently reduce the number of higher-order mutants. In this technique, only the positions of defs and uses are considered as locations to seed the mutation. 

\subsubsection*{Synergistic Effects}
\synPartitioning~/~\synPartitioningWitness: data-flow analysis identifies a set of def-use relations, steering mutation analysis to associate higher-order mutations with those relations. 

\subsubsection*{Inter-Analysis Workflow}
\workCascade

\subsubsection*{Mapping-Function Interpretation Structure}
\structDF: the results of data-flow analysis consist of relations between defs and uses to be considered to seed the mutations. 

\subsubsection*{Mapping-Function Mechanics}
\mechanicAssociation: program locations marked as defs and uses with data-flow analysis are directly used for mutation analysis.

\subsection{PROPR: Property-Based Automatic Program Repair \cite{gissurarson_propr_2022}}

\subsubsection*{Summary}
This technique addresses automatic program repair (APR) by combining (in a feedback loop) property-based testing (i.e., dynamic analysis for generating test cases and identifying the ones that violate properties, if any), fault localization (i.e., dynamic analysis for evaluating coverage against failing and non-failing test cases, respectively, to detect the likely-faulty instructions), a fix-suggestion technique based on the Haskell compiler (in turn based on a static analysis with respect to the program AST and the likely-faulty instructions).  

\subsubsection*{Synergistic Effects}
\begin{itemize}
    \item \synTraversal~/~\synTraversalSeedInputs:  property-based testing provides allow for fault localization based on the identified test cases. 
    \item \synTraversal~/~\synTraversalSeedSink: fault localization allows for fix-detection to work on the instructions identified as likely-faulty.
    \item \synTraversal~/~\synTraversalTransform: fix-detection generates a variant of the problem by patching the original problem.
\end{itemize}

\subsubsection*{Inter-Analysis Workflow}
\workFeedback: a feedback loop of property-based testing, which feeds fault localization, which in turn feeds static-analysis-based fix selection, which feeds again property-based testing to evaluate/accept current fixes and progress to identify further fixes.

\subsubsection*{Mapping-Function Interpretation Structure}
\begin{itemize}
    \item \structProgram:  property-based testing provides passing and failing test cases related to the program under analysis.
    \item \structCFG: fault localization identifies likely-faulty instructions.
    \item \structProgram: Static-analysis-based fix selection selects fixes for the program.
\end{itemize}
\subsubsection*{Mapping-Function Mechanics}
\mechanicAssociation: Test cases and fixes are associated with the target program as a whole.

\mechanicML: Spectrum-based fault localization computes metrics/statistics to measure the risk of instructions to be faulty, based on the frequencies with which they are executed in passing and failing test cases, respectively.

\subsection{Toward optimal mc/dc test case generation \cite{godboley_toward_2021}}

\subsubsection*{Summary}
This paper aims to provide MC/DC coverage for bounded programs. The technique relies on static analysis to produce the MC/DC sequences for the decision points in the program. Then it relies on symbolic execution with constraint solving to produce test inputs, and on backward symbolic analysis with interpolants to discard infeasible MC/DC sequences. 

\subsubsection*{Synergistic Effects}
\synPartitioning~/~\synPartitioningCoverage: The synergy is to optimize  MC/DC coverage.

\subsubsection*{Inter-Analysis Workflow}
\workFeedback: between symbolic execution and backward symbolic analysis (after the initial step that computes the MC/DC sequences).

\subsubsection*{Mapping-Function Interpretation Structure}
\structCFG: MC/DC sequences, symbolic states and interpolants associated with program instructions and decision points.

\subsubsection*{Mapping-Function Mechanics}
\mechanicAssociation: MC/DC sequences and symbolic states associates with program instructions and decision points.

\mechanicInterpolation: backward symbolic analysis is used to compute interpolants to discard infeasible MC/DC sequences.

\subsection{Proving memory safety of floating-point computations by combining static and dynamic program analysis \cite{godefroid_proving_2010}}

\subsubsection*{Summary}
This technique combines dynamic symbolic execution with a lightweight path-insensitive "may" static analysis of Floating-Point-computation-dependent instructions, such that symbolic execution can consider this program-semantic information to steer the symbolic analysis algorithm. 

\subsubsection*{Synergistic Effects}
\synIntepretability~/~\synIntepretabilityEntities: static analysis adds semantics to the Floating-Point-computation-dependent instructions in the program. 

\subsubsection*{Inter-Analysis Workflow}
\workCascade:  between static analysis (to mark and characterize Floating-Point-computation-dependent program parts) and dynamic symbolic execution to generate test cases and prove safety. 

\subsubsection*{Mapping-Function Interpretation Structure}
\structCFG: static analysis computes and propagates the information on the Floating-Point-computation-dependent instructions.

\subsubsection*{Mapping-Function Mechanics}
\mechanicAssociation: static analysis computes and propagate the information on the Floating-Point-computation-dependent instructions.

\subsection{Automatic generation of oracles for exceptional behaviors \cite{goffi_automatic_2016}}

\subsubsection*{Summary}
This technique statically processes Javadoc "throws" comments according to the Toradocu approach, and uses that information to add exception-related oracles to test cases generated with EvoSuite and Randoop, to cover more bugs and filter out false alarms.

\subsubsection*{Synergistic Effects}
\synAlarms~/~\synAlarmsStatic: the identified oracles allows for avoiding  that the generated test cases signal false alarms upon observing exceptions that, however, represent the correct behavior according to the Javadoc specifications. 

\subsubsection*{Inter-Analysis Workflow}
\workCascade: the analysis of the Javadoc allows for enriching test suites with oracles.

\subsubsection*{Mapping-Function Interpretation Structure}
\structCG: the oracles extracted from Javadoc comments are associated with the class methods. 

\subsubsection*{Mapping-Function Mechanics}
\mechanicSummary: The processing of Javadoc comments is further refined in the form of logic-formula-like contracts for the class methods.

\subsection{Can We Predict the Quality of Spectrum-based Fault Localization? \cite{golagha_can_2020}}

\subsubsection*{Summary}
The paper combines static and dynamic derived software metrics to construct a prediction model that is effective in the task of fault localization.

\subsubsection*{Synergistic Effects}
\synFeature: the technique aims to obtain reliable predictions of faultiness by exploiting static and dynamic software metrics in combination.

\subsubsection*{Inter-Analysis Workflow}
\workSidebyside: the technique computes static and dynamic software metrics separately, and then integrates the metrics in a feature vector to train a multi-class classifier of the faultiness levels of the class.

\subsubsection*{Mapping-Function Interpretation Structure}
\structModules: the metrics are seen as features associated with Java classes.

\subsubsection*{Mapping-Function Mechanics}
\mechanicML:the metrics are the results of numerically summarizing the execution of static and dynamic analysis algorithms. 

\subsection{Comparing and Combining File-based Selection and Similarity-based Prioritization towards Regression Test Orchestration \cite{greca_comparing_2022}}

\subsubsection*{Summary}
The paper investigates a combination of test case selection (TCS) and test case prioritization (TSP) approaches. To do so, it uses two existing tools for test case selection (which uses dynamic analysis to extract dependencies and determine relevant tests to be re-run) and test case prioritization (which uses static analysis, in particular it uses the string representation of the test cases to determine the prioritization) executed separately and combines their results in \workSidebyside style, to get selected and ordered test cases.

\subsubsection*{Synergistic Effects}
\synFeature: the metrics from TCS and TCP are exploited as integrated features to produce the final test cases to be executed in the given order.

\subsubsection*{Inter-Analysis Workflow}
\workSidebyside: they use two existing tools for test case selection and test case prioritization executed separately, and combine their results.

\subsubsection*{Mapping-Function Interpretation Structure}
\structModules: the results of the analysis are associated with test-case modules.
Both TCS and TCP associate properties (i.e., selection marks and rank numbers, respectively) to the test cases.

\subsubsection*{Mapping-Function Mechanics}
\mechanicML: the metrics from TCS and TCP are numerically summarized out of the results of dynamic and static analysis, respectively. 

\subsection{Shooting from the heap: Ultra-scalable static analysis with heap snapshots \cite{grech_shooting_2018}}

\subsubsection*{Summary}
The paper proposes a technique which first dynamically obtains a heap state (snapshot)  and then performs a static analysis of the program behavior based on the snapshot. The goal is to improve upon the efficiency of static analysis by focusing on a specific heap state (snapshot).

\subsubsection*{Synergistic Effects}
\synRewrite~/~\synRewriteConcrete: some relations used during static analysis (e.g., the FieldPointsTo relation and the CallGraphEdge relation) are rewritten based on the results that were observed during dynamic analysis. 

\subsubsection*{Inter-Analysis Workflow}
\workCascade: the proposed technique combines the heap state (snapshot) computed with dynamic analysis with static analysis of the program behavior.

\subsubsection*{Mapping-Function Interpretation Structure}
\structProgram:the information from multiple heap dumps is merged into facts that associate with the program under analysis. The facts relate to object field values (values an object’s fields can point to), static field values (values a class’s static fields can point to), and array content values (values an array’s contents can point to).

\subsubsection*{Mapping-Function Mechanics}
\mechanicMerging: the information from multiple heap dumps is merged to generate a generalized view of the dynamically observed snapshots.

\subsection{Preventing database deadlocks in applications \cite{grechanik_preventing_2013}}

\subsubsection*{Summary}
The paper proposes a technique for preventing database deadlocks. 
It combines static analysis, which detects hold-and-wait cycles that specify how resources (e.g., database tables) are held in contention during executions of SQL statements, with runtime monitoring, which uses the Petri net models synthesized in the static analysis phase to automatically prevent database deadlocks at runtime.

\subsubsection*{Synergistic Effects}
\synIntepretability~/~\synIntepretabilityAPI: the static analysis synthesizes a Petri Net model that adds locking-related semantics to the database APIs (i.e., the database transactions) executed by the application.

\subsubsection*{Inter-Analysis Workflow}
\workCascade: the static analysis synthesize a Petri Net model of the application's transactions, and feeds the model to the runtime monitoring stage.

\subsubsection*{Mapping-Function Interpretation Structure}
\structProgram: the result of the static analysis phase is a Petri Net model of the application's transactions, which allows for detecting the possible hold-and-wait cycles and is the input for runtime monitoring.

\subsubsection*{Mapping-Function Mechanics}
\mechanicAssociation: the Petri Net model derived in the static analysis phase is associated with the application to be monitored at runtime.

\subsection{SYNERGY: A new algorithm for property checking \cite{gulavani_synergy_2006}}

\subsubsection*{Summary}
The work proposes a technique for identifying whether or not a given program satisfies a specified safety property, modeled as an error state (abort) in the program. To do this, it uses a feedback loop in which it uses directed testing (in the style of concolic execution) to try to identify counter-example test cases, and backward reachability analysis to try to find a proof that the error state is unreachable. 

\subsubsection*{Synergistic Effects}

\synPartitioning~/~\synPartitioningWitness: the safety property represents an abstract state for which the technique aims to identify a witness, which can be either a counter-example or a proof.

\subsubsection*{Inter-Analysis Workflow}
\workFeedback: the technique alternates between direct testing and reachability analysis. 

\subsubsection*{Mapping-Function Interpretation Structure}
\structCFG: the computed test cases and reachability conditions are mapped to the program's instructions. 

\subsubsection*{Mapping-Function Mechanics}
\mechanicConstraint: is exploited to generate test cases and determine unreachable states.

\subsection{PredRacer: Predictively Detecting Data Races in Android Applications \cite{guo_predracer_2024}}

\subsubsection*{Summary}
The paper presents a technique for detecting data races in Android applications: 
\begin{itemize}
    \item It first refers to dynamic analysis to capture execution traces of an Android application and events that may potentially participate in data races,
    \item Then it statically reorders the events within the trace based on partial orders by analyzing the relations among tasks, threads, and callback methods in the application, trying to identify orderings that may result in data races with high probability. It can possibly result in false alarms. 
\end{itemize}

\subsubsection*{Synergistic Effects}
\synIntepretability~/~\synIntepretabilityEntities: the information collected dynamically allows for associating the threads with information about events that may potentially participate in data races.

\subsubsection*{Inter-Analysis Workflow}
\workCascade: the information collected dynamically, is then analyzed to identify potential data races.

\subsubsection*{Mapping-Function Interpretation Structure}
\structModules: the information collected dynamically is associated with the threads in the program.

\subsubsection*{Mapping-Function Mechanics}
\mechanicAssociation: the information collected dynamically is associated with the threads in the program.

\subsection{Hybrid Slicing: Integrating Dynamic Information with Static Analysis \cite{gupta_hybrid_1997}}

\subsubsection*{Summary}
This work applies program slicing for focusing the relevant parts of a program during debugging.  It aims to produce a hybrid slice that is more precise than the static slice and less costly than the dynamic slice. Hybrid slicing integrates dynamic information from a specific execution into a static slice analysis. 
The technique uses
dynamic information to more accurately predict control flow and thus
eliminate some of the paths that could not have been involved in the
specific execution. The particular dynamic information exploited is breakpointing information and dynamic procedure call and return information.
This information is integrated into a static slicing analysis to more accurately estimate the potential paths taken by the program. 
The breakpoint information consists of breakpoint positions in the code
that are encountered as well as breakpoint positions that are not encountered. For interprocedural slicing, procedure calls and returns are used,
including the code position of the call sites.

\subsubsection*{Synergistic Effects}
\synPartitioning~/~\synPartitioningWitness: the considered breakpoints and the possible procedure calls represent the partitions of interest, which get associated with witnessing execution data in the dynamic analysis phase.

\subsubsection*{Inter-Analysis Workflow}
\workCascade: the information on breakpoints that are executed or not executed, and executed call sites and calls, is fed to the slicing algorithm.

\subsubsection*{Mapping-Function Interpretation Structure}
\structCFG: the breakpoint information is mapped to corresponding instructions in the code.

\structCG: the dynamic call information is mapped to corresponding calls and call sites in the call graph.

\subsubsection*{Mapping-Function Mechanics}
\mechanicAssociation: the breakpoints and call sites are associated with the property of having been executed or not executed during the dynamic analysis.

\subsection{Constraint-based behavioral consistency of evolving software systems \cite{hahnle_constraint-based_2018}}

\subsubsection*{Summary}
The paper discusses how behavioral consistency (between anticipated and observed behavior)  in software systems can be captured by combining symbolic execution and runtime monitoring (augmented with specification mining).

\subsubsection*{Synergistic Effects}
\synIntepretability~/~\synIntepretabilityOracle: the model learned via symbolic execution and specification mining is instrumented in the program, allowing for runtime monitoring to interpret the execution results in order to detect anomalies. 

\subsubsection*{Inter-Analysis Workflow}
\workFeedback: runtime monitoring finds anomalous execution that triggers symbolic execution to update (in the style of specification mining) the constraint-based global model. In turn this allows for instrumenting the program for runtime monitoring of the anomalies.

\subsubsection*{Mapping-Function Interpretation Structure}
\structProgram: the results from runtime monitoring and symbolic execution allow for learning (according to specification mining) a model of the program under analysis, which in turn allows for identifying anomalies.

\subsubsection*{Mapping-Function Mechanics}
\mechanicMining.

\subsection{Strong higher order mutation-based test data generation \cite{harman_strong_2011}}

\subsubsection*{Summary}
SHOM is a mutation-based test data generation approach that combines Dynamic Symbolic Execution (DSE) and Search Based Software Testing (SBST). DSE is used to construct test inputs that weakly kill mutants, while SBST is used to propagate the infection of the state to an output, i.e., to strongly kill the mutant.

\subsubsection*{Synergistic Effects}
\synTraversal~/~\synTraversalSeedInputs. 

\subsubsection*{Inter-Analysis Workflow}
\workCascade. 

\subsubsection*{Mapping-Function Interpretation Structure}
\structPaths. 

\subsubsection*{Mapping-Function Mechanics}
\mechanicAssociation. 

\subsection{Python API Misuse Mining and Classification Based on Hybrid Analysis and Attention Mechanism \cite{he_python_2023}}

\subsubsection*{Summary}
The HatPAM approach aims to determine whether some Python APIs are misused. It performs a change analysis over the repository of the code, plus a dataflow analysis, and trains a deep-learning model. Here we refer to the combination of change analysis and dataflow analysis.

\subsubsection*{Synergistic Effects}
\synFlow.  

\subsubsection*{Inter-Analysis Workflow}
\workSidebyside. 

\subsubsection*{Mapping-Function Interpretation Structure}
\structDF. 

\subsubsection*{Mapping-Function Mechanics}
\mechanicML: a matrix with metrics is passed to the deep-learning model.  

\subsection{Profiting from Unit Tests for Integration Testing \cite{holling_profiting_2016}}

\subsubsection*{Summary}
The paper presents OUTFIT, an approach to integration testing of cascades of two components. First, KLEE is used to build tests for the downstream component, then, KLEE is again used to build tests for the integrated components by constraining symbolic execution through the upstream component with the path conditions of the downstream tests.

\subsubsection*{Synergistic Effects}
\synTraversal~/~\synTraversalSeedSink. 

\subsubsection*{Inter-Analysis Workflow}
\workCascade. 

\subsubsection*{Mapping-Function Interpretation Structure}
\structPaths. 

\subsubsection*{Mapping-Function Mechanics}
\mechanicAssociation. 

\subsection{Athena: A framework to automatically generate security test oracle via extracting policies from source code and intended software behaviour \cite{homaei_athena_2019}}

\subsubsection*{Summary}
Athena is a combined static/dynamic approach to extract oracles about the security policies of a program. Dynamic analysis over test inputs generated from user scenarios tracks the activity of a set of gates, i.e., points where the program interacts with the environment. The interprocedural CFG and the DFG are decorated by the security properties calculated by the static and dynamic analyses.

\subsubsection*{Synergistic Effects}
\synFlow. 

\subsubsection*{Inter-Analysis Workflow}
\workSidebyside. 

\subsubsection*{Mapping-Function Interpretation Structure}
\structDF. 

\subsubsection*{Mapping-Function Mechanics}
\mechanicMerging. 

\subsection{Revealing injection vulnerabilities by leveraging existing tests \cite{hough_revealing_2020}}

\subsubsection*{Summary}

RIVULET is a combined dynamic/dynamic tool for detecting several injection vulnerabilities in Java code. A dynamic taint analysis is performed based on tests provided by users to determine whether a potentially vulnerable source-sink flow exists. Then, the detected tests are perturbed and rerun with a suitable white-box approach to elicit the vulnerability.

\subsubsection*{Synergistic Effects}
\synTraversal~/~\synTraversalSeedInputs. 

\subsubsection*{Inter-Analysis Workflow}
\workCascade. 

\subsubsection*{Mapping-Function Interpretation Structure}
\structDF. 

\subsubsection*{Mapping-Function Mechanics}
\mechanicAssociation. 

\subsection{Hybrid learning: Interface generation through static; dynamic; and symbolic analysis \cite{howar_hybrid_2013}}

\subsubsection*{Summary}
The paper reports the {\sc X-Psyco} approach to infer interface automata of software components (Java classes). The automaton is learned with the $L^*$ algorithm, but several techniques (concrete execution, symbolic execution, static analysis with partial order reduction) are integrated in the teacher part of the algorithm to make everything efficient.

\subsubsection*{Synergistic Effects}
\synRefine~/~\synRefineIncorporate , \synRefinePrune 

\subsubsection*{Inter-Analysis Workflow}
\workFeedback. 

\subsubsection*{Mapping-Function Interpretation Structure}
\structProgram: the learner passes to the teacher the whole inferred automaton. 

\subsubsection*{Mapping-Function Mechanics}
\mechanicMining. 

\subsection{A Highly Scalable; Hybrid; Cross-Platform Timing Analysis Framework Providing Accurate Differential Throughput Estimation via Instruction-Level Tracing \cite{hsu_highly_2023}}

\subsubsection*{Summary}
MCAD is a tool for differential throughput estimation, i.e., prediction of performance impact of software changes. It combines a dynamic analysis based on QEMU, that collects instruction traces along with other dynamic information, with a static throughput estimation.

\subsubsection*{Synergistic Effects}
\synIntepretability~/~\synIntepretabilityEntities: static throughput analysis is usually limited as it can reliably analyze only a single basic block. Dynamic analysis provides static analysis with feasible, long instruction sequences. 

\subsubsection*{Inter-Analysis Workflow}
\workCascade. 

\subsubsection*{Mapping-Function Interpretation Structure}
\structPaths. 

\subsubsection*{Mapping-Function Mechanics}
\mechanicAssociation. 

\subsection{A Semantics-Based Hybrid Approach on Binary Code Similarity Comparison \cite{hu_semantics-based_2021}}

\subsubsection*{Summary}
The presented {\sc BinMatch} approach detects whether two pieces of binary code are semantically similar. It first performs dynamic analysis based on testing of the first piece of binary code, monitoring several usage information, then runs the other piece of binary code with the same inputs and execution context, extracting the 'semantic signature' of the two pieces of code on similar executions, and then comparing those signatures for determining whether the execution respects the same usage.

\subsubsection*{Synergistic Effects}
\synPartitioning~/~\synPartitioningDirect. 

\subsubsection*{Inter-Analysis Workflow}
\workCascade. 

\subsubsection*{Mapping-Function Interpretation Structure}
\structCG. 

\subsubsection*{Mapping-Function Mechanics}
\mechanicAssociation. 

\subsection{ACHyb: A hybrid analysis approach to detect kernel access control vulnerabilities \cite{hu_achyb_2021}}

\subsubsection*{Summary}
AcHyb is a hybrid static/dynamic tool for detecting access control vulnerabilities in operating systems. It first applies static analysis to identify the potentially vulnerable paths and then applies dynamic analysis to further reduce the false positives.

\subsubsection*{Synergistic Effects}
\synAlarms~/~\synAlarmsDynamic: dynamic analysis confirms the alarm paths detected by static analysis. 

\subsubsection*{Inter-Analysis Workflow}
\workCascade. 

\subsubsection*{Mapping-Function Interpretation Structure}
\structPaths. 

\subsubsection*{Mapping-Function Mechanics}
\mechanicAssociation. 

\subsection{Discovering Likely Program Invariants for Persistent Memory \cite{huang_discovering_2024}}

\subsubsection*{Summary}
Persistent memory, that can retain data after power loss, needs explicit synchronization operation to ensure that in case of power losses the memory content stays consistent. To infer such persistent memory invariant a combined static/dynamic approach is presented: Static analysis is used to calculate control and data dependencies and instrument the program. Dynamic analysis, based on tests and on the instrumentation, calculates a set of properties that read/write memory operations have during execution.

\subsubsection*{Synergistic Effects}
\synTraversal~/~\synTraversalSeedSink. 

\subsubsection*{Inter-Analysis Workflow}
\workCascade. 

\subsubsection*{Mapping-Function Interpretation Structure}
\structDF. 

\subsubsection*{Mapping-Function Mechanics}
\mechanicAssociation. 

\subsection{Utilization of Dependence and Weight to Improve Fault Localization Method of Regression Test Cases \cite{hui_utilization_2017}}

\subsubsection*{Summary}
A fault localization method is proposed that calculates coverage targets from program dependence graph and uses them in the style of spectrum-based fault localization.

\subsubsection*{Synergistic Effects}
\synTraversal~/~\synTraversalSeedSink. 

\subsubsection*{Inter-Analysis Workflow}
\workCascade. 

\subsubsection*{Mapping-Function Interpretation Structure}
\structDF. 

\subsubsection*{Mapping-Function Mechanics}
\mechanicAssociation. 

\subsection{Testing of data-centric and event-based dynamic service compositions \cite{hummer_testing_2013}}

\subsubsection*{Summary}
A technique is presented to test compositions of services. A first stage calculates the dataflow dependencies between the services that participate in a given composition. A second stage synthesizes a set of test cases to satisfy all the relevant dependencies.

\subsubsection*{Synergistic Effects}
\synPartitioning~/~\synPartitioningCoverage. 

\subsubsection*{Inter-Analysis Workflow}
\workCascade. 

\subsubsection*{Mapping-Function Interpretation Structure}
\structDF. 

\subsubsection*{Mapping-Function Mechanics}
\mechanicAssociation. 

\subsection{Efficient testing of different loop paths \cite{huster_efficient_2015}}

\subsubsection*{Summary}
The paper introduces a technique for testing programs with loops. Its first stage performs static analysis identifying equivalence classes of loop paths, and determines a set of representative loop paths sufficient to cover each equivalence class. The second stage uses symbolic execution to build a test for each path. 

\subsubsection*{Synergistic Effects}
\synPartitioning~/~\synPartitioningDirect. 

\subsubsection*{Inter-Analysis Workflow}
\workCascade. 

\subsubsection*{Mapping-Function Interpretation Structure}
\structPaths. 

\subsubsection*{Mapping-Function Mechanics}
\mechanicAssociation. 

\subsection{Using robustness testing to handle incomplete verification results when combining verification and testing techniques \cite{huster_using_2017}}

\subsubsection*{Summary}
The paper augments an approach based on Hoare-style verification (Code Contracts) complemented with automated tests on unverified contracts (Pex), by adding robustness tests over contracts that depend on unverified contracts, in order to inject further robustness checks and avoid propagation of possible errors.

\subsubsection*{Synergistic Effects}
\synAlarms~/~\synAlarmsDynamic. 

\subsubsection*{Inter-Analysis Workflow}
\workCascade. 

\subsubsection*{Mapping-Function Interpretation Structure}
\structDF. 

\subsubsection*{Mapping-Function Mechanics}
\mechanicAssociation. 

\subsection{Scalable and scope-bounded software verification in Varvel \cite{ivancic_scalable_2015}}

\subsubsection*{Summary}
Varvel is a software model checker for C programs. To enhance scalability it abstracts the call environment of the checked C function, and some of the functions this may transitively invoke. The result of this abstraction are (likely) pre/postconditions and function stubs that are used in place of the removed code to improve the successive model checking.

\subsubsection*{Synergistic Effects}
\synIntepretability (\synIntepretabilityEntities). 

\subsubsection*{Inter-Analysis Workflow}
\workCascade. 

\subsubsection*{Mapping-Function Interpretation Structure}
\structCG. 

\subsubsection*{Mapping-Function Mechanics}
\mechanicSummary. 

\subsection{Test oracle assessment and improvement \cite{jahangirova_test_2016}}

\subsubsection*{Summary}
This work addresses improvement of test oracles (specifically, runtime assertions) by combining test case generation (that they use to investigate false positives from the oracles) and mutation testing (that they use to investigate false negatives from the oracles). By doing so they then improve the oracles, in order to reduce the incidence of both false positives and false negatives.

\subsubsection*{Synergistic Effects}
\synTraversal~/~\synTraversalTransform: the programs are transformed by inserting fabricated branch that, if executed during search-based testing (executed against the original programs or the mutants), indicate false-positive or false-negative outcomes of the current oracles. 

\subsubsection*{Inter-Analysis Workflow}
\workSidebyside: the test execution sessions to  reveal false-positive or false-negative outcomes of current oracles, respectively, are independently executed and provide complementary results to the analysts in charge of refining the assertions.

\subsubsection*{Mapping-Function Interpretation Structure}
\structCFG: the test cases that reveal false-positive or false-negative outcomes of current oracles are associated with the assertions rendered as instructions in the program.

\subsubsection*{Mapping-Function Mechanics}
\mechanicAssociation: the test cases that reveal false-positive or false-negative outcomes of current oracles are associated with the assertions.

\subsection{Generating classified parallel unit tests \cite{jannesari_generating_2014}}

\subsubsection*{Summary}
AutoRT is a  unit test generator for testing races in parallel programs which uses both dynamic and static approaches for program analysis.
For a given program the algorithm considers all possible method pairs as candidates for parallelism-oriented unit testing, where a test case with consist of running those methods in separate threads.
The algorithm identifies relevant method pairs in two independent analyses:
1) A static analysis identifies a candidate set of method pairs such that two methods access the same variables.
2) A dynamic analysis identifies the candidate set to method pairs as two methods that were observed to run in parallel.

\subsubsection*{Synergistic Effects}
\synReports: two analysis stage indicate method pairs that can potentially lead to races, and the final result is computed by integrating their respective results.

\subsubsection*{Inter-Analysis Workflow}
\workSidebyside: the two analysis stages independently identify a candidate set of method pairs each, and then the results are intersected to identify the final  set of candidate method pairs to be considered in the test cases.

\subsubsection*{Mapping-Function Interpretation Structure}
\structCG: the results are provided with reference to the methods of the program under test.

\subsubsection*{Mapping-Function Mechanics}
\mechanicAssociation: the results of both analysis stages are associated with the methods of the program under test.

\subsection{Test data generation and feasible path analysis \cite{jasper_test_1994}}

\subsubsection*{Summary}
This work describes a technique (called TSDT) to generate test cases that satisfy MCDC coverage for Ada programs. TSDT starts from existing test cases and analyzes the program under test to compute MCDC coverage and missing MCDC elements. Then this information is further processed based on symbolic analysis (weakest preconditions) to compute test cases that improve MCDC coverage.

\subsubsection*{Synergistic Effects}
\synPartitioning~/~\synPartitioningCoverage: the overall technique aims at addressing the MCDC coverage criterion.

\subsubsection*{Inter-Analysis Workflow}
\workCascade: the coverage analysis stage computes MCDC coverage and missing MCDC elements, and is followed by the test generation stage. 

\subsubsection*{Mapping-Function Interpretation Structure}
\structCFG: covered and missing MCDC are interpreted with reference to the decision points in the program.

\subsubsection*{Mapping-Function Mechanics}
\mechanicAssociation: covered and missing are associated with to corresponding decision points in the program.

\subsection{An automatic approach of domain test data generation \cite{jeng_automatic_1999}}

\subsubsection*{Summary}
The paper combines static and dynamic approaches for test data generation for boundary testing of program branch conditions (called ON and OFF points of a branch condition in the paper). The program starts from a seed input, and determines the corresponding path condition in similar fashion as concolic execution. Then it uses a constraint solving method based on hill-climbing to find the ON and OFF points of the predicates in the path condition. 
The second stage uses program slicing and adapts the concolic method of the first step to reason on the ON and OFF points missed in the previous stage.

\subsubsection*{Synergistic Effects}

\synPartitioning~/~\synPartitioningCoverage: the synergy relates with incrementally satisfying the ON and OFF points of the branch conditions in the program.

\subsubsection*{Inter-Analysis Workflow}
\workCascade: concolic execution is followed by program slicing and then again by concolic execution.

\subsubsection*{Mapping-Function Interpretation Structure}
\structCFG: the analysis stages propagate coverage data and test cases related to the ON and OFF points of the program branch conditions.

\subsubsection*{Mapping-Function Mechanics}
\mechanicAssociation: coverage data and test cases are associated with the ON and OFF points of the program branch conditions.

\subsection{F3: Fault localization for field failures \cite{jin_f3_2013}}

\subsubsection*{Summary}
This paper presents F3 a novel technique that builds on BugRedux and extends it with support for fault localization. The technique relies on KLEE (symbolic execution) to identify test cases that lead to a crash (a stack trace available from a crash that occurred during execution in the field). Then test cases are used to perform profiling with the goal of performing statistical fault localization.

\subsubsection*{Synergistic Effects}
\synPartitioning~/~\synPartitioningDirect: the results from BugRedux direct the fault-localization stage to focus on specific (passing and failing)  program traces. 

\subsubsection*{Inter-Analysis Workflow}
\workCascade: the results from BugRedux are fed to the fault localizations stage.

\subsubsection*{Mapping-Function Interpretation Structure}
\structProgram: The results from BugRedux are test cases associated with the program under test.

\subsubsection*{Mapping-Function Mechanics}
\mechanicAssociation. 

\subsection{Efficient runtime-enforcement techniques for Policy weaving \cite{joiner_efficient_2014}}

\subsubsection*{Summary} 
The paper focuses on policy-weaving systems. Policy weaving is a program-transformation technique that rewrites a program so that it is guaranteed to be safe with respect to a security policy. To do this, it combines static analysis (to identify points in the program at which policy violations might occur) and runtime checks (inserted at such points to monitor policy states and prevent violations from occurring).
At runtime, they either suppress the execution of violating statements, or dynamically instrument generated code that is not available to the static analysis. 

\subsubsection*{Synergistic Effects}
\synTraversal~/~\synTraversalTransform: the static analysis stage allows for augmenting the program with runtime checks, to monitor policy states and prevent violations from occurring.

\subsubsection*{Inter-Analysis Workflow}
\workCascade: static analysis identifies points in the program at which policy violations may occur, and inserts runtime checks  at such points. At runtime, the transformed programs implements a dynamic analysis that monitors the program states, aiming to prevent violations from occurring.

\subsubsection*{Mapping-Function Interpretation Structure}
\structCFG: instructions to be transformed, associated with runtime checking code.

\subsubsection*{Mapping-Function Mechanics}
\mechanicAssociation: instructions to be transformed, associated with runtime checking code.

\subsection{Addressing the regression test problem with change impact analysis for Ada \cite{jones_addressing_2016}}

\subsubsection*{Summary}
The paper introduces an approach to solving the test selection problem for regression testing given a combination of both static and dynamic data for a program and a change-set.
The technique consists in:
\begin{itemize}
    \item identifying the difference between the original and modified source code. This phase is a static analysis that identifies changes at the interface, package and function/procedural levels. The changes as a whole are called a change-set.
    \item They build a dependency graph of the Ada source code, by merging both static data (in particular dependencies on other ADA packages either at the type or use level) and dynamic data (in particular subprograms invoked as observed from the test cases, at a (previous) execution of the existing test baseline). 
    \item Given the change-set and the dependency graph, a subset of the test suite that is affected by the changes in the change-set is extracted.
\end{itemize}

\subsubsection*{Synergistic Effects}
\synFlow: static and  information are combined to represent the possible execution flows through functions, procedures and packages.

\subsubsection*{Inter-Analysis Workflow}
\workSidebyside: static information on dependencies with respect to the change set is merged with  dynamic dependencies with respect to the test cases.  

\subsubsection*{Mapping-Function Interpretation Structure}
\structModules: the dependency and execution information refer to call and use relations at unit/package levels.

\subsubsection*{Mapping-Function Mechanics}
\mechanicAssociation: the dependency and execution information is associated with corresponding calls and use relations at unit/package levels.

\subsection{PathWalker: A dynamic symbolic execution tool based on LLVM byte code instrumentation \cite{jun-xian_pathwalker_2015}}

\subsubsection*{Summary}
This work presents a technique to extend concolic execution to C programs that take data structures as input. First a static analysis  inspects the types of the  data-structure inputs and recursively unfolds their fields, allowing concolic execution to synthesize a test driver that associate the fields with primitive inputs of the test driver. Then, the technique relies on  standard concolic execution with respect to the primitive inputs.

\subsubsection*{Synergistic Effects}
\synTraversal~/~\synTraversalTransform: the static analysis stage augments the target program with the test driver that handles the fields of the data-structure inputs, then concolic execution can thus traverse program paths that depend on the data-structure inputs.

\subsubsection*{Inter-Analysis Workflow}
\workCascade: static analysis generate the test driver and then concolic execution generates test cases.

\subsubsection*{Mapping-Function Interpretation Structure}
\structCG: the test driver is associated with a target function of a program under analysis.

\subsubsection*{Mapping-Function Mechanics}
\mechanicAssociation: the results of statically analyzing the types of the inputs are fed to the concolic execution stage, where they are exploited to seed suitable symbolic inputs for concolic execution.

\subsection{Lightweight state capturing for automated testing of multithreaded programs \cite{kahkonen_lightweight_2014}}

\subsubsection*{Summary}
The paper combines dynamic symbolic execution with a state-matching approach, in order to avoid the exploration of execution paths that correspond to the same states multiple times. The state-matching stage receives the information on the states explored with DSE, internally models (with a Petri net) the already explored interleavings. Based on such models, it can feed DSE with decisions on whether or not a next-state-candidate should be explored.

\subsubsection*{Synergistic Effects}
\synPartitioning~/~\synPartitioningCoverage: the information that DSE passes to the state-matching stage represents the executed paths, while the information fed by the state-matching stage indicates which program paths that DSE is computing are valid candidates for further exploration. 

\subsubsection*{Inter-Analysis Workflow}
\workFeedback: the state-matching stage receives the information on the states explored with DSE, internally models (with a Petri net) the explored interleavings, and feeds DSE back with decisions on whether or not a next-state-candidate shold be explored.

\subsubsection*{Mapping-Function Interpretation Structure}
\structPaths: DSE data and state-matching decisions are associated with the program paths being explored.

\subsubsection*{Mapping-Function Mechanics}
\mechanicAssociation: DSE data and state-matching decisions are associated with the program paths being explored.

\subsection{A change impact analysis tool: Integration between static and dynamic analysis techniques \cite{kama_change_2015}}

\subsubsection*{Summary}
This paper combines static and dynamic analysis techniques for impact analysis, aiming at providing reliable estimation on potential impacted classes.
The static analysis part consists in identifying classes directly and indirectly  affected by the changes in the requirements. The subsequent dynamic analysis analyzes the method execution paths extracted from the classes identified in the previous stage, to remove false-detected impacted classes.

\subsubsection*{Synergistic Effects}
\synRefine~/~\synRefinePrune: the synergy with dynamic analysis allows to eliminate false impacted classes from the impact-analysis model.

\subsubsection*{Inter-Analysis Workflow}
\workCascade: static analysis maps change requests to impacted classes by considering the tracking data between requirements and classes, and the dependencies detected in the code of the classes. The dynamic analysis collects data on the method execution paths of the classes reported by the static analysis step, aiming to eliminate false impacted classes.

\subsubsection*{Mapping-Function Interpretation Structure}
\structModules: the static analysis computes information on the dependencies between the classes in the software under test. 

\subsubsection*{Mapping-Function Mechanics}
\mechanicAssociation: the information computed in the static analysis phase is associated with the classes in the software under test. 

\subsection{Effective and efficient API misuse detection via exception propagation and search-based testing \cite{kechagia_effective_2019}}

\subsubsection*{Summary}
The paper aims at identifying violations of usage constraints of APIs.  
To do this, it combines static exception propagation analysis with automatic search-based test case generation.
The main idea is to focus the search space of the automatic test case generator (EvoSuite) on candidate misuses (crash-prone api-call locations), i.e., method calls that might throw exceptions at runtime. 
Then, traditional code coverage heuristics and the previously identified
candidate misuses are used for focusing the automatic test suite
generator EvoSuite towards the generation of test cases
that trigger the candidates’ (propagated) exceptions.

\subsubsection*{Synergistic Effects}
\synPartitioning~/~\synPartitioningCoverage: the candidate misuses computed with static exception propagation analysis are used as coverage heuristics (along with traditional code coverage) for focusing the automatic test suite generator EvoSuite.

\subsubsection*{Inter-Analysis Workflow}
\workCascade: static exception propagation analysis is used to compute candidate misuses (crash-prone API-call locations) and this information is fed to the search-based testing stage where the candidate misuses are used as coverage goals.

\subsubsection*{Mapping-Function Interpretation Structure}
\structCG: the candidate misuses are mapped at the call sites of crash-prone API-call locations.

\subsubsection*{Mapping-Function Mechanics}
\mechanicAssociation: the candidate misuses are associated with the corresponding call sites of crash-prone API-call locations.

\subsection{Lightweight and modular resource leak verification \cite{kellogg_lightweight_2021}}

\subsubsection*{Summary}
The goal of a leak detector for a Java-like language is to ensure
that required methods (such as close()) are called on all relevant
objects. They deem this a must-call property.
The paper proposes a static analysis defined as accumulation analysis (which in turn is a special-case of typestate analysis) consisting in attaching a finite-state machine (FSM) to each relevant object. To improve the accumulation analysis, they first run an intra-procedural dataflow
analysis stage for alias tracking.

\subsubsection*{Synergistic Effects}
\synIntepretability~/~\synIntepretabilityEntities: the alias analysis stage compute information to improve the precision of the following accumulation analysis stage.

\subsubsection*{Inter-Analysis Workflow}
\workCascade: the accumulation analysis stage uses the information computed with alias analysis.

\subsubsection*{Mapping-Function Interpretation Structure}
\structCFG: alias analysis annotates alias semantics for the instructions in the programs. 

\subsubsection*{Mapping-Function Mechanics}
\mechanicAssociation: alias analysis associates alias semantics with the corresponding instructions in the programs. 

\subsection{Investigating ARIMA models of software system quality \cite{khoshgoftaar_investigating_1995}}

\subsubsection*{Summary}
This works follows the classic approach of combining static software metrics (number of operators, number of operands, number of executable statements, number of times the control flow crosses itself, fan-out, fan-in) and the historical information on the number of known faults, 
collected during testing and during in-field executions, to build a prediction model for the number of faults in the next build of the program.

\subsubsection*{Synergistic Effects}
\synFeature: the prediction model exploits the integrated set of metrics as independent variables.

\subsubsection*{Inter-Analysis Workflow}
\workSidebyside: classic approach to join static software metrics and faultiness data derived from static analysis and concrete execution, respectively.

\subsubsection*{Mapping-Function Interpretation Structure}
\structProgram: the metrics are associated with the a build of the target program.

\subsubsection*{Mapping-Function Mechanics}
\mechanicML: the metrics are summarized out of the results of static analysis and concrete executions, and used in prediction models.

\subsection{A First Look at the Inheritance-Induced Redundant Test Execution \cite{kim_first_2024}}

\subsubsection*{Summary}
The paper proposes a hybrid approach that combines static and dynamic analysis to identify and locate inheritance-induced redundant test cases.
First, static analysis is applied to extract the inheritance hierarchy in test classes, and thus identify test cases candidates that can be redundantly executed.
Then, dynamic analysis is applied while executing the test cases to compute code coverage and perform oracle analysis, to filter the candidates that are truly redundant as they result in identical coverage and same oracle results.

\subsubsection*{Synergistic Effects}
\synAlarms~/~\synAlarmsDynamic: dynamic analysis confirms the test cases that are signaled as truly redundant, avoiding false alarms on redundant test cases.

\subsubsection*{Inter-Analysis Workflow}
\workCascade: static analysis identifies redundant test cases candidates, and then they further analyze those candidate test cases in the dynamic analysis stage.

\subsubsection*{Mapping-Function Interpretation Structure}
\structModules: in this cases the test cases are seen as modules of the software under analysis.

\subsubsection*{Mapping-Function Mechanics}
\mechanicAssociation: the redundancy reports are associated with the test cases.

\subsection{Combining weak and strong mutation for a noninterpretive Java mutation system \cite{kim_combining_2013}}

\subsubsection*{Summary}
This work designs a (strong) mutation testing system optimized in efficiency by combining weak mutation analysis and strong mutation analysis. After generating the mutants, they first compute the killed mutants according to weak mutation analysis (which is efficient). Then, strong mutation is conducted only for the mutants reported as weakly killed by the first stage.

\subsubsection*{Synergistic Effects}
\synPartitioning~/~\synPartitioningWitness: the overall technique marks weakly-killed mutants, and then focus the strong mutation phases only on the mutants that were witnessed by the weak mutation phase.

\subsubsection*{Inter-Analysis Workflow}
\workCascade: the results of weak mutation analysis are fed to strong mutation analysis, in order to improve efficiency.

\subsubsection*{Mapping-Function Interpretation Structure}
\structProgram: the weakly killed mutants are associated with the program under analysis.

\subsubsection*{Mapping-Function Mechanics}
\mechanicAssociation: the weakly killed mutants are associated with the program under analysis.

\subsection{Precise Concolic Unit Testing of C Programs using Extended Units and Symbolic Alarm Filtering \cite{kim_precise_2018}}

\subsubsection*{Summary}
Automatically generated unit test drivers/stubs for concolic testing may lead to raising false alarms because they over-approximate the real execution contexts of a target function f, allowing for infeasible executions of f as well.
The proposed technique constructs an extended unit of f that consists of f and the functions that are closely relevant to f. The relevant functions are identified with dynamic analysis of a set of test cases, as their execution led to calling f with high probability. 
Then they provide realistic execution contexts for the function f. 
They filter out false alarms by checking feasibility of the  symbolic execution path of each alarm with respect to to f's symbolic calling contexts obtained by symbolic execution of the closely related predecessor functions.

\subsubsection*{Synergistic Effects}
\synAlarms~/~\synAlarmsStatic: the execution contexts are analyzed to refuse the alarms that are not feasible in any of the considered execution contexts.

\subsubsection*{Inter-Analysis Workflow}
\workCascade: after detecting each alarm with concolic execution, they apply the analysis of the possible execution contexts to confirm or refuse the alarm. 

\subsubsection*{Mapping-Function Interpretation Structure}
\structPaths: the alarms are associated with the path condition of the corresponding program path.

\subsubsection*{Mapping-Function Mechanics}
\mechanicAssociation: the alarms are associated with the path condition of the corresponding program path.

\subsection{A metric for evaluating residual complexity in software \cite{krisper_metric_2017}}

\subsubsection*{Summary}
The paper proposes a new metric for evaluating the complexity (called \emph{residual complexity}) of a software which is derived from the combination of a statically derived complexity metrics (e.g., the cyclomatic complexity) and a dynamically derived test coverage metric.

\subsubsection*{Synergistic Effects}
\synFeature: the residual complexity is computed by integrating the two features of the methods in a class, extracted with static and dynamic analysis, respectively.

\subsubsection*{Inter-Analysis Workflow}
\workSidebyside: the two base metrics are computed independently, and the results are combined to derive the residual complexity. 

\subsubsection*{Mapping-Function Interpretation Structure}
\structCG: the base metrics are associated with the methods in the program (and allow for computing the residual complexity of the classes that include the methods).

\subsubsection*{Mapping-Function Mechanics}
\mechanicML: the results of the sided analysis stages are summarized as function-level software metrics.

\subsection{Static detection of unsafe component loadings \cite{kwon_static_2012}}

\subsubsection*{Summary}
This paper presents a static analysis of binaries, aiming at detecting software-component-loading errors.
The techniques combines a first stage static analysis in which it performs backward program slicing, starting from the component loading call sites. A context-sensitive executable slice is computed  for each execution context of the relevant call sites.
Once each slice is obtained it is then executed (the paper says \emph{emulated}) to compute what components may be loaded at the relevant program locations.
This combination of slicing and emulation aims to achieve scalability and precision.

\subsubsection*{Synergistic Effects}
\synTraversal~/~\synTraversalTransform: the first stage relies on program slice to make the analysis of the second stage address component loading-related errors related to that call-site. 

\subsubsection*{Inter-Analysis Workflow}
\workCascade: the first stage performs program slicing, then the second stages runs the slices to detect component loading-related errors via dynamic analysis. 

\subsubsection*{Mapping-Function Interpretation Structure}
\structCG: the slices from the first analysis stage are associates with the call sites where the second stage aims to focus the dynamic analysis.

\subsubsection*{Mapping-Function Mechanics}
\mechanicAssociation: each slice is associated with a call site to be dynamically analyzed.

\subsection{Staged static techniques to efficiently implement array copy semantics in a MATLAB JIT compiler \cite{lameed_staged_2011}}

\subsubsection*{Summary}
This work proposes a combination of compile-time static analyses to guarantee the copy semantics of array assignments for Matlab programs, while  minimizing the number of copies and copy checks.
The first phase applies very simple and inexpensive flow-insensitive analysis to identify read-only arrays, which then do not need to be copied.
The next phases apply variations of classic forward and backward data flow analysis, to locate all places where an array update requires a copy (forward necessary copy analysis) and then determine the best location for copies (backward copy placement analysis).

\subsubsection*{Synergistic Effects}

\synIntepretability~/~\synIntepretabilityEntities: the analysis incrementally compute the copy semantics of array assignments, with each phase relying on the semantics computed by the previous phases.

\subsubsection*{Inter-Analysis Workflow}
\workCascade: the static analysis phases are executed sequentially. 

\subsubsection*{Mapping-Function Interpretation Structure}

\structCFG: the later phases report the copy locations of the variables to be copied. 

\subsubsection*{Mapping-Function Mechanics}
\mechanicAssociation: the program variables are associated with the corresponding copy locations.

\subsection{Model extraction and test generation from JUnit test suites \cite{lamela_seijas_model_2018}}

\subsubsection*{Summary}
The technique, called James, combines control-dependency and data-dependency information derived from executing an existing test suite to infer state machine models that represent the behaviour of a target Web Service. The model is then used for test generation in the style of property-based testing.

\subsubsection*{Synergistic Effects}
\synFlow: James combines control-dependency and data-dependency information derived via dynamic analysis. 

\synIntepretability~/~\synIntepretabilityOracle: the model  built with dynamic analysis (as above) enables generating and monitoring test cases in the style of property-based testing.

\subsubsection*{Inter-Analysis Workflow}
\workSidebyside: James executes dynamic analysis to collect both control-dependency and data-dependency information out of the execution of a given test suite. Then it combines does dependency to build a model, in the style of specification mining. The model is lately used to generate test cases in property-based testing style.

\subsubsection*{Mapping-Function Interpretation Structure}
\structProgram: the model built by James represents the behavior of the program under analysis.

\subsubsection*{Mapping-Function Mechanics}
\mechanicMining: James combines control-dependency and data-dependency information, to build a model of the behavior of the program under analysis.

\subsection{Difuzer: Uncovering Suspicious Hidden Sensitive Operations in Android Apps \cite{samhi2022difuzer}}

\subsubsection*{Summary}
Difuzer is a tool for determining suspicious hidden sensitive operations in Android apps. This tool cascades two static analysis stages. The first one is an analysis of the control flow graph that determines a set of potential sensitive operations (if constructs conditions). This information is transferred to the second stage, a taint analysis that determines whether the sensitive operations, treated as taint sources, can propagate suspicious data to conditional statements, and trigger logic bombs. Difuzer further finalizes the findings by using a SVM classifiers that pinpoints the taint analysis alarms that most likely correpond to actual logic bomb cases.

\subsubsection*{Synergistic Effects}
\synAlarms~/~\synAlarmsStatic: The second stage taint analysis is used to evaluate if the potential sensitive operations identified during the first stage analysis can propagate suspicious data to conditional statements, and trigger logic bombs.

\subsubsection*{Inter-Analysis Workflow}
\workCascade: The if constructs conditions identified during the first stage analysis are provided to the second analysis. There is no feedback between the tow stages.

\subsubsection*{Interpretation Structures of Mapping Functions}
\structCFG: The information about the relevant sensitive operations (if constructs conditions) are identified on the control flow graph that is used during the taint analysis phase.

\subsubsection*{Mechanics of Mapping Functions}
\mechanicAssociation: Identified constructs in the control flow are provided as is to the second stage analysis (taint analysis).

\subsection{NIVAnalyzer: a Tool for Automatically Detecting and Verifying Next-Intent Vulnerabilities in Android Apps \cite{Tang:NIVAnalyzer:2017}}

\subsubsection*{Summary}
This paper presents NIVAnalyzer, which automatically detects and verifies next-intent vulnerabilities (NIV). It is composed by two modules. The NIV discovery module conducts static intent flow analysis, which is designed to track the target intent instance and check whether it meets all the features of NIV. Meanwhile, it generates relevant information to guide the vulnerability exploitation. The NIV exploitation module installs the vulnerable app on the Android emulator and then automatically constructs a test project which includes test cases to exploit the NIV and simulate the possible login process. With the log information collected from the Android emulator during the exploitation, it is able to see whether the vulnerability is successfully exploited.

In details:
Uses static analysis to check whether the given app contains next-intent vulnerabilities (NIV). First scans the disassembled code to find key points (instructions that retrieve the target intent from another intent). For each methods with key points, it derives its CFG and obtains all the possible paths that starts from the key point and reach an endpoint (a leaf of the CFG).
Then the intent-flow analysis takes the paths as an input and perform forward analysis to check whether this path would pass the target intent to a sink method. The sink methods are defined as inter-component communication methods. If the target intent can reach a sink method, the analysis continues to perform backward analysis to check where the proxy intent is coming from to be able to manipulate the potentially vulnerable method.
Finally, dynamic analysis is performed by creating test cases that execute the potentially vulnerable code and automatically generate logs that highlight if a NIV is found.

\subsubsection*{Synergistic Effects}

\synAlarms~/~\synAlarmsStatic: All the performed analyses have the goal of identifying potential NIV in the analyzed app starting from a less precise analysis (static) to a more precise dynamic analysis that woks in taint analysis fashion.

\subsubsection*{Inter-Analysis Workflow}
\workCascade: Each analysis stage passes their results to the following stage.

\subsubsection*{Interpretation Structures of Mapping Functions}
\begin{itemize}
    \item \structCFG: First to second stage analysis. Uses static analysis to check whether the given app contains next-intent vulnerabilities (NIV).
    First scans the disassembled code to find key points (instructions that retrieve the target intent from another intent). For each methods with key points, it derives its CFG and obtains all the possible paths that starts from the key point and reach an endpoint (a leaf of the CFG).
    Then the intent-flow analysis takes the paths as an input and perform forward analysis to check whether this path would pass the target intent to a sink method. The sink methods are defined as inter-component communication methods. If the target intent can reach a sink method, the analysis continues to perform backward analysis to check where the proxy intent is coming from to be able to manipulate the potentially vulnerable method.
    \item \structCG: Second to third stage analysis. Dynamic analysis is performed by creating test cases that execute the potentially vulnerable code and automatically generate logs that highlight if a NIV is found.
\end{itemize}

\subsubsection*{Mechanics of Mapping Functions}
\mechanicAssociation: The information derived from the various analysis stages are provided as-is to the following stage.

\subsection{ConfDiagnoser: An Automated Configuration Error Diagnosis Tool for Java Software~\cite{zhang2013confdiagnoser}}
The ConfDiagnoser technique diagnoses configuration errors in Java software. It exploits a first static analysis stage based on program slicing in which it computes the dependency between the branch conditions in the program and the configuration parameters of the program. This information is then associated with the branch conditions, and passed to the second stage analysis. 
The second stage is a dynamic analysis that users can activate to diagnose the root cause of a program crash. This stage uses the information from the first stage to instrument the program branch conditions that depend on configuration parameters. Then, when activated, it monitors which branch conditions get executed differently than in normal executions, signals the associated configuration parameters as configuration errors.

\subsubsection*{Synergistic Effects}
\synIntepretability~/~\synIntepretabilityArtifacts: The static analysis stage associates program locations with corresponding dependencies on configuration files, and then the second stage runtime analysis exploits those dependencies to diagnose failing executions due configuration issues.

\subsubsection*{Inter-Analysis Workflow}
\workCascade: The first stage statically derived slice (based on branch conditions) is provided to the second stage dynamic analysis. There is no feedback.

\subsubsection*{Interpretation Structures of Mapping Functions}
\structCFG: The static slice derived during the first stage static analysis is represented by marking the relevant branch conditions on the CFG. This enriched CFG is then passed to the second stage analysis.

\subsubsection*{Mechanics of Mapping Functions}
\mechanicAssociation: The results of each analysis stage are provided as-is to the next stage analysis.

\subsection{The software model checker Blast: Applications to software engineering~\cite{beyer:blast:isttt:2007}}

Described in Example 2.3 in the paper.

\subsubsection*{Synergistic Effects}
\synAlarms~/~\synAlarmsStatic,
\synAlarms~/~\synAlarmsDynamic

\subsubsection*{Inter-Analysis Workflow}
\workFeedback

\subsubsection*{Interpretation Structures of Mapping Functions}
\structProgram,
\structPaths,
\structCG

\subsubsection*{Mechanics of Mapping Functions}
\mechanicInterpolation,
\mechanicConstraint,
\mechanicAssociation

\subsection{SLAM2: Static driver verification with under 4\% false alarms~\cite{ball:slam:fmcad:2010}}

Described in Example 2.3 in the paper.

\subsubsection*{Synergistic Effects}
\synAlarms~/~\synAlarmsStatic,
\synAlarms~/~\synAlarmsDynamic

\subsubsection*{Inter-Analysis Workflow}
\workFeedback

\subsubsection*{Interpretation Structures of Mapping Functions}
\structProgram,
\structPaths,
\structCG

\subsubsection*{Mechanics of Mapping Functions}
\mechanicInterpolation,
\mechanicConstraint,
\mechanicAssociation

\subsection{Directed automated random testing~\cite{godefroid:dart:pldi:2005}}

Described in Example 2.2 in the paper.

\subsubsection*{Synergistic Effects}
\synPartitioning~/~\synPartitioningDirect,
\synRewrite~/~\synRewriteConcrete

\subsubsection*{Inter-Analysis Workflow}
\workFeedback

\subsubsection*{Interpretation Structures of Mapping Functions}
\structPaths,
\structCG

\subsubsection*{Mechanics of Mapping Functions}
\mechanicConstraint,
\mechanicAssociation

\subsection{CUTE: a concolic unit testing engine for C~\cite{sen:cute:esec:2005}}

Described in Example 2.2 in the paper.

\subsubsection*{Synergistic Effects}
\synPartitioning~/~\synPartitioningDirect,
\synRewrite~/~\synRewriteConcrete

\subsubsection*{Inter-Analysis Workflow}
\workFeedback

\subsubsection*{Interpretation Structures of Mapping Functions}
\structPaths,
\structCG

\subsubsection*{Mechanics of Mapping Functions}
\mechanicConstraint,
\mechanicAssociation

\subsection{Check'n'crash: combining static checking and testing~\cite{csallner:check-n-crash:icse:2005}}

Described in Example 2.1 in the paper.

\subsubsection*{Synergistic Effects}
\synAlarms~/~\synAlarmsDynamic

\subsubsection*{Inter-Analysis Workflow}
\workCascade

\subsubsection*{Interpretation Structures of Mapping Functions}
\structCG

\subsubsection*{Mechanics of Mapping Functions}
\mechanicConstraint

\subsection{Combined Constraint-Based Analysis for Efficient Software Regression Detection in Evolving Programs \cite{le_combined_2013}}

\subsubsection*{Summary}
This paper proposes an approach for effective detection of software regression in evolving programs.
It introduces a concolic-based approach to detect the regression bugs. Instead of individually performing concolic testing on both old and new programs, it suggests combining constraints extracted from both programs, thus ensuring the detection of any regression error. They propose two versions of the algorithm, the first one simply tries to combine all constraints; since this is inefficient, they propose an evolution of it that takes into account only solvable constraints. The evolved version generates a test case for each path condition of the original program (always in concolic execution style), following this it will use each test case to find its corresponding path conditions in the original and evolved program versions. Based on the retrieved path conditions it keeps generating relevant constraints to generate suitable test cases, that are used to derive more path conditions.

\subsubsection*{Synergistic Effects}
\synPartitioning~/~\synPartitioningDirect

The aim of concolic execution is to increase the efficiency and the effectiveness of traditional symbolic execution in exploring the path space of a program, usually for the sake of generating test cases that execute the program paths explored thereby.
In this specific case, the technique leverages it to solve path conditions that are derived from the combination of path conditions of two versions of the same program; but the underlying technique is basically the same. 

\subsubsection*{Inter-Analysis Workflow}
\workFeedback.

Each time a new test case is generated it can be used to derive new constraints that are used to potentially generate new combinations of path conditions, that will once again lead to the creation of new test cases. This is the classic concolic execution behavior.

\subsubsection*{Interpretation Structures of Mapping Functions}
\begin{itemize}
    \item Runtime monitoring → Symbolic: \structPaths.

    Execution traces associated to corresponding program paths.

    \item Symbolic → Runtime monitoring: \structCG.

    Symbolic execution feeds inputs to further execute the target function. synthesized via constraint solving, for executing the target function(s).
    
\end{itemize}

\subsubsection*{Mechanics of Mapping Functions}
\begin{itemize}
    \item Runtime monitoring → Symbolic: \mechanicAssociation.

    Execution traces associated to corresponding program paths.

    \item Symbolic → Runtime monitoring: \mechanicConstraint.

    Constraint solving is used to identify input values that satisfy the path conditions that symbolic execution identified as executability conditions of yet-unvisited program paths.
\end{itemize}

\subsection{HybriDroid: Static analysis framework for android hybrid applications \cite{lee_hybridroid_2016}}

\subsubsection*{Summary}
This paper presents a static analysis technique that analyzes Android hybrid apps by constructing call graphs for both Java and JavaScript, and bridging the two results to provide the call graph of hybrid programs that include both JavaScript and Java code. They consider the bridge communication mechanism to call Java methods from JavaScript programs.

\subsubsection*{Synergistic Effects}
\synIntepretability~/~\synIntepretabilityEntities

Combining the two analysis stages allow for improve the interpretability as the final call graph represents the class in both Java and JavaScript code.

\subsubsection*{Inter-Analysis Workflow}
\workCascade.

The result achieved for the Java parts can be cascaded in the call graph constructed while analyzing the JavaScript parts. 

\subsubsection*{Interpretation Structures of Mapping Functions}
\structCG.

The Java analysis propagates the call graph of the Java methods that can be called from JavaScript.

\subsubsection*{Mechanics of Mapping Functions}
\mechanicAssociation.

The Java analysis associates the call graph of the Java methods with the calls of those methods.

\subsection{A sealant for inter-app security holes in android \cite{lee_sealant_2017}}

\subsubsection*{Summary}
SEALANT is an integrated technique that monitors and protects ICC (Inter-component communication) paths through which Android inter-app attacks can take place. 
SEALANT recognizes each instance of ICC as a relation between a sender, a receiver, and an intent. When an intent
from a sender component matches an intent that can be received by a receiver component, SEALANT reports an ICC relation. SEALANT builds an ICC graph in which vertices are components and edges are the ICC relations. It then extracts all possible vulnerable ICC paths in the ICC graph and monitors them at runtime. When an instance of ICC matches one of the extracted vulnerable paths, SEALANT may block it based on the user’s choice.
SEALANT’s combination of static and dynamic analysis improves upon existing techniques in automatically identifying the vulnerable ICC paths between a set of apps, monitoring each instance of ICC to detect potential attacks, and empowering end-users to stop the attacks.

\subsubsection*{Synergistic Effects}
\synAlarms~/~\synAlarmsDynamic

The static analyzer identifies the ICC paths connecting two apps via intent and that may expose potential security vulnerabilities. The runtime monitoring is able to detect which paths are being activated at runtime so that they can be reported and eventually blocked by the user.

\subsubsection*{Inter-Analysis Workflow}
\workCascade.

The list of ICC paths identified with the static analyzer is passed to the runtime monitoring to monitor their execution.

\subsubsection*{Interpretation Structures of Mapping Functions}
\structPaths.

The static analyzer produces a set of paths that needs to be monitored by the runtime monitoring component.

\subsubsection*{Mechanics of Mapping Functions}
\mechanicAssociation.

Statically derived ICC paths associated to corresponding executed ICC paths.

\subsection{FlakeRepro: automated and efficient reproduction of concurrency-related flaky tests \cite{leesatapornwongsa_flakerepro_2022}}

\subsubsection*{Summary}
FlakeRepro is a tool which aims at reproducing failures of concurrency-related flaky tests. A static analysis stage, which uses data- and control-flow analysis named FlakeAnalyzer, determines a set of critical memory accesses that may affect the outcome of the test. Next, a dynamic stage, named FlakeFinder, systematically explores the interleavings of the critical memory accesses identified by the first stage, in search for an interleaving that makes the test fail.

\subsubsection*{Synergistic Effects}
\synAlarms~/~\synAlarmsDynamic

During the static analysis phase, FlakeRepro starts from the failed assertion, and identifies the relevant parts of the code that lead to the fail by performing data- and control-flow slicing. After FlakeAnalyzer identifies a set of critical accesses that may affect the test outcome, FlakeFinder dynamically explores the space of interleavings of critical accesses to find a buggy interleaving that causes the test failure.

\subsubsection*{Inter-Analysis Workflow}
\workCascade.

The results of the static analysis part of the technique are provided to the second part without any additional feedback from it.

\subsubsection*{Interpretation Structures of Mapping Functions}
\structCFG.

The static analysis part identifies the statements in the code related to critical accesses that may affect the test outcome. These critical accesses locations are dynamically monitored to find buggy interleavings.

\subsubsection*{Mechanics of Mapping Functions}
\mechanicAssociation.

The critical access locations identified in the code are the ones that are dynamically monitored.

\subsection{Calculating source line level energy information for Android applications \cite{li_calculating_2013}}

\subsubsection*{Summary}
The paper presents a new approach that provides developers with source line level energy information.
It works as follows: While measuring the energy consumption of a smartphone at hardware level, the approach uses efficient path profiling to identify which parts of the application are executing and correlates these paths with the measured energy. Then, the approach statically analyzes the paths to identify and adjust for high-energy events, such as thread switching, before applying robust regression analysis to calculate each source line's energy consumption.
In detail:
In the first step, a tester executes an instrumented version of the program that tracks the CFG path executed while a power meter records the power usage. Power meter data and executed instructions carry a timestamp to associate to each other. This process is reiterated for several tests.

In the second step, the paths are statically analyzed to adjust the corresponding energy measurements to account for special API invocations, tail energy, and interleaving threads. 

In the third step, linear regression analysis is used in order to calculate each source line's energy consumption. In detail, the technique sets up the equations E = mX, where E is the adjusted power measurements, X represents the path traversals with each row representing a frequency vector of bytecodes present in the measured path. Then the technique solves for the coefficients m by means of regression to determine the energy consumed by the specific instructions in the path segment.

\subsubsection*{Synergistic Effects}
\synRefine~/~\synRefinePrune

The precision of a program model can be also refined by pruning away some portions of the model that were initially included for the sake of guaranteeing conservative results, but could be then identified as irrelevant based on the information conveyed from the partner analysis stage.

The dynamic analysis step detects the energy used during the execution of a certain path, this information (which cannot be obtained statically), is then used by the static analyzer to further refine the energy consumption measurements by adjusting it for high-energy events. This information is finally used to derive a metric that expresses the energy consumption of a line of code, which could not have been compute without either static or dynamic analysis.

\subsubsection*{Inter-Analysis Workflow}
\workCascade.

Each step is executed in sequence without any sort of feedback.

\subsubsection*{Interpretation Structures of Mapping Functions}
\structPaths.

The dynamic analysis part tracks the paths executed and the associated energy usages.
These paths are then statically analyzed to adjust energy measurements to account for special API invocations, tail energy, and interleaving threads.
The paths with the adjusted power measurements are the used to predict the line level energy consumption by making use of linear regression.

\subsubsection*{Mechanics of Mapping Functions}
\mechanicAssociation.

The paths identified during the dynamic analysis are provided as-is to the static analysis part that can use them to map them directly to corresponding instructions in a CFG.

\subsection{View-based maintenance of graphical user interfaces \cite{li_view-based_2008}}

\subsubsection*{Summary}
This paper investigates the combination of a hybrid dynamic and static approach to allow for view-based maintenance of GUIs. Dynamic analysis reconstructs object relationships, providing a concrete context in which maintenance can be performed. Static checking restricts that only changes in the design view which can meaningfully be translated back to source are allowed.
In detail:
A dynamic model is created by manually executing the program, making sure that a dynamic view the developer is interested in, is rendered. The model is created by intercepting the creation of a widget, the changes to properties of a widget and the containment relationship between container widgets and their child widgets.
The second step is triggered when a programmer wants to edit the code of a widget from the dynamic view obtained by the dynamic model. The first thing that happens is the determination of the change context by identifying the least ancestor of the affected widget that is a custom widget. In other words, starting from the widget in the component tree, they walk up the tree until they encounter a custom widget. Then, the located custom widget class is identified from the runtime model.
Next, the located custom class is analyzed using a traditional GUI editor static recovery technique that returns the static design model of the class.
Finally the sub-tree of the dynamic model rooted at the change context is compared to the design model of its class. In particular, if the top-down tree-to-tree comparison matches it means that the change can be performed because the dynamic view matches the static implementation.

\subsubsection*{Synergistic Effects}
\synRewrite~/~\synRewriteSimulator

Once the dynamic model is obtained, the programmer selects a code item to edit from within the derived dynamic view. From this the affected custom widget class is identified and analyzed using static analysis to derive the design model of the class.

\subsubsection*{Inter-Analysis Workflow}
\workCascade.

Once the dynamic model is obtained, the programmer selects a code item to edit from within the derived dynamic view. From this the affected custom widget class is identified and analyzed using static analysis to derive the design model of the class. There is a clear order to the execution and there is no feedback provided between the techniques.

\subsubsection*{Interpretation Structures of Mapping Functions}
\structGUI.

The dynamic and static design models (their sub-trees) are the structures which are compared to determine if a change can be performed.

\subsubsection*{Mechanics of Mapping Functions}
\mechanicAssociation.

The two design models are directly compared against each other.

\subsection{Human-Machine Collaborative Testing for Android Applications \cite{li_human-machine_2023}}

\subsubsection*{Summary}
This paper proposes an approach that combines classic static program analysis and crowdsourced testing to implement human-machine collaborative testing for Android applications.
The presented technique employs the static analysis technique to model the possible GUI window sequences into window transition graphs (WTG). Then, a depth-first search algorithm is used to traverse the WTG and generate the testing task (test path) lists.
In the testing process, they recommend these tasks for testers and adjust the task priority based on user feedback to optimize collaborative testing efficiency and effectiveness.
There is no ``strict'' interaction but:
For each testing task, the corresponding activities are extracted, and their names are used by a test automation tool (Appium) to obtain the GUI element corresponding to each activity. The GUI will then be screenshotted and shown to the testers to help them in identifying the testing objective.

\subsubsection*{Synergistic Effects}
\synPartitioning~/~\synPartitioningDirect

The activities which are represented as code, are executed to obtain their concrete representation in the GUI so that it can be screeshotted and shown to testers.

\subsubsection*{Inter-Analysis Workflow}
\workCascade.

The two stages of the technique (activity identification and activity execution) are sequential.

\subsubsection*{Interpretation Structures of Mapping Functions}
\structGUI.

The first technique identifies the GUI activities statically, while the second technique executes them.

\subsubsection*{Mechanics of Mapping Functions}
\mechanicAssociation.

The identified activities are used directly by the second stage of the technique.

\subsection{Fault localization to detect co-change fixing locations \cite{li_fault_2022}}

\subsubsection*{Summary}
The paper proposes a novel Deep Learning-based Fault Localization approach that aims to locate co-change fixing locations within one or multiple methods. To do this it extracts method-level and statement-level features that are then used to train some machine learning models with the goals of localizing co-change fixing locations. The features are extracted statically (e.g. method name, structure (AST)...) and dynamically (stack trace of a test case faulty execution, execution trace, etc...). 

\subsubsection*{Synergistic Effects}
\synFeature.

The synergistic effect between the static and dynamic analysis arises from the joining together of the computed feature vectors, as this allows for the machine learning algorithm at hand to consider a richer set of possible correlations.

\subsubsection*{Inter-Analysis Workflow}
\workSidebyside.

The statically and dynamically derived features are extracted independently from each other without any sequentiality.

\subsubsection*{Interpretation Structures of Mapping Functions}
\structCFG.

The technique provides an indicator of a faulty statement.

\subsubsection*{Mechanics of Mapping Functions}
\mechanicML.

The features becomes part of the decision model internal to the classification technique.

\subsection{Mining android app usages for generating actionable GUI-based execution scenarios \cite{linares-vasquez_mining_2015}}

\subsubsection*{Summary}
The framework presented in the paper aims at mining models capable of generating feasible and fully replayable scenarios reflecting either natural user behavior or uncommon usages for a given app.

The technique consists of several stages:
\begin{itemize}
    \item developers/testers use the app naturally; the event logs representing scenarios executed by the developers/testers are recorded. Each log line represents an action that has been performed on a specific part of the GUI which does not contain information such as Activity, GUI component, window, etc.;
    \item The Data collector replays the logs in a ripping mode in order to dynamically collect the GUI information related to the event. These will be stored in a tuple containing: Activity, window, GUI component, action performed and component class;
    \item the source code of the app and the event sequences are mined to build a vocabulary of feasible events. The event sequences are used to augment the dynamically built vocabulary from user event streams. To achieve this, GUI components are extracted from decompiled APKs before links between these components to activities, windows, and actions/gestures are constructed.
    \item language models are derived using the vocabulary of feasible events;
    \item the models are used to Generate event sequences;
    \item the sequences are validated on the target device (infeasible events are removed for generating actionable scenarios)
\end{itemize}

The relevant stages for our survey are the first three.

\subsubsection*{Synergistic Effects}
\begin{itemize}
    \item Stage 1 $\to$ Stage 2: \synPartitioning~/~\synPartitioningDirect.
    
The logged events direct the replay of the execution in ripping-mode to collect GUI items and actions.
The sequence of events saved as logging info during user execution is used as an input by a replay tool that extracts more informative GUI data.

    \item Stage 2 $\to$ Stage 3: \synGUI.
    
    The statically derived event sequences are used to augment the dynamically built vocabulary from user event streams.
\end{itemize}
\subsubsection*{Inter-Analysis Workflow}
\begin{itemize}
    \item Stage 1 $\to$ Stage 2: \workCascade.

    The first stage needs to be executed before the second stage, there is no feedback.

    \item Stage 2 $\to$ Stage 3: \workSidebyside.
    
    These two stages can be executed in an independent order since they do not use the output from the other.

\end{itemize}
\subsubsection*{Interpretation Structures of Mapping Functions}
\begin{itemize}
    \item Stage 1 $\to$ Stage 2: \structProgram.
    
    The sequence of events saved as logging info for the program. Then, during user execution is used as an input by a replay tool that extracts GUI information.

    \item Stage 2 $\to$ Stage 3: \structGUI.
    
    The vocabularies of GUI event tuples derived statically and dynamically are merged. 
\end{itemize}

\subsubsection*{Mechanics of Mapping Functions}
\mechanicAssociation.

For both stages the extracted information are used as is when passing them to the next stage or when combining them.

\subsection{MARVIN: Efficient and Comprehensive Mobile App Classification through Static and Dynamic Analysis \cite{lindorfer_marvin_2015}}

\subsubsection*{Summary}
Described in Example 2.4 in the paper.

\subsubsection*{Synergistic Effects}
\synFeature.

\subsubsection*{Inter-Analysis Workflow}
\workSidebyside.

\subsubsection*{Interpretation Structures of Mapping Functions}
\structProgram.

\subsubsection*{Mechanics of Mapping Functions}
\mechanicML.

\subsection{Essential or Excessive? MINDAEXT: Measuring Data Minimization Practices among Browser Extensions \cite{ling_essential_2024}}

\subsubsection*{Summary}
The paper presents MINDAExT, a framework that automatically examine end-to-end data minimization practices in browser extensions by description text analysis and hybrid program analysis techniques.

MINDAExT consists of two parts: one aims at determining the set of minimized personal data essential to each extension’s purpose/functionality, the second one identifies the set of personal data collected by extensions.
Only the second part is relevant to our survey since the first one relies on NLP to derive the set of essential personal data and no static/dynamic analysis is performed.

To identify the set of personal data collected by extensions, two analyses are performed: 1) code feature analysis on privacy-related APIs in source code, and 2) runtime analysis on dynamically loaded UI pages and network traffic.

During code feature analysis, the JavaScript code of the extensions is parsed into a Abstract Syntax Trees (ASTs). Then, the ASTs are traversed using depth-first search (DFS) and all nodes in the type of CallExpression related to function calls are identified. If the function name of these nodes matches with any API in the privacy-related APIs list provided by the browser, the associated personal data types are added to the set of collected personal data.

During dynamic runtime analysis, user actions are simulated and the analysis traverses extension UIs by depth-first search. Collected personal data can be obtained in two ways:
\begin{itemize}
    \item for each UI, the tool interacts with all page elements and observe if any collected personal data is requested by checking the HTML for keyword-matching rules.
    \item while the extension is executed, HTTP(S) requests are captured. From the HTTP requests, the tool extracts all the URLs, key-value pairs in query strings, request bodies, and POST request forms. Then, it identifies the collected and transmitted data types by the same keyword-matching rules as what is used in UI analysis.
\end{itemize}

\subsubsection*{Synergistic Effects}
\synReports.

Some collected personal data can only be obtained statically while others dynamically and they are merged together.

\subsubsection*{Inter-Analysis Workflow}
\workSidebyside.

The static and dynamic analyses are not dependent on each other.

\subsubsection*{Interpretation Structures of Mapping Functions}
\structProgram.

The collected personal data comes from different analyses and at the end they are simply pooled together as a set of personal data used by the application.

\subsubsection*{Mechanics of Mapping Functions}
\mechanicAssociation.

The collected personal data are merged as is once extracted by the techniques.

\subsection{A General Framework to Detect Design Patterns by Combining Static and Dynamic Analysis Techniques \cite{liu_general_2021}}

\subsubsection*{Summary}
The paper proposes a framework for the automatic detection of design patterns by combining static and dynamic analysis.
The first step of the technique consists in the creation of a specification for each design patterns that needs to be discovered in the software. The specification contains elements such as: the role of the design pattern, a mapping from the roles to their values, the structural constraints, a function that identifies a set of pattern instance invocations from the method call set of a pattern instance, a function to check the behavioral constraints of all invocations of a pattern instance.

Static analysis takes as input the source code and return a set of candidate pattern instances by considering only the information extracted from the source code. A pattern instance is represented as a tuple of the participants (classes and methods) each representing a particular role.

Because candidate pattern instances detected by existing static tools may be incomplete, some roles needed to perform the behavioral constraint checking are missing. Software execution data is used to complete the candidates in an exhaustive way.

For each complete candidate pattern instance and its execution data, the behavioral constraints given in the specification are checked to verify whether they are satisfied with respect to all invocations of a pattern instance. A candidate pattern instance is valid if there exists at least one pattern instance invocation that satisfies all behavioral constraints, otherwise, it is not valid according to the execution data.

\subsubsection*{Synergistic Effects}

\begin{itemize}
\item \synRefine~/~\synRefineIncorporate

The precision of a program model can be refined by incorporating new details in the model, aiming to make the abstraction less coarse.

The candidate pattern instances detected by the first stage static analysis may be incomplete so, in the second stage, software execution data is used to complete the candidates in an exhaustive way.

\item \synPartitioning~/~\synPartitioningWitness

The dynamic analysis stage is performed with the goal of validating if a certain potential pattern detected statically is effectively found.

\end{itemize}

\subsubsection*{Inter-Analysis Workflow}
\workCascade.

The static analysis used to detect pattern instances is the first step of the technique. The dynamic analysis which completes and validate the pattern is the second part.

\subsubsection*{Interpretation Structures of Mapping Functions}
\structCG~\&~\structModules.

Design patterns are detected at the class and method level.

\subsubsection*{Mechanics of Mapping Functions}
\mechanicAssociation.

The partial patterns identified in the static step are verified dynamically.

\subsection{PROMAL: Precise Window Transition Graphs for Android via Synergy of Program Analysis and Machine Learning \cite{liu_promal_2022}}

\subsubsection*{Summary}
The paper presents an hybrid analysis that synergistically combines static analysis, dynamic analysis, and machine learning to construct a precise Window Transition Graph (WTG). It first applies static analysis to build a static WTG and then applies dynamic analysis to verify the transitions in the static WTG.
Due to the coverage issues of dynamic analysis and over-approximation of the static analysis, it is expected that a substantial amount of transitions cannot be verified by the dynamic analysis. For these unverified transitions, it uses a machine learning technique, window transition prediction, that leverages the features for the unverified transitions (e.g., screenshots and text) to predict which transition to include.

\subsubsection*{Synergistic Effects}
\synRefine~/~\synRefinePrune

The precision of a program model can be also refined by pruning away some portions of the model that were initially included for the sake of guaranteeing conservative results, but could be then identified as irrelevant based on the information conveyed from the partner analysis stage.

The dynamic analysis result is used to confirm the transitions between GUI activities that are sure to be feasible. 

\subsubsection*{Inter-Analysis Workflow}
\workCascade.

Static analysis (Window Transition Graph building) is the pre-requisite for the dynamic analysis step which is responsible for identifying feasible transitions.

\subsubsection*{Interpretation Structures of Mapping Functions}
\structGUI.

The statically and dynamically discovered GUI elements and transitions are modeled into a Window Transition Graph which is then compared to identify the transitions that can effectively be executed.

\subsubsection*{Mechanics of Mapping Functions}
\mechanicAssociation.

The Window Transition Graphs, derived from the two analysis phases, are combined as-is.

\subsection{Efficient online cycle detection technique combining with Steensgaard points-to information \cite{liu_efficient_2016}}

\subsubsection*{Summary}
This paper proposes bootstrapping as a way to improve cycle detection predictability of pointer analysis. The main idea is to run a sequence of increasingly precise pointer analyses to feed into the next more precise analysis to improve the efficiency of the latter analysis. 
They use Steensgaard's fast unification algorithm as the bootstrap, and devise a new cycle detection method for Andersen's inclusion-based flow-insensitive, context-insensitive analysis.

\subsubsection*{Synergistic Effects}
\synIntepretability~/~\synIntepretabilityEntities

Each step provides additional knowledge about pointers and pointer aliases in the program.

\subsubsection*{Inter-Analysis Workflow}
\workCascade.

The considered pointer analyses are executed in sequence.

\subsubsection*{Interpretation Structures of Mapping Functions}
\structProgram.

The flow-insensitive, context-insensitive point-to information is associated with the program variables of the entire program.

\subsubsection*{Mechanics of Mapping Functions}
\mechanicAssociation.

The flow-insensitive, context-insensitive point-to information is associated with the program variables of the entire program.

\subsection{TDroid: Exposing app switching attacks in android with control flow specialization \cite{liu_tdroid_2018}}

\subsubsection*{Summary}
The paper introduces TDroid, an approach to detecting app switching attacks. The challenge lies in how to handle a plethora of input-dependent branch predicates (forming an exponential number of paths) that control the execution of the code responsible for launching such attacks.
TDroid tackles this challenge by combining static and dynamic analysis to analyze an app without producing any false positives.

TDroid first performs a static “Pre-Analysis” to look for all suspicious startActivity() calls that may lead to app switching attacks. TDroid then handles each suspicious startActivity() call separately.

In the second phase, TDroid constructs runnable slices starting from the identified startActivity() calls using “Static Backward Slicing".

Finally, it uses dynamic analysis to execute these slices (repackaged APKs) on an Android phone or emulator to expose their malicious GUIs.

\subsubsection*{Synergistic Effects}
\begin{itemize}
    \item Stage 1 $\to$ 2: \synPartitioning~/~\synPartitioningDirect.

    The startActivity() calls are identified and used as the starting goals for the backward slicing technique.
    
    \item Stage 2 $\to$ 3: \synTraversal~/~\synTraversalTransform\ and \synAlarms~/~\synAlarmsDynamic.

    The slices extracted with static analysis and repackaged into APKs are executed with the goal of identifying if they lead to suspicious behaviors (app switching attacks).
\end{itemize}

\subsubsection*{Inter-Analysis Workflow}
\workCascade.

Each stage of the analysis technique provides its output to the next stage.

\subsubsection*{Interpretation Structures of Mapping Functions}
\begin{itemize}
    \item Stage 1 $\to$ 2: \structCG.
    
    The startActivity() calls identified during the first static analysis phase are passed to the second static analysis phase.
    
    \item Stage 2 $\to$ 3: \structProgram.
    
    The runnable slices repackaged into APKs identified during the second static analysis phase are executed during the dynamic analysis phase.
\end{itemize}

\subsubsection*{Mechanics of Mapping Functions}
\mechanicAssociation.

The call graph is passed to the stage 2, and similarly the executable slices are passed to stage 3.

\subsection{IPA: Improving predictive analysis with pointer analysis \cite{liu_ipa_2016}}

\subsubsection*{Summary}
This paper proposes a predictive analysis (analysis that starts with the trace of an execution and mutates the schedule order of the trace to ``predict'' the executions that expose the hidden races) that allows changing the accessed locations. They solve the challenge through a combination of predictive analysis and pointer analysis.

\subsubsection*{Synergistic Effects}
\synIntepretability~/~\synIntepretabilityEntities

The pointer analysis provides information about pointers in the program, such that predictive analysis can use that  information to better interpret the program semantics.

\subsubsection*{Inter-Analysis Workflow}
\workCascade.

Pointer analysis feeds information to predictive analysis.

\subsubsection*{Interpretation Structures of Mapping Functions}
\structCFG.

The work consider flow sensitive pointer analysis that associate pointer-access information to the program instructions in the trace being considered.

\subsubsection*{Mechanics of Mapping Functions}
\mechanicAssociation.

They associate pointer-access information to the program instructions in the trace being considered.

\subsection{VD-Guard: DMA Guided Fuzzing for Hypervisor Virtual Device \cite{liu_vd-guard_2023}}

\subsubsection*{Summary}
This paper proposes a hybrid method to detect vulnerabilities in virtual devices called VD-GUARD. It first leverages static control flow analysis to track call chains from various data entry points of virtual devices (MMIO/PIO functions) to the critical dispatcher points (DMA functions). Then it generates seeds that can trigger this call chain via static analysis and limited fuzzing test. Finally, VD-GUARD takes these seeds as input and employs DMA guided fuzzing to discover DMA related vulnerabilities.

The first step of the technique extracts the suspicious memory operations by locating the memory regions which are registered with the MMIO/PIO flag and the corresponding memory operations.
The memory operations in the memory regions registered with MMIO/PIO flag are the suspicious memory operations (i.e., MMIO/PIO functions), and are the entry point of the call chain.

In the second step, the call graph is built and the technique search the call chain which originates from the extracted MMIO/PIO functions and lead to the DMA operations.
Indirect function calls (i.e., function pointer) hinder the construction of the call graph by static analysis since the actual callee function is uncertain. To solve this, when a function pointer is encountered during the construction process of the call graph, we look for all its assignment statements and add the corresponding assigned function to the call graph.
The call graph is then visited in DFS (Depth-First Search) fashion. During the searching process, when a call chain is identified as containing the function call to the DMA operations, each node in the call chain is marked and the call chain information is stored. In the subsequent searching process, if the direct successors of current node are these marked nodes, the predecessors in the stored call chain  are replaced with the current node and its predecessors and mark these nodes as well.
The result of this visit is the set of call chains originating from the MMIO/PIO functions that lead to the DMA operations.

In the third step, an initial seed that could trigger the collected call chains is constructed. To do this, the technique builds a PDG (Program Dependency Graph) for each call chain identified in the previous search. The nodes of the PDG are the statements of the functions in the call chain, and the edges are the control and data flow dependencies. 
Using this information an initial seed can be partially constructed, however, the static analysis method has its limitations in handling all the constraints.

In the fourth step, the technique uses the seed constructed by the static method before as the initial seed. Instead of triggering a crash, the limited fuzzing process continues until the call chain is triggered. The fuzzing result is the initial seed of the next stage.

In the fifth step, dynamic testing is performed with fuzzing using the seed derived in the previous step as the starting point. The end goal is identifying vulnerabilities in virtual devices.

\subsubsection*{Synergistic Effects}

The technique is composed of several stages; each has its own synergy:

\begin{itemize}
\item Stage 1 → Stage 2: \synPartitioning~/~\synPartitioningDirect. 

Discover potential alarm locations. The stage 1 technique statically extracts suspicious memory operations by locating the memory regions which are registered with the MMIO/PIO flag and the corresponding memory operations. The stage 2 technique searches the call graph for the call chain which originates from the extracted MMIO/PIO functions and leads to the DMA operations.

\item Stage 2 → Stage 3: \synTraversal~/~\synTraversalSeedSink.

The call chains identified  by stage 2 technique are used as a starting point to construct a partial initial seed that can trigger them.

\item Stage 3 → Stage 4: \synAlarms~/~\synAlarmsDynamic. 

Add test cases to incrementally confirm the alarms. The stage 4 technique uses the seed constructed by stage 3 technique as the initial seed of a limited fuzzing process that continues until the call chain is triggered.

\item Stage 4 → Stage 5: \synAlarms~/~\synAlarmsDynamic. 

Add test cases to incrementally confirm the alarms. In the final stage, dynamic testing is performed with fuzzing using the seed derived in the stage 4 technique as the starting point. The end goal is identifying vulnerabilities in virtual devices.

\end{itemize}

\subsubsection*{Inter-Analysis Workflow}
\workCascade.

Each phase is executed one after another without any sort of feedback.

\subsubsection*{Interpretation Structures of Mapping Functions}
\begin{itemize}
    \item Stage 1 → Stage 2: \structCG.
    
    The memory operations are identified in the call graph and provided to the second stage so that the call chain can be built.

    \item Stage 2 → Stage 3: \structCG.
    
    The call chain derived in step 2 is used by step 3 to build a PDG.

    \item Stage 3 → Stage 4: \structProgram.
    
    The seed derived purely from static analysis is used as the first seed of a limited fuzzing process that has the goal of covering the whole call chain.

    \item Stage 4 → Stage 5: \structProgram.
    
    The seed that covers the whole call chain derived dynamically is used as the initial seed of the fuzzing process that has the goal of locating vulnerabilities.
\end{itemize}

\subsubsection*{Mechanics of Mapping Functions}
\mechanicAssociation.

All the information exchanged between different stages are provided as is.

\subsection{RATA: Rapid atomic type analysis by abstract interpretation - Application to JavaScript optimization \cite{logozzo_rata_2010}}

\subsubsection*{Summary}
This paper introduces RATA, a static analysis based on abstract interpretation for the  inference of atomic types in JavaScript programs. RATA is a combination of 
numerical invariant inference with intervals and kinds, and variation analysis for type refinement.

\subsubsection*{Synergistic Effects}
\synIntepretability~/~\synIntepretabilityEntities

Interval analysis and kind analysis provide information on numerical invariant inference for each program point. Then variation analysis exploits this semantic information with its algorithm.

\subsubsection*{Inter-Analysis Workflow}
\workCascade.

Interval analysis and kind analysis feed inferences to variation analysis.

\subsubsection*{Interpretation Structures of Mapping Functions}
\structCFG.

The numerical invariant analysis (interval analysis and kind analysis)  infers invariants for each program point.

\subsubsection*{Mechanics of Mapping Functions}
\mechanicAssociation.

The numerical invariant analysis (interval analysis and kind analysis) associates the inferred invariants to the corresponding program points.

\subsection{A Two-Phase Approach for Conditional Floating-Point Verification \cite{lohar_two-phase_2021}}

\subsubsection*{Summary}
The paper proposes a framework for combining different static and dynamic analyses that allows to automatically prove the absence or detect the presence of large floating-point roundoff errors or the special values NaN and Infinity.
In particular, the proposed framework consists of:
\begin{itemize}
    \item a first static analysis stage in which abstract interpretation is used to infer a sound over-approximation of the kernel ranges;
    \item a second dynamic analysis stage that uses fuzzing to monitor the lower and upper bounds of the kernel ranges seen during concrete executions, enlarging the statically-inferred intervals upon observing values larger than upper bounds or smaller than lower bounds;
    \item a third static analysis stage is performed to automatically prove the absence of special values by using the sound floating-point roundoﬀ analysis tool Daisy. \item For the warnings generated by Daisy, i.e., the warning about special values that can potentially occur, they use the CBMC bounded model checker, to discard  spurious warnings.
    
\end{itemize}

\subsubsection*{Synergistic Effects}
\begin{itemize}
    \item \synIntepretability~/~\synIntepretabilityEntities

    Abstract interpretation and fuzzing estimate kernel ranges. Then, the sound floating-point roundoff analysis tool Daisy exploits these ranges while proving the absence of special values.

    \item \synAlarms~/~\synAlarmsStatic
    
    CBMC can discard spurious warnings identified by Daisy. 
\end{itemize}

\subsubsection*{Inter-Analysis Workflow}
\workCascade.

Results of abstract interpretation get refined during fuzzing and are then fed to Daisy and CBMC.

\subsubsection*{Interpretation Structures of Mapping Functions}
\structCFG. 

Abstract interpretation and fuzzing annotate the estimated kernel ranges per program point. Daisy annotates the possibly problematic intermediate expressions per program point. 

\subsubsection*{Mechanics of Mapping Functions}
\begin{itemize}
    \item \mechanicAssociation.

    Abstract interpretation and fuzzing associate the estimated kernel ranges to the corresponding program points. Daisy associates the possibly problematic intermediate expressions to the corresponding program points. 
    
    \item \mechanicConstraint.
    
    CBMC models the program as formulas and uses constraint solving to generate counter examples (if any within the specified bounds). 
\end{itemize}

\subsection{Software Bug Localization Based on Key Range Invariants \cite{ma_software_2018}}

\subsubsection*{Summary}
This paper proposes a software bug localization method based on key range invariants.
In the first stage, the key variables are identified by monitoring the execution of test suites.
Then bug localization is done by inferring  range invariants for 
the values of key variables in successful test case, and monitoring the violations of the invariants for the failing test cases. 

\subsubsection*{Synergistic Effects}
\synIntepretability~/~\synIntepretabilityOracle

Bug localization is done by inferring  range invariants for the values of key variables in successful test case, and monitoring the violations of the invariants for the failing test cases. 

\subsubsection*{Inter-Analysis Workflow}
\workCascade.

Throughout identification of key variables, invariant inferences and bug localization.

\subsubsection*{Interpretation Structures of Mapping Functions}
\structCFG.

The range invariants are associated with the instructions in the program.

\subsubsection*{Mechanics of Mapping Functions}
\mechanicInvariants. 

The results of monitoring the passing test cases is post processed by computing 
likely range invariants to be used in the subsequent analysis stage.

\subsection{Practical static analysis of JavaScript applications in the presence of frameworks and libraries \cite{madsen_practical_2013}}

\subsubsection*{Summary}
This paper proposes a technique which combines points-to flow-insensitive analysis with another static analysis (called use analysis) to improve call graph resolution and other static deductions for JavaScript code that interface with JavaScript libraries. The use analysis allows for recovering necessary information about the structure of objects returned from libraries, when those objects are passed into callbacks declared within the application. The recovered information allow for resolving links from the JavaScript code to the objects of underlying HTML page, thus refining the results of the analysis. 

\subsubsection*{Synergistic Effects}
\synIntepretability~/~\synIntepretabilityEntities.

The use analysis allows for recovering necessary information about the structure of objects returned from the JavaScript libraries invoked in the programs. the recovered information allows for refining points-to deductions.

\subsubsection*{Inter-Analysis Workflow}
\workFeedback.

To improve call graph resolution, the inferences made with point-to analysis are refined (unification) based on the inferences made with use analysis, iterating until the fixpoint.  

\subsubsection*{Interpretation Structures of Mapping Functions}
\structCG.

The two analysis share data by associating information (points-to relations and object-structure information) with the call sites in the program.

\subsubsection*{Mechanics of Mapping Functions}
\mechanicAssociation.

The points-to relations and object-structure information are associated with the corresponding call sites at which is relevant to share them across the points-to analysis and use analysis, respectively.

\subsection{EvoDroid: Segmented evolutionary testing of Android apps \cite{mahmood_evodroid_2014}}

\subsubsection*{Summary}
The paper presents EvoDroid, a search-based test generator tool for Android that aims at maximizing code coverage.
Compared to other tools it leverages statically derived models to aid the evolutionary test generation.

From the source code, EvoDroid extracts two types of models, representing the app's external interfaces and internal behaviors, to automatically generate the tests: Interface Model (IM) and Call Graph Model (CGM).

IM provides a representation of the app's external interfaces and in particular the ways in which an app can be exercised, e.g., the inputs and events available on various screens to generate tests that are valid for those screens. 
The IM is obtained by extracting information contained in the configuration files and meta-data included in Android APK. The IM is built by listing all Android components (e.g., Activities, Services) and then parsing the layout file of each Activity.
EvoDroid uses the IM to determine the structure of individuals, i.e., the input and event types that are coupled together. 

CGM is an extended representation of the app's call graph that includes information about the implicit call relationships caused by events. A particular use case follows a certain path through the CGM. EvoDroid uses CGM to (1) determine segments (the parts of the code that can be searched independently), and (2) evaluate the fitness of different test cases, based on the paths they cover through the CGM, thus guiding the search.

Using these two models, EvoDroid employs a step-wise evolutionary test generation algorithm called segmented evolutionary testing. It aims to find test cases covering as many unique CGM paths from the starting node of an app to all its leaf nodes. In doing so, it logically breaks up each path into segments. It uses heuristics to search for a set of inputs and sequence of events to incrementally cover the segments. By carefully composing the test cases covering each segment into system test cases covering an entire path in the CGM, EvoDroid is able to promote the genetic makeup of good individuals in the search. The test cases are evaluated based on a fitness function that rewards code coverage and uniqueness of the covered path. The focus of EvoDroid is on generating test cases that maximize code coverage, not on whether the test cases have passed or failed.

\subsubsection*{Synergistic Effects}
\begin{itemize}
    \item Stage 1 $\to$ Stage 3: \synIntepretability~/~\synIntepretabilityEntities
    
    The Interface Model, which is an abstraction of the GUI which provides a representation of the app's external interfaces (e.g., the inputs and events available on various screens to generate tests that are valid for those screens) is provided to the search-based technique to build appropriate individuals.
    
    \item Stage 2 $\to$ Stage 3: \synPartitioning~/~\synPartitioningCoverage
    
    The search-based technique aims at executing as many unique Call Graph Model paths from the starting node of an app to all its leaf nodes.
\end{itemize}

\subsubsection*{Inter-Analysis Workflow}
\workCascade.

The two static analysis stages are preemptive to the dynamic analysis stage.

\subsubsection*{Interpretation Structures of Mapping Functions}
\begin{itemize}
    \item Stage 1 $\to$ Stage 3: \structGUI.
    
    The Interface Model, which is an abstraction of the GUI, is provided to the search-based technique to build appropriate individuals.
    
    \item Stage 2 $\to$ Stage 3: \structCG.
    
    The Call Graph Model is passed to the search-based technique to aid the evolutionary algorithm.

\end{itemize}

\subsubsection*{Mechanics of Mapping Functions}
\mechanicAssociation.

In both cases the models are provided as-they-are to the search-based technique.

\subsection{Sapienz: Multi-objective automated testing for android applications \cite{mao_sapienz_2016}}

\subsubsection*{Summary}
Sapienz combines static string analysis (to extract the strings used within an Android app, as those string can be meaningful for testing the app) with dynamic analysis of both the GUI screens and widgets executed at run time, to support search-based testing of Android apps. 

\subsubsection*{Synergistic Effects}
\begin{itemize}
    \item \synGUI.

    The results about string analysis and GUI screens can be sided while generating next-offspring test cases, aiming to achieving effective offsprings.

    \item \synPartitioning~/~\synPartitioningCoverage

    The results about coverage are tracked at the level of GUI screens and  widgets, in order to compute fitness during search-based testing. (Occasionally, if source code is available, also traditional code coverage is exploited.)
\end{itemize}
 
\subsubsection*{Inter-Analysis Workflow}
\begin{itemize}
    \item \workSidebyside.

    Both static string analysis and dynamic GUI screen feed information to the search engine for generating next offsprings (where the test cases can get further engineered with sequences of actions based on the knowledge of the executed screens and possible strings for filling the text fields therein).

    \item \workCascade.

    The search engine work with multi-level coverage analysis for making decisions on fitness of the test suites.
\end{itemize}

\subsubsection*{Interpretation Structures of Mapping Functions}
\begin{itemize}
    \item \structProgram. 

    The results of string analysis are associated at the program-level with the Android app under test. 

    \item \structGUI.
    
    The data about executed screens and executed widgets relate with the screens that comprise the GUI of the android app under test.
\end{itemize}

\subsubsection*{Mechanics of Mapping Functions}
\mechanicAssociation.

The mapping functions above work by associating information with the app under test (relevant strings) and the GUI screens therein. 

\subsection{Pinpointing and exploiting opportunities for enhancing data reuse \cite{marin_pinpointing_2008}}

\subsubsection*{Summary}

This paper describes a technique to analyze data-locality issues in binary programs by separately collecting two type of metrics, related to temporal reuse and spatial reuse, respectively. They then present both types of metrics with integrated visualizations to software engineers. Temporal reuse metrics pinpoint scarce-locality issues based on dynamic-analysis-based measurements of cache misses. Spatial reuse metrics are computed with static analysis of access patterns to quantify spatial reuse, by identifying \emph{related references} (references that access the
same data arrays with the same stride) via detailed analysis of the binaries. The HPCToolkit allows engineers to visualize the integrated metrics, to pinpoint program issues and possible solutions.

\subsubsection*{Synergistic Effects}
\synReports.

The technique integrate the results of both temporal reuse analysis and spatial reuse analysis in an integrated reporting, to assist software engineers to pinpoint program issues and possible solutions.

\subsubsection*{Inter-Analysis Workflow}
\workSidebyside. 

Temporal reuse analysis and spatial reuse analysis are performed separately. Then their results are reported (visualized) in integrated fashion.

\subsubsection*{Interpretation Structures of Mapping Functions}
\structCFG.

The reported metrics are associated to statements, loops, arrays and variables in the program.

\subsubsection*{Mechanics of Mapping Functions}
\mechanicML.

The results of both the static and the dynamic analysis are numerically processed to report temporal-reuse and spatial-reuse metrics.

\subsection{KATCH: High-coverage testing of software patches \cite{10.1145/2491411.2491438}}

\subsubsection*{Summary}
This paper addresses the problem of bugs introduced with code patches by using a technique for automatically testing them.
The technique proposed by the paper works in similar fashion to concolic execution in which an initial seed (input) is used to explore a path and then it is iteratively modified with the goal of executing the target (the patched code).

\subsubsection*{Synergistic Effects}
\synPartitioning~/~\synPartitioningDirect

The aim of concolic execution is to increase the efficiency and the effectiveness of traditional symbolic execution in exploring the path space of a program, usually for the sake of generating test cases that execute the program paths explored thereby.
In this specific case, the technique leverages it with the goal of executing the part of the code containing the patch; but the underlying technique is basically the same. 

\subsubsection*{Inter-Analysis Workflow}
\workFeedback.

Each time a new seed is used it can be used to derive new constraints that are used to potentially generate new combinations of path conditions, that will once again lead to the potential creation of new seeds. In particular, the constraint that is less distant from the target is used to derive a new seed to further the exploration, until the target is executed. This is all in concolic execution fashion.

\subsubsection*{Interpretation Structures of Mapping Functions}
\begin{itemize}
    \item Runtime monitoring → Symbolic: \structPaths.
    
    Execution traces associated to corresponding program paths.

    \item Symbolic → Runtime monitoring: \structCG.
    
    Symbolic execution feeds inputs to further execute the target function. synthesized via constraint solving, for executing the target function(s).
\end{itemize}

\subsubsection*{Mechanics of Mapping Functions}
\begin{itemize}
    \item Runtime monitoring → Symbolic: \mechanicAssociation.
    
    Execution traces associated to corresponding program paths.

    \item Symbolic → Runtime monitoring: \mechanicConstraint.
    
    Constraint solving is used to identify input values that satisfy the path conditions that symbolic execution identified as executability conditions of yet-unvisited program paths.
\end{itemize}

\subsection{Determining microservice boundaries: A case study using static and dynamic software analysis \cite{matias_determining_2020}}

\subsubsection*{Summary}
The paper describes a systematic approach for refactoring systems to microservice architectures that uses static analysis to determine the system's structure and dynamic analysis to understand its actual behavior.

During the static analysis stage, software artifacts are analyzed and the collected information used to build a graph-like model of the system, representing components as nodes and the dependencies between them as edges. Components and dependencies can be of different types, and identifying them will depend on the used programming languages, frameworks and environments. For example, components can refer to classes, packages or modules, and dependencies to imports or method calls. Each edge is assigned a weight to represent the strength of the dependency. This is a function of the number and quality of connections between the two components. The weight of edges after static analysis can, for example, be the sum of the number of imports and method calls between its two components.

In the dynamic analysis stage, the system is monitored at runtime to gather operational data, which is analyzed to identify how the dependencies are exercised during execution, and gain an understanding of how the system is actually used. Such information is used to compute a new weight for each edge of the graph. The final weight values are a function of the static and dynamic weights, and are a measure for
how the components in the system are mutually bound. The underlying assumption is that a high amount of interaction between two components correlates with belonging to a common bounded context. Including them in different microservices would imply higher costs in latency and in maintaining resilience and fault tolerance.

Finally, clustering is performed with the goal of identifying decomposition suggestions. A graph of the service composition will support identifying different clusters of components. The nodes connected by the edges with higher weight values will be grouped to form clusters of relatively high cohesion. These clusters will depend on each other through edges with low weight values, representing relatively low coupling. The clusters can, therefore, be used to determine a set of possible service cuts.

\subsubsection*{Synergistic Effects}
\synRefine~/~\synRefineIncorporate

The static analysis stage builds a graph-like model of the system representing the components and their dependencies and assigning to each edge a weight based on strength of the dependency between two components. The dynamic analysis stage refines the generated graph model by generating new weights for the dependencies based on their usage.

\subsubsection*{Inter-Analysis Workflow}
\workCascade.

The static analysis stage that identifies the relevant features to build a model is followed by the dynamic analysis stage.

\subsubsection*{Interpretation Structures of Mapping Functions}
\structModules.

The quality data used to determine a refactoring is computed at the component level.

\subsubsection*{Mechanics of Mapping Functions}
\mechanicAssociation.

The relevant features derived during the static analysis stage are provided as-is to the dynamic analysis stage.

\subsection{Input domain reduction through irrelevant variable removal and its effect on local, global, and hybrid search-based structural test data generation \cite{mcminn_input_2012}}

\subsubsection*{Summary}
This paper proposes a static dependence analysis derived from program slicing that can be used to support search space reduction when performing Search-Based Software Testing. 
The static analysis stage determines which are the input variables that affects each objective (i.e. branches). In the second stage, a search-based algorithm is used to try to satisfy all objectives by focusing specifically on the inputs associated to each objective.

\subsubsection*{Synergistic Effects}
\synTraversal~/~\synTraversalSeedInputs

The static analysis stage reduces the program inputs for the search-based second stage; that is, only the relevant variables (found in the first stage) for each branch will be used during the search-based analysis step.

\subsubsection*{Inter-Analysis Workflow}
\workCascade.

The static analysis stage is followed by the search-base testing stage without feedback between them.

\subsubsection*{Interpretation Structures of Mapping Functions}
\structCFG

The variables relevant for a branch objective are provided to the search-based technique.

\subsubsection*{Mechanics of Mapping Functions}
\mechanicAssociation.

The variables relevant for a branch objective are provided as-is to the search-based technique.

\subsection{Employing user profiles to test a new version of a GUI component in its context of use \cite{memon_employing_2006}}

\subsubsection*{Summary}
This paper presents a technique to test the new version of GUI components, by using the previous GUI version as test reference for test inputs and test oracle. First, dynamic analysis (profiling) is used to capture event-level information and the GUI states while the previous version of the GUI is being used in the field.  
Second, a static analysis of the new version of the GUI components generate a mapping between the events of the new components and the previous versions. Third, the new components are tested by using the profiled event sequences as test inputs and both the profiled GUI states and the mapping to check oracles (comparing GUI states for corresponding events).

\subsubsection*{Synergistic Effects}
\begin{itemize}
    \item \synPartitioning~/~\synPartitioningDirect
    
    The data from profiling guide the testing stage to progress on executing specific event sequences where the oracles are available.

    \item \synIntepretability~/~\synIntepretabilityOracle
    
    The event mapping and profiled GUI states allow for interpreting the execution semantics via suitable oracles based on the relation with previous versions of the components under test.
\end{itemize}

\subsubsection*{Inter-Analysis Workflow}
\workCascade.

Both the data collected during profiling and the statically generated event mappings are fed to the testing stage, where they are used for synthesizing test inputs and test oracles.

\subsubsection*{Interpretation Structures of Mapping Functions}
\structGUI.

Both event-profiling and event-mapping data are associated with GUI entities (events and event sequences) for being exploited during testing.

\subsubsection*{Mechanics of Mapping Functions}
\mechanicAssociation.

Both event-profiling and event-mapping data are associated with corresponding events and event sequences.

\subsection{Measurement and tracing methods for timing analysis: Independently and in combination with modelling methods \cite{merriam_measurement_2013}}

\subsubsection*{Summary}
This work combines profiling of timing properties of the software at runtime, with static worst-case execution analysis, in synergistic fashion.
Static analysis provides information on loop bounds that can be used to add specific test cases to the measurement test suite used for profiling. 
Timing properties measured during profiling and data dependencies observed therein may improve the precision of the estimation models used for worst-case execution analysis.

\subsubsection*{Synergistic Effects}
\begin{itemize}
    \item \synPartitioning~/~\synPartitioningDirect

    Static analysis provides information on loop bounds that can be used to direct the profiling stage to progress on relevant execution cases. 

    \item \synRefine~/~\synRefineIncorporate
    
    The results from profiling allow for improving the precision of the estimation models used for worst-case execution analysis.
\end{itemize}

\subsubsection*{Inter-Analysis Workflow}
\workFeedback.

The static information on loop bounds allows for improving profiling, and the data from profiling allow for improving worst-case execution analysis.

\subsubsection*{Interpretation Structures of Mapping Functions}
\structCFG.

The static analysis (worst-case execution analysis) and dynamic analysis (profiling) layers exchange information for code entities, e.g., loops and instructions.

\subsubsection*{Mechanics of Mapping Functions}
\mechanicAssociation.

The information is associated with corresponding code entities to be used on demand in other analysis stages.

\subsection{Log-based slicing for system-level test cases \cite{messaoudi_log-based_2021}}

\subsubsection*{Summary}
This paper proposes an approach which automatically decomposes a complex system test case into separate test case slices. The idea is to use test case execution logs, obtained from past regression testing sessions, to identify “hidden” dependencies in the slices generated by static slicing. Since logs include run-time information about the system under test, they can be used to extract access and usage of global resources and refine the slices generated by static slicing.

\subsubsection*{Synergistic Effects}
\synIntepretability~/~\synIntepretabilityEntities

Some hidden dependencies in the program can only be derived dynamically (e.g. resources accessed and the actions performed). This information is extracted from logs and used to aid the slicing of system test cases with the goal of obtaining smaller and more focused test cases.

\subsubsection*{Inter-Analysis Workflow}
\workCascade.

The log traces dynamically obtained from regression testing are used by the second stage static analysis with the goal of obtaining smaller test cases from the initial system test cases.

\subsubsection*{Interpretation Structures of Mapping Functions}
\structProgram.

The second stage analysis requires the logs obtained from system tests that requires the whole system (program).

\subsubsection*{Mechanics of Mapping Functions}
\mechanicAssociation.

The log traces dynamically obtained from regression testing are provided as is to the second stage static analysis.

\subsection{Reducing combinatorics in GUI testing of android applications \cite{mirzaei_reducing_2016}}

\subsubsection*{Summary}
This paper presents the technique TrimDroid. It relies on static program analysis for identifying the interactions (control- and data-flow dependencies) among the widgets and actions available on the Android app under test. The set of interacting widgets become candidates for t-way combinatorial testing.
Then, TrimDroid enumerates (and generates) the test sequences according to relevant criteria, by using the Alloy solver after representing the possible integrations as a model in Alloy.

\subsubsection*{Synergistic Effects}
\synPartitioning~/~\synPartitioningCoverage

Identifying the interactions and extracting the corresponding test sequences via Alloy allows for addressing the testing process based on relevant criteria.

\subsubsection*{Inter-Analysis Workflow}
\workCascade.

TrimDroid first relies on static program analysis for identifying the interactions among the widgets, and then it enumerates the test sequences  via Alloy.

\subsubsection*{Interpretation Structures of Mapping Functions}
\structGUI.

The results of the former, static analysis stage are propagated as dependencies between widgets and GUI actions.

\subsubsection*{Mechanics of Mapping Functions}
\mechanicAssociation.

The results of the former, static analysis stage associate with widgets and entities of the GUI. 

\subsection{Fuzzing Class Specifications \cite{molina_fuzzing_2022}}

\subsubsection*{Summary}
The paper presents SPECFUZZER, a novel technique that combines grammar-based fuzzing, dynamic invariant detection with Daikon, and mutation analysis, to automatically produce class specifications. The technique consists of several stages:
\begin{itemize}
    \item Invariant generation: A static analyzer takes as input a class and creates a grammar expressing the language of candidate assertions for the class. This grammar is then exploited with grammar-based fuzzing to generate candidate assertions that denote method preconditions, postconditions, and class invariants. 
    \item Test case generation: Test cases are generated for each analyzed class by using Randoop.
    \item Invariant validation: A Dynamic Invariant Detector (Daikon) is used to evaluate the plausibility of the candidate assertions produced by the fuzzer. The dynamic invariant detector takes as input the test cases produced by the test generator, and the set of assertions produced by the fuzzer. This component instruments the program with the assertions generated by the fuzzer and runs the tests to verify which assertions hold across the observed executions. The resulting assertions are reported as likely invariants.
    \item Mutation-analysis-based invariant selection: An Invariant Selector based on mutation analysis partition the assertions that were deemed valid, grouping together similar assertions, and taking a single representative from each partition. To do this, it relies on mutation analysis to detect which of the valid invariants can reveal faults and which fault each reveal. 
    The rationale for the mutation based partition is that assertions that kill different mutants are non-equivalent (or, alternatively, that assertions that kill the same mutants are ``similar''); the rationale for ranking assertions according to the number of failures is that assertions that are falsified a greater number of times are ``stronger''.
    This component reports a subset of the likely assertions it receives as input, ranking the invariants by the number of failures in corresponding code assertions. The Invariant Selector reduces the number of reported assertions.
\end{itemize}

\subsubsection*{Synergistic Effects}
\synPartitioning~/~\synPartitioningWitness

The invariants generated in the first stage provide a set of partitions that are crosschecked with witness test cases and witness mutants in the next steps.

\subsubsection*{Inter-Analysis Workflow}
\workCascade.

Invariant generation and test cases are cascaded to Daikon, and the results further cascade into the mutation analysis stage.

\subsubsection*{Interpretation Structures of Mapping Functions}
\begin{itemize}
    \item \structModules.
    
    Test cases and some invariants are associated at the class level.

    \item \structCG.
    
    Some invariants are associated with class methods.
\end{itemize}

\subsubsection*{Mechanics of Mapping Functions}
\begin{itemize}
    \item \mechanicAssociation.

    Test cases and statically produced invariants are associated with the class and the methods therein.

    \item \mechanicInvariants.
    
    The results of dynamic analysis are further processed to select likely invariants. The results of mutation analysis are again further process to select likely invariants.
\end{itemize}

\subsection{Boosting Grey-box Fuzzing for Connected Autonomous Vehicle Systems \cite{moukahal_boosting_2021}}

\subsubsection*{Summary}
This paper presents a hybrid fuzz testing framework (VulFuzz++) that unites the efficiency of fuzzing and the precision of concolic execution to provide the automotive industry with a reliable security testing tool. VulFuzz++ offloads most of the exploration process to the vulnerability-oriented fuzzer (VulFuzz) explicitly designed for automotive systems. When the fuzzer stops exploring different paths, VulFuzz++ examines the untraversed branches and prioritizes them based on their potential to expose vulnerabilities. It utilizes a tailored, targeted concolic engine that limits the symbolic exploration to only specific functions. When the concolic engine discovers new system inputs, testing is handed over again to the fuzzer to perform a quick and efficient evaluation of the newly explored region.

\subsubsection*{Synergistic Effects}
\synPartitioning~/~\synPartitioningCoverage

The technique consists of first stage fuzzer that explores the analyzed target until it cannot discover any new paths, once this happens a concolic engine is used to further explore specific parts of the systems which contains uncovered branches. Once new inputs are discovered by the concolic engine, they are provided once again to the fuzzer to further the exploration.

\subsubsection*{Inter-Analysis Workflow}
\workFeedback.

The fuzzer explore the target system until it halts, then a concolic engine is run with the goal of covering unexplored branches, following this once new relevant inputs are discovered the testing is once again handed over to the fuzzer.

\subsubsection*{Interpretation Structures of Mapping Functions}
\structCG.

The inputs on which the fuzzer works on and that are used as initial seed for the concolic engine are mapped to functions/methods and consequently are call graph entities.

\subsubsection*{Mechanics of Mapping Functions}
\mechanicAssociation.

The functions and respective input values are provided as-is to the following analysis stage.

\subsection{Effective static deadlock detection \cite{naik_effective_2009}}

\subsubsection*{Summary}
This work proposes a cascade-composition-style combination of static analysis phases, each of which approximates a different necessary condition for a deadlock. The sequential phases are based on call-graph analysis, may-alias analysis, thread-escape analysis, may-happen-in-parallel analysis and approximated must-alias analysis.

\subsubsection*{Synergistic Effects}
\synAlarms~/~\synAlarmsStatic

Each analysis phase verifies further necessary conditions for deadlock, up to eventually confirm only the deadlocks that satisfy all necessary conditions.

\subsubsection*{Inter-Analysis Workflow}
\workCascade.

The approach consists of sequential phases of static analysis, with each phase addressing a different property necessary for the existence of deadlocks.

\subsubsection*{Interpretation Structures of Mapping Functions}
\structModules.

Namely, threads in this case; each static-analysis phase devises properties to the threads in the program, and the properties from a phase are inputs for deciding the properties to be checked in the next phase.

\subsubsection*{Mechanics of Mapping Functions}
\mechanicAssociation.

Each static-analysis phase associates new properties to the threads in the program.

\subsection{Improving spectral-based fault localization using static analysis \cite{neelofar_improving_2017}}

\subsubsection*{Summary}
The paper introduces a weighting technique by combining static and dynamic program analysis. Static analysis is performed to categorize program statements into different classes and giving them weights based on the likelihood of being buggy statement. Statements are finally ranked on the basis of the weights computed by statements' categorization (static analysis) and scores computed by SBFL metrics (dynamic analysis).

\subsubsection*{Synergistic Effects}
\synReports.

The dynamic and static technique both computes a probability for a certain statement to be buggy. Then their probability scores are combined to obtain a single bug probability score.

\subsubsection*{Inter-Analysis Workflow}
\workSidebyside.

The two program analysis techniques can be executed independently from each other. There is no sequentiality.

\subsubsection*{Interpretation Structures of Mapping Functions}
\structCFG.

The scores that are computed and then combined by the two analysis techniques are calculated at the statement level.

\subsubsection*{Mechanics of Mapping Functions}
\mechanicAssociation.

The mapping of the scores obtained from the two techniques is done by association on the relevant statement.

\subsection{Multiple Mutation Testing for Timed Finite State Machine with Timed Guards and Timeouts \cite{nguena_timo_multiple_2019}}

\subsubsection*{Summary}
The traditional model-based testing approach considers a fault domain as the universe of all machines with a given number of states and input-output alphabet while mutation-based approaches define a list of mutants to kill with a test suite. 

This work develops a mutation testing technique for real-time systems, after a step in which it represents the fault domain for the system under test with timed finite state machines with timed guards and timeouts (TFSM-TG). Thus, the seeded faults consist of fault-seeded versions of the  TFSM-TG. By means of this approach the test generation step, which aims to generate a test suite that covers all the relevant mutants, avoids the one-by-one enumeration of the mutants and is based on constraint solving.

\subsubsection*{Synergistic Effects}
\synPartitioning~/~\synPartitioningCoverage

The first phase produces the model (TFSM-TG) of the fault domain of the program, which guides the test generation phase to produce test suites which can be deemed complete with respect to the considered model-based mutation coverage criterion.

\subsubsection*{Inter-Analysis Workflow}
\workCascade.

Cascade composition between the static analysis phase that generates the model (TFSM-TG) to represent the fault domain of the program, and the step that generate complete test suites (complete with respect to the considered model-based mutation coverage criterion) by exploiting the fault-domain model with constraint solving.

\subsubsection*{Interpretation Structures of Mapping Functions}
\structProgram.

The first phase produces the model (TFSM-TG) of the fault domain of the program, which is shared to the second phase by associating it with the program under test. The final test suite is associated with the program as well.

\subsubsection*{Mechanics of Mapping Functions}
\mechanicAssociation.

The first phase produces the model (TFSM-TG) of the fault domain of the program, which is shared to the second phase by associating it with the program under test.

\subsection{Using dynamic analysis to generate disjunctive invariants \cite{nguyen_using_2014}}

\subsubsection*{Summary}
This work proposes a technique to generate disjunctive invariants by combining two analysis stages: a stage of dynamic analysis that infers candidate (likely) conjunctive invariants by using a variant of existing geometric hull approaches, and a step of bounded software model checking that statically validate the invariants discarding false invariants. 

\subsubsection*{Synergistic Effects}
\synPartitioning~/~\synPartitioningWitness

The inferred invariants represent relevant state space partitions, associated with witnesses in the form of formal proofs generated with model checking. 

\subsubsection*{Inter-Analysis Workflow}
\workCascade.

Cascade between a stage of dynamic analysis that infers candidate (likely) conjunctive invariants, and a step of bounded model checking that validate the invariants to discard false invariants. 

\subsubsection*{Interpretation Structures of Mapping Functions}
\structCFG. 

The first analysis stage infers invariants associated to program locations.

\subsubsection*{Mechanics of Mapping Functions}
\mechanicInvariants.

The first analysis stage post-processes the results of dynamic analysis to infer likely invariants.

\subsection{Badger: Complexity analysis with fuzzing and symbolic execution \cite{noller_badger_2018}}

\subsubsection*{Summary}
The paper describes Badger, a new hybrid approach for complexity analysis with the goal of discovering vulnerabilities which occur when the worst-case time or space complexity of an application is significantly higher than the average case. Badger uses fuzz testing to generate a diverse set of inputs that aim to increase not only coverage but also a resource-related cost associated with each path. Since fuzzing may fail to execute deep program paths due to its limited knowledge about the conditions that influence these paths, the analysis is complemented with a concolic execution which is also customized to search for paths that increase the resource-related cost. Symbolic execution is particularly good at generating inputs that satisfy various program conditions but by itself suffers from path explosion. Therefore, Badger uses fuzzing and concolic execution in tandem, to leverage their benefits and overcome their weaknesses.

\subsubsection*{Synergistic Effects}
\synPartitioning~/~\synPartitioningCoverage

The technique consists of first stage fuzzer that explores the analyzed target until it cannot discover any new paths, once this happens a concolic engine is used to further explore specific parts of the system which contains uncovered branches. Once new inputs are discovered by the concolic engine, they are provided once again to the fuzzer to further the exploration.

\subsubsection*{Inter-Analysis Workflow}
\workFeedback.

The fuzzer explores the target system until it halts, then a concolic engine is run with the goal of covering unexplored branches, following this once new relevant inputs are discovered the testing is once again handed over to the fuzzer.

\subsubsection*{Interpretation Structures of Mapping Functions}
\structCG.

The inputs on which the fuzzer works on and that are used as initial seed for the concolic engine are mapped to functions/methods and consequently are call graph entities.

\subsubsection*{Mechanics of Mapping Functions}
\mechanicAssociation.

The functions and respective input values are provided as-is to the following analysis stage.

\subsection{Hydiff: Hybrid differential software analysis \cite{noller_hydiff_2020}}

\subsubsection*{Summary}
The paper presents HyDiff; the first hybrid approach for differential software analysis. HyDiff integrates and extends two very successful testing techniques: Feedback-directed greybox fuzzing for efficient program testing and shadow symbolic execution for systematic program exploration. HyDiff extends greybox fuzzing with divergence-driven feedback based on novel cost metrics that also take into account the control flow graph of the program. Furthermore HyDiff extends shadow symbolic execution by applying four-way forking in a systematic exploration and still having the ability to incorporate concrete inputs in the analysis. HyDiff applies divergence revealing heuristics based on resource consumption and control-flow information to efficiently guide the symbolic exploration, which allows its efficient usage beyond regression testing applications. The authors introduce differential metrics such as output, decision and cost difference, as well as patch distance, to assist the fuzzing and symbolic execution components in maximizing the execution divergence.

\subsubsection*{Synergistic Effects}
\synPartitioning~/~\synPartitioningCoverage

The technique consists of first stage fuzzer that explores the analyzed target until it cannot discover any new paths, once this happens a concolic engine is used to further explore specific parts of the system which contains uncovered branches. Once new inputs are discovered by the concolic engine, they are provided once again to the fuzzer to further the exploration.

\subsubsection*{Inter-Analysis Workflow}
\workFeedback.

The fuzzer explores the target system until it halts, then a concolic engine is run with the goal of covering unexplored branches, following this once new relevant inputs are discovered the testing is once again handed over to the fuzzer.

\subsubsection*{Interpretation Structures of Mapping Functions}
\structCG.

The inputs on which the fuzzer works on and that are used as initial seed for the concolic engine are mapped to functions/methods and consequently are call graph entities.

\subsubsection*{Mechanics of Mapping Functions}
\mechanicAssociation.

The functions and respective input values are provided as-is to the following analysis stage.

 \subsection{An extensible approach for taming the challenges of JavaScript dead code elimination \cite{Obbink:SANER:2018}}

\subsubsection*{Summary}
This paper presents an approach for improving JavaScript dead code elimination by merging statically generated and dynamically generated call graphs. Static call graphs offer broad coverage but suffer from over-approximation, while dynamic call graphs are precise but incomplete. The technique constructs both graphs independently, static analysis extracts a conservative call graph from the entire program. Dynamic analysis collects runtime call edges based on actual execution traces. The two graphs are then merged to produce a richer call graph that compensates for the weaknesses of each input. The resulting combined call graph supports more accurate dead code elimination.

\subsubsection*{Synergistic Effects}
\synFlow:
Static and dynamic call graph information provide complementary knowledge: the static graph fills gaps left by the dynamic graph, while the dynamic graph removes many spurious static edges. The integration refines program-flow information.

\subsubsection*{Inter-Analysis Workflow}
\workSidebyside: The two call graphs are generated independently from one another. Only after both analyses complete are they merged; no ordering, refinement, or feedback cycle exists between the techniques.

\subsubsection*{Mapping-Function Interpretation Structure}
\structCG:
Both analyses produce call graphs expressed using nodes (functions) and edges (calls), and the mapping function aligns these entities during merging.

\subsubsection*{Mapping-Function Mechanics}
\mechanicMerging:
The merged call graph is formed by directly associating corresponding nodes and edges from both graphs, unifying them.

\subsection{Detecting DOM-sourced cross-site scripting in browser extensions \cite{Pan:ICSME:2017}}

\subsubsection*{Summary}
This paper uses a combination of static analysis and symbolic execution to detect DOM-based XSS vulnerabilities in the Greasemonkey browser extension environment. The static component analyzes extension code to locate points where DOM elements may influence security-sensitive operations. Dynamic symbolic execution is then used to explore possible values and flows of these DOM-derived inputs. By jointly considering syntactic patterns and symbolic behaviors, the technique can detect feasible XSS injections while eliminating unrealistic ones.

\subsubsection*{Synergistic Effects}
\synAlarms~/~\synAlarmsDynamic:
Static analysis identifies potentially dangerous flows but tends to over-approximate. The dynamic symbolic execution then concretizes and filters these candidates, validating whether an actual exploit scenario exists.

\subsubsection*{Inter-Analysis Workflow}
\workCascade:
Static analysis first identifies potential vulnerable functions in a Greasemonkey script. The dynamic symbolic execution then focuses on these identified potentially vulnerable scripts.

\subsubsection*{Mapping-Function Interpretation Structure}
\structModules:
The mapping occurs across units of code (scripts, handlers, modules) where static warnings are checked against execution paths derived symbolically.

\subsubsection*{Mapping-Function Mechanics}
\mechanicAssociation:
The dynamic symbolic executor simply takes the static scripts defined as vulnerable by the static analysis and checks them symbolically.

\subsection{Accelerating JavaScript static analysis via dynamic shortcuts \cite{Park:JMESECSFSE:2021}}

\subsubsection*{Summary}
This paper proposes a hybrid static–dynamic technique to accelerate costly JavaScript static analysis. During program execution, dynamic analysis collects concrete values for variables when possible. These observed values are inserted as “dynamic shortcuts” in the static abstract interpreter. Whenever dynamic concrete information is available, the static analysis uses it to skip expensive abstract computations; when it is not available, the static analysis proceeds normally with abstract values. The technique substantially reduces analysis cost while preserving soundness under the specified assumptions.

\subsubsection*{Synergistic Effects}
\synRewrite~/~\synRewriteConcrete: The static analysis is strengthened by the concrete runtime observations.

\subsubsection*{Inter-Analysis Workflow}
\workFeedback: The analysis alternates between dynamic and static based on the information available during the program analysis.

\subsubsection*{Mapping-Function Interpretation Structure}
\structCFG, \structCG:
Dynamic values are associated with the program’s control points and program functions

\subsubsection*{Mapping-Function Mechanics}
\mechanicMerging:
Concrete dynamic states are merged for the corresponding control points and program functions to provide possible replacements for the abstract states.

\subsection{ConflictJS: Finding and understanding conflicts between JavaScript libraries \cite{Patra:PICSE:2018}}

\subsubsection*{Summary}
This paper introduces a dynamic-analysis–based technique for detecting conflicts between JavaScript libraries. The approach uses two dynamic analyses:

Dynamic write logging: Each library is executed in isolation to observe which memory locations, object properties, or global variables it writes to. Conflicts are hypothesized when multiple libraries modify the same locations.

Dynamic test generation: Once potential conflict points are identified, the system generates and executes test cases that combine the libraries in various ways to elicit diverging behaviors. These tests help confirm whether the suspected conflicts are true or false conflicts.

\subsubsection*{Synergistic Effects}
\synAlarms~/~\synAlarmsDynamic:
The first dynamic phase produces many potential conflicts based on overlapping writes. The second dynamic phase uses targeted testing to determine which suspected conflicts manifest at runtime, filtering out false alarms.

\subsubsection*{Inter-Analysis Workflow}
\workCascade:
Dynamic execution of libraries produces candidate conflicts, and these candidates are then further examined through dynamic test-case generation. The second stage directly depends on the output of the first.

\subsubsection*{Mapping-Function Interpretation Structure}
\structModules:
Modules constitute the main structural entities. Their write sets and subsequent dynamic behaviors are mapped across modules to determine conflicts.

\subsubsection*{Mapping-Function Mechanics}
\mechanicAssociation:
The write sets of libraries are directly compared to identify overlaps, and the test generator directly uses these overlaps to produce tests.

\subsection{Specifying Callback Control Flow of Mobile Apps Using Finite Automata \cite{Perez:TSE:2019}}

\subsubsection*{Summary}
This paper introduces a technique for constructing Callback Control Flow Automata (CCFA) by integrating three statically generated models:
\begin{itemize}
\item the Windows Transition Graph (WTG) capturing GUI window relationships,
\item interprocedural control-flow graphs (ICFGs) connecting callback methods, and
\item Predicate Callback Summaries (PCSs) encoding conditions affecting callback invocations.
\end{itemize}
By combining these graphs side-by-side, the technique produces a finite automaton representing feasible callback sequences within an app. This unified model improves developers’ understanding of callback-driven behavior and supports downstream analyses such as testing or verification.

\subsubsection*{Synergistic Effects}
\synFlow:
Each static model captures a different facet of callback behavior. Integrating them enriches the global program-flow representation, enabling a more accurate automaton.

\subsubsection*{Inter-Analysis Workflow}
\workSidebyside:
The WTG, ICFG, and PCS are produced independently and combined afterward. They coexist without ordering or feedback.

\subsubsection*{Mapping-Function Interpretation Structure}
\structCG.
Callbacks, invocation edges, and conditional transitions from the various models form the structural units aligned in the automaton.

\subsubsection*{Mapping-Function Mechanics}
\mechanicMining:
The CCFA is effectively mined from the combination of the multiple static graphs, translating program-flow information into automaton transitions.

\subsection{Detection, assessment and mitigation of vulnerabilities in open source dependencies \cite{ponta2020detection}}

\subsubsection*{Summary}
This paper presents a pipeline that combines static and dynamic techniques to understand and mitigate vulnerabilities introduced through third-party dependencies. The workflow is:

Static diffing: Identify all code fragments in the third-party dependency modified by a vulnerability-fixing commit.

Static reachability analysis: Determine which parts of the dependent program can reach the modified code in the third-party dependency.

Dynamic execution: Run the dependent program’s existing test suite to identify the dynamically reachable modified code of the dependency.

Static expansion: From dynamically reached points, apply static analysis again to find all code potentially reachable from those dynamic entry points.

The combination of these analysis stages narrows down which vulnerable code portions are actually relevant to the application and helps guide mitigation or patching.

\subsubsection*{Synergistic Effects}
\synAlarms~/~\synAlarmsDynamic:
The dynamic execution indicates which parts of the dependency are exercised. Static reachability refines the results, ensuring that only genuinely reachable (or relevant) vulnerable code is reported

\subsubsection*{Inter-Analysis Workflow}
\workCascade: detection of vulnerable library calls, dynamic analysis to identify reachable statements, static analysis to identify reachable statements from the dynamically executed states.

\subsubsection*{Mapping-Function Interpretation Structure}
\structCFG: vulnerable and reachable statements

\subsubsection*{Mapping-Function Mechanics}
\mechanicAssociation:
reachability information associated to reachable statements.

\subsection{Feature location using probabilistic ranking of methods based on execution scenarios and information retrieval \cite{poshyvanyk2007feature}}

\subsubsection*{Summary}
This paper proposes a hybrid feature-location technique that integrates static information-retrieval–based ranking with dynamic execution profiling. The static component applies Latent Semantic Indexing (LSI) to source code, comments, and documentation in order to generate feature–method similarity scores. LSI produces a vector-space model in which each method is represented as a weighted textual feature vector, enabling probabilistic ranking based on semantic proximity to a feature description.
The dynamic component, Scenario-based Probabilistic Ranking (SPR), collects program-execution traces for scenarios known to be related to the feature under investigation. SPR analyzes the frequency and temporal concentration of method-execution events in these traces to assign likelihood scores indicating whether a method is involved in implementing the feature.
To combine the analyses, the approach computes a unified score that integrates the LSI-based semantic likelihood with the SPR-based behavioral probability. A normalization and confidence-weighting scheme is used to reconcile heterogeneous scoring scales, allowing both techniques to contribute proportionally. The resulting ranking offers a unified view that reflects both textual semantics and runtime behavior, improving the accuracy and interpretability of feature-localization results.

\subsubsection*{Synergistic Effects}
\synReports: The static LSI model and the dynamic SPR model compute each a score which is then merged into a single report through a simple mathematical formula.

\subsubsection*{Inter-Analysis Workflow}
\workSidebyside: The static and dynamic analyses operate independently: LSI extracts textual semantics while SPR derives behavioral probabilities. Their results are then combined without directing or constraining one another.

\subsubsection*{Mapping-Function Interpretation Structure}
\structCG:
Both techniques ultimately assign relevance scores to methods, and the combination is performed over these structural units.

\subsubsection*{Mapping-Function Mechanics}
\mechanicML:
The two ranked lists are merged using probabilistic confidence weighting.

\subsection{Statically checking API protocol conformance with mined multi-object specifications \cite{pradel2012statically}}

\subsubsection*{Summary}
This paper presents a two-phase approach that first mines API usage specifications dynamically and then statically checks program conformance against the mined model. In the first phase, dynamic traces of program executions are collected to observe object interactions and API-usage patterns. From these traces, the technique infers multi-object specifications, represented as state machines describing valid object states and the transitions allowed by correct API protocols.
In the second phase, a static analysis examines the source code of the program to ensure that all API usages adhere to the mined state machine. The static checker analyzes call sequences, dataflow relations, and possible state transitions to detect whether code paths violate protocol constraints.
By leveraging mined specifications rather than manually written ones, the approach provides developers with actionable API-usage rules learned from real executions, while the static checker enforces these inferred rules across all possible \structPaths:

\subsubsection*{Synergistic Effects}
\synIntepretability~/~\synIntepretabilityEntities:
Dynamic mining reveals the semantics of valid API behaviors, while static checking ensures those inferred semantics are respected across the program.

\subsubsection*{Inter-Analysis Workflow}
\workCascade:
The mined state machine is produced by dynamic analysis and then used as input to the static checker.

\subsubsection*{Mapping-Function Interpretation Structure}
\structModules:
Specifications describe behaviors over objects and their interactions, which correspond to classes and their method protocols.

\subsubsection*{Mapping-Function Mechanics}
\mechanicMining:
Dynamic traces are mined into behavioral specifications that are then applied directly during static checking.

\subsection{Making malory behave maliciously: targeted fuzzing of android execution environments \cite{Rasthofer:2017:ICST}}

\subsubsection*{Summary}
This paper applies fuzzing to test a given target in an Android app. For depended APIs, it collects return values by solving constraints collected via data-flow analysis, and then feeds those return values to the fuzzing process, where they are used instead of executing the APIs.

\subsubsection*{Synergistic Effects}
\synTraversal~/~\synTraversalSeedInputs:
Input values computed by static analysis stages are passed to the fuzzing stage for improving the ability to reach the given target.

\subsubsection*{Inter-Analysis Workflow}
\workCascade:
Static analysis produces data-flow constraints, which it solves to provide API-return-values that drive the dynamic fuzzing strategy.

\subsubsection*{Mapping-Function Interpretation Structure}
\structCG: The statically derived values given by the value providers are used by the fuzzer as substitute to the return values to the API calls that are not available.

\subsubsection*{Mapping-Function Mechanics}
\mechanicConstraint: For the symbolic value provider, the values are obtained through constraint solving.

\subsection{RRF: A Race Reproduction Framework for Use in Debugging Process-Level Races \cite{Rattanasuksun:RRF:2016}}

\subsubsection*{Summary}
RRF is a hybrid framework designed to reproduce process-level race conditions in programs. Given a bug report and the programs under test, the framework first records system-call traces through dynamic executions, tracking every operation that may access shared resources implicated in the reported race (e.g. \texttt{open}, \texttt{read}, \texttt{execve}, etc.). The identified events are then mapped back to the corresponding source-level system-call invocations, reconstructing the code locations responsible for the interleaving. To force the race to manifest, RRF then modifies the program under test by injecting \texttt{sleep} statements at key locations. The instrumented program is then executed once again, to observe the new behavior and attempt the identification of the race. This approach allows developers to consistently reproduce subtle race-condition bugs that would otherwise be difficult or impossible to trigger reliably.

\subsubsection*{Synergistic Effects}
\synIntepretability~/~\synIntepretabilityEntities. Through the use of the first and the second analysis stage, they interpret the semantics of which code statements result in system calls.

\subsubsection*{Inter-Analysis Workflow}
\workCascade.

\subsubsection*{Mapping-Function Interpretation Structure}
\structCG. The source code instructions are related to the execution of system calls.

\subsubsection*{Mapping-Function Mechanics}
\mechanicAssociation.

\subsection{Practical dynamic reconstruction of control flow graphs \cite{Rimsa:SPE:2020}}

\subsubsection*{Summary}
This work introduces a dynamic technique for reconstructing control-flow graphs (CFGs) from binary executables and integrates these results with statically generated CFGs. Dynamic execution is instrumented to observe taken branches, indirect jumps, and previously unobserved control transfers, allowing the implicit discovery of actual runtime control structures, including dynamically computed jumps that static analysis often misses.
Once the dynamic CFG fragments are collected, they are merged with statically derived CFGs to form a more complete and precise representation of program control flow. The merged CFG benefits from static coverage of unreachable or rarely executed paths and from dynamic fidelity for computed or obfuscated flows.

\subsubsection*{Synergistic Effects}
\synFlow:
Dynamic traces refine and extend the static CFG, while static CFGs provide structure where dynamic information is incomplete.

\subsubsection*{Inter-Analysis Workflow}
\workSidebyside:
Static and dynamic CFGs are generated independently and later merged.

\subsubsection*{Mapping-Function Interpretation Structure}
\structCFG:
Basic blocks, edges, and jump instructions constitute the central mapping structures.

\subsubsection*{Mapping-Function Mechanics}
\mechanicMerging:
Static and dynamic CFG elements are combined by matching blocks and edges across both representations.

\subsection{Approach for solving the feature location problem by measuring the component modification impact \cite{Rohatgi:IET:2009}}

\subsubsection*{Summary}
This paper presents a feature-location approach based on correlating runtime execution traces with statically derived class-dependency graphs. Initially, the program is instrumented and then executed to record which classes and methods are used for a specific program feature. Following this, the code is statically analyzed to build a class dependency graph capturing relations such as inheritance, aggregation, and method invocation between program components, while at the same time applying a filter so that the resulting graph only contains the classes that have been executed for the feature under examination. Finally, an impact analysis is performed, where various metrics are computed for the remaining set of classes and used to rank them. The intuition is that classes that are dynamically involved in a feature and statically influential are more likely to be core feature components. The final output of the technique is made up of multiple rankings for the identified class components, one for each proposed metric.

\subsubsection*{Synergistic Effects}
\begin{enumerate}
\item Dynamic $\rightarrow$ Static: \synTraversal~/~\synTraversalSeedSink. The classes and methods that are actually executed for a specific feature are used to restrict the second analysis stage so that the computed class dependency graph only features the relevant classes.
\item Static $\rightarrow$ Impact: \synTraversal~/~\synTraversalSeedSink. The restricted class dependency graph computed by the first technique is provided to the impact analysis so that the latter can calculate metrics only on the classes that are actually relevant for the feature under analysis.
\end{enumerate}

\subsubsection*{Inter-Analysis Workflow}
\begin{enumerate}
\item Dynamic $\rightarrow$ Static: \workCascade. The dynamic analysis outputs cascade directly to the static class dependency graph analysis stage, without the possibility of feedback.
\item Static $\rightarrow$ Impact: \workCascade. The restricted class dependency graph is provided to the impact analysis for metric calculation in a cascading fashion.
\end{enumerate}

\subsubsection*{Mapping-Function Interpretation Structure}
\begin{enumerate}
\item Dynamic $\rightarrow$ Static: \structModules. The communication between the partner analysis stages occurs at the class level, indicating the ones that have been executed during the former to the latter.
\item Static $\rightarrow$ Impact: \structModules. The class dependency graph stores information related to the class structures of the program under examination.
\end{enumerate}

\subsubsection*{Mapping-Function Mechanics}
\begin{enumerate}
\item Dynamic $\rightarrow$ Static: \mechanicAssociation.
\item Static $\rightarrow$ Impact: \mechanicAssociation.
\end{enumerate}

\subsection{FSCT: A new fuzzy search strategy in concolic testing \cite{sabbaghi2019fsct}}

\subsubsection*{Summary}
FSCT introduces a concolic testing strategy that augments classic symbolic execution with fuzzy search heuristics. First, a static analysis produces an augmented control-flow graph (CFG) in which edges carry weights reflecting branch difficulty or coverage priority. A coverage table is initialized to track the exploration status of program branches.
During execution, concolic analysis generates symbolic path constraints and concrete executions. After each run, infeasible paths are recorded, and the fuzzy expert system analyzes branch weights, past coverage, and the set of infeasible or unproductive paths to select promising new branches as targets for subsequent concolic searches.
This feedback loop between static CFG weighting and dynamic concolic results continually updates the path-exploration strategy, reducing time spent on infeasible paths and steering the symbolic executor toward high-value program regions.

\subsubsection*{Synergistic Effects}
\synPartitioning~/~\synPartitioningCoverage:
Static path models are refined using dynamic infeasibility information, helping eliminate unrealizable paths and improving exploration to prioritize unexplored state-space regions.

\subsubsection*{Inter-Analysis Workflow}
\workFeedback:
Dynamic concolic results inform and update the next iteration’s control-flow-edges weights, aimed to branch selection during concolic execution.

\subsubsection*{Mapping-Function Interpretation Structure}
\begin{itemize}
\item \structCFG:
An analysis stage computes control-flow-edges weights, aimed to branch selection during concolic execution.

\item \structPaths: The other analysis stage provides information on infeasible paths, to allow for refining the weights
\end{itemize}

\subsubsection*{Mapping-Function Mechanics}
\mechanicAssociation.

\subsection{PATDroid: Permission-Aware GUI Testing of Android \cite{Sadeghi:JMFSE:2017}}

\subsubsection*{Summary}
PATDroid integrates dynamic GUI exploration with data-flow analysis to identify the minimum number of permission combinations for a given application (called the App Under Test or AUT) that should be tested for each of the test cases in a test suite, deployed as part of another application called Test Harness App (THA), which is a common pattern in Android application testing. The technique employs three analysis stages: the THA Analyzer, the App Analyzer, and the AUT Analyzer. First, the THA Analyzer is used to statically analyze the THA through the use of a data flow analysis, with the goal of identifying which widgets are exercised by which test; in this case, only interactions are considered, while mere information lookups are ignored. At the same time, the App Analyzer instruments the AUT and executes the AUT/THA combination to dynamically extract the entry points that are invoked by each of the tests in the THA, through the use of injected log statements. These combinations are then used by the AUT Analyzer, which performs further static analysis on the AUT aimed at extracting which entry points and widgets are invoked by each test and which permissions are required for the proper execution, through the use of an inter-procedural fixed-point and data-flow analysis. In the end, the Interaction Detector (which is not an analysis step) combines the results of the THA Analyzer and the AUT Analyzer through output matching to then determine the minimum set of permissions required to test the AUT under all possible conditions.

\subsubsection*{Synergistic Effects}
\begin{enumerate}
\item App Analyzer $\rightarrow$ AUT Analyzer: \synRefine~/~\synRefineIncorporate. The App Analyzer constructs a mapping between the tests executed by the THA and the entry points invoked in the AUT, which is then provided to the partner analysis stage to augment it with the addition of information related to the widgets and permissions that are required during the execution of each entry point.
\item THA Analyzer + AUT Analyzer: \synFlow. The pairings of test to widget obtained by the THA Analyzer and the richer information computed by the AUT Analyzer are combined together to obtain more information on the permissions required by the program flows.
\end{enumerate}

\subsubsection*{Inter-Analysis Workflow}
\begin{enumerate}
\item App Analyzer $\rightarrow$ AUT Analyzer: \workCascade. The results of the dynamic App Analyzer are provided to the AUT Analyzer without the possibility of feedback.
\item THA Analyzer + AUT Analyzer: \workSidebyside. The THA Analyzer and AUT Analyzer are executed in side-by-side fashion, with the results being combined at the end.
\end{enumerate}

\subsubsection*{Mapping-Function Interpretation Structure}
\begin{enumerate}
\item App Analyzer $\rightarrow$ AUT Analyzer: \structCG. The communication between the two partner analysis stages occurs through the pairing of test and entry point, which correspond to functions in Android.
\item THA Analyzer + AUT Analyzer: \structCG~+ \structGUI. The integration of the results from the two partner analysis stages occurs both at the level of tests and entry points, which correspond to functions in Android, but also leveraging information on GUI elements (namely, widgets).
\end{enumerate}

\subsubsection*{Mapping-Function Mechanics}
\begin{enumerate}
\item App Analyzer $\rightarrow$ AUT Analyzer: \mechanicAssociation.
\item THA Analyzer + AUT Analyzer: \mechanicAssociation.
\end{enumerate}

\subsection{Mining understandable state machine models from embedded code \cite{Said:ESE:2020}}

\subsubsection*{Summary}
The paper proposes an approach focused on a semi-automatic approach focused on extraction of state machines from the source code of embedded programs. The process is divided into multiple phases. Initially, a static analysis is used to extract the state variables, i.e. variables that are deemed relevant as they hold state for the embedded system, according to a set of criteria (including but not limited to global variables or cases of read-before-write). The set of variables identified by the technique can then be inspected by human operators, who can decide to filter the given results to focus the next steps on a subset of relevant variables. The second analysis stage focuses on extracting all possible paths through the targeted program through the use of a slightly modified concolic execution, as the user can interactively restrict the explored space by specifying additional constraints to be provided to the symbolic engine. The concolic execution paths are then used to extract only atomic conditions, meaning conditions that are made up of only a boolean symbolic variable or the application of a relational operator between symbolic expressions, which represent the output of the technique together with the full path examination of the concolic engine. The user is once again involved as they can pick which atomic conditions should be considered for the next stages of the technique. Finally, the third analysis stage is executed, with the goal of generating the state machine. This is done through a static analysis step where both states and transitions are extracted based on examination of both the paths enumerated by the concolic engine and the atomic conditions chosen by the user. The transition guards are then minimized to ease the comprehension of the final result.

\subsubsection*{Synergistic Effects}
\begin{enumerate}
\item State Variable Extraction $\rightarrow$ Path Enumeration: \synTraversal~/~\synTraversalSeedSink. The state variables as determined by the first analysis stage (and possibly filtered by a human operator) are provided to the second analysis stage to restrict the computation of conditions only for the relevant variables.
\item Path Enumeration $\rightarrow$ State Machine Generation: \synTraversal~/~\synTraversalSeedSink. The atomic conditions computed by the partner analysis stage are provided as an input to the state machine generation to restrict its analysis ability only to the ones that have been deemed relevant.
\end{enumerate}

\subsubsection*{Inter-Analysis Workflow}
\begin{enumerate}
\item State Variable Extraction $\rightarrow$ Path Enumeration: \workCascade. The set of variables to inspect is provided to the partner analysis stage without the possibility of feedback.
\item Path Enumeration $\rightarrow$ State Machine Generation: \workCascade. The paths examined by the concolic execution along with the corresponding conditions are given to the partner analysis sage in cascading fashion.
\end{enumerate}

\subsubsection*{Mapping-Function Interpretation Structure}
\begin{enumerate}
\item State Variable Extraction $\rightarrow$ Path Enumeration: \structDF. The state variables to be considered are used as the communication structure between the two analysis stages.
\item Path Enumeration $\rightarrow$ State Machine Generation: \structPaths. The communication between the two partner analysis stages occurs at the level of program paths and associated conditions.
\end{enumerate}

\subsubsection*{Mapping-Function Mechanics}
\begin{enumerate}
\item State Variable Extraction $\rightarrow$ Path Enumeration: \mechanicAssociation.
\item Path Enumeration $\rightarrow$ State Machine Generation: \mechanicAssociation.
\end{enumerate}

\subsection{Exploiting program dependencies for scalable multiple-path symbolic execution \cite{santelices_exploiting_2010}}

\subsubsection*{Summary}
This paper presents a new technique, called Symbolic Program Decomposition that exploits
results from data flow analysis to optimize the symbolic analysis.

\subsubsection*{Synergistic Effects}
\synIntepretability~/~\synIntepretabilityEntities: the results of data flow analysis allows the symbolic program decomposition to interpret the relations between definitions and uses of the program.

\subsubsection*{Inter-Analysis Workflow}
\workCascade

\subsubsection*{Mapping-Function Interpretation Structure}
\structDF: The communication between the two partner analysis stages occurs based on the data flow relations that are determined.

\subsubsection*{Mapping-Function Mechanics}
\mechanicAssociation

\subsection{Software Analysis Method for Assessing Software Sustainability \cite{saputri_software_2020}}

\subsubsection*{Summary}
This paper combines static analysis, dynamic analysis, and machine learning to evaluate software sustainability. The analyzed technique is split into four phases: the building phase (focused on preparing the source code and scenarios for analysis), the profiling phase (focused on executing static and dynamic analysis to gather sustainability metrics), the evaluation phase (where ML classifiers are used to estimate some measurements), and the visualization phase (where a mapping between scenarios and the source code is generated for traceability purposes). For the purposes of this survey, we will focus on the profiling phase as it is the part of the technique where the analysis stage occurs. In this phase, various metrics are computed through static and dynamic analysis, which are executed essentially in parallel. these metrics are then integrated to allow the following stages to reason on them and determine the sustainability of the target program.

\subsubsection*{Synergistic Effects}
\synReports. The analysis stages produce various reports in the forms of metrics, which are then combined together.

\subsubsection*{Inter-Analysis Workflow}
\workSidebyside. The static and dynamic analysis stages are executed in parallel in side-by-side fashion.

\subsubsection*{Mapping-Function Interpretation Structure}
\structCG. The metrics operate on the call-graph entities.

\subsubsection*{Mapping-Function Mechanics}
\mechanicAssociation + \mechanicML. The communication mechanic changes depending on the targeted metric.

\subsection{Worst-case execution time analysis for a Java processor \cite{schoeberl_worst-case_2010}}

\subsubsection*{Summary}
The paper introduces a  technique that combines data flow analysis (to infer valid callees  and receiver objects of method calls in the Java program under analysis) and Worst Case Execution Time analysis in the JVM context. 

\subsubsection*{Synergistic Effects}
\synIntepretability~/~\synIntepretabilityEntities.
Data-flow information is collected about valid callees and provided to the Worst Case Execution Time analysis of choice (model checking or implicit path enumeration technique).

\subsubsection*{Inter-Analysis Workflow}
\workCascade: The results of data flow analysis are used for WCET analysis.

\subsubsection*{Mapping-Function Interpretation Structure}
\structCG: The provided information augments the semantics of method calls by indicating valid callees and receiver objects.  

\subsubsection*{Mapping-Function Mechanics}
\mechanicAssociation

\subsection{Combining static analysis and state transition graphs for verification of event-condition-action systems in the RERS 2012 and 2013 challenges \cite{schordan_combining_2014}}

\subsubsection*{Summary}
The paper presents a multi-step technique aimed at verifying event-condition-action systems, considering a subcategory of problems presented at the RERS 2012 and 2013 challenges. The end goal of the proposed technique is to verify the given assertions through reachability analysis, and this is done by making use of five different analysis stages. The first analysis step (ASA) is an approximating interval analysis executed statically, with the goal of tracking the interval of values that a given variable can be set to. At the end of the analysis, an additional step is performed where every variable whose interval size is equal to 1 is marked as a constant. The second analysis step (IBAV) takes the output of the first as input and performs a verification of the assertions based on the computed interval information. In this instance, every constant variable is replaced with its corresponding constant value, which allows for the evaluation of some of the conditions. At the same time, the output of ASA is also given to the third analysis stage (STGGen), which leverages it along with the control flow graph of the program under test to generate a state-transition graph (the STG). In particular, the operation is essentially a loop that starts from the initial state of the program and then gradually expands the generated graph by computing stages and edges until the possibility space is exhausted. Finally, the STG is provided as an input to the fourth analysis stage (SBAV), which behaves similarly to the second, as the goal is still verification of assertions. Finally, the technique also sees a final fifth step based on verification of LTL formulas, but it is not relevant for the purposes of this survey.

\subsubsection*{Synergistic Effects}
\begin{enumerate}
\item ASA $\rightarrow$ IBAV: \synRewrite~/~\synRewriteConcrete. The intervals outputted by ASA are used to rewrite the variables in the assertions that have been deemed constant with their corresponding values.
\item ASA $\rightarrow$ STGGen: \synRewrite~/~\synRewriteConcrete. The intervals outputted by ASA are used to rewrite the variables in the program that have been deemed constant with their corresponding values, to allow for this information to be employed by the partner analysis stage.
\item STGGen $\rightarrow$ SBAV: \synIntepretability~/~\synIntepretabilityOracle. The STG computed by the first analysis stage is provided to the second, which can then reason on both states and the transitions associated to them.
\end{enumerate}

\subsubsection*{Inter-Analysis Workflow}
\begin{enumerate}
\item ASA $\rightarrow$ IBAV: \workCascade.
\item ASA $\rightarrow$ STGGen: \workCascade.
\item STGGen $\rightarrow$ SBAV: \workCascade.
\end{enumerate}

\subsubsection*{Mapping-Function Interpretation Structure}
\begin{enumerate}
\item ASA $\rightarrow$ IBAV: \structDF. The communication occurs at the level of program variables.
\item ASA $\rightarrow$ STGGen: \structDF. The communication occurs at the level of program variables.
\item STGGen $\rightarrow$ SBAV: \structProgram. The communication between the partner analysis stages occurs through a state machine model of the target program.
\end{enumerate}

\subsubsection*{Mapping-Function Mechanics}
\begin{enumerate}
\item ASA $\rightarrow$ IBAV: \mechanicAssociation.
\item ASA $\rightarrow$ STGGen: \mechanicAssociation.
\item STGGen $\rightarrow$ SBAV: \mechanicAssociation.
\end{enumerate}

\subsection{A Combinatorial Testing-Based Approach to Fault Localization \cite{sh_ghandehari_combinatorial_2020}}

\subsubsection*{Summary}
The paper proposes a technique called BEN, with the goal of performing fault localization through the results of combinatorial testing. In particular, BEN is made up of two phases that are executed sequentially. In the first phase, the results of a previous execution of combinatorial testing are used as a starting point to determine the combination of inputs that are likely to cause failures. This is done through a dynamic analysis approach which determines combinations that are deemed potentially suspicious, followed by a ranking. As part of the process, test cases are generated and then executed to determine which combinations actually lead to failures. Following this, the results of this step are used in the second phase, where the ranking information of the input combinations are used to generate an additional test case. This test case is executed, with its results being used to determine which statements inside the module under test cause the failure. Finally, an ad-hoc metric is used to rank the statements.

\subsubsection*{Synergistic Effects}
\synPartitioning~/~\synPartitioningWitness. The first technique identifies witnesses representing combinations that lead to failure, dividing the failure space in partitions based on the ranking. This information is then provided to the partner analysis stage to rank the statements of the identified partitions.

\subsubsection*{Inter-Analysis Workflow}
\workCascade. The first dynamic stage's results are given directly to the second dynamic stage, without callbacks.

\subsubsection*{Mapping-Function Interpretation Structure}
\structModules. The information from the first technique can be used in the second phase to test any software module, from a single method to an entire class to wider structures.

\subsubsection*{Mapping-Function Mechanics}
\mechanicAssociation. The results are propagated with identity-style mechanics.

\subsection{Self-hiding behavior in Android apps: Detection and characterization \cite{shan_self-hiding_2018}}

\subsubsection*{Summary}
This paper proposes an analysis framework to analyze Android applications and determine which of them employ SHB, i.e. self-hiding behavior, to hide certain actions they perform. This analysis is done statically, by combining a series of analysis steps in both cascading and side-by-side fashion, to then identify potential malware. In particular, the proposed approach is to first extract information in the form of the call graph and aliases by analyzing an Android application bytecode, through the use of Soot. From this preliminary step, a series of different analysis stages are executed side-by-side: SAPI Analysis (the control flow and call graph are analyzed to identify calls to certain APIs, followed by backwards dataflow analysis to identify the values of the parameters of the calls), PAPI Analysis (done via a variant of taint analysis), UD Analysis (through analysis of whether the callbacks are invoked in response to conscious user action), AF Analysis (simple control flow analysis to check if the finish method is invoked during Activity creation), Attribute Analysis (checking the values that are used to invoke certain API methods). The results of all these analysis steps are used by detection rules, which flag certain SHBs if the corresponding conditions are identified.

\subsubsection*{Synergistic Effects}
\synReports: The various analysis steps all gather different information that is then merged to provide supporting evidence towards the determination of whether an application is potentially malware due to SHBs or not.

\subsubsection*{Inter-Analysis Workflow}
\workSidebyside: The various analysis stages are executed essentially in parallel and are then merged at the end through the use of detection rules into a report.

\subsubsection*{Mapping-Function Interpretation Structure}
\structCG: The results of the analysis stage relate the methods in the app under test. 

\subsubsection*{Mapping-Function Mechanics}
\mechanicAssociation.

\subsection{Fault localization and repair for Java runtime exceptions \cite{sinha_fault_2009}}

\subsubsection*{Summary}
An approach is presented to detect the faulty instruction of a Java program that generated a runtime exception, and to determine the other statements that may be involved in the fault, to assist program repair. Fault detection is performed by feeding by dynamic information from the failed execution (typically stack traces) to a static backwards dataflow analysis. To determine the candidate faulty statements, other forward and backwards static analyses are performed based on the result of the former, to identify instructions involved with the fault.

\subsubsection*{Synergistic Effects}
\synTraversal~/~\synTraversalSeedSink

\subsubsection*{Inter-Analysis Workflow}
\workCascade

\subsubsection*{Mapping-Function Interpretation Structure}
\structDF: instructions on which the target exception depends, according to a possible type of null-values-related dependencies

\subsubsection*{Mapping-Function Mechanics}
\mechanicAssociation

\subsection{GrammarForge: Learning Program Input Grammars for Fuzz Testing \cite{sochor_grammarforge_2025}}

\subsubsection*{Summary}
This paper presents an approach to infer a grammar from a parser for it. In the initial analysis stage, the control flow graph of the parser is analyzed to construct a initial grammar structure, where the terminal symbols are replaced by suitable placeholders. Then, in the next analysis stage, random strings from the grammar structure are selected, and placeholders are randomly replaced with terminal symbols. The parser is queried with the resulting strings, and if the string belongs to the language the grammar is refined by assigning the terminal symbols to the placeholders. 

\subsubsection*{Synergistic Effects}
\synIntepretabilityEntities: the grammar synthesized in the first stage conveys semantic of the input structures, which is exploited in the second stage.

\subsubsection*{Inter-Analysis Workflow}
\workCascade

\subsubsection*{Mapping-Function Interpretation Structure}
\structProgram: the grammar synthesized in the first stage conveys semantic of the input structures of the program.

\subsubsection*{Mapping-Function Mechanics}
\mechanicAssociation

\subsection{CEMENT: On the Use of Evolutionary Coupling Between Tests and Code Units. A Case Study on Fault Localization \cite{sohn_cement_2022}}

\subsubsection*{Summary}
This paper proposes CEMENT, a static technique for fault localization by correlating tests to the regions of code that are relevant to test execution, based on the difference of time between the commit of a code change and the commit of a test. This technique is used to perform fault localization, combined with another fault-localization technique from literature by  integrating their respective results side-by-side.

\subsubsection*{Synergistic Effects}
\synReports

\subsubsection*{Inter-Analysis Workflow}
\workSidebyside

\subsubsection*{Mapping-Function Interpretation Structure}
\structCFG

\subsubsection*{Mapping-Function Mechanics}
\mechanicAssociation

\subsection{ITree: Efficiently discovering high-coverage configurations using interaction trees \cite{song_itree_2014}}

\subsubsection*{Summary}
This paper proposes a combination between pairwise testing and coverage analysis that allows for selecting an efficient (less combinations, stronger than simply pairwise) set of combinations.

\subsubsection*{Synergistic Effects}

\synIntepretability~/~\synIntepretabilityArtifacts: from runtime analysis to combinatorial testing,

\synPartitioning~/~\synPartitioningDirect: from combinatorial testing to runtime analysis

\subsubsection*{Inter-Analysis Workflow}
\workFeedback: at each iteration, pairwise testing selects a set of configuration-option pairs to be considered, in order to the improve the knowledge on the relevant combinations. Then, at test execution time, the coverage is tracked at runtime and the coverage data are further elaborated with machine learning (clustering, decision trees) to determine the relevant configuration-option pairs to refine the knowledge on the combinations to be tested. 

\subsubsection*{Mapping-Function Interpretation Structure}

\structProgram: program configurations to be tested for runtime analysis, or relevant for combinatorial testing according to the analysis of the coverage data. 

\subsubsection*{Mapping-Function Mechanics}
\mechanicML: from runtime analysis to combinatorial testing,

\mechanicAssociation: from combinatorial testing to runtime analysis

\subsection{TATOO: Testing and analysis tool for object-oriented software \cite{souter_tatoo_2001}}

\subsubsection*{Summary}
A stage of static points-to analysis to construct the annotated points-to escape (APE) graph, and a second stage that generates test cases based on the APE graph.

\subsubsection*{Synergistic Effects}
\synIntepretability~/~\synIntepretabilityEntities

\subsubsection*{Inter-Analysis Workflow}
\workCascade

\subsubsection*{Mapping-Function Interpretation Structure}
\structProgram

\subsubsection*{Mapping-Function Mechanics}
\mechanicAssociation

\subsection{Crystallizer: A Hybrid Path Analysis Framework to Aid in Uncovering Deserialization Vulnerabilities \cite{srivastava_crystallizer_2023}}

\subsubsection*{Summary}
Crystallizer presents a novel hybrid framework to automatically uncover deserialization vulnerabilities by combining static and dynamic analyses. The intuition is to first over-approximate possible payloads through static analysis (to constrain the search space). Then; it uses dynamic analysis to instantiate concrete payloads as a proof-of-concept of a vulnerability (giving the analyst concrete examples of possible attacks).
In particular:
\begin{itemize}
    \item A) A gadget graph is statically derived by taking as input a library and information about trigger gadgets. This information is used to automatically infer which methods in a library can be used as entry points.
    \item B) Candidate gadgets that may use arbitrary objects are dynamically inferred by executing an instrumented code.
    \item C) A set of static filters validates if the candidate gadgets use arbitrary objects and thus can be considered sinks.
    \item D) The final dynamic analysis step consists in identifying if there exists an input payload that exercises a gadget chain (that ends in a sink) when passed to a deserialization entry point.
\end{itemize}

\subsubsection*{Synergistic Effects}
\begin{itemize}
    \item \synRefine~/~\synRefineIncorporate: A $\to$ B. The gadget graph is enriched with information about which are the potential sinks.
    \item \synRefine~/~\synRefineIncorporate: B $\to$ C. The gadget graph is pruned from the potential sinks that are not relevant.
    \item \synAlarms~/~\synAlarmsDynamic: C $\to$ D. The identified paths in the gadget graph are dynamically analyzed to discover if they can effectively be instantiated, and consequently a vulnerability (payload) may be identified.
\end{itemize}

\subsubsection*{Inter-Analysis Workflow}
\begin{itemize}
    \item \workCascade: A $\to$ B.
    \item \workCascade: B $\to$ C.
    \item \workCascade: C $\to$ D.
\end{itemize}

\subsubsection*{Mapping-Function Interpretation Structure}
\begin{itemize}
    \item \structCG: A $\to$ B: Entry points are method invocations.
    \item \structCFG: B $\to$ C: Potential sinks are statements in the code that use the input gadget in some way.
    \item \structPaths: C $\to$ D: For each entry point and sink a path is searched using a Djikstra-like algorithm and subsequently an input is tried to be found that executes the specific path.
\end{itemize}

\subsubsection*{Mapping-Function Mechanics}
\begin{itemize}
    \item \mechanicAssociation.
    \item \mechanicAssociation.
    \item \mechanicAssociation.
\end{itemize}

\subsection{Experience Report: Automated System Level Regression Test Prioritization Using Multiple Factors \cite{ulrich_experience_2016}}

\subsubsection*{Summary}
The paper proposes a new method of determining an effective ordering of regression test cases. The tool generates an efficient order to run the cases in an existing test suite by using expected or observed test duration and combining priorities of multiple factors associated with test cases; including previous fault detection success; interval since last executed; and modifications to the code tested.

\subsubsection*{Synergistic Effects}
\synFeature: merge previous fault detection success, interval since last executed, and modifications to the code tested priority metrics.

\subsubsection*{Inter-Analysis Workflow}
\workSidebyside: the analyses are performed independently from each other.

\subsubsection*{Mapping-Function Interpretation Structure}
\structCG: Test cases.

\subsubsection*{Mapping-Function Mechanics}
\mechanicML: Test priority metric.

\subsection{Combining symbolic execution and model checking for data flow testing \cite{su_combining_2015}}

\subsubsection*{Summary}
This paper presents a combined approach to automatically generate data flow-based test data. The proposed approach synergistically combines two techniques: dynamic symbolic execution (DSE) and counterexample-guided abstraction refinement-based model checking (CEGAR).

The basic workflow of the combined DSE-CEGAR approach is the following:
\begin{itemize}
    \item The static analysis is used to find def-use pairs and their cut points from the program under test.
    \item The DSE-based approach is first used to cover as many feasible pairs as possible (within a time bound on each pair).
    \item After the DSE-based testing, for the remaining uncovered pairs, the CEGAR-based approach is used to identify infeasible pairs and cover new feasible pairs (which have not yet been covered in DSE) within a time bound.
\end{itemize}
The process is iterative in the sense that, after a first pass, the time bounds for DSE and CEGAR are increased and the technique is executed again to uncover new feasible and infeasible pairs. This is done up until the saturation of the time budget.
  
\subsubsection*{Synergistic Effects}
\begin{itemize}
    \item \synPartitioning~/~\synPartitioningCoverage: The DSE technique aims at covering all possible def-use pairs.
    \item \synAlarms~/~\synAlarmsStatic\ +\ \synPartitioning~/~\synPartitioningCoverage: The CEGAR technique aims at removing infeasible def-use pairs in conjunction with trying to cover additional feasible def-use pairs.
\end{itemize}

\subsubsection*{Inter-Analysis Workflow}
\workCascade: The techniques are executed one after the other.

\subsubsection*{Mapping-Function Interpretation Structure}
\begin{itemize}
    \item \structDF: The DSE approach aims at covering as many def-use pairs as possible.
    \item \structDF: The CEGAR approach aims at covering as many def-use pairs as possible and exclusing infeasible ones.
\end{itemize}

\subsubsection*{Mapping-Function Mechanics}
\mechanicAssociation.

\subsection{TLV: Abstraction through testing; learning; and validation \cite{sun_tlv_2015}}

\subsubsection*{Summary}
The paper presents an automatic approach, called TLV, which combines testing, learning, and validation, to constructing an abstraction of a given class. TLV has three phases: learning, validation and refinement that are described in the following:
\begin{itemize}
    \item In the learning phase, the technique apply automatic testing techniques to generate, inexpensively, sample behavior of the class, which consists of sequences of method calls. Tests are generated randomly with Randoop.
    \item In the validation phase, symbolic execution is used to validate the abstraction so that the abstraction is guaranteed to be correct and accurate. In particular, Symbolic PathFinder is used to discharge proof obligations by being executed against an instrumented code, which includes precondition and postcondition (assertions) that can be used to verify that no exceptions are raised.
    \item In the refinement phase, the abstraction is model checked to identify and prune spurious counterexamples.
\end{itemize}
If the abstraction is found to be too abstract by the model checker, it is refined and then the technique restarts from the learning phase. The iterative process ends when a correct and accurate abstraction is constructed.

\subsubsection*{Synergistic Effects}
\begin{itemize}
    \item \synRefine~/~\synRefinePrune~/~\synRefineIncorporate: Symbolic execution is used to refine the previously obtained abstract model. The results of the symbolic execution may prune or incorporate additional elements in the abstract model.
    \item \synRefine~/~\synRefinePrune: The model checker identifies spurious counterexamples and consequently prunes the model.
\end{itemize}

\subsubsection*{Inter-Analysis Workflow}
\workFeedback: The technique is iterative in the sense that the abstraction may need to be refined after it has been constructed and requires further iterations of the technique.

\subsubsection*{Mapping-Function Interpretation Structure}
\structModules: All the techniques reason on the class that needs to be abstracted.

\subsubsection*{Mapping-Function Mechanics}
\begin{itemize}
    \item \mechanicAssociation: with instrumentation.
    \item \mechanicAssociation.
\end{itemize}

\subsection{ComboRT: A New Approach for Generating Regression Test Cases for Evolving Programs \cite{sun_combort_2016}}

\subsubsection*{Summary}
ComboRT is a technique for regression testing. It first performs a change impact analysis technique to determine a ranked list of impacted methods from a set of changed classes. Then, it selects test cases based on the coverage information, and ranks the selected test cases based on the probability of the methods to be impacted.

\subsubsection*{Synergistic Effects}
\synFlow: Change impact analysis and test coverage analysis merge their outputs to produce ranking information.

\subsubsection*{Inter-Analysis Workflow}
\workSidebyside: Change impact analysis and test coverage analysis are performed side by side.

\subsubsection*{Mapping-Function Interpretation Structure}
\structCG: The analyses share impact and coverage information about the methods of the program.

\subsubsection*{Mapping-Function Mechanics}
\mechanicAssociation.

\subsection{GPTScan: Detecting Logic Vulnerabilities in Smart Contracts by Combining GPT with Program Analysis \cite{sun_gptscan_2024}}

\subsubsection*{Summary}
GPTScan is a hybrid tool that combines static analysis and large language models to determine vulnerabilities in smart contracts. A first static analysis stage determines the candidate impacted function. A second GPT stage determines the relevant scenarios/properties (functionalities containing a vulnerable code pattern), and pinpoints them based on  the variables/functions involved in the pattern. Finally, a third static analysis stage confirms whether the candidate vulnerability exists.

\subsubsection*{Synergistic Effects}
First stage $\to$ second stage: \synTraversal~/~\synTraversalSeedSink: The first stage provides information to the second stage about the code entities to analyze, thus restricts the scope of the second stage's~GPT analysis. 
Second stage $\to$ third stage: \synTraversal~/~\synTraversalSeedSink: The second stage provides information to the third stage about which methods may be affected by a vulnerability, thus restricts the scope of the third stage's~static analysis. 

\subsubsection*{Inter-Analysis Workflow}
\workCascade. 

\subsubsection*{Mapping-Function Interpretation Structure}
\structCG, \structDF. 

\subsubsection*{Mapping-Function Mechanics}
\mechanicAssociation. 

\subsection{Software Numerical Instability Detection and Diagnosis by Combining Stochastic and Infinite-Precision Testing \cite{tang_software_2017}}

\subsubsection*{Summary}
This paper introduces a toolchain to detect numerical instabilities in software. A set of tools perform suitable transformations on the program to check, yielding versions of the program with different degrees of precision. Then, the transformed programs are randomly tested to determine whether numerical instabilities are present.

\subsubsection*{Synergistic Effects}
\synTraversal~/~\synTraversalTransform. 

\subsubsection*{Inter-Analysis Workflow}
\workCascade. 

\subsubsection*{Mapping-Function Interpretation Structure}
\structProgram. 

\subsubsection*{Mapping-Function Mechanics}
\mechanicAssociation. 

\subsection{Separate abstract interpretation for control-flow analysis \cite{tang_separate_1994}}

\subsubsection*{Summary}
A combined static/static analysis technique is presented that exploits (1) type and effect analysis to information on the method signatures in separately-compiled programs, and (2) abstract interpretation to execute the target analysis (e.g, control flow analysis).

\subsubsection*{Synergistic Effects}
\synIntepretability~/~\synIntepretabilityEntities. 

\subsubsection*{Inter-Analysis Workflow}
\workCascade. 

\subsubsection*{Mapping-Function Interpretation Structure}
\structCG. 

\subsubsection*{Mapping-Function Mechanics}
\mechanicAssociation. 

\subsection{Automated System Testing of Dynamic Web Applications \cite{tanida_automated_2013}}

\subsubsection*{Summary}
An approach to system testing of web application is proposed. First, dynamic web crawling is exploited to build a model of the navigation behavior of the web application, and then model checking is used to check this model for various properties of interest.

\subsubsection*{Synergistic Effects}
\synIntepretability~/~\synIntepretabilityEntities. 

\subsubsection*{Inter-Analysis Workflow}
\workCascade. 

\subsubsection*{Mapping-Function Interpretation Structure}
\structGUI. 

\subsubsection*{Mapping-Function Mechanics}
\mechanicAssociation.

\subsection{Automatic performance prediction of multithreaded programs: a simulation approach \cite{Tarvo:ASE:2018}}

\subsubsection*{Summary}
Static and dynamic analysis are used to build a probabilistic model of the program. Then, the model is simulated to predict the program's performance.

\subsubsection*{Synergistic Effects}
\synFlow. 

\subsubsection*{Inter-Analysis Workflow}
\workSidebyside. 

\subsubsection*{Mapping-Function Interpretation Structure}
\structCG. 

\subsubsection*{Mapping-Function Mechanics}
\mechanicMining\ (to build the probabilistic call graph). 

\subsection{Demand-driven Information Flow Analysis of WebView in Android Hybrid Apps \cite{Tiwari:ISSRE:2023}}

\subsubsection*{Summary}
The paper reports the static analysis technique {\sc IwanDroid} to perform information flow analysis of hybrid Android apps, aimed at detecting integrity or confidentiality violations. Here ``hybrid apps'' means apps that contain both native (Java) code and web (JavaScript) code that interoperate.

\subsubsection*{Synergistic Effects}
\synIntepretability~/~\synIntepretabilityEntities: There are three stages in the analysis. The synergistic effects of the first stage towards the second is to determine the invocation relations between the Java and the JavaScript code (``bridge'' objects and methods). The effect of the second stage towards the third is to summarize some procedural call: The second stage calculates the summaries of the calls to the bridge methods. Stage 3 is the information flow analysis. In both cases the synergistic effect is  \synIntepretability~/~\synIntepretabilityEntities. 

\subsubsection*{Inter-Analysis Workflow}
\workCascade: Each stage provides its output to the next one. 

\subsubsection*{Mapping-Function Interpretation Structure}
Stage 1 $\to$ 2: \structCG; Stage 2 $\to$ 3: \structCG.

\subsubsection*{Mapping-Function Mechanics}
Stage 1 $\to$ 2: \mechanicAssociation; Stage 2 $\to$ 3: \mechanicSummary. The extended call graph is passed to stage 2, and similarly the function summaries are passed to stage 3.

\subsection{A hybrid approach for safe memory management in C \cite{Tlili:AMAST:2008}}

\subsubsection*{Summary}
A hybrid program analysis is presented for detecting violations of safe memory accesses in C programs. An enhanced type and effect system must be used to annotate the program, and static (undecidable) type analysis determines violations. Dynamic analysis based on runtime monitoring determines whether violations not caught by static analysis arise at runtime.

\subsubsection*{Synergistic Effects}
\synAlarms~/~\synAlarmsDynamic: The type and effect system is undecidable and imprecise, thus it may generate false alarms. Runtime monitoring is used to check the feasibility of the generated alarms. Note that the second stage instruments the program for runtime monitoring, but does not by itself generate tests or in any way perform execution of the program (the tests must be provided externally), so it is debatable whether it is a combined technique. 

\subsubsection*{Inter-Analysis Workflow}
\workCascade: The static analysis provides input to the dynamic one. No feedback is generated. 

\subsubsection*{Mapping-Function Interpretation Structure}
\structPaths: The static analysis produces as result a set of (possibly infeasible) error paths. The second stage instruments the code to insert runtime monitors that detect violations of memory safety properties along the detected paths. 

\subsubsection*{Mapping-Function Mechanics}
\mechanicAssociation: The paths, with the associated information (which error they potentially trigger) are directly passed to the instrumentation phase. 

\subsection{Hybrid security analysis of web Javascript code via dynamic partial evaluation \cite{Tripp:ISSTA:2014}}

\subsubsection*{Summary}
This paper presents JSA, a combined static-dynamic vulnerability analysis of JavaScript code run client-side (in-browser). The vulnerabilities are DOM-based XSS and open redirect. The analysis starts from a call graph of the JavaScript code, detects the source-sink dataflow pairs in the call graph by means of dataflow analysis, and for each pair, if the source flows into the sink, a string propagation analysis is performed to refine this taint analysis. The abstract string that is propagated in this analysis is obtained by partially evaluating from a dynamic oracle that partially concretizes the abstract strings based on segments of the concrete string values observed at runtime.

\subsubsection*{Synergistic Effects}
\synAlarms~/~\synAlarmsStatic: The string analysis refines the alarms identified with taint analysis. It refines the information about whether a source-sink reachable pair determined by taint analysis might be feasible, lowering the number of false alarms. 

\synRewrite~/~\synRewriteConcrete: the string analysis phase works by incrementally rewriting  abstract states from taint analysis with respect to the concrete strings observed at runtime: Partially concretizing the abstract string values used as seeds in the string analysis does indeed enable the string analysis that refines taint analysis. 

\subsubsection*{Inter-Analysis Workflow}
\workCascade: taint analysis feeds string analysis, dynamic analysis feeds string analysis. There is no feedback. 

\subsubsection*{Mapping-Function Interpretation Structure}
\begin{itemize}
    \item \structCG: Taint analysis produces a call graph annotated with source-sink pairs from taint analysis (Call graph entities).
    \item \structCG: The dynamic analysis produces a set of input strings associated to the possible input points (Call graph entities). Note that the call graph is considered as an input of the taint analysis.
\end{itemize}

\subsubsection*{Mapping-Function Mechanics}
\begin{itemize}
    \item \mechanicAssociation: for the mapping between taint analysis and string analysis;
    \item \mechanicAssociation: for the mapping between dynamic analysis and string analysis: The values of dynamic analysis are used to partially concretize the abstract strings used as seeds for the string analysis.
\end{itemize}

\subsection{Slimming javascript applications: An approach for removing unused functions from javascript libraries \cite{Vasquez:IST:2019}}

\subsubsection*{Summary}
The paper presents a simple combined static-dynamic analysis to determine dead code in JavaScript bundles. Static analysis, performed with an off-the-shelf tool (Uglify) determines the unused modules. Dynamic analysis is used to suggest to users the possibly unused functions in the used modules.

\subsubsection*{Synergistic Effects}
\synReports: The synergistic effect of the dynamic analysis is to refine the coarse-grain results of the static analysis. Note that the results of the dynamic analysis are unsound (a function that is reported as unused might be actually used). 

\subsubsection*{Inter-Analysis Workflow}
\workSidebyside: The two analyses are independent (neither uses the results of the other) and their results are merged together. 

\subsubsection*{Mapping-Function Interpretation Structure}
\structModules: The modular structure of the program. 

\subsubsection*{Mapping-Function Mechanics}
\mechanicAssociation: Static analysis lists the unused modules, dynamic analysis lists the (possibly) unused functions, and the two sources of information are merged. 

\subsection{Mutation-inspired symbolic execution for software testing \cite{ValleGomez:IETSW:2022}}

\subsubsection*{Summary}
This paper presents Naive MISE (Mutation-Inspired Symbolic Execution), a way to combine mutation testing and dynamic symbolic execution. As mutants are generated, dynamic symbolic execution is naively applied on them to generate tests, hoping to kill more of them than by just running dynamic symbolic execution on the original program.

\subsubsection*{Synergistic Effects}
\synPartitioning~/~\synPartitioningCoverage: This is done by steering dynamic symbolic execution towards regions in the ``mutant space'' so it achieves a better mutation coverage. 

\subsubsection*{Inter-Analysis Workflow}
\workCascade: The generated mutants are given as input to dynamic symbolic execution. There is no feedback from dynamic symbolic execution to mutation testing other than coverage. 

\subsubsection*{Mapping-Function Interpretation Structure}
\structProgram: the program suitably mutated. 

\subsubsection*{Mapping-Function Mechanics}
\mechanicAssociation. 

 \subsection{The Impact of Program Reduction on Automated Program Repair \cite{Vidziunas:ICSME:2024}}

\subsubsection*{Summary}
The effect of several kinds of reductions on automated program repair is assessed through a series of experiments. A program repair tool, TBar, is considered, and different combinations of reductions are applied to it based on the results of dynamic slicing of the source code: Either reduction of the analyzed code, or reduction of the test suite used to validate the generated patch, or reduction of the list of suspicious statements used to localize the fault, or a combination of the above.

\subsubsection*{Synergistic Effects}
\synTraversal~/~\synTraversalTransform: This is achieved by applying program slicing  to the program before program repairing.

\synTraversal~/~\synTraversalSeedInputs: This is achieved by applying reduction of the test suite to the program before program repairing.

\synTraversal~/~\synTraversalSeedSink: This is achieved by applying reduction of the list of suspicious statements used to localize the fault.

\subsubsection*{Inter-Analysis Workflow}
\workCascade: Slicing is applied to the program before program repairing, and the results of slicing affect how program repairing is performed. 

\subsubsection*{Mapping-Function Interpretation Structure}
\begin{itemize}
    \item \structProgram: suitably sliced, in the case of program reduction;
   \item \structCG: set of input states, tests, in the case of tests reduction;
   \item \structCFG: set of statements, in the case of suspicious statements reduction.
\end{itemize}

\subsubsection*{Mapping-Function Mechanics}
\mechanicAssociation. 

\subsection{Growing a test corpus with bonsai fuzzing \cite{Vikram:ICSE:2021}}

\subsubsection*{Summary}
The paper introduces the technique of \emph{bonsai fuzzing} for generating test inputs as, e.g., strings conforming to a grammar. The objective is to produce test strings that are short and understandable. The approach is based on establishing a hierarchy of fuzzers, ordered by bounds on the size of the output they can produce. The fuzzers that can produce the smallest outputs provide the seeds to the less-constrained fuzzers. The topmost fuzzer is the one that produces the least-constrained outputs---the ones of interest. Experiments show that bonsai fuzzing has an advantage over directly fuzzing and reducing the outputs of the fuzzer.

\subsubsection*{Synergistic Effects}
\synTraversal~/~\synTraversalSeedInputs: Achieved by seeding a (guided) random search so its converges more easily to a result. 

\subsubsection*{Inter-Analysis Workflow}
\workCascade: Lattice-shaped hierarchy. Each fuzzer feeds its ``neighbours'' in the space of the size-bounds. 

\subsubsection*{Mapping-Function Interpretation Structure}
\structCG: inputs associated to the input points. 

\subsubsection*{Mapping-Function Mechanics}
\mechanicAssociation. 

\subsection{A low-overhead value-tracking approach to information flow security \cite{Vorobyov:SEFM:2012}}

\subsubsection*{Summary}
This paper presents a hybrid static-dynamic program analysis technique to determine at runtime whether secure values may flow to unsecure sinks. The static phase analyzes the program and instruments it with statements that track values during execution and assertions that will fail if security violations occur at run time.

\subsubsection*{Synergistic Effects}
\synAlarms~/~\synAlarmsDynamic: By enabling the runtime analysis to track the flow of values and detect alarms (we may say that static analysis ``builds'' a part of dynamic analysis). 

\subsubsection*{Inter-Analysis Workflow}
\workCascade: The static analysis produces an output that is passed to the dynamic analysis. 

\subsubsection*{Mapping-Function Interpretation Structure}
\structCFG: program statements, instrumented. 

\subsubsection*{Mapping-Function Mechanics}
\mechanicAssociation. 

\subsection{Combodroid: Generating high-quality test inputs for android apps via use case combinations \cite{wang_combodroid_2020}}

\subsubsection*{Summary}
Combodroid is a technique that monitors sequence of events in an Android app, and from it builds new sequences of events to exercise the application. A first stage performs a depth-first exploration of the app GUI states to mine an execution trace, from which automata are inferred. The transitions between automata states (corresponding to stable GUI states, with transitions labelled by application methods) are then combined to determine new sequences of actions, thus new tests. To obtain meaningful combinations, data dependency relations are calculated between the methods of two consecutive actions.

\subsubsection*{Synergistic Effects}
\synPartitioning~/~\synPartitioningCoverage. 

\subsubsection*{Inter-Analysis Workflow}
\workCascade. 

\subsubsection*{Mapping-Function Interpretation Structure}
\structProgram. 

\subsubsection*{Mapping-Function Mechanics}
\mechanicMining. 

\subsection{String Test Data Generation for Java Programs  \cite{wang_string_2022}}

\subsubsection*{Summary}
JustinStr is a test case generator for programs that accept as input, and manipulate, strings. First, slicing is used to determine the path along which the program manipulates the input string, and the sequence of API calls that are used to manipulate the input string is isolated. Then, the sequence is mutated to elicit also exceptions. Then, each API call is (statically) mapped to a regular expression, and according to the determined sequences of API calls the regex are combined to build a regex characterizing the input parameter.

\subsubsection*{Synergistic Effects}
\synIntepretability~/~\synIntepretabilityEntities. 

\subsubsection*{Inter-Analysis Workflow}
\workCascade. 

\subsubsection*{Mapping-Function Interpretation Structure}
\structCG. 

\subsubsection*{Mapping-Function Mechanics}
\mechanicMining. 

\subsection{Automatic Detection, Validation, and Repair of Race Conditions in Interrupt-Driven Embedded Software \cite{Wang:TSE:2022}}

\subsubsection*{Summary}
The paper introduces SDRacer, a technique that combines static analysis, guided symbolic execution and dynamic validation via virtualization, to detect and fix races between embedded software and interrupt routines. First, a lightweight static analysis determines potential pairs of racy events (read or write of variables). Then, guided symbolic execution is used to exclude unfeasible event pairs and to construct inputs that cover the feasible ones. Next, the races are dynamically validated by executing them on a virtual platform that allows to control the scheduling of the interrupts. Finally, the emitted races are passed to a stage that proposes, based on the kind of race, a bug fix.

\subsubsection*{Synergistic Effects}
\begin{itemize}
    \item \synAlarms~/~\synAlarmsStatic: Symbolic execution excludes some, but not all, of the false alarms.
    \item \synAlarms~/~\synAlarmsDynamic: Dynamic analysis validates the alarms produced by the previous stage.
\end{itemize}

\subsubsection*{Inter-Analysis Workflow}
\workCascade: as reported in Figure 2 in the paper. 

\subsubsection*{Mapping-Function Interpretation Structure}
\begin{itemize}
    \item \structPaths: For static analysis $\to$ guided symbolic execution
    \item \structPaths: For guided symbolic execution
\end{itemize}

\subsubsection*{Mapping-Function Mechanics}
\begin{itemize}
    \item \mechanicAssociation: For static analysis $\to$ guided symbolic execution
    \item \mechanicConstraint: For guided symbolic execution $\to$ dynamic analysis 
\end{itemize}

\subsection{Could I Have a Stack Trace to Examine the Dependency Conflict Issue? \cite{wang_could_2019}}

\subsubsection*{Summary}
The paper describes {\sc Riddle}, a test generator aimed at producing test that highlight problems due to conflicting dependencies versions. {\sc Riddle} performs static analysis to produce a set of potential stack traces, then it uses EvoSuite's search to find tests that hit the stack traces.

\subsubsection*{Synergistic Effects}
\synPartitioning~/~\synPartitioningCoverage. 

\subsubsection*{Inter-Analysis Workflow}
\workCascade. 

\subsubsection*{Mapping-Function Interpretation Structure}
\structCG. 

\subsubsection*{Mapping-Function Mechanics}
\mechanicAssociation. 

\subsection{Practical blended taint analysis for JavaScript \cite{Wei:ISSTA:2013}}

\subsubsection*{Summary}
The paper describes a combined static/dynamic taint analysis for the dynamic language JavaScript, aimed at showing that the combination of static analysis with dynamic analysis can overcome the limits of static analysis when applied to dynamic languages. The analysis demonstrates a general static/dynamic analysis framework for JavaScript (and for dynamic languages in general) where a first dynamic phase selects and executes a set of tests with high method coverage, then static analysis is performed on each dynamic trace from test execution, and the collected results are integrated.  

\subsubsection*{Synergistic Effects}
\synIntepretability~/~\synIntepretabilityEntities: Dynamic analysis obviates for the inability of static analysis of determining the information flow statically. 

\subsubsection*{Inter-Analysis Workflow}
\workCascade: The static analysis follows the dynamic one. 

\subsubsection*{Mapping-Function Interpretation Structure}
\structPaths. 

\subsubsection*{Mapping-Function Mechanics}
\mechanicAssociation. 

\subsection{Architecture-independent dynamic information flow tracking \cite{Whelan:CC:2013}}

\subsubsection*{Summary}
This paper describes {\sc Pirate}, a tool to perform dynamic taint analysis of binary code. This tool achieves independence on the ISA by analyzing LLVM bitcode. The authors demonstrate that it is possible to analyze multiple ISAs by defining models for 29 LLVM bitcode instructions, and by defining a static analysis for the more complex instructions that have a C model.

\subsubsection*{Synergistic Effects}
\synIntepretability~/~\synIntepretabilityEntities. 

\subsubsection*{Inter-Analysis Workflow}
\workCascade: The static analysis is followed by the dynamic analysis. 

\subsubsection*{Mapping-Function Interpretation Structure}
\structCFG: bitcode instructions. 

\subsubsection*{Mapping-Function Mechanics}
\mechanicAssociation. 

\subsection{LeanBin: Harnessing Lifting and Recompilation to Debloat Binaries \cite{LeanBin:Wodiany:ASE:2024}}

\subsubsection*{Summary}
LeanBin is a binary debloating tool based on lifting and recompilation: A binary is debloated by determining the part of it that is effectively executed, lifting it to C source code, and recompiling the obtained source. LeanBin is structured as a combined dynamic/static tool: Dynamic trace analysis, run over a test suite, explores basic blocks, indirect branches and a part of the direct branches, determining a first version of the program's CFG. Then, static analysis recovers unexplored direct branches, refines the CFG, and disassembles the code.

\subsubsection*{Synergistic Effects}
\synPartitioning~/~\synPartitioningCoverage: The dynamic analysis seeds the static analysis with feasible CFG entities, to make the static analysis faster / more precise in constructing the final CFG. 

\subsubsection*{Inter-Analysis Workflow}
\workCascade: The dynamic analysis'~output is the input of the static analysis. 

\subsubsection*{Mapping-Function Interpretation Structure}
\structCFG. 

\subsubsection*{Mapping-Function Mechanics}
\mechanicAssociation: The static analysis completes/refines the information calculated by the dynamic analysis. 

\subsection{S-Looper: Automatic summarization for multipath string loops \cite{Xie:ISSTA:2015}}
\subsubsection*{Summary}
S-Looper is a technique for summarizing loops that perform string traversals. It first identifies, from the CFG of the program, the induction variables of the loop, possibly pruning the non-induction variables. Second, for every path in the loop body a static analysis infers the subrange of the string traversed by the path. Third, a string constraint component (possibly based on symbolic execution) generates a string constraint for every path in the loop body. The string constraints are conjoined to obtain the loop summary.

\subsubsection*{Synergistic Effects}
\synIntepretability~/~\synIntepretabilityEntities. 

\subsubsection*{Inter-Analysis Workflow}
\workCascade. 

\subsubsection*{Mapping-Function Interpretation Structure}
\begin{itemize}
    \item \structCFG: stage 1 $\to$ 2
    \item \structDF: stage 1 $\to$ 2
    \item \structPaths: stage 2 $\to$ 3
\end{itemize}

\subsubsection*{Mapping-Function Mechanics}
\mechanicAssociation. 

\subsection{Every Mutation Should Be Rewarded: Boosting Fault Localization with Mutated Predicates \cite{Xu:ICSME:2020}}

\subsubsection*{Summary}
The paper introduces {\sc Flip}, a fault localization technique that combines spectrum-based fault localization (SBFL) and mutation-based fault localization (MBFL). Spectrum-based fault localization is used to determine an initial ranking of the program statements'~suspiciousness. Then, the predicates that are considered more suspicious are mutated and tests are rerun on them. Slicing is applied to determine the statements that affect the outcome of the predicates. Finally, the ranking is adjusted based on the result of mutation analysis.

\subsubsection*{Synergistic Effects}
\synPartitioning~/~\synPartitioningCoverage: SBFL focuses the mutation analysis on the part of the program that is more likely to contain the faulty statement.

\subsubsection*{Inter-Analysis Workflow}
\workCascade. 

\subsubsection*{Mapping-Function Interpretation Structure}
\structCFG: statements, predicates, with ranking. SBFL determines a suspiciousness ranking of CFG entities that is exploited by MBFL to focus its analysis. 

\subsubsection*{Mapping-Function Mechanics}
\mechanicAssociation. 

\subsection{SWAT4J: Generating System Call Allowlist for Java Container Attack Surface Reduction \cite{Xu:SANER:2024}}

\subsubsection*{Summary}
SWAT4J analyzes Java programs run in a Docker container and determines which operating system calls the program under analysis may perform. This way it is possible to configure Docker to disallow the other system calls, thus reducing the attack surface of the container application. SWAT4J combines static analysis and dynamic analysis to detect the system calls the Java application may perform.

\subsubsection*{Synergistic Effects}
\synReports: SWAT4J combines the results of different analyses for a same problem, that are able to detect different outcomes with different precision (the paper says, e.g., that static analysis is not able to determine the system calls during the container startup). 

\subsubsection*{Inter-Analysis Workflow}
\workSidebyside: The outputs of the two analyses are merged together. 

\subsubsection*{Mapping-Function Interpretation Structure}
\structCFG: statements. 

\subsubsection*{Mapping-Function Mechanics}
\mechanicAssociation. 

\subsection{CSFL: Fault Localization on Real Software Bugs Based on the Combination of Context and Spectrum \cite{Yan:SETTA:2021}}

\subsubsection*{Summary}
The authors introduce an approach to fault localization that complements spectrum-based fault localization (SBFL) with context information. The approach first calculates the suspiciousness ranking of the statements using a standard spectrum-based approach. Then, it performs a static data dependency analysis, and for each suspicious statement it adds to its score the score of the statements it depends on.

\subsubsection*{Synergistic Effects}
\synIntepretability~/~\synIntepretabilityEntities:
The second stage computes a suspiciousness score for the statements, by exploiting the computed data dependency information along with the suspiciousness scores calculated by the first stage. 

\subsubsection*{Inter-Analysis Workflow}
\workCascade: The output of the SBFL analysis is sent to the data dependency analysis. 

\subsubsection*{Mapping-Function Interpretation Structure}
\structCFG: Statements with suspiciousness ranking. 

\subsubsection*{Mapping-Function Mechanics}
\mechanicML. 

\subsection{Multi-Objective Software Defect Prediction via Multi-Source Uncertain Information Fusion and Multi-Task Multi-View Learning \cite{yang_multi-objective_2024}}

\subsubsection*{Summary}
The proposed technique estimates the defect proneness of a software module by considering the output of a number of static analysis tools (like PMD or CppCheck) and of code metrics, that are used to train a deep neural model to infer the module defect rate.

\subsubsection*{Synergistic Effects}
\synFeature. 

\subsubsection*{Inter-Analysis Workflow}
\workSidebyside. 

\subsubsection*{Mapping-Function Interpretation Structure}
\structProgram. 

\subsubsection*{Mapping-Function Mechanics}
\mechanicML. 

\subsection{PermDroid: Automatically testing permission-related behaviour of Android applications \cite{10.1145/3533767.3534221}}

\subsubsection*{Summary}
PermDroid aims to test the behavior of Android apps under different set of permissions, at the purpose of detecting permission-related bugs. To this ends, it combines static analysis and dynamic GUI testing of the app. First, static analysis constructs a state transition graph (STG) of the app, where all the states performing API calls relevant to the declared permissions are marked as \emph{active}. Then, the app's GUI is dynamically explored based on the STG until all the active states are covered or all the reachable screens have been explored.

\subsubsection*{Synergistic Effects}
\synPartitioning~/~\synPartitioningCoverage: Steering the dynamic analysis towards relevant states (the active ones) and determining the termination condition of the dynamic analysis (coverage). 

\subsubsection*{Inter-Analysis Workflow}
\workCascade: Static analysis provides input to the dynamic one. 

\subsubsection*{Mapping-Function Interpretation Structure}
\structCG. 

\subsubsection*{Mapping-Function Mechanics}
\mechanicAssociation: The state model is used to guide the dynamic exploration of the app's~UI. 

\subsection{Efficient and precise dynamic slicing for client-side Javascript programs \cite{Ye:2016:SANER}}

\subsubsection*{Summary}
JS-Slicer is a dynamic slicer for JavaScript program run in a browser. It constructs a program dependence graph (PDG) with three kind of dependencies: Data, control, and DOM. Control dependency is captured by a hybrid static/dynamic analysis: Static analysis captures intra-procedural control dependency, dynamic analysis captures inter-procedural control dependency.

\subsubsection*{Synergistic Effects}
\synFlow. 

\subsubsection*{Inter-Analysis Workflow}
\workSidebyside. 

\subsubsection*{Mapping-Function Interpretation Structure}
\begin{itemize}
    \item \structCFG , 
    \item \structDF ,
    \item \structCG .
\end{itemize}

\subsubsection*{Mapping-Function Mechanics}
\mechanicAssociation. 

\subsection{Efficient loop-extended model checking of data structure methods \cite{Yi:ASEA:2011}}

\subsubsection*{Summary}
The paper presents LEMC, a software model checking approach that is focused on limiting state space explosion in the presence of loops that scan data structures. To this end, LEMC combines dynamic symbolic execution and static analysis. By means of dynamic symbolic execution, information is gathered on a concrete execution of a path traversing the loop, which is used to statically check and prune the input states that are expected to follow paths with similar loop-traversing behavior. In this way, a path analyzed with  DSE reveals the partition of similar paths, and allows for steering the analysis process towards inputs that explore a different partition.

\subsubsection*{Synergistic Effects}
\synPartitioning~/~\synPartitioningCoverage. 

\subsubsection*{Inter-Analysis Workflow}
\workCascade: The dynamic analysis feeds the static one. 

\subsubsection*{Mapping-Function Interpretation Structure}
\structPaths. 

\subsubsection*{Mapping-Function Mechanics}
\mechanicAssociation. 

\subsection{FRIES: Fuzzing Rust Library Interactions via Efficient Ecosystem-Guided Target Generation \cite{yin_fries_2024}}

\subsubsection*{Summary}
FRIES generates and fuzzes Rust programs with the aim of checking Rust libraries. The first stage of the technique calculates a weighted API dependency graph by means of static analysis and uses this graph to determine relevant fuzz targets (i.e., programs that use the API), the second stage is a fuzz tool that tries to discover bugs in the driver.

\subsubsection*{Synergistic Effects}
\synPartitioning~/~\synPartitioningCoverage 

\subsubsection*{Inter-Analysis Workflow}
\workCascade. 

\subsubsection*{Mapping-Function Interpretation Structure}
\structCG.  

\subsubsection*{Mapping-Function Mechanics}
\mechanicAssociation. 

\subsection{Combining Static Concurrency Analysis with Symbolic Execution \cite{Young:TSE:1988}}

\subsubsection*{Summary}
This paper combines symbolic execution and static interleaving analysis for concurrent programs. Static analysis determines a set of feasible task interleavings, that are sent to symbolic execution to prune unexecutable paths. Concurrency analysis reduces the number of interleavings that must be considered by the symbolic executor, while symbolic execution reduces the number of spurious error reports produced by the concurrency analysis algorithm.

\subsubsection*{Synergistic Effects}
\synIntepretability~/~\synIntepretabilityEntities: Static analysis allows to determine a (super)set of the feasible paths that the symbolic analysis will explore. 

\subsubsection*{Inter-Analysis Workflow}
\workCascade. 

\subsubsection*{Mapping-Function Interpretation Structure}
\structPaths. 

\subsubsection*{Mapping-Function Mechanics}
\mechanicAssociation. 

\subsection{Symbolic string verification: Combining string analysis and size analysis \cite{Yu:TACAS:2009}}

\subsubsection*{Summary}
String analysis and size analysis are performed on a program that manipulates strings. Abstract states are represented as string and string length automata, and statements as post-image operators on these automata. An abstract interpretation framework (with widening) is used to determine an over-approximation of the possible string values and string sizes at the different program points.

\subsubsection*{Synergistic Effects}
\synReports: Here we have two sources, string values and string sizes. 

\subsubsection*{Inter-Analysis Workflow}
\workSidebyside: We have two abstract interpretations (on string values and on string sizes) that are performed in parallel on the same program. 

\subsubsection*{Mapping-Function Interpretation Structure}
\structDF: The analysis reports the possible values/lengths of the string-typed variables. 

\subsubsection*{Mapping-Function Mechanics}
\mechanicAssociation. 

\subsection{ConTesa: Directed test suite augmentation for concurrent software \cite{Yu:TSE:2020}}

\subsubsection*{Summary}
An approach for the augmentation of regression test suites in concurrent settings. First, a static analysis determines a set of interleaving schedules that covers a number of def-use pairs, as new targets for the augmentation. Then, regression tests with random thread schedules are used to cover as many targets as possible. Finally, symbolic execution is used to generate new tests and cover more interleavings.

\subsubsection*{Synergistic Effects}
\synPartitioning~/~\synPartitioningCoverage: The static analysis provides targets to quantify coverage of the subsequent random testing stage, and to steer the subsequent symbolic execution stage.

\subsubsection*{Inter-Analysis Workflow}
\workCascade. 

\subsubsection*{Mapping-Function Interpretation Structure}
\structDF: def-use pairs. 

\subsubsection*{Mapping-Function Mechanics}
\mechanicAssociation. 

\subsection{DESCRY: Reproducing system-level concurrency failures \cite{yu_descry_2017}}

\subsubsection*{Summary}
DESCRY is a tool to reproduce interprocess concurrency failures from log data. Static analysis derives from logs of failures which processes and which statements (goals) are involved in the failure. Then symbolic execution synthesizes candidate inputs from sequences of goals. If the inputs do not trigger the failure, dynamic analysis of the execution trace is used to synthesize alternative schedules.

\subsubsection*{Synergistic Effects}
\synPartitioning~/~\synPartitioningDirect: Static analysis calculates sequences of goals as objectives for steering symbolic execution. Similarly, execution of tests produced by symbolic execution provides schedules which are used by dynamic analysis to produce alternative schedules. 

\subsubsection*{Inter-Analysis Workflow}
\workCascade. 

\subsubsection*{Mapping-Function Interpretation Structure}
\structPaths. 

\subsubsection*{Mapping-Function Mechanics}
\mechanicAssociation. 

\subsection{Discover deeper bugs with dynamic symbolic execution and coverage-based fuzz testing \cite{zhang_discover_2018}}

\subsubsection*{Summary}
BREACHER is a tool that combines dynamic symbolic execution and fuzz testing to test programs, a technique called ``hybrid testing'' and implemented in previous tools (e.g., Driller). Fuzzing is used to cover as quickly as possible. Information from the fuzzer is used to discover paths to uncovered branches upon saturation of fuzz testing. Dynamic symbolic execution then builds inputs that are used to seed further fuzzing. {\sc Breacher} extends previous work by improving over some issues of the general approach (path explosion, etc.)

\subsubsection*{Synergistic Effects}
\begin{itemize}
    \item \synPartitioning~/~\synPartitioningCoverage: Fuzzing $\to$ dynamic symbolic execution. \item \synTraversal~/~\synTraversalSeedInputs: Dynamic symbolic execution $\to$ fuzzing.
\end{itemize}

\subsubsection*{Inter-Analysis Workflow}
\workFeedback. 

\subsubsection*{Mapping-Function Interpretation Structure}
\begin{itemize}
    \item \structPaths: Fuzzing $\to$ dynamic symbolic execution.
    \item \structCG: Dynamic symbolic execution $\to$ fuzzing (namely, program inputs). 
\end{itemize}

\subsubsection*{Mapping-Function Mechanics}
\mechanicAssociation. 

\subsection{Detecting infeasible paths via mining branch correlations \cite{zhang_detecting_2012}}

\subsubsection*{Summary}
The paper proposes an approach to detect infeasible \structPaths. The approach combines a dynamic analysis to determine feasible traces and data mining to determine patterns of sequences of branch predicates evaluations along the feasible \structPaths. A second static analysis step queries the data miner to determine whether the statically analyzed path has a sequence of branch predicates evaluations that matches the knowledge base or not.

\subsubsection*{Synergistic Effects}
\synIntepretability~/~\synIntepretabilityEntities. 

\subsubsection*{Inter-Analysis Workflow}
\workCascade. 

\subsubsection*{Mapping-Function Interpretation Structure}
\structCFG. 

\subsubsection*{Mapping-Function Mechanics}
\mechanicML. 

\subsection{Hybrid Regression Test Selection by Integrating File and Method Dependences \cite{zhang_hybrid_2024}}

\subsubsection*{Summary}
JcgEks is a technique for regression testing selection (RTS). It combines static analysis and dynamic analysis of dependencies to determine a safe and precise subset of tests that must be rerun upon code change. 

\subsubsection*{Synergistic Effects}
\synFlow: Static and dynamic analysis calculate at different granularities the dependencies that are used to determine the test classes. 

\subsubsection*{Inter-Analysis Workflow}
\workSidebyside. 

\subsubsection*{Mapping-Function Interpretation Structure}
\structModules. 

\subsubsection*{Mapping-Function Mechanics}
\mechanicAssociation. 

\subsection{Comparing and Combining Analysis-Based and Learning-Based Regression Test Selection \cite{zhang_comparing_2022}}

\subsubsection*{Summary}
The proposed hybrid technique aims at performing regression test selection by combining program analysis and machine learning. A set of tests is selected by performing a program-analysis-based regression test selection procedure (either Ekstazi or STARTS). Then, this set is reduced by further selecting tests through a ML-based regression set procedure.

\subsubsection*{Synergistic Effects}
\synTraversal~/~\synReports. The two analysis produce their results for each test case, which can be selected/non-selected according the analysis-based regression test selection procedure, and is assigned a score by the ML-based stage. Then the final selectin derives from considering both results, yielding the selected test cases also scored more than a given score threshold.  

\subsubsection*{Inter-Analysis Workflow}
\workSidebyside. The two analyses are both executed on the same test cases modules, though (implementation wise) the ML-based stage can applied only to the test cases selected with the analysis-based regression test selection procedure.   

\subsubsection*{Mapping-Function Interpretation Structure}
\structProgram. Test cases to be re-executed against the program under test. 

\subsubsection*{Mapping-Function Mechanics}
\mechanicAssociation. 

\subsection{Hybrid Regression Test Selection \cite{zhang_hybrid_2018}}

\subsubsection*{Summary}
HyRTS is a technique for regression testing selection (RTS). It combines analysis of dependencies at two different granularities to determine a safe and precise subset of tests that must be rerun upon code change.

\subsubsection*{Synergistic Effects}
\synFlow: The analyses at different granularities are performed in parallel. 

\subsubsection*{Inter-Analysis Workflow}
\workSidebyside. 

\subsubsection*{Mapping-Function Interpretation Structure}
\structModules. 

\subsubsection*{Mapping-Function Mechanics}
\mechanicAssociation. 

\subsection{Runtime prevention of concurrency related type-state violations in multithreaded applications \cite{zhang_runtime_2014}}

\subsubsection*{Summary}
The paper proposes a combined static-dynamic analysis to detect potential violations of correct sequences of method calls, described as type-checking state machines as inputs, in concurrent programs. A static control flow analysis determines whether a method call may be followed by another method call in the same thread. This information is exploited by a dynamic analysis to determine whether a future violation of the correct sequences of method calls may happen.

\subsubsection*{Synergistic Effects}
\synIntepretability~/~\synIntepretabilityEntities. 

\subsubsection*{Inter-Analysis Workflow}
\workCascade. 

\subsubsection*{Mapping-Function Interpretation Structure}
\structCG. 

\subsubsection*{Mapping-Function Mechanics}
\mechanicAssociation. 

\subsection{Combined static and dynamic automated test generation \cite{zhang_combined_2011}}

\subsubsection*{Summary}
Palus is a test generator for object-oriented (Java) programs, that generates tests setting up input objects through their interface methods. It combines a dynamic analysis to build a call sequence model, and a static analysis to enrich the model with method dependence information. This model is passed to a random test generator that uses it to incrementally create method sequences.

\subsubsection*{Synergistic Effects}
\begin{itemize}
    \item \synIntepretability~/~\synIntepretabilityEntities: Dynamic analysis $\to$ random and static analysis $\to$ random, 
    \item \synFlow: Static analysis and dynamic analysis. 
\end{itemize}

\subsubsection*{Inter-Analysis Workflow}
\begin{itemize}
    \item \workCascade: Dynamic analysis $\to$ random and static analysis $\to$ random.
    \item \workSidebyside: Static analysis and dynamic analysis. 
\end{itemize}

\subsubsection*{Mapping-Function Interpretation Structure}
\structModules. 

\subsubsection*{Mapping-Function Mechanics}
\mechanicAssociation. 

\subsection{Adaptive Tracing and Fault Injection based Fault Diagnosis for Open Source Server Software \cite{zhang_adaptive_2023}}

\subsubsection*{Summary}
The proposed technique aims at performing fault diagnosis of software systems. First, mutation analysis is performed on the software, and from this a set of mutants is obtained. Then, from a runtime analysis of the execution of the mutants a graph neural network is trained that, in a third round of runtime analysis over the non-mutated system, is used to determine whether the executions are faulty or not.

\subsubsection*{Synergistic Effects}
\synTraversal~/~\synTraversalTransform. 

\subsubsection*{Inter-Analysis Workflow}
\workCascade: mutation analysis $\to$ runtime analysis. 

\subsubsection*{Mapping-Function Interpretation Structure}
\structCFG. 

\subsubsection*{Mapping-Function Mechanics}
\mechanicML. 

\subsection{Heuristic guided selective path exploration for loop structure in coverage testing \cite{zhang_heuristic_2017}}

\subsubsection*{Summary}
A technique is presented aimed at generating a test covering a given instruction in a program, that is based on symbolic execution but can manage loops. In presence of some loop structure the technique generates partial paths up to the loop entry that are fed a concrete test generator. The concrete tests are prioritized based on a branching distance metric w.r.t. the target, so that the loop is traversed on a heuristically chosen path.

\subsubsection*{Synergistic Effects}
\begin{itemize}
    \item \synPartitioning~/~\synPartitioningDirect: Symbolic execution $\to$ testing;
    \item \synPartitioning~/~\synPartitioningDirect: Testing $\to$ symbolic execution. 
\end{itemize}

\subsubsection*{Inter-Analysis Workflow}
\workFeedback. 

\subsubsection*{Mapping-Function Interpretation Structure}
\begin{itemize}
    \item \structPaths: Symbolic execution $\to$ testing;
    \item \structPaths: Testing $\to$ symbolic execution. 
\end{itemize}

\subsubsection*{Mapping-Function Mechanics}
\begin{itemize}
    \item \mechanicAssociation: Symbolic execution $\to$ testing ;
    \item \mechanicConstraint: testing $\to$ symbolic execution .
\end{itemize}

\subsection{AndroidLeaker: A hybrid checker for collusive leak in android applications \cite{zhang_androidleaker_2017}}

\subsubsection*{Summary}
AndroidLeaker is a hybrid static/dynamic taint analysis tool. Static analysis is used to check the information leak in the individual applications and dynamic checking at runtime is responsible for preventing the information leak caused by cooperation of multiple applications.

\subsubsection*{Synergistic Effects}
\synIntepretability~/~\synIntepretabilityEntities: for a set of program entities the static analysis provides to the dynamic analysis information about the private information that can be potentially disclosed by external messages. 

\subsubsection*{Inter-Analysis Workflow}
\workCascade. 

\subsubsection*{Mapping-Function Interpretation Structure}
\structDF. 

\subsubsection*{Mapping-Function Mechanics}
\mechanicAssociation. 

\subsection{Boundary value analysis in automatic white-box test generation \cite{zhang_boundary_2016}}

\subsubsection*{Summary}
The paper presents a technique that extends white box testing based on symbolic execution with boundary value testing. A first stage, namely test generation based on symbolic execution, feeds a second stage with path conditions; the second stage uses a black box technique (AETG) to enrich the path condition so it generates a set of boundary values tests.

\subsubsection*{Synergistic Effects}
\synPartitioning~/~\synPartitioningDirect. 

\subsubsection*{Inter-Analysis Workflow}
\workCascade. 

\subsubsection*{Mapping-Function Interpretation Structure}
\structPaths. 

\subsubsection*{Mapping-Function Mechanics}
\mechanicAssociation. 

\subsection{Test Generation for Programs with Binary Tree Structure as Input \cite{zhao_test_2015}}

\subsubsection*{Summary}
The paper proposes an approach to generate binary tree data structures as test inputs that hit a given program path. The technique is based on genetic algorithms (for determining the shape of the tree) and constraint solving based on symbolic execution (for determining the values stored in the nodes of the tree). A genetic algorithm is used to evolve a population of binary trees for (one, a number of) iterations. Then, the population is passed to symbolic execution that determines whether, for any of the trees, some values exist that cover the program path. In the negative case, a fitness value based on branch distance is calculated for the tree, and the population is evolved again.

\subsubsection*{Synergistic Effects}
\begin{itemize}
    \item \synTraversal~/~\synTraversalSeedInputs: Genetic algorithm $\to$ symbolic execution;
    \item \synPartitioning~/~\synPartitioningCoverage: Symbolic execution $\to$ genetic algorithm.
\end{itemize}

\subsubsection*{Inter-Analysis Workflow}
\workFeedback. 

\subsubsection*{Mapping-Function Interpretation Structure}
\begin{itemize}
    \item \structCG: Genetic algorithm $\to$ symbolic execution;
    \item \structPaths: Symbolic execution $\to$ genetic algorithm.
\end{itemize}

\subsubsection*{Mapping-Function Mechanics}
\begin{itemize}
    \item \mechanicAssociation: Genetic algorithm $\to$ symbolic execution;
    \item \mechanicML: Symbolic execution $\to$ genetic algorithm.
\end{itemize}

\subsection{A Framework for Scanning Privacy Information based on Static Analysis \cite{zhao_framework_2022}}

\subsubsection*{Summary}
An approach is proposed to determine privacy information in a program's code. A first step ingests privacy policy documents in a language model (long-short term memory) to detect privacy-related words. A second step, based on static code analysis, determines the methods and the variables whose names can in some way be associated to the privacy-related words.

\subsubsection*{Synergistic Effects}
\synIntepretability~/~\synIntepretabilityArtifacts 

\subsubsection*{Inter-Analysis Workflow}
\workCascade. 

\subsubsection*{Mapping-Function Interpretation Structure}
\structCFG. 

\subsubsection*{Mapping-Function Mechanics}
\mechanicML. 

\subsection{Minerva: browser API fuzzing with dynamic mod-ref analysis \cite{zhou_minerva_2022}}

\subsubsection*{Summary}
{\sc Minerva} is a fuzzing tool for browser API. To determine interesting API interactions it performs a first, dynamic, analysis of the browser code and builds an API interference graph. Then, fuzzing exploits this graph to generate test cases that are aware of interference relations between APIs.

\subsubsection*{Synergistic Effects}
\synPartitioning~/~\synPartitioningCoverage: By analyzing the implementation of the browser APIs the upstream, dynamic analysis provides the downstream analysis semantic information about the relevant interactions between different APIs. 

\subsubsection*{Inter-Analysis Workflow}
\workCascade. 

\subsubsection*{Mapping-Function Interpretation Structure}
\structCG. 

\subsubsection*{Mapping-Function Mechanics}
\mechanicAssociation. 

\subsection{Dynamic Generation of Python Bindings for HPC Kernels \cite{zhu_dynamic_2021}}

\subsubsection*{Summary}
{\sc WayOut} generates automatically language bindings from Python rapid-prototyping-oriented programs to C/C++ functions of kernels. It leverages static analysis to analyze kernel headers and thus generate binding templates, and then dynamic analysis to determine types at runtime and instantiate the binding templates to ad-hoc bindings.

\subsubsection*{Synergistic Effects}
\synRewrite~/~\synIntepretabilityEntities. 

\subsubsection*{Inter-Analysis Workflow}
\workCascade. 

\subsubsection*{Mapping-Function Interpretation Structure}
\structCG. 

\subsubsection*{Mapping-Function Mechanics}
\mechanicAssociation.

\end{document}